\documentclass{aa}

\usepackage{graphicx}	% Including figure files
\usepackage{amsmath}	% Advanced maths commands
\usepackage{amssymb}	% Extra maths symbols
\usepackage{newtxtext,newtxmath}
\usepackage{multirow}
\usepackage{lscape}
\usepackage{makecell}
\usepackage{subfigure}
\usepackage{float}
\usepackage{newtxtext,newtxmath}
\usepackage{cellspace} 
\usepackage[export]{adjustbox}
\usepackage{placeins}
\usepackage{hyperref}
\usepackage{txfonts}
\newcommand{\nustar}{\textit{NuSTAR}\xspace}
\newcommand{\chandra}{\textit{Chandra}\xspace}
\makeatletter
\renewcommand*\aa@institutefont{\footnotesize}  % smaller affiliations
\makeatother

\begin{document}

   \title{The origin of the apparent ultraluminous X-ray source deficit in NGC 3256}
 
   \author{N. Torres-Alb\`a\inst{1, 2}
          \and
          P. G. Boorman\inst{3} 
          \and
          A. S. Evans\inst{1,4}
          \and
          J. Nagarajan-Swenson\inst{1,4}
          \and
          S. T. Linden\inst{5,6}
          \and
          G. C. Privon\inst{1,4,7}
          \and
          L. Abalo\inst{8}
          \and
          L. Armus\inst{9}
          \and
          L. Barcos-Mu\~{n}oz\inst{1,4}
          \and
          T. Bohn\inst{10}
          \and
          M. Brightman\inst{11}
          \and
          V. Charmandaris\inst{13,14,15}
          \and
          T. D\'iaz-Santos\inst{13,14}
          \and
          H. P. Earnshaw\inst{11}
          \and
          H. Inami\inst{12}
          \and
          M. A. Johnstone\inst{1}
          \and
          D. Kakkad\inst{16}
          \and
          L. Lenki\'c\inst{9}
          \and
          C. M. Lofaro\inst{13,14}
          \and
          M. A. Malkan\inst{17}
          \and
          M. J. Middleton\inst{18}
          \and
          C. Ricci\inst{19,20,21}
          \and
           M. S\'anchez-Garc\'ia\inst{22}
          \and
          D. Stern\inst{23}
          \and
          E. Treister\inst{24}
          \and
          D. J. Walton\inst{16}
          }

   \institute{
            Department of Astronomy, University of Virginia, P.O. Box 400325, Charlottesville, VA 22904, USA \newline \email{ntorres@ice.csic.es}
        \and 
            Department of Physics and Astronomy, Clemson University, Kinard Lab of Physics, Clemson, SC 29634-0978, USA
        \and
            Max Planck Institute for Extraterrestrial Physics, Giessenbachstrasse, 85748 Garching, Germany
        \and
            National Radio Astronomy Observatory, 520 Edgemont Road, Charlottesville, VA 22903, USA
        \and
            Steward Observatory, University of Arizona, 933 North Cherry Avenue, Tucson, AZ 85721, USA
        \and
            School of Physics and Astronomy, Rochester Institute of Technology, 84 Lomb Memorial Drive, Rochester, NY 14623, USA
        \and 
            Department of Astronomy, University of Florida, P.O. Box 112055, Gainesville, FL 32611, USA
        \and
            Huygens-Kamerlingh Onnes Laboratory, Leiden University, Postbus 9504, 2300 RA Leiden, The Netherlands
        \and
            IPAC, California Institute of Technology, 1200 E. California Boulevard, Pasadena, CA 91125, USA
        \and
            Research Center for Space and Cosmic Evolution, Ehime University, Bunkyo-cho 2-5, Matsuyama, Ehime 790-8577, Japan
        \and
            Cahill Center for Astronomy and Astrophysics, California Institute of Technology, 1200 E California Blvd, Pasadena CA 91101
        \and
            Hiroshima Astrophysical Science Center, Hiroshima University, 13-1 Kagamiyama, Higashi-Hiroshima, Hiroshima 739-8526, Japan
        \and
            Department of Physics, University of Crete, 70013 Heraklion, Greece
        \and
            Institute of Astrophysics, Foundation for Research and Technology - Hellas (FORTH), Heraklion, GR-70013, Greece
        \and
            School of Sciences, European University Cyprus, Diogenes street, Engomi, 1516 Nicosia, Cyprus
        \and
            Centre for Astrophysics Research, Department of Physics, Astronomy and Mathematics, University of Hertfordshire, Hatfield, AL10 9AB, UK
        \and
            Department of Physics and Astronomy, University of California, Los Angeles, 430 Portola Plaza, Los Angeles, CA  90095-1547, USA
        \and
            Physics and Astronomy, University of Southampton, Southampton, Hampshire SO17 1BJ, UK
        \and
            Department of Astronomy, University of Geneva, ch. d’Ecogia 16,1290 Versoix, Switzerland
        \and    
            Instituto de Estudios Astrof\'isicos, Facultad de Ingenier\'ia y Ciencias, Universidad Diego Portales, Av. Ej\'ercito Libertador 441, Santiago,Chile
        \and
            Kavli Institute for Astronomy and Astrophysics, Peking University,Beijing 100871, China
        \and
            Centro de Astrobiolog\'ia (CAB), CSIC-INTA, Camino Bajo del Castillo s/n, 28692, Villanueva de la Ca\~{n}ada, Madrid, Spain
        \and
            Jet Propulsion Laboratory, California Institute of Technology, 4800 Oak Grove Drive, Pasadena, CA 91109, USA
        \and
            Instituto de Alta Investigaci\'on, Universidad de Tarapac\'a, Casilla 7D, Arica, Chile
        }

   \date{Received XX, XX; accepted XX, XX}

 \abstract{}{}{}{}{} 
% 5 {} token are mandatory
 
  \abstract
   {Ultraluminous X-ray (ULX) sources are bright, off-nuclear X-ray point sources, mostly associated with high-mass X-ray binaries (HMXBs), with a compact object accreting from a donor star at super-Eddington luminosities. Although their HMXB nature implies a connection to star formation processes, previous works have found that luminous infrared galaxies (LIRGs) appear to have a deficit of ULXs compared to their very high star formation rates (SFRs).}
   {Here, we analyze three \chandra observations of NGC 3256, a merging LIRG with SFR$\sim$50-60~M$_\odot$~yr$^{-1}$. Based on local scaling relations, it could be expected to host between 100-120 individual ULXs, which could have a non-negligible impact on the energetics and feedback processes of the galaxy. We aim to characterize all its detected X-ray point sources to determine if there is a ULX deficit in this galaxy. We also aim to explore the potential impact of host-galaxy obscuration and metallicity in said deficit.}
   {We characterized the luminosity of each X-ray point source in each \chandra observation, and used all observations to determine their potential ULX nature and the fraction of transient ULXs, as well as construct a ULX luminosity function for this galaxy.}
   {We find that the galaxy hosts 10$-$15 ULXs, making it one of the most ULX-rich galaxies known to date. While a comparison with the ULX/SFR ratio of nearby galaxies predicted as many as $\sim$100~ULXs, we show that, when accounting for its very high metallicity, NGC~3256 does not have a ULX deficit; and may even present an excess.  While the high obscuration in LIRGs has previously been invoked to explain the ULX deficit, we conclude that it is not required to explain the ULX to SFR ratio in NGC~3256. If obscuration still plays a role, then LIRGs may actually be significantly more efficient at producing ULXs than normal galaxies (of the same metallicity), in disagreement with previous results.}
   {}


   \maketitle
    \nolinenumbers
%
%-------------------------------------------------------------------

\section{Introduction}

Ultraluminous X-ray sources (ULXs) are defined as off-nuclear point sources with X-ray luminosities of $L_x > 10^{39}$~erg~s$^{-1}$ \citep[e.g.,][]{Kaaret2017,King2023}, which can only be achieved as a result of accretion onto a compact object. These sources have garnered the interest of the community since their discovery \citep[e.g.,][]{Long1981,Fabbiano1987}, given that $10^{39}$~erg~s$^{-1}$ is the Eddington luminosity for a 10~$M_\odot$ black hole. Sources with a significantly higher luminosity must therefore have a significantly higher mass (i.e. potential intermediate-mass black holes, IMBHs), or be stellar mass objects accreting at super-Eddington rates. 

Intermediate-mass black holes bridge the gap between stellar and supermassive black holes, offering fundamental information about black hole seeding mechanisms and growth \citep[see e.g.,][]{Greene2020}. Very few are considered reliable candidates, and these tend to have masses in the range of $10^5-10^6$~M$_\odot$ \citep[e.g.,][]{Chilingarian2018,Zaw2020}. Candidate ULX IMBHs, however, have candidate black hole masses on the order of $\sim100-1000$~M$_\odot$ \citep[e.g., HLX-1 in ESO 243-49, M82 X-1, NGC 2276-3c, and NGC 5252 X-1;][]{Farrell2009,Davis2011,Pasham2014,Mezcua2015,Mezcua2018}, making them representative of the low-end tail of the IMBH mass distribution, and potential seeds for supermassive black holes (SMBHs).

\cite{Bachetti2014} identified, for the first time, the ULX M82 X-2 as a high-mass X-ray binary (HMXB), with an accreting neutron star and a donor star. Since then, a select number of ULXs \citep[out of $\sim 1800$ candidates known to date,][]{Walton2022} have been firmly identified as neutron stars thanks to the detection of pulsations \citep[e.g.,][]{Furst2016,Israel2017,Carpano2018,Sathyaprakash2019,Rodriguez-Castillo2020,Pintore2025,Ducci2025}, which further supported the notion that the majority of the population may comprise stellar-mass compact objects. An X-ray binary origin of ULXs is also supported by the correlation found between the number of ULXs in nearby, star-forming galaxies and their star formation rate (SFR); with an average of 2 ULXs per unit SFR \cite[e.g.,][]{Mapelli2010}. If ULXs have a binary origin, they are the ideal site to study super-Eddington accretion. The extreme X-ray luminosity of some sources \citep[$>10^{41}$~erg~s$^{-1}$; e.g.,][]{Davis2011,Sutton2012,Amato2025} may be more difficult to explain without an IMBH. However, ULX-1 in NGC~5907 has been reported to reach such luminosities, despite its confirmed pulsar nature \citep[e.g.,][]{Israel2017,Furst2017}; and accreting neutron-star models invoking geometrical beaming manage to achieve them \citep[e.g.,][]{King2009,Dauser2017}. As such, ULXs and their nature remain a topic of debate.

Luminous infrared galaxies (LIRGs) have infrared luminosities of $L_{\rm IR}>10^{11}$~L$_\odot$ and SFRs that easily surpass $10$~M$_\odot$~yr$^{-1}$. In the local Universe, they are often the result of mergers \citep[e.g.,][]{Sanders1996,Larson2016} and, at the peak of star formation in the Universe (z$\sim2-3$), they dominate the infrared luminosity density function \citep[e.g.,][]{Caputi2007,Magnelli2011}. At an expected rate of 2 ULXs per unit SFR, LIRGs should be the ideal sites to search for ULXs. However, the most extensive study to date \citep[focusing on 17 systems;][]{Luangtip2015} from the Great Observatories All-Sky LIRG Survey \citep[GOALS,][]{Armus2009} found that the rate of ULXs per unit SFR was only 0.2, a factor of 10 lower than expected.

LIRGs are inherently gas-rich and dusty \citep[e.g.,][]{Downes1998,Herrero-Illana2019,Chartab2022}, which leads to high levels of extinction \citep[A$_{\rm V}$ up to 20~mag, e.g.,][]{Piqueras-Lopez2013}, even in the X-ray regime \citep[e.g.,][]{Iwasawa2011,Torres-Alba2018} and, particularly, in their nuclear regions \citep[$>10^{23}-10^{24}$~cm$^{-2}$, e.g.,][]{Ricci2017mnras,Ricci2021}. The high levels of extinction make observing sources with a luminosity of $\sim10^{39}$~erg~s$^{-1}$ extremely challenging, and thus \cite{Luangtip2015} propose this as the dominant explanation for the deficit of ULXs in local LIRGs. However, they also suggest two other potential contributors to the deficit.

First, they suggest the observed dependence of ULX abundance on metallicity. The high SFRs of LIRGs lead to high supernovae rates, which in turn leads to higher-than-solar metallicities \citep[see e.g.,][who find super-solar gas metallicities in the inner few kiloparsecs of NGC 3256]{Rich2012}. Previous works indicate that metallicity does have an impact on the abundance of ULXs in galaxies \cite[e.g.,][]{Mapelli2011,Prestwich2013,Lehmer2021}, with a higher metallicity leading to reduced formation of high-mass stars, which results in a decrease in the detected number of ULXs. 

The second effect could be the age of the starbursts within these systems. That is, if a significant fraction of the star formation is younger than $\sim10$~Myr \citep{Linden2010}, not enough stars may have died and turned into accreting, compact objects. Estimates of star formation histories (SFHs) in local LIRGs are consistent with a significant fraction of their recent star-formation activity occurring within the last 10~Myr \citep[e.g.,][Nagarajan-Swenson in prep.]{Cortijo-Ferrero2017,Pereira-Santaella2015}.  Although \cite{Luangtip2015} argue that neither of these effects alone is sufficient to fully explain the ULX deficit in LIRGs, it is unclear how much the combination of both can account for. For example, an in-depth study of the HMXB population in another LIRG, NGC~7552, concludes the opposite; that there likely is an intrinsic lack of X-ray binaries, rather than an obscured population \citep{West2023}.

Another factor that could be at play is angular resolution. LIRGs have SFRs high enough that some could have hundreds of ULXs within them, if the ULX to SFR ratio holds. Furthermore, their SFRs tend to be highly concentrated in the nuclear regions, within a radius of  $\lesssim 1$~kpc \citep[e.g.,][]{Haan2013,Song2021}. Given these two facts, and at distances of up to $\sim50-60$~Mpc \citep[as is the case for previous LIRG studies;][]{Luangtip2015}, it is very likely that part of the population is unresolved, even with \textit{Chandra}. In such instances, multiple ULXs could be mistakenly counted as a single source, resulting in reduced counts and higher luminosities.

In this work, we investigate the potential ULX deficit of NGC~3256 and its causes. The galaxy is classified as a LIRG \citep{Armus2009}, and a late-stage merger \citep{Stierwalt2013}. Its high SFR, 49$-$61~M$_\odot$~yr$^{-1}$ \citep{Diaz-Santos2017,Michiyama2020A}, implies it could contain up to $\sim100$ ULXs, according to the scaling relations of $\sim2$~ULXs per unit SFR mentioned above. It is also among the closest 20 LIRGs, making it ideal for high-resolution studies. Previous works report different numbers of X-ray point sources \citep[from 4 to 14, e.g.,][]{Lira2002,Liu2011,Lehmer2015,Walton2022,Tranin2024}, but generally include the nuclei, and miss point sources due to using only one or two of three available \chandra observations.

Here, we present a comprehensive study of the detectable ULX population in NGC 3256 across all available \chandra epochs, including their spectral properties and their epoch-by-epoch variability, tracing timescales of a few years to a decade. In Sec. \ref{sec:obs_data}, we discuss the properties of the available observations and their data reduction, as well as the point source association with NGC~3256, and cross-match against previous studies. In Sec. \ref{sec:analysis}, we discuss the analysis methodology followed throughout the paper. In Sec. \ref{sec:results}, we present the results obtained from the spectral fitting. We discuss their interpretation, in the context of previous studies, in Sec.~\ref{sec:discussion}, and summarize our conclusions in Sec. \ref{sec:conclusions}.

\section{NGC~3256}\label{sec:SFR}

NGC~3256 has an infrared luminosity of ${\rm log L}_{\rm IR}$~=~11.64~L$_\odot$ \citep{Armus2009}. It has two distinguishable nuclei \citep[angular separation of $\sim5$'';][]{Sakamoto2014}, the southernmost of which is obscured at optical wavelengths \citep[e.g.,][]{Kotilainen1996}. It is an ideal candidate for ULX studies in LIRGs due to its close distance to us, $\sim45$~Mpc, at which the \chandra resolution, $\sim0.5$'', corresponds to a physical scale of $\sim100$~pc. This places NGC~3256 among the closest 20 LIRGs, and the one with the highest SFR among them. 

From IR estimations \cite{Diaz-Santos2017} estimate an SFR~=~61.7~$\pm$~1.3 ~M$_\odot$~yr$^{-1}$. However, the nuclei in NGC 3256, particularly the southern one, have long been considered to potentially host heavily buried active galactic nuclei (AGN), based on radio and X-ray observations \cite[e.g.][]{Neff2003,Jenkins2004,Sakamoto2014}. If the galaxy indeed hosts one (or two) AGN, this would likely result in the far-IR determinations of SFR being overestimated. It is unclear whether this is the case, since despite the significant \textit{NuSTAR} detection of NGC~3256 \citep{Lehmer2015,Yamada2021}, and the nuclear outflows \citep[][Evans et al. in prep.]{Emonts2014,Pereira-Santaella2015,Bohn2024}, its energetics and spectral properties remain consistent with an extreme starburst. 

A recent estimate, based on a combination of MUSE and ALMA \citep[Atacama Large Millimeter/sub-millimeter Array,][]{Wootten2009} data, combining hydrogen recombination lines in optical, millimeter, and free-free emission, reports SFR~=~49~$\pm$~2~M$_\odot$~yr$^{-1}$, where the contributions from nuclear and disk-wide starbursts are $\sim$34\% and $\sim$66\%, respectively \citep{Michiyama2020A}. This implies an estimated SFR of $\sim32$~M$_\odot$~yr$^{-1}$ in the non-nuclear regions. A ULX is, by definition, an off-nuclear point source\footnote{Galactic nuclei likely contain ULX-like sources, particularly in galaxies with strong nuclear starbursts. However, they are by definition excluded from consideration due to the impossibility of disentangling them from SMBH accretion.}. Since we only consider X-ray point sources that are off-nuclear, considering the nuclear SFR biases the calculation of the ULX ratio. However, the estimate of the nuclear versus off-nuclear SFR here comes from ALMA+MUSE data, while throughout this work we compare to previous works, such as \cite{Luangtip2015} and \cite{Lehmer2021}, who use IR-derived SFRs. As such, we report results for both ends of the possible SFR determinations, SFR~=~32$-$62 ~M$_\odot$~yr$^{-1}$. Since we do not have an IR-based estimate for the nuclear SFR contribution, 62 ~M$_\odot$~yr$^{-1}$ should be viewed as an upper limit to the off-nuclear SFR.

\section{Observations and data reduction}\label{sec:obs_data}

NGC 3256 has been observed with \chandra three times, for a total of 71.3~ks. Details on each individual observation can be found in Table \ref{tab:observations}. Each observation was reduced individually, following the standard \chandra data analysis procedure, as described by the CIAO analysis guides\footnote{https://cxc.cfa.harvard.edu/ciao/}. We used CIAO version 4.17 to carry out the analysis. Throughout the paper, we use as cosmological constants $H_{\rm 0}$ = 70 km s$^{-1}$ Mpc$^{-1}$, $\Omega_{\Lambda}$ = 0.73, and $\Omega _{\rm M}$ = 0.27.

\begin{table}[t]
\centering
\caption{Properties of the \chandra observations used in this work.}
\label{tab:observations}
\begin{tabular}{Sc Sc Sc Sc}
OBSID & Date & Detector & Exp. time [ks] \\
(1) & (2) & (3) & (4) \\
\hline\hline
835 & 2000-01-05 & ACIS-S & 28.2 \\
3569 & 	2003-05-23 & ACIS-S & 27.6 \\
16026 & 2014-08-27 & ACIS-S & 15.6 \\\hline
\end{tabular}
\footnotesize{ \newline \\
{\it Columns:} 1) Observation ID. 2) Date in which the observation was taken. 3) Detector used in the observation. 4) Total exposure time of the observation.}
\end{table}

\subsection{Point source detection}\label{sec:point_source_detection}

\begin{figure*}
        \centering
        \includegraphics[scale=0.5]{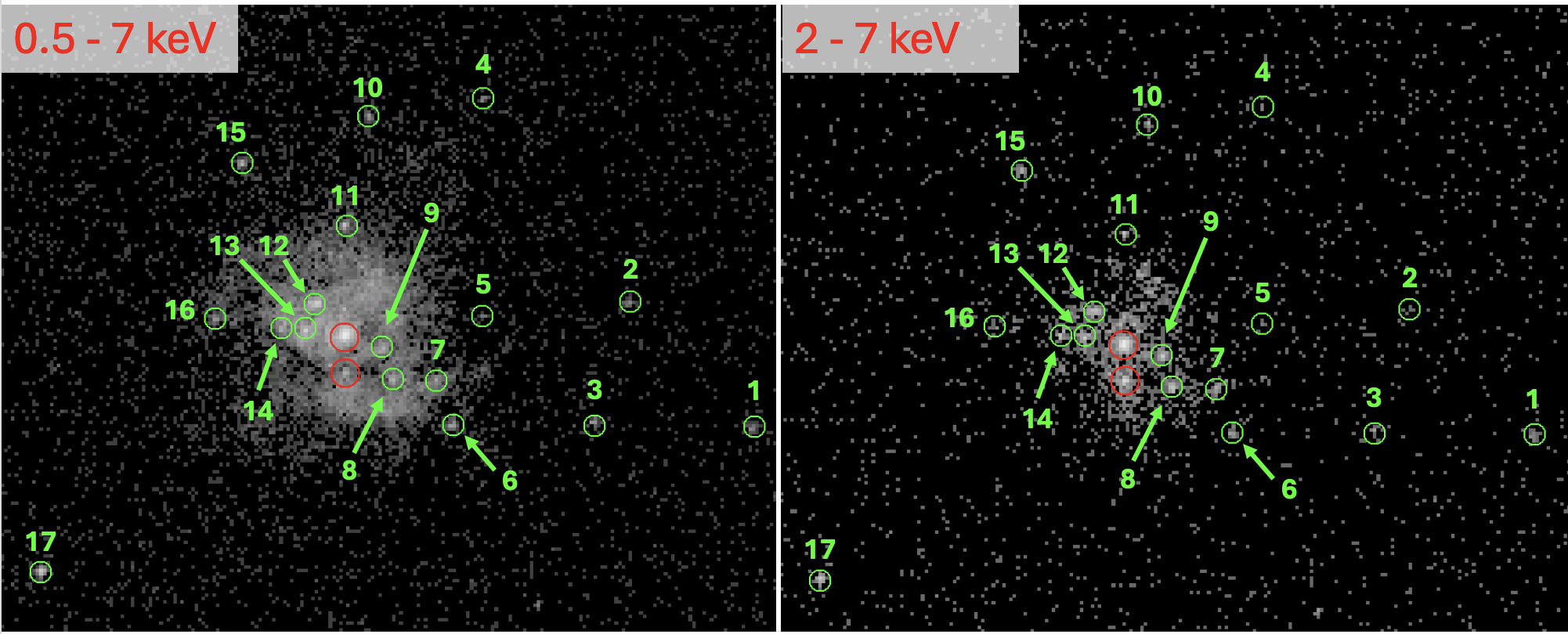}
        \caption{Stacked \textit{Chandra} images of NGC 3256, in the 0.5$-$7~keV (full, left panel) and 2$-$7~keV (hard, right panel) bands. All considered point sources are marked with green circles, and numbered according to the data shown in Table \ref{tab:point_sources}, in order of increasing right ascension. The two red circles correspond to the nuclei of the galaxy. North is up and East is left in all images.} \label{fig:chandra_images}
\end{figure*}

\begin{figure*}
        \centering
        \includegraphics[width=\textwidth]{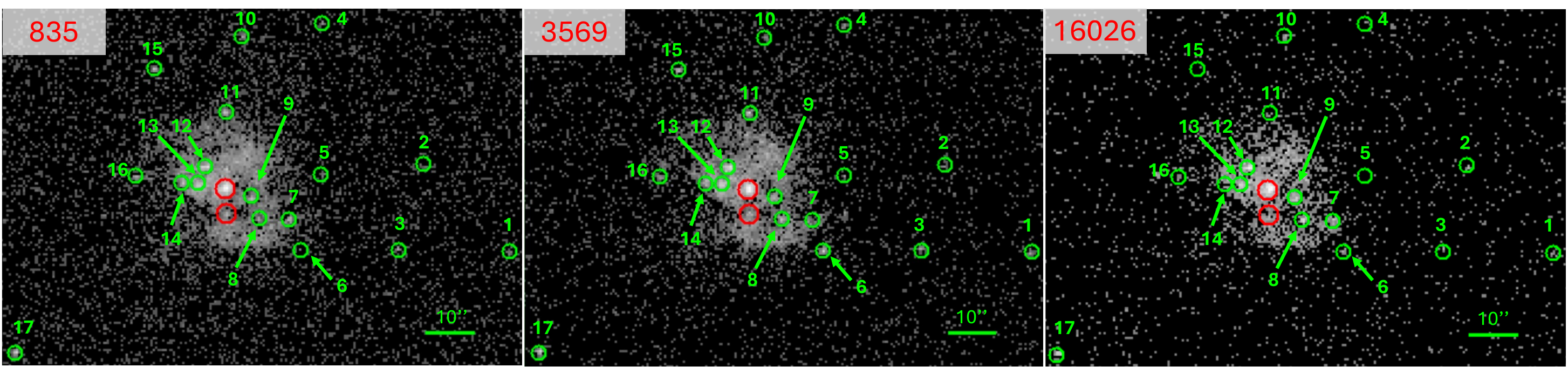}
        \caption{Individual \textit{Chandra} observations of NGC 3256, in the 0.5$-$7~keV (full) band. All considered point sources are marked with green circles, and numbered according to the data shown in Table \ref{tab:point_sources}, in order of increasing right ascension. The two red circles correspond to the nuclei of the galaxy. The 10'' angular scale bar corresponds to a physical scale of $\sim$2.2~kpc. North is up and east is left in all images.} \label{fig:chandra_obsid}
\end{figure*}

We reprocessed each observation individually using the CIAO task \texttt{chandra\_repro}, and then stacked all observations, using \texttt{merge\_obs}, in order to increase the signal-to-noise and detect as many point sources as possible. We generated two broadband stacked images: 0.5$-$7~keV (full band) and 2$-$7~keV (hard band). The full band covers the range in which \chandra is most sensitive, which maximizes our chances of detecting fainter ULXs. In the hard band, the prominent diffuse X-ray emission of NGC 3256, associated with the intense starburst, is significantly reduced, which makes it easier to isolate point sources. The two images are shown in Fig. \ref{fig:chandra_images}, and Fig.~\ref{fig:chandra_obsid} shows the three individual observations separately, in the full band.

 \begin{figure*}
\sidecaption
  \includegraphics[scale=0.2755]{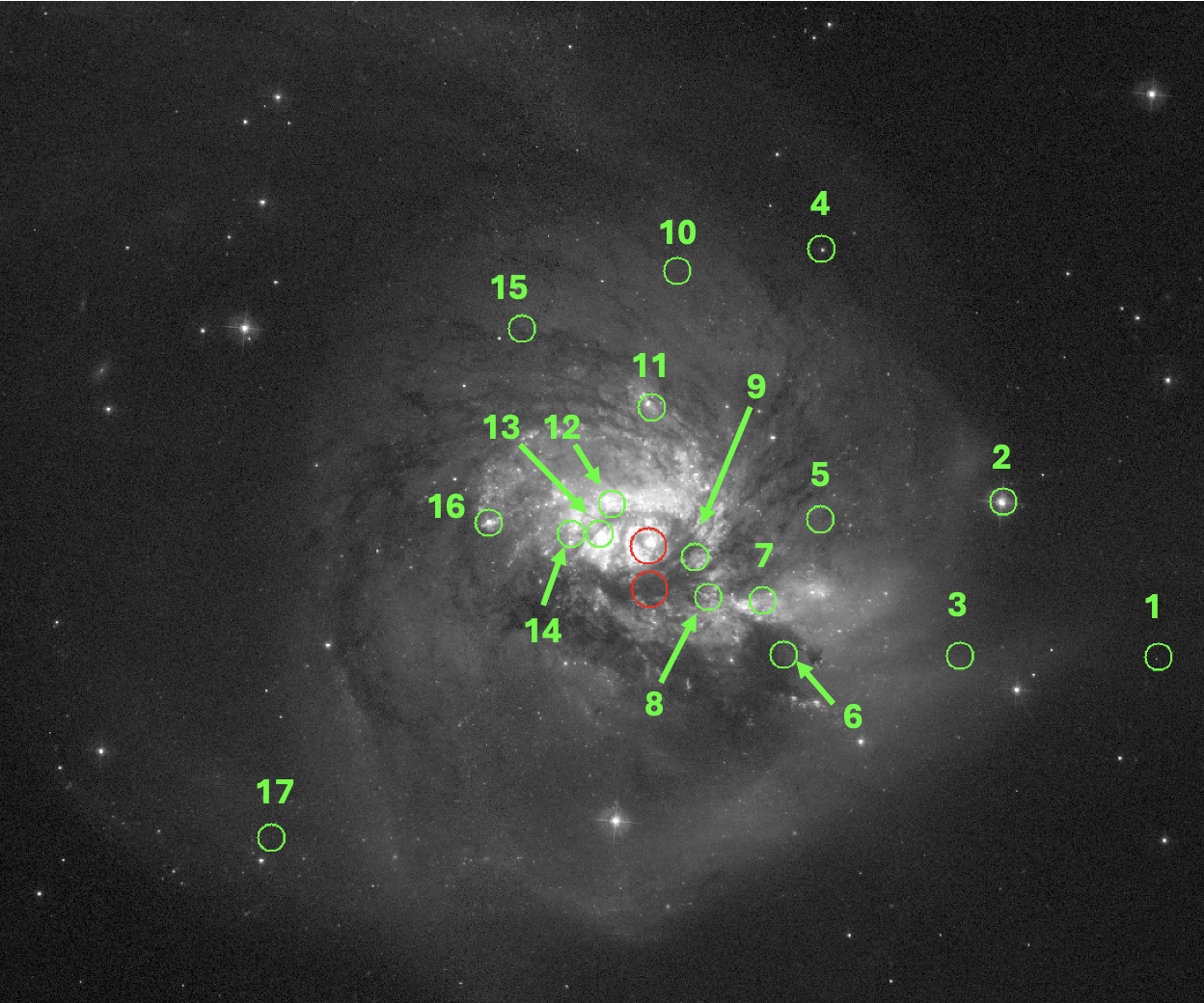}
   \includegraphics[scale=0.29]{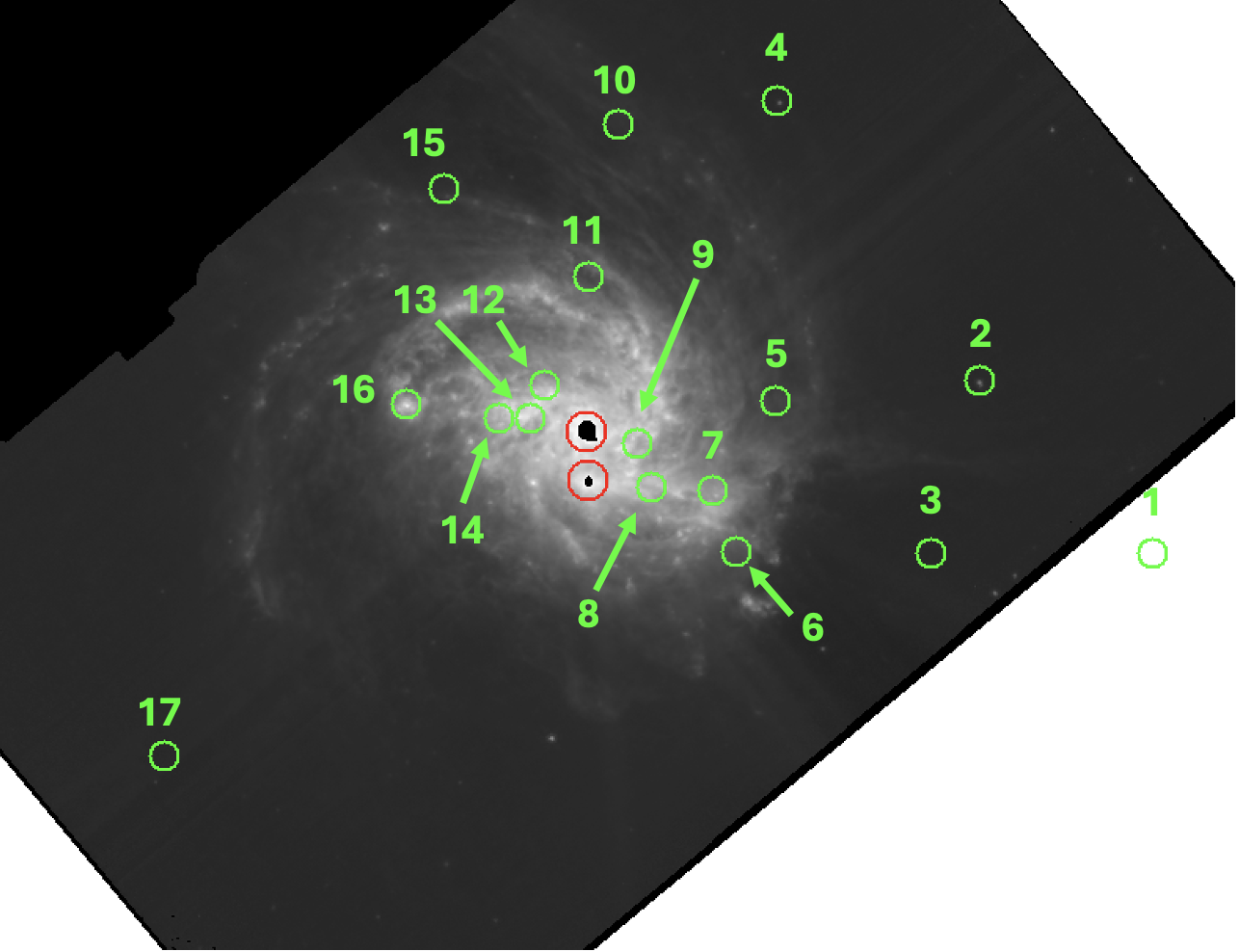}
     \caption{\textit{HST} ACS/WFC F435W (left) and \textit{JWST} MIRI F770W images (right) of NGC 3256. The green circles correspond to the \chandra-detected point sources listed in Table \ref{tab:point_sources}, PS1-17. The red circles correspond to the two nuclei in the galaxy. The \textit{JWST} image has been masked at the nuclei to prevent issues with saturation. North is up and East is left in all images.}
     \label{fig:MW}
\end{figure*}

We ran the point source detection tool \texttt{wavdetect} on both the full band and the hard band images. We used default parameters, except for \texttt{scales}, which corresponds to the different pixel scales for which a wavelet transform is performed. Due to the extended emission in the galaxy being non-uniform, we fixed \texttt{scales}=``1'', rather than the default \texttt{scales}=``1 2 4 8 16''. This minimized the detection of extended regions as point sources. We consider all point sources (either in the hard or the full bands) that overlap with the visible optical extension of NGC 3256 as per its \textit{Hubble Space Telescope (HST)} observation; see Fig.~\ref{fig:MW}. A total of 17 point sources were detected, which are listed in Table \ref{tab:point_sources}, together with the positions of both galactic nuclei. {Table \ref{tab:point_sources} reports net counts and detection significance for each point source, both in the full and hard bands, obtained following the method of \cite{Li1983}.}

\begin{table*}[t]
\scriptsize
\centering
\caption{X-ray point sources identified in NGC 3256, in order of RA.}
\label{tab:point_sources}
\begin{tabular}{Sc Sc Sc Sc Sc Sc Sc Sc Sc Sc Sc Sc}
PS & RA & DEC & F/H & $\sigma_F$/$\sigma_H$& CSC &  Lira & Liu & Lehmer & Walton & Tranin & Class\\
 &  &  &  & & & +2002 & +2011 & +2015 & +2022 & +2024 & \\
(1) & (2) & (3) & (4) &  (5) & (6) & (7) & (8) & (9) & (10) & (11) & (12)\\
\hline\hline
PS1 & 10:27:45.95 & -43:54:26.62 & 38.9/16.9 & 12.4/7.9 & 2CXO J102745.9$-$435426 & $-$ & $-$ & $-$ & $-$ & $-$ &BQ? \\
PS2 & 10:27:47.55 & -43:54:09.20 & 30.2/4.6 & 9.9/2.8 & 2CXO J102747.5$-$435409 & $-$ & $-$ & $-$ & $-$ & $-$ & * \\
PS3 & 10:27:48.01 & -43:54:26.34 & 43.4/10.9 & 12.9/5.9 & 2CXO J102748.0$-$435425 & 1 &  $-$ & $-$ & $-$ & $-$ \\
PS4 & 10:27:49.45 & -43:53:40.90 & 19.7/2.7 & 6.9/1.8  & $-$ & $-$ & $-$ & $-$ & $-$ & $-$ & * \\
PS5 & 10:27:49.46 & -43:54:11.22 & 13.3/3.6 & 4.9/2.3 & $-$ & $-$ & $-$ & $-$ & $-$ & $-$ \\
PS6 & 10:27:49.84 & -43:54:26.38 & 73.4/16.3 & 14.1/6.8 & 2CXO J102749.8$-$435426 & $-$ & NGC3256-X15 & $-$ & Included & Included\\
PS7 & 10:27:50.05 & -43:54:20.29 & 119.3/31.0 & 12.1/8.2 & $-$ & 2 & NGC3256-X12 & NGC 3256 1 & $-$ & $-$\\
PS8 & 10:27:50.60 & -43:54:19.97 & 97.1/64.4 & 7.8/12.1 & 2CXO J102750.6$-$435420 & $-$ & NGC3256-X7  & NGC 3256 2 & $-$ & Included &\\
PS9 & 10:27:50.75 & -43:54:15.52 & 144.0/61.3 & 10.5/11.2 & 2CXO J102750.7$-$435415 & 3 & NGC3256-X8 & NGC 3256 3 & $-$& $-$\\
PS10 & 10:27:50.93 & -43:53:43.39 & 49.2/18.7 & 12.2/8.0 & $-$ & 4 & $-$ & $-$ & $-$ & $-$ \\
PS11 & 10:27:51.21 & -43:53:58.55 & 123.8/26.7 & 14.8/8.8  & 2CXO J102751.2$-$435358 & 5 & NGC3256-X14 & $-$ & Included & Included\\
South & 10:27:51.22 & -43:54:19.15 & $-$ & $-$& 2CXO J102751.2$-$435418 & 7 & NGC3256-X9 & $-$ & $-$ & $-$ \\
North & 10:27:51.24 & -43:54:14.21 & $-$ & $-$ & 2CXO J102751.2$-$435413 & 8 & $-$ & NGC 3256 4 & $-$ & $-$  \\
PS12 & 10:27:51.62 & -43:54:09.43 & 391.6/69.6 & 20.0/10.1 & 2CXO J102751.5$-$435409 & 9 & NGC3256-X3 & NGC 3256 5 & $-$ & Included\\
PS13 & 10:27:51.74 & -43:54:12.89 & 301.2/50.3 & 13.2/6.4 & 2CXO J102751.7$-$435413 & 10 & NGC3256-X2 & NGC 3256 6 & $-$ & $-$ \\
PS14 & 10:27:52.05 & -43:54:12.89 & 13.3/31.4 & 0.9/5.4  & $-$ & 11 & NGC3256-X4 & $-$ & $-$ & Included\\
PS15 & 10:27:52.55 & -43:53:49.82 & 89.7/26.7 & 18.3/10.3 & 2CXO J102752.5$-$435349 & 12 & NGC3256-X19 & $-$ & Included & Included\\
PS16 & 10:27:52.90 & -43:54:11.57 & 36.7/7.1 & 6.0/3.5 &$-$ & 13 & $-$ & $-$ & $-$ & $-$\\
PS17 & 10:27:55.16 & -43:54:46.99 & 175.2/51.7 & 31.4/16.0 & 2CXO J102755.1$-$435446 & 14 & NGC3256-X10 & NGC 3256 7 & Included & Included & BQ? \\
\hline
\end{tabular}
\footnotesize{ \newline \\
{\it Columns:} 1) Point Source (PS) number. 2, 3) Right ascension and declination, as per the centroid of the \textit{Chandra} point source. {4) Full (S; $0.5-7$~keV) and Hard (H;$2-7$~keV) band net counts detected in the stacked observation. 5) Detection significance in the Full and Hard bands in the stacked observation, following \cite{Li1983}.} 6) CSC name, if the source is present in the catalog \citep{Evans_csc2024}. 7) Corresponding point source name if detected by \cite{Lira2002} in ObsID 835. 8) Corresponding point source name if detected by \cite{Liu2011} in ObsIDs 835 or 3569. 9) Corresponding point source name if detected by \cite{Lehmer2015} in OBSID 16026. 10,11). Wether the sources are included, with their CSC name, in the \cite{Walton2022} and \cite{Tranin2024} catalogs, respectively.
12) \textbf{Source class, if potentially not associated to NGC~3256.} Sources marked as `*' are confirmed stars. Sources marked as `BQ?' are potential background quasars. See Sec. \ref{sec:point_source_association}.  \newline
The table includes the positions and counterparts to both galactic nuclei, labeled as `South' and `North'.}
\end{table*}

Table \ref{tab:point_sources} also includes a comparison between the point sources recovered in this work, previous studies \citep{Lira2002,Liu2011,Lehmer2015}, and the \textit{Chandra} Source Catalog \citep[CSC v2.2.,][]{Evans_csc2024}. We also include two recent ULX catalogs, \cite{Walton2022} and \cite{Tranin2024}, both based on the CSC v2.0\footnote{{We note that the CSC does not include observations taken before January 29 2000 \citep[i.e., ObsID 835;][]{Evans_csc2024}, which explains the lower number of sources for the catalog-based works.}}. Overall, we detect a larger number of sources than previous works, which we attribute to the use of multiple observations.

\cite{Lira2002} analyzed only the first \chandra observation (ObsID 835), as it was the only one available at the time. We recover 13 of their 14 point sources, although two correspond to the nuclei. The exception is the source labeled ``6'' in their work which, after performing the stacking, appears to be extended rather than point-like. \cite{Liu2011} use both ObsIDs 835 and 3569, but a stacking analysis is not performed prior to detecting sources. This likely explains why they do not detect all our point sources. We note that, while their numbering goes up to X19, they only report 18 point sources. Of these, we recover 11. Out of the ones we do not recover, sources X11, X13, X16 and X17 are consistent with diffuse emission instead. Source X6 is a point source, but falls outside of the extension of the galaxy (roughly 2' further from the center of the galaxy than PS1). Source X18 is located even further away, and we do not detect any X-ray emission above background levels near that position. \cite{Lehmer2015} focus on the study of the nuclei, using a 2014 \textit{Nuclear Spectroscopic Telescope Array} \citep[\nustar,][]{Harrison2013} observation. As such, they choose the simultaneous \chandra observation (ObsID 16026), which is shorter and also taken at a point where the degradation of the \chandra effective area impacted sensitivity in the soft band, compared to the previous observations. Due to this fact, they detect a lower number of point sources than the other two works, only 7 (one of which is a nucleus), all of which we recover. The two ULX catalogs \citep{Walton2022,Tranin2024} detect 4 and 7 ULXs respectively, which we also recover. 

Previous works have tended to consider the two nuclei among the X-ray point sources, although here we opt to exclude them from the numbering convention, due to their different nature. Analysis of the nuclear X-ray spectral emission can be found in multiple works \citep{Lira2002,Lehmer2015,Ohyama2015}, which remain inconclusive regarding the presence of an AGN in either of them, as the extreme starburst is sufficient to explain the detected emission. The debate extends to other wavelengths, and remains even in the analysis of \textit{JWST} data \citep[e.g.,][]{Bohn2024,Gonzalez-Martin2025}. Although the nuclei may contain a number of ULXs, they cannot be spatially resolved with the current observations. As such, we exclude their analysis from this work.

\subsection{Point source association with NGC 3256}\label{sec:point_source_association}

Although the X-ray point sources listed in Table~\ref{tab:point_sources} overlap spatially with the optical extent of the galaxy\footnote{As determined by the diffuse HST emission; see Fig.~\ref{fig:MW}.}, not all of them are necessarily associated with NGC 3256. Point sources located near the edge of the galaxy (PS1, PS2, PS4, PS17), particularly those that do not follow the distribution of material in the spiral arms, have a higher likelihood of being unrelated foreground (e.g., stars), or background objects (e.g., quasars). Fig~\ref{fig:MW} shows \textit{HST} and \textit{James Webb Space Telescope} \citep[\textit{JWST},][]{Gardner2006} images of NGC~3256, from \cite{Kim2013} and \cite{Linden2024} respectively, with the \chandra point source regions overlaid.

We look for potential optical counterparts in publicly available optical data, in order to ascertain the potential non-association of some of the point sources with NGC~3256. We use the \textit{Gaia} DR3 \citep{Gaia2023}, as well as MUSE \citep[Multi Unit Spectroscopic Explorer,][]{Bacon2010} data. PS2 and PS4 have \textit{Gaia} counterparts (Gaia DR3 5367700730400924416 and Gaia DR3 5367700528540910208, respectively), but only PS4 has a parallax measurement firmly placing it within our galaxy (1.28$\pm$0.14 mas). PS2, with a \textit{Gaia} parallax 0.27$\pm$0.45 mas, is bright and star-like in the \textit{HST} images and well detected by MUSE. We use the Penalized Pixel Fitting Routine \citep[pPXF][]{Cappellari2017,Cappellari2023}, as well as the empirical determinations of the calcium triplet of \cite{Cenarro2001} to measure a redshift by means of fitting said triplet. 

\begin{figure}
        \centering
        \includegraphics[scale=0.13]{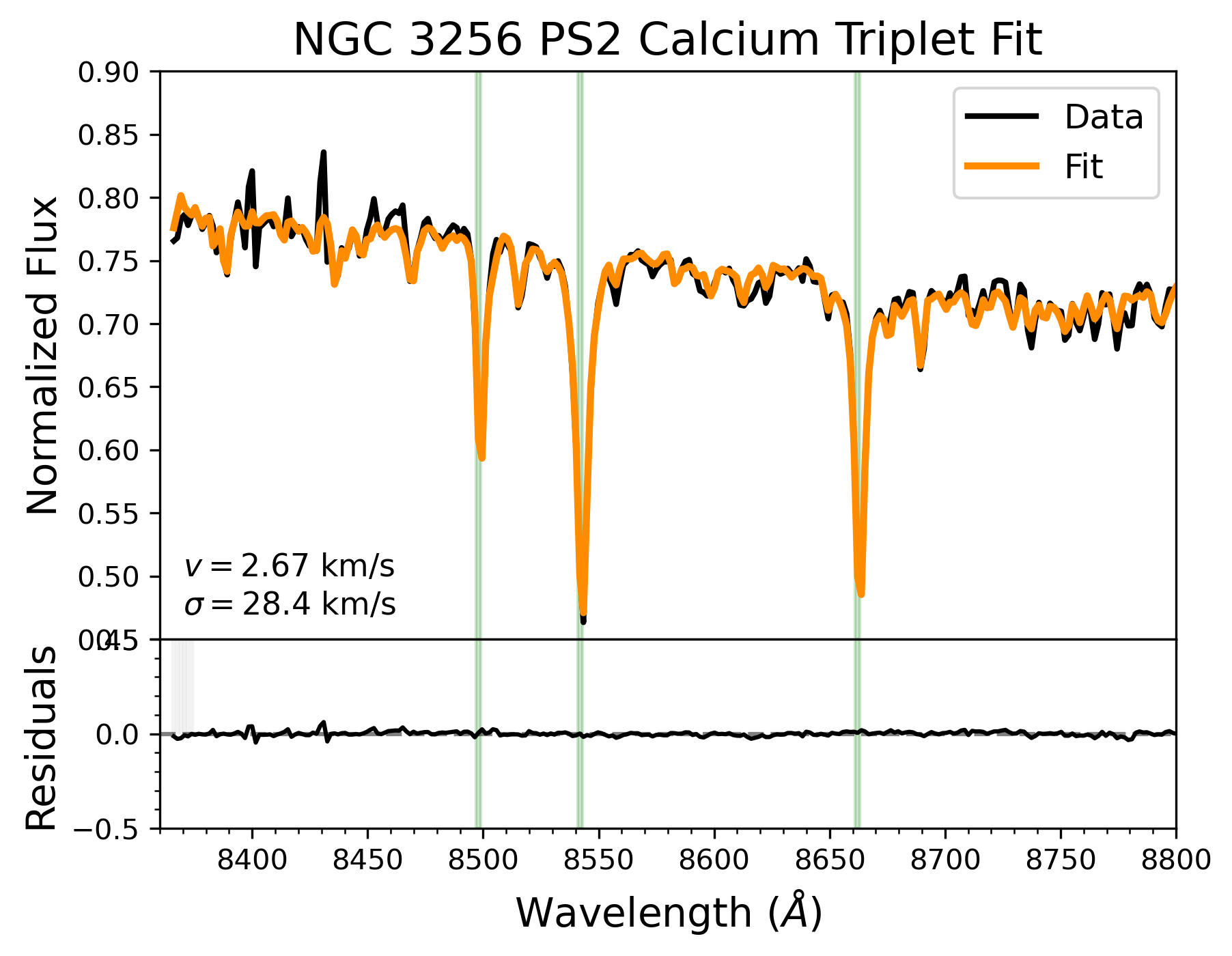}
        \caption{Calcium triplet absorption feature fit for PS2, which yields a redshift z$\ll$0.00055. The low value, incompatible with a cosmological redshift, implies PS2 is a foreground star rather than a ULX associated with NGC~3256.} \label{fig:calcium_triplet}
\end{figure}

In Fig. \ref{fig:calcium_triplet} we present a fit to the calcium triplet absorption feature of PS2, taken from the MUSE data. It yields a redshift measurement $<10^{-5}$ (i.e. not a cosmological redshift), or more accurately, a velocity measurement of $v$ = $3\pm45$~km/s; placing the source within the Milky Way and confirming its stellar nature. PS1 also has an optical counterpart in the \textit{HST} image (point-source like), but it is too faint to be detected by either \textit{Gaia} or MUSE. PS17 lacks an optical counterpart, even in the \textit{HST} data\footnote{PS5, PS6, PS10 and PS15 also lack clear optical counterparts. However, this does not affect the potential ULX classification of any source.}.

In order to calculate how many background quasars are expected in the stacked observation, we simulated a point source emission under optimal circumstances for detection: the \chandra on-axis effective area of the year 2000 observation, but with an observing time of 71.3~ks. This provides an upper limit for the detectability of background quasars, given how the later observations will have lower effective area below 2~keV. We assumed a power-law emission with photon index $\Gamma$~=~$1.8$ \citep[average for AGN;][]{Nandra1994} and absorption $N_{\rm H}$~=~$8.91\times10^{20}$~cm$^{-2}$, to account for the gas in our own Galaxy \citep{Kalberla2005}. We then adjusted the normalization until a 71.3~ks exposure produces 10~cts in the 0.5$-$7~keV band. This, once more, certifies that we provide a robust upper limit, given that the lowest-count source in our sample (PS5) has a total of 30 cts in the stacked observation. 

The flux of such a model, in the 2$-$10~keV band, is $\sim7\times10^{-16}$~erg~s$^{-1}$~cm$^{-2}$. We can then compare this flux with the $2-10$~keV log(N) $-$ log(S) distribution of \cite{Ueda2014}, obtaining a density of AGN above the given flux threshold of $\sim2700$~deg$^{-2}$. The extension of NGC 3256 is $\sim1$', which results in an expected number of detectable background AGN of 2.33. This is consistent with the furthest point sources from the center of the galaxy, PS1 and PS17, potentially being background AGN\footnote{However, we note that both sources are reached by tidal tails stemming from NGC~3256, and thus remain potentially associated with the galaxy.}.

\subsection{Chandra spectral extraction}\label{sec:spectral:extraction}

We extracted spectra for all point sources labeled in Table~\ref{tab:point_sources}, regardless of their association with NGC 3256, for all three available observations separately. We excluded the nuclei from the analysis, as the potential AGN activity in the galaxy is not the focus of this work; the analysis of the stacked nuclei spectra will be presented in Evans et. al. (in prep.).

The source spectra were extracted using a circular region with a radius of 1.5'', which corresponds to a physical scale of $\sim300$~pc. This size was selected to enclose the majority of the \chandra PSF\footnote{The on-axis \chandra PSF encloses $\sim90$\% of the energy fraction within 1.5'', https://cxc.cfa.harvard.edu/proposer/POG}, while excluding most of the contamination from the diffuse emission in the galaxy. We also used a localized background for each point source to account for the extended emission in which a number of them are embedded. In order to ensure a background that is as representative as possible of the ULX extraction region, we use an annulus when the point sources are isolated, and multiple circular background regions (on different sides of the point source) when they have other ULXs nearby (e.g., PS12, PS13, and PS14). This leaves us with a total of 50 source spectra to fit, since PS5 has zero counts in ObsID~16026.

\section{X-ray spectral analysis}\label{sec:analysis}

For each source, we fit each spectrum separately (for the three available observations) using three different models, listed below:
\begin{align}
& \texttt{phabs*powerlaw} \\
& \texttt{phabs*zphabs*powerlaw}\\
& \texttt{phabs*diskpbb}, 
\end{align}
where \texttt{phabs} accounts for the obscuration within our own Galaxy in the direction of NGC 3256, measured to be $N_{\rm H,gal}$~=~$8.91\times10^{20}$~cm$^{-2}$ \citep{HI4PI2016}. The value of this obscuration is fixed, and not varied within the fits. The first model is thus a simple power law, while the second model accounts for local obscuration in the environment of each source (\texttt{zphabs}); likely expected given the amount of soft X-ray emission associated with star-forming processes. The final model is a multiple blackbody disk model, with disk temperature proportional to $r^{-p}$. In this expression, $r$ is the radius and $p$ is a free parameter. This model is commonly used for X-ray binaries \citep[e.g.,][]{Mineshige1994,Watarai2000,Kubota2005}. We opt to not include more model variations (e.g., local obscuration on the \texttt{diskpbb} model, additional thermal components) given the data quality of most of the observations (with the lowest having $\sim10$~cts). The three models used provide a realistic description of the potential range of luminosities for all considered point sources, which is the main objective of this paper.

Each spectra was binned using the HEASoft tool \texttt{ftgrouppha}, and the optimal binning scheme of \cite{Kaastra2016}. This binning is sufficient to fit the data using Cash statistics \citep{Cash1979}, as described by \cite{Buchner2023}. Fitting was carried out using the Bayesian X-ray Analysis software package \citep[BXA,][]{Buchner2014,Buchner2016bxa}, using the nested sampling package \texttt{ULTRANEST} \citep{Buchner2016ultranest,Buchner2021UltraNest}. 

Model (1) has two free parameters: photon index ($\Gamma$) and normalization ($norm$). We assigned a uniform prior to the former and a log-uniform prior to the latter, in the ranges of $\Gamma\in\left[-5,10\right]$ and ${\rm log} \left(norm\right)\in\left[[-11,-2\right]$.

Model (2) has an additional free parameter, obscuration along the line of sight, $N_{\rm H}$. We assign it a log-uniform prior in the range ${\rm log}\left(N_{\rm H}\right)\in\left[-3,2\right]$, in units of $10^{22}$~cm$^{-2}$. The extent of our priors for most parameters is far broader than the expected and commonly observed range. However, due to the low number of counts in some of the spectra, allowing full freedom for photon index values for model (2) resulted in unrealistic luminosity values (i.e. $10^{45}$~erg~s$^{-1}$). This is due to the degeneracy between photon index and local obscuration. In order to avoid exploration of infeasible regions of luminosity parameter space, we limit the photon index to vary between $\in\left[1,3\right]$ for model (2). 

Model (3) has three free parameters: the temperature of the inner disk radius, $T_{\rm in}$; the exponent of the radial dependence of the disk temperature, $p$; and normalization. We assign $p$ a uniform prior in the range $p\in\left[0.5,1\right]$, and $T_{\rm in}$ and normalization a log-uniform prior, in the ranges $T_{\rm in}\in\left[0.1,10\right]$~keV and $log \left(norm\right)\in\left[[-11,-2\right]$.

\section{Results}\label{sec:results}

\begin{figure}
        \centering
        \includegraphics[scale=0.5]{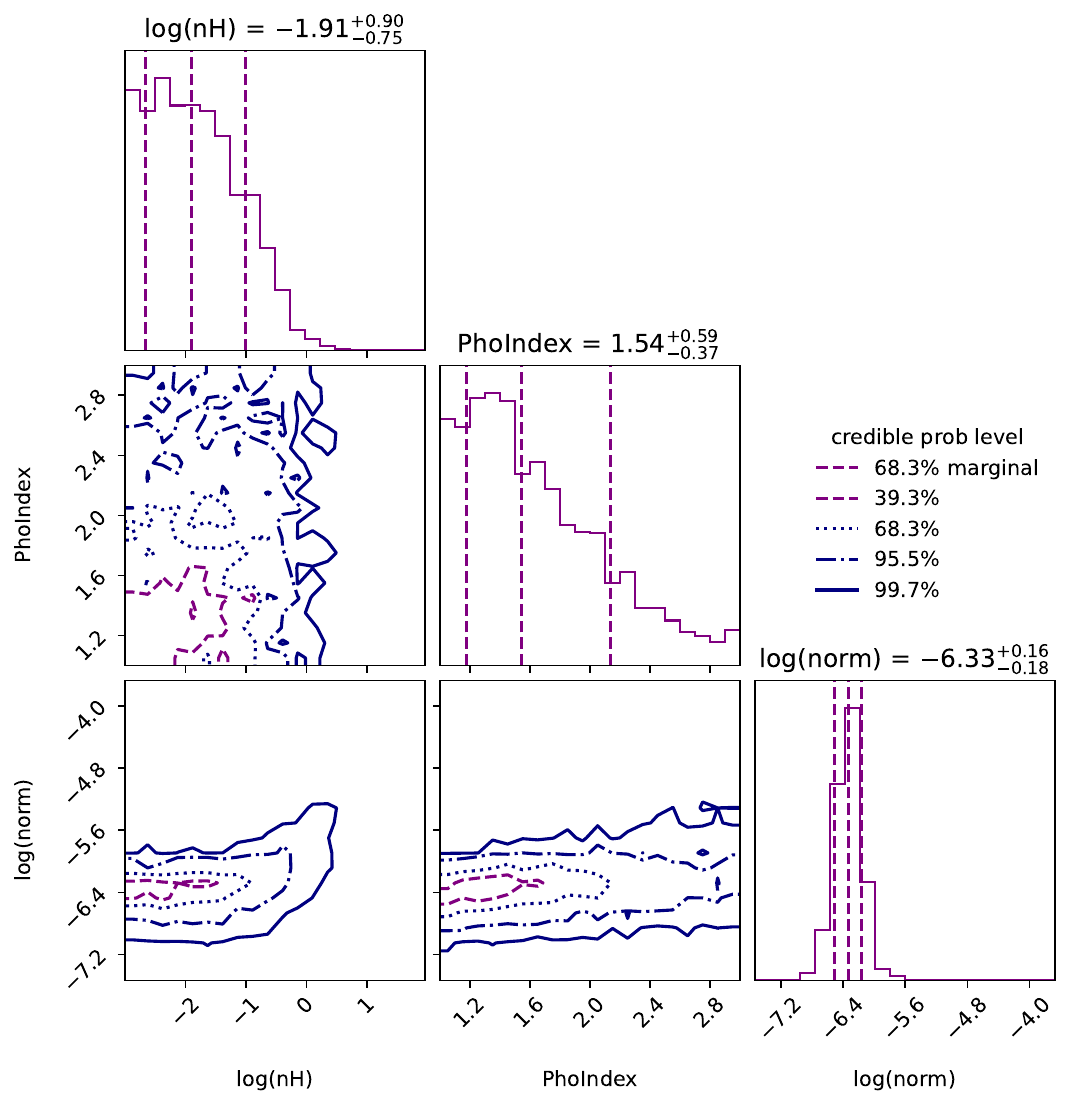}
        \caption{Corner plot showing the posterior distribution for all parameters in the fit of PS1, Obsid 835, for model (2). As can be seen, log(nH), in units of 10$^{22}$~cm$^{-2}$, is unconstrained on the lower boundary, $\Gamma$ (PhoIndex) is unconstrained on both ends of the parameter space, and log(norm) is well-constrained.} \label{fig:corner_plot_example}
\end{figure}

In this section we report the results of the spectral fitting described in Sect. \ref{sec:analysis}. Table \ref{tab:fit_medians} reports the best fit values of all parameters for each point source, for each observation (as well as the 90\% credible interval derived from the posterior distribution). On certain occasions, the posterior does not converge to zero before reaching one end (or both) of the considered parameter space. In such instances, the 90\% credible interval, which corresponds to 90\% of the values in the posterior, does not necessarily have a significant statistical meaning. That is, if the parameter space had been broader, the posterior would have reached further down, making the error range for the parameter potentially larger (or, at the very least, compatible with the hard parameter boundary). This is exemplified in Fig. \ref{fig:corner_plot_example}, corresponding to a corner plot for a \texttt{powerlaw} fit (model 1) to the first observation of PS1.

The posterior distribution for all relevant fit parameters (replacing log(norm) for a luminosity distribution) is shown in Figs. \ref{fig:app:posteriors_123} to \ref{fig:app:posteriors_1617}. The figures also display the median of the posterior distribution for each parameter, as well as the 90\% credible interval. When the credible interval is almost at the boundary of the plot (or goes beyond it), the measurement should be interpreted as an upper/lower limit (or no constraint at all, in cases where the credible interval is not bound in either direction). These are also indicated, for all parameters, with the letter `u' in Tables \ref{tab:fit_medians} and \ref{tab:luminosity}. This is particularly significant for the posterior distributions of luminosity: a ULX with a luminosity compatible with the lower threshold of the plot ($L_{\rm 0.3-10 keV}$=$10^{37}$~erg~s$^{-1}$) is compatible with a non-detection during that epoch. For context, our simulations show that a source with luminosity log$L\rm_x >38.1$~erg~s$^{-1}$ will be detected in the stacked observation with $>5$~cts with a 90\% probability\footnote{Assuming an isolated point source.}.

For each point source, PS1$-$PS17, and each model used, Figs. \ref{fig:app:spectra_panel1} to \ref{fig:app:spectra_panel3} also report the percent of the posterior above the ULX luminosity threshold, $10^{39}$~erg~s$^{-1}$. These numbers, together with the median of the luminosity, are reported in Table \ref{tab:luminosity}. Reported luminosities are intrinsic, corrected for obscuration.

We imposed two different thresholds to determine whether a point source is classified as a ULX or not: a less conservative threshold, considering a point source to be a ULX if in at least one observation $\geq$50\% of the posterior is above the ULX luminosity threshold; and a more conservative threshold, considering a point source to be a ULX if in at least one observation $\geq$90\% of the posterior is above the ULX threshold. Table \ref{tab:isitULX} reports the ULX determinations considering all models. It also reports whether a source that is determined as a ULX in at least one observation also has at least one observation compatible with a non-detection. We call this behavior ``transient ULX'' (but see Sec.~\ref{sec:transients} for a discussion on our definition of transient ULX).

According to the simple \texttt{powerlaw} model, 13 sources make the 50\% ULX cut, and 12 make the 90\% cut. For the absorbed \texttt{powerlaw}, these are 15 and 13 ULXs respectively, and 13 and 12 ULXs for \texttt{diskpbb}. One could argue that the inclusion of absorption is realistic in a gas-rich/dusty galaxy such as a LIRG, but the inclusion of the other two models guarantees a more conservative estimate, which allows one to determine the minimum number of detected ULXs in the galaxy: 12-13, depending on the chosen threshold. However, the association of PS1 and PS17 with the galaxy is doubtful, bringing the number of firmly determined ULXs down to 10: PS6, PS7, PS8, PS9, PS10, PS11, PS12, PS13, PS15, and PS16. Additionally, PS3 would be considered a ULX at a 50\% confidence threshold (except for the \texttt{diskpbb} model), PS14 would be considered a ULX according to the absorbed \texttt{powerlaw} model (at 90\% confidence) and the \texttt{diskpbb} model (at 50\% confidence)\footnote{Moreover, the simple  \texttt{powerlaw} model, the only one that does not meet any ULX thresholds, is statistically disfavored. See Sect.~\ref{sec:model_comparison}}, and PS1 and PS17 remain potential ULXs. 

\begin{table*}[t]
\scriptsize
\centering
\caption{ULX determinations.}
\label{tab:isitULX}
\renewcommand{\arraystretch}{1.2}
\begin{tabular}{l|ccc|ccc|ccc|c}
\hline\hline
 &  \multicolumn{3}{c}{pow} & \multicolumn{3}{c}{abspowlim} & \multicolumn{3}{c}{diskpbb} & \\
&  ULX-50\% & ULX-90\% & Transient & ULX-50\% & ULX-90\% & Transient & ULX-50\% & ULX-90\% & Transient & Class.\\ \hline
\noalign{\vspace{2pt}}
PS1$^{a}$ & YES & YES & NO & YES & YES & NO & YES & NO & NO &   Unk.\\
PS2 & NO & NO & NO & NO & NO & NO & NO & NO & NO & Star\\
PS3 & YES & NO & NO & YES & NO & NO & NO & NO & NO & Unk.\\
PS4 & NO & NO & NO & NO & NO & NO & NO & NO & NO & Star\\
PS5 & NO & NO & NO & YES & NO & YES & NO & NO & NO & Unk.\\
PS6 & YES & YES & YES & YES & YES & YES & YES & YES & YES & ULX\\
PS7 & YES & YES & YES & YES & YES & NO & YES & YES & NO & ULX\\
PS8 & YES & YES & NO & YES & YES & NO & YES & YES & NO & ULX\\
PS9 & YES & YES & NO & YES & YES & NO & YES & YES & NO & ULX\\
PS10 & YES & YES & NO & YES & YES & NO & YES & YES & NO & ULX\\
PS11 & YES & YES & YES & YES & YES & YES & YES & YES & YES & ULX\\
PS12 & YES & YES & NO & YES & YES & NO & YES & YES & NO & ULX\\
PS13 & YES & YES & YES & YES & YES & NO & YES & YES & NO & ULX\\
PS14 & NO & NO & NO & YES & YES & YES & YES & NO & NO & Unk.\\
PS15 & YES & YES & YES & YES & YES & YES & YES & YES & YES & ULX\\
PS16 & YES & YES & YES & YES & YES & YES & YES & NO & YES & ULX\\
PS17$^{a}$ & YES & YES & NO & YES & YES & NO & YES & YES & NO & Unk. \\
\hline
\hline
\end{tabular}
\footnotesize{\newline \\ 
{\it Columns:} 1) Point Source number. 2,5,8) For the  \texttt{powerlaw}, absorbed  \texttt{powerlaw} and \texttt{diskpbb} models respectively; is $\geq$50\% of the luminosity posterior for at least one observation above the ULX threshold. 3,6,9) For the  \texttt{powerlaw}, absorbed  \texttt{powerlaw} and \texttt{diskpbb} models respectively; is $\geq$90\% of the luminosity posterior for at least one observation above the ULX threshold. 4,7,10) For the  \texttt{powerlaw}, absorbed \texttt{powerlaw} and \texttt{diskpbb} models respectively; is there at least one observation above the ULX threshold, and one non-detection. 11) Final source classification (see Sect.~\ref{sec:point_source_comments} for details). Sources labeled as 'Unk.' remain ULX candidates. $^{a}$: Suspected background quasar. \newline}
\end{table*}

\subsection{Point source properties}\label{sec:point_source_comments}

{In this section, we comment on each source's potential ULX classification, as well as their variability properties. We compare against the variability flags listed in the CSC v2.2 when the source is listed in the catalog. ObsID 835 is not included in the CSC. The CSC variability results therefore rely only on ObsIDs 3569 and 16026.}

PS1 and PS17 would be considered non-transient ULXs, but may not actually be associated with the host galaxy. Both are also compatible with remaining non-variable in terms of luminosity (i.e., a source that is firmly detected in all observations with 90\% luminosity credible intervals that overlap). {PS17 is flagged as non-variable in the CSC v2.2. PS1 is flagged as variable
with an inter-observation variability index of 0.95, but only
in the 0.5$-$1.2~keV band. Our results show consistent luminosity between observations, within errors, when considering the
full band.}

PS2 and PS4 are, as discussed previously, confirmed to be stars within our own Galaxy. We still opt to report the values obtained from our fitting in terms of their X-ray luminosity. We note that the reported values are lower limits, calculated with the highest parallax measurement possible given by the \textit{Gaia} data, as reported in Sec. \ref{sec:point_source_association}. PS2 has no upper bound on distance, and therefore no upper bound on potential X-ray luminosity. PS4 has a highest possible parallax of 1.14 mas, corresponding to a distance of 704 pc. The lowest possible parallax, 1.42 mas, corresponds to a distance of 877~pc. Its luminosity could therefore be $\sim$55\% higher than reported. {Neither source is variable in X-rays, but only PS2 is listed in the CSC v2.2, where it is also flagged as non-variable.}

{PS3 only reaches the lower threshold for ULX classification, having 50\% of the posterior above the limit. In the case of the \texttt{diskpbb} model, even this more lenient threshold is not met. We see no signs of flux variability, which agrees with the CSC v2.2 results, in which it is flagged as non-variable.}

PS5 is undetected in ObsID 16026, to the degree that not a single detected count is within its extraction region. As such, we do not carry out a fit to the data, and its last observation should be considered a non-detection. The first observation is also compatible with a non-detection in all models, and only the absorbed  {powerlaw} model results in it potentially being a ULX ($\geq$50\% threshold only). {PS5 is not listed in the CSC v2.2.}

PS6, PS11, PS15, and PS16 are transient ULX candidates according to all models, being undetected in one observation, while being classified as ULXs in the others. For PS6 and PS16, some of the detected observations/models only meet the 50\% ULX threshold. {PS6, PS11 and PS15 are all flagged as variable in the CSC v2.2, with an inter-observation variability probability of 1. PS16 is not listed in the CSC v2.2.}.

PS7 and PS13 are ULXs in at least one observation. They both appear to be variable and, in one observation, one of the models is compatible with a non-detection. Due to the other two models disagreeing, and most of the posterior remaining in the detectability range even for the non-detection model, we opt to not classify these as transients. {PS13 is flagged as variable in the CSC v2.2, with an inter-observation variability probability of 0.99 in the full band (only 0.44 in the hard band). PS7 is not listed in the CSC v2.2.}.

PS8 and PS12 are confirmed, bright ULXs in all observations, with at least one observation with $L_{\rm 0.3-10 kev}>10^{40}$~erg~s$^{-1}$. PS9 would fall in the same category for the absorbed  \texttt{powerlaw} model, although the other two models place the posterior medians below $10^{40}$~erg~s$^{-1}$. {PS8 is flagged as variable in the CSC v2.2, with an inter-observation variability index of 0.9, but only in the hard band. Our results show consistent luminosity between observations, within errors, but we only consider the full band. PS12 and PS9 are consistent with no variability, both in our results and in the CSC v2.2.}

PS10 is classified firmly as a ULX by all models in the first observation, but only meets the 50\% threshold for the other two (or only the second, depending on model used). It is not classified as a transient ULX, since the source is firmly detected despite potentially dropping below the ULX threshold. We note all measurements remain compatible within 90\% confidence. {PS10 is not listed in the CSC v2.2.}

PS14 is a peculiar case. The point source is deeply embedded in a star-forming region (see Fig.\ref{fig:chandra_images}), which results in a very high background. As such, almost all models are compatible with a non-detection for all observations. Despite that, the source still meets the ULX threshold at 50\% for diskpbb, and at 90\% for the absorbed  \texttt{powerlaw} model, and with median luminosity $L_{\rm 3-10 kev}\geq 10^{40}$~erg~s$^{-1}$. The latter makes sense, as a high level of obscuration is expected given its environment. However, due to the disparity in results, we cannot firmly claim the source as a ULX. {PS14 is not listed in the CSC v2.2.}

\subsection{Model comparison}\label{sec:model_comparison}

\begin{figure*}[t]
        \centering
        \includegraphics[scale=0.5]{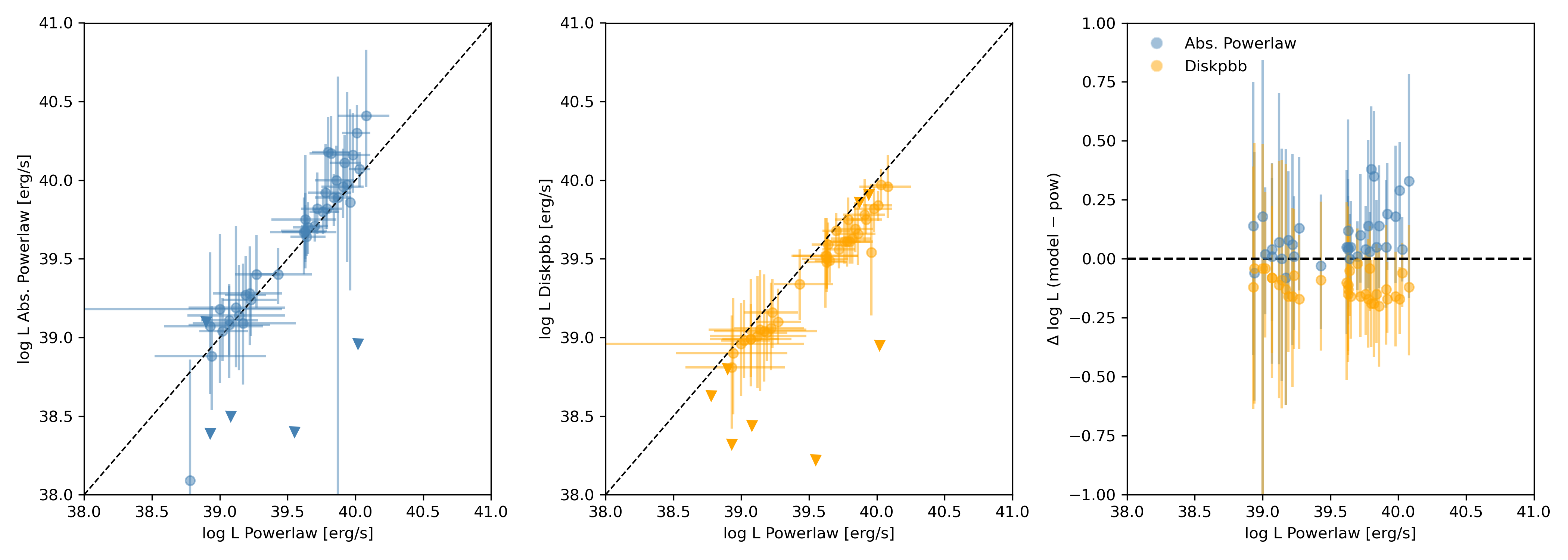}
        \caption{Comparison of the luminosity determined for each observation of each point source with the absorbed \texttt{powerlaw} model (left) and the \texttt{diskpbb} model (center), against the simple \texttt{powerlaw} model. The right panel shows the typical log difference between model determinations.} \label{fig:model_comp}
\end{figure*}

We performed model comparison using the Bayesian evidence ($Z$, the likelihood integrated over the prior volume, which quantifies how well a model explains the data while penalizing model complexity) computed by the nested sampling algorithm implemented in BXA. For each source and observation, we compared the two more ``complex'' models (absorbed \texttt{powerlaw} and \texttt{diskpbb}, which have an additional free parameter) to the simplest model used, \texttt{powerlaw}, by evaluating their relative evidences.

We computed relative log-evidences, $\Delta{\rm log}Z$~=~${\rm log}Z-{\rm log}Z_{\rm pow}$. These were used to estimate Bayes factors between models, assuming uniform model priors. We adopted a threshold corresponding to a Bayes factor of 30 (i.e., $\Delta{\rm log}Z$~=~${\rm log}30$) to identify strongly disfavored models, following the Jeffreys scale \citep{Jeffreys196}. Models with lower evidence below this threshold were considered statistically ruled out. We note, however, that no universal Bayes-factor threshold exists, as the relationship between Bayes factors and false-selection rates depends on the competing models, prior assumptions, and data quality. Simulations of Chandra spectra by \cite{Buchner2014} showed that, for the models and signal-to-noise ratios tested, a Bayes factor of $\sim$10 was already sufficient to achieve a false-selection rate below 1\%. However, their simulations start at a minimum of 100~cts per observation, which is above the signal for most of our sample. As such, we adopt the traditional threshold of 30, as a more conservative approach.

Based on the methodology above, the complex model is preferred over the simple  \texttt{powerlaw} for either one or two observations, for 10 point sources: PS6 (ObsID 3569), PS7 (ObsID 835), PS8 (ObsID 3569, 16026), PS9 (ObsID 835, 3569), PS12 (ObsID 16026), PS13 (ObsID 835, 3569),  PS14 (ObsID 835), PS15 (ObsID 835), PS16 (ObsID 835, 3569), PS17 (ObsID 16026). With the exception of PS10 and PS11, all confirmed ULXs in the galaxy have at least one epoch in which they are better fit with a more ``curved'' spectral model than a simple  \texttt{powerlaw}. However, determining which model is preferred between an obscured powerlaw (which implies higher intrinsic luminosity) and a disk black body model (which implies lower intrinsic luminosity) is not possible for any observation. The latter model is favored in high signal-to-noise spectral analysis of \nustar data for bright ULXs \cite[e.g.,][]{Rana2015,Pintore2018b}, but obscuration cannot be ruled out in the dusty environments of LIRGs.

Figure \ref{fig:model_comp} shows the typical differences in luminosity determination between the three models used. As can be perceived, compared to the simple \texttt{powerlaw} model, the absorbed \texttt{powerlaw} model favors higher luminosity values, by$\sim$0.2~dex. The \texttt{diskpbb} model yields slightly lower luminosities instead, similarly by $\sim$0.2~dex.

For the sources/observations for which a simple \texttt{powerlaw} is a worse fit than the complex model, it is thus not possible to ascertain whether their true luminosity is higher or lower, given how the other models predict opposite trends. PS14 is a slight exception to this fact, since for ObsID 835 both the absorbed \texttt{powerlaw} and \texttt{diskpbb} models predict a ULX luminosity (at 90\% and 50\% thresholds, respectively), while the statistically discarded \texttt{powerlaw} model does not. 

\subsection{Population properties}

In this section, we use the luminosity posteriors of each observation of each ULX to characterize the global behavior of the ULX population in NGC~3256. In Sec.~\ref{sec:photonindex}, we derive the distribution of photon indexes; and in Sec.~\ref{sec:lum_function}, we derive the X-ray ULX luminosity function of NGC~3256.

\subsubsection{Photon index as a function of luminosity}\label{sec:photonindex}

\begin{figure}
        \centering
        \includegraphics[scale=0.5]{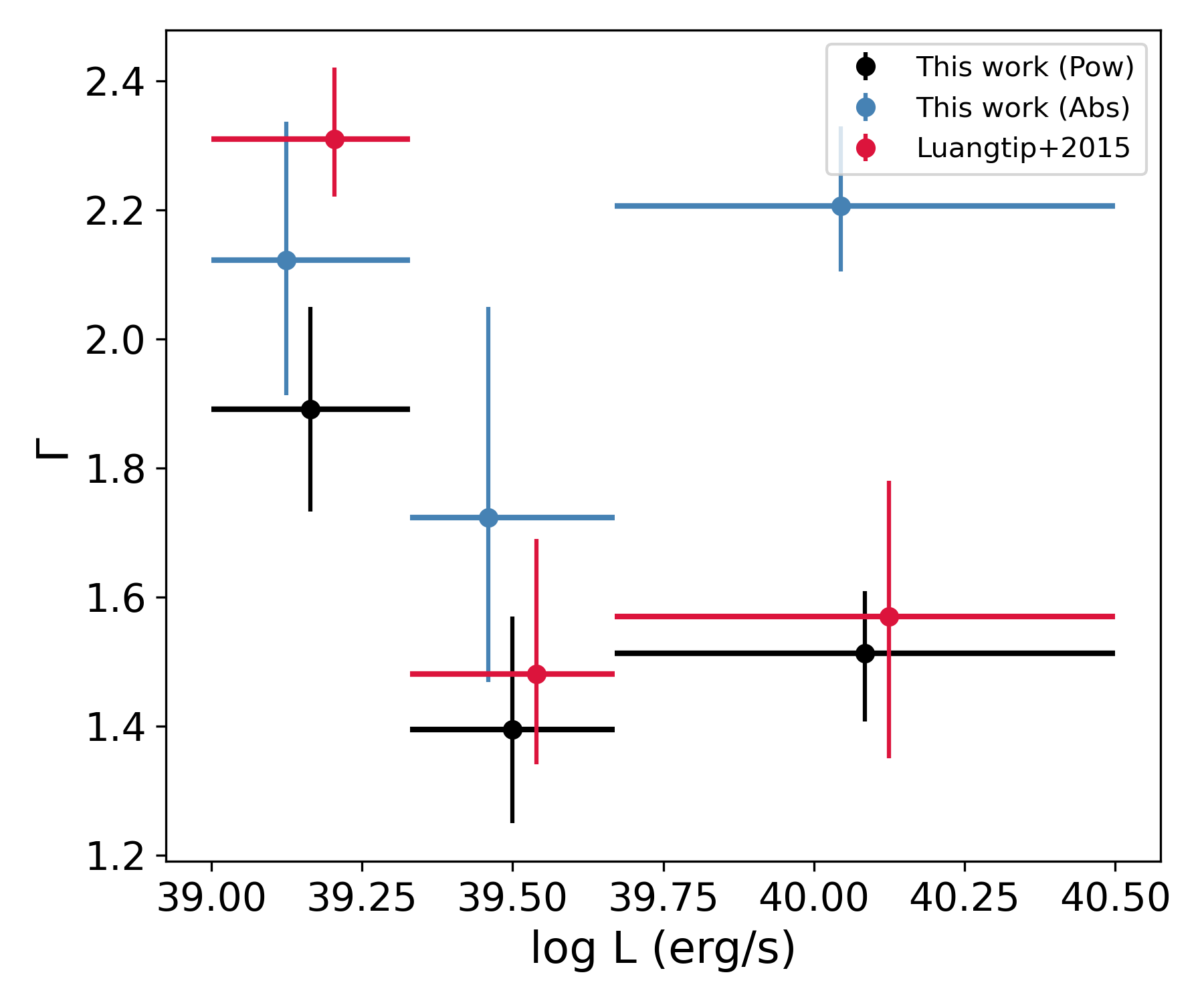}
        \caption{Median photon index of our ULX sample per luminosity bin (39<log $L_{\rm X}$<39.33, 39.33<log $L_{\rm X}$<39.67, log $L_{\rm X}$>39.37), for both the \texttt{powerlaw} (black) and absorbed \texttt{powerlaw} (blue) models. The higher luminosity bin extends up to the highest median luminosity value of the sample, for the absorbed \texttt{powerlaw} model. Results from our work are compared to the stacking analysis fits performed by \cite{Luangtip2015}. Bin centroids are the same, but datapoints have been shifted by $\pm0.04$ for the absorbed  \texttt{powerlaw} model and the \cite{Luangtip2015} dataset, for easier visualization.} \label{fig:gamma_vs_lum}
\end{figure}

The most extensive previous work on ULXs in LIRGs fit a sample of 17 nearby LIRGs \citep{Luangtip2015}. They opted to separate the detected point sources into three luminosity bins: low luminosity (39 $\leq~\rm log\frac{L_{x}}{\rm erg~s^{-1}}~\leq$ 39.33, 25 sources); medium
luminosity (39.33 $\leq~\rm log\frac{L_{x}}{\rm erg~s^{-1}}~\leq$ 39.67, 21 sources) and high luminosity (39.67 $\leq~\rm log\frac{L_{x}}{\rm erg~s^{-1}}~\leq$ 40, 6 sources), in order to study their spectra. That is, instead of individual source fits, they fit a stacked spectrum in each luminosity bin. They obtained luminosity estimates using the web interface version of
the Portable Interactive Multi-Mission Simulator (WEBPIMMS)\footnote{https://heasarc.gsfc.nasa.gov/cgi-bin/Tools/w3pimms/w3pimms.pl}, extrapolating from the detected count-rates, using an absorbed \texttt{powerlaw} model with fixed $\Gamma$~=~$2$ and $N_{\rm H}$~=~1.5$\times 10^{21}$~cm$^{-2}$. These values are only used for the preliminary determination of luminosity, in order to separate sources into three luminosity bins. The spectra in each bin are then stacked, and the resulting three spectra (one per bin) are fit with an obscured power law model. $\Gamma$ and $N_{\rm H}$ are then left free to vary. The photon index of the fit of each luminosity bin is shown in Fig.~\ref{fig:gamma_vs_lum}, and compared to the results of our analysis.

To perform this comparison, we split our sample into the same three luminosity bins, using the median value of the posterior. We consider each observation of a ULX an independent occurrence of a ULX in a LIRG, excluding PS2 and PS4. This approach allows us to improve our statistics, as well as average out the effects of variability. This results in 11, 8, and 15 detections in the low, medium, and high-luminosity bins, respectively, for the \texttt{powerlaw} model; 10, 3 and 21  for the absorbed \texttt{powerlaw} model; and 12, 6, and 16, respectively, for the \texttt{diskpbb} model.

After splitting the sources, we used the hierarchical Bayesian framework implemented in \texttt{posterior stacker}\footnote{https://github.com/JohannesBuchner/PosteriorStacker} \citep{Baronchelli} to combine the individual source posteriors within each bin. Rather than averaging or co-adding the posteriors, this method assumes that the parameters of all sources in a given bin are drawn from an underlying parent distribution, and infers its shape from the data. It then yields an estimate of the parent distribution of the parameter of interest within each luminosity bin, while propagating uncertainties and avoiding biases that would arise from directly stacking or averaging individual posteriors.

As seen in Fig.~\ref{fig:gamma_vs_lum}, we recover similar behavior to \cite{Luangtip2015} for the \texttt{powerlaw} model, although inconsistent within errors in the lowest luminosity bin, where our fits yield lower $\Gamma$. However, a transition into harder spectra is seen, as is common in ULXs at $\gtrsim 3\times10^{39}$~erg~s$^{-1}$, the boundary of the lowest luminosity bin \citep[e.g.,][]{Gladstone2009,Sutton2013}. On the other hand, the absorbed \texttt{powerlaw} model is consistent in the first two luminosity bins; but goes back to a very soft state at the highest luminosity bin. This may simply be an effect of our low-count statistics, in which soft photon indices are easily permitted under unconstrained obscuration. 

Compared to the work of \cite{Luangtip2015}, we have a lower number of detections in the explored range (34 in total), but we have more than double the detections in the high-luminosity bin. This remains true even when considering in-an-out scattering per bin, as the \cite{Luangtip2015} results would be 6$\pm$2, while our model with the lowest number of detections in that bin would contain 15$\pm$3. This could be due to a variety of reasons (i.e., intrinsically higher luminosities, lack of resolution), which are explored in Sec.~\ref{sec:ulx_deficit_origin}.

\subsubsection{Luminosity function}\label{sec:lum_function}

\begin{figure}[h!]
        \centering
        \includegraphics[scale=0.50]{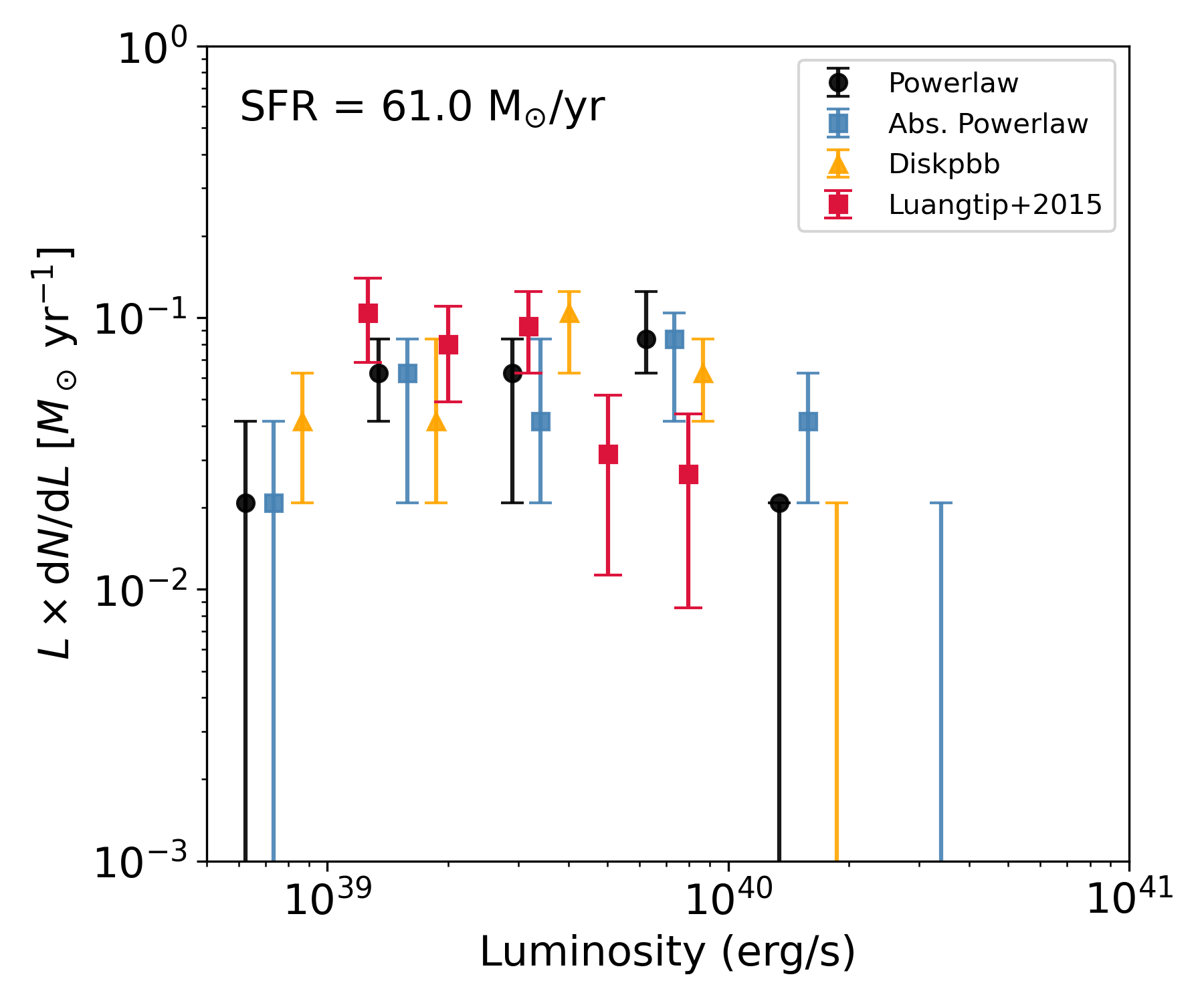}
         \caption{ULX luminosity function for the point sources in NGC 3256, normalized for star-formation rate, for all three models used in this work. {Data points for different models are offset for clarity}. Results are compared to the luminosity function derived by \cite{Luangtip2015}.} \label{fig:lum_function1}
\end{figure}

\begin{figure*}[h!]
        \centering
        \includegraphics[scale=0.50]{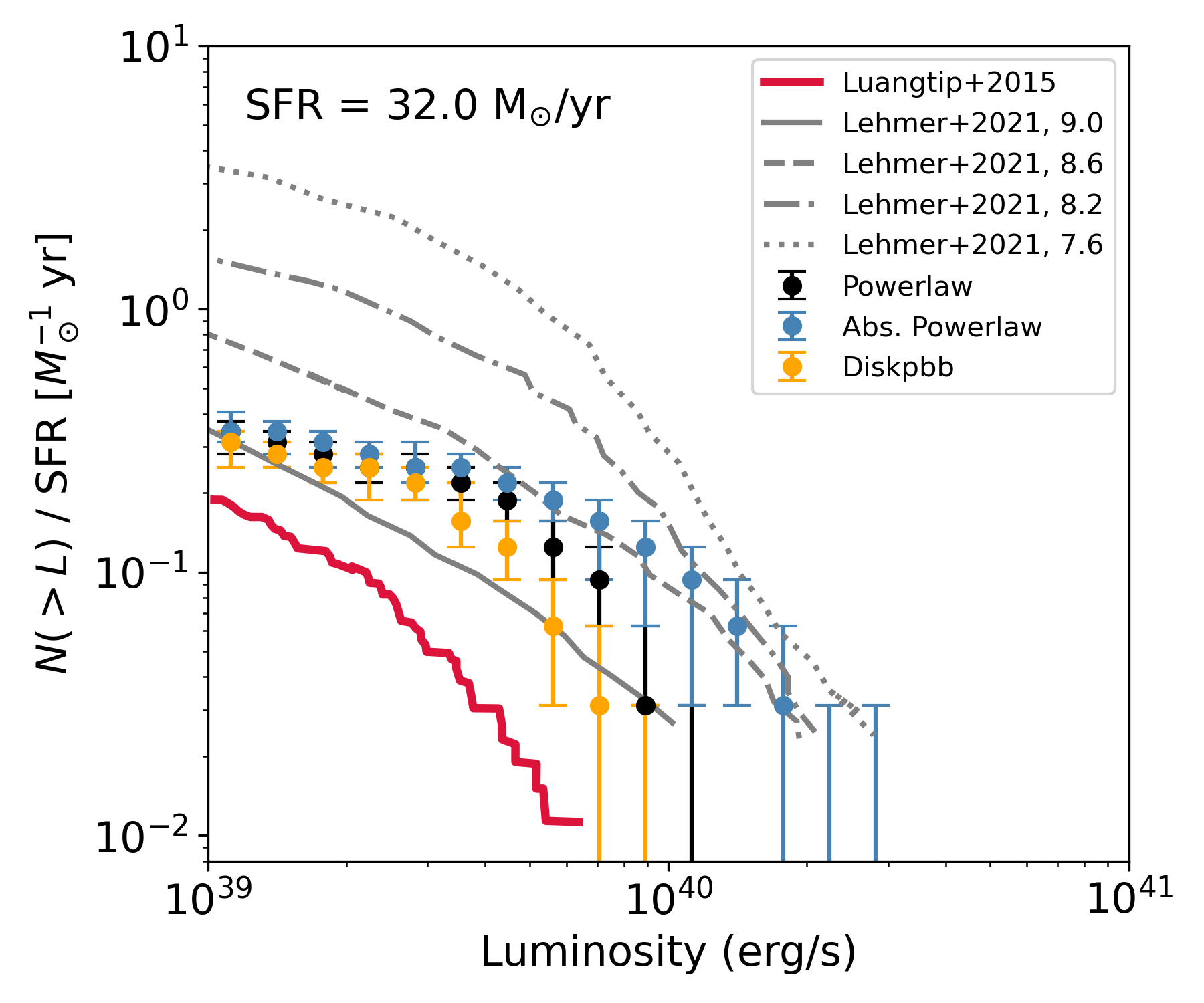}
        \includegraphics[scale=0.50]{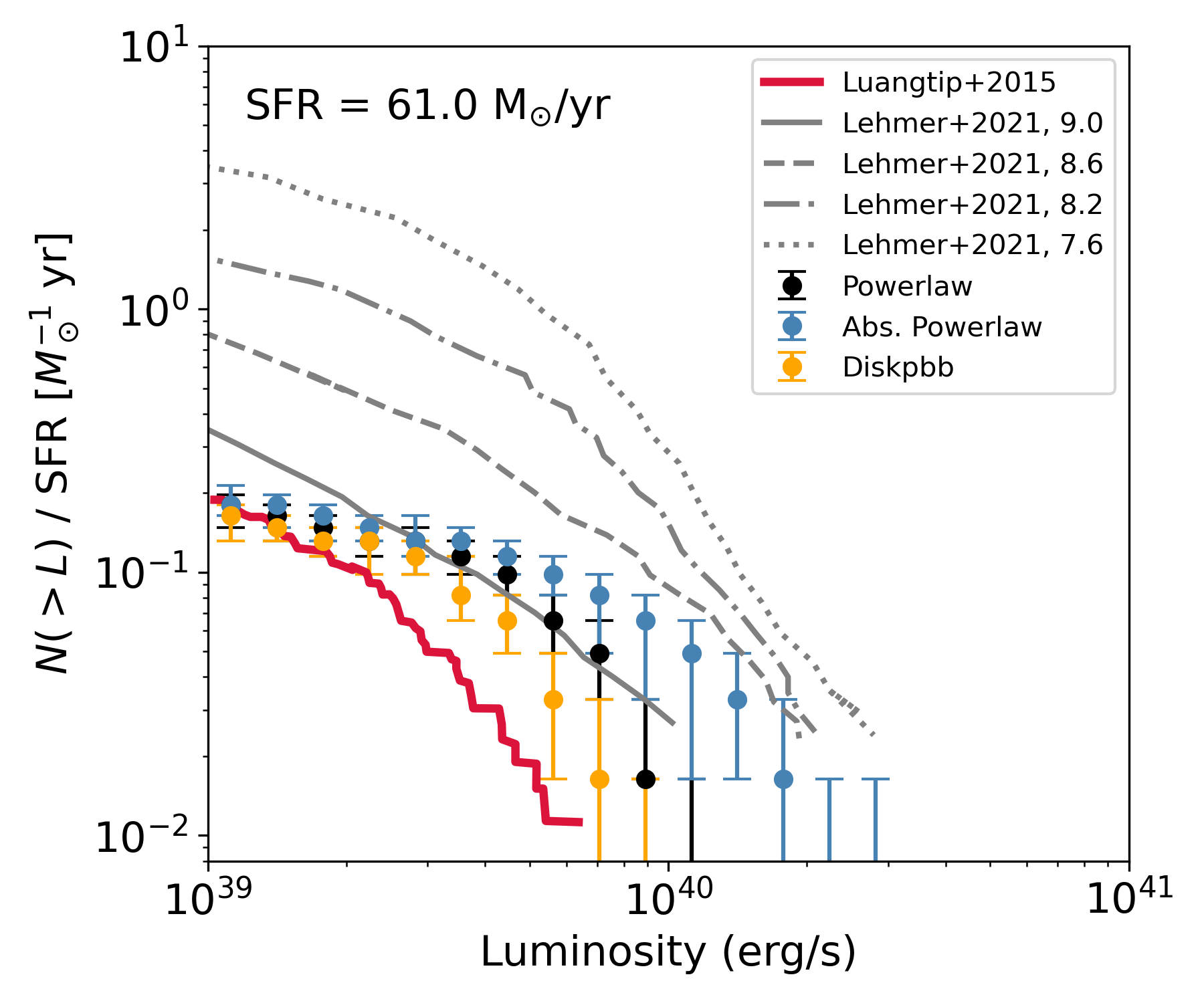}
        \caption{Cumulative luminosity function, normalized by two different SFRs (left, 32~M$_\odot$~yr$^{-1}$, from \cite{Michiyama2020A}; right, 61~M$_\odot$~yr$^{-1}$, from \cite{Diaz-Santos2017}), for all three models used in this work. Results are compared to previous work by \cite{Luangtip2015}, and \cite{Lehmer2021}. For the latter, we plot results from four different metallicity bins, corresponding to: 12 + log(O/H) = 8.7$-$9.2 (solid), 8.6$-$8.7 (dashed), 8.1$-$8.4 (dash-dotted), 7.0$-$8.1 (dotted).} \label{fig:lum_function2}
\end{figure*}

{In order to minimize variability effects, we constructed the luminosity function by selecting one of the three observations at random for every source, and then drawing one luminosity value from that epoch’s posterior distribution. This process was repeated 3000 times, randomly iterating between epochs, to sample all possible states of the ULX. We report the median and 16th–84th percentile range across realizations as our luminosity function and its uncertainty. We exclude PS2 and PS4, which are confirmed foreground stars.}

Figure \ref{fig:lum_function1} shows the luminosity function, normalized by the SFR. We use the highest of the two SFRs discussed in Sec.~\ref{sec:ulx_rate}, and thus showing a lower limit for the luminosity function. All values shown would be higher by a factor of $\sim$2 if we were to take the lowest possible SFR. Results are compared to those of \cite{Luangtip2015} who obtain a compatible value at $\sim10^{39}$~erg~s$^{-1}$. At higher luminosities, however, our luminosity function sits above theirs, particularly for the powerlaw models. This echoes the behavior discussed in Sec.~\ref{sec:photonindex}.

The difference becomes more evident when constructing the cumulative luminosity function (Fig.~\ref{fig:lum_function2}). We compare our results to those of \cite{Luangtip2015}, as well as to the luminosity functions of HMXBs derived by \cite{Lehmer2021}, which are split into different metallicity bins. Our luminosity function is flatter at low luminosities, and has a cutoff at higher luminosities (See Sec.~\ref{sec:ulx_deficit_origin} for a discussion).

We opt to show the cumulative luminosity function for the two values at the extremes of the literature determinations of the star-formation rate of the galaxy (see Sec.~\ref{sec:SFR}). While a SFR~=~61~$M_\odot$~yr$^{-1}$ is more consistent with the results of \cite{Luangtip2015}, that determination includes the nuclei, which are excluded from our analysis due to the standard definition of ULX. The true curve is therefore likely above that shown in the right panel of Figure~\ref{fig:lum_function2}. The two extremes of our luminosity function lay above and below the highest metallicity HMXB luminosity functions obtained by \cite{Lehmer2021}. 

\section{Discussion} \label{sec:discussion}

In this section we discuss the results presented in the previous section. We determine the rate of ULXs NGC 3256 in Sec.~\ref{sec:ulx_rate}, and discuss the origin of the ULX deficit in Sec.~\ref{sec:ulx_deficit_origin}. We also comment on the fraction of transient ULXs detected in the galaxy (Sec.~\ref{sec:transients}).

\subsection{The ULX Ratio in NGC 3256}\label{sec:ulx_rate}

Depending on the model and threshold chosen (see Sect. \ref{sec:results}), we identify between 12$-$15 ULXs in NGC~3256. However, given that PS1 and PS17 are near the outskirts of the galaxy, and we expect up to two background quasars in a field this size (see Sect. \ref{sec:point_source_association}), a more conservative estimate is that we detect at least 10 ULXs in NGC~3256, and up to 15 at most. Regardless of the final count, NGC~3256 is one of the galaxies with the highest number of ULXs detected \citep[exceeded by the Cartwheel galaxy, with 16$-$29 ULXs, depending on the work considered;][]{Gao2003,Wolter2004,Salvaggio2023}. Apart from the Cartwheel, the Antennae galaxy is reported to potentially have 13 ULXs \citep{Clark2005,Zezas2006}, and Arp~299 up to 14 \citep{Anastasopoulou2016}, making them comparable to, and potentially higher than, our estimate for NGC~3256.

Despite that, and even when considering the maximum number of ULXs reported here, it still presents a significant ULX deficit when compared to its star-formation rate \citep[based on the correlation by e.g.,][which predicts a rate of 2 ULXs per unit SFR]{Mapelli2010}. Based on the two estimates presented in Sec.~\ref{sec:SFR}, and our range of potential ULXs, the ULX to SFR ratio of NGC~3256 is in the range of 0.16$-$0.46 ULXs per unit SFR.

The first study of ULX populations in LIRGs, carried out by \cite{Luangtip2015} obtained a rate of 0.2 per unit SFR, far from the 2 reported for lower-SFR galaxies \citep{Mapelli2010}. This factor of $\sim$10 difference was referred to as the ULX deficit in LIRGs. NGC~3256 is in reasonable agreement with the results for other LIRGs (within a factor of $\sim$2), and thus presents the same ``deficit''. 

\subsection{The origin of the ULX deficit}\label{sec:ulx_deficit_origin}

As originally discussed by \cite{Luangtip2015}, there are likely multiple factors contributing to the ULX deficit in LIRGs. We consider obscuration, metallicity, and lack of angular resolution (i.e. source confusion). We also discuss potential biases caused by different estimators of SFR and metallicity.

The age of the starburst may also play a role in any potential deficits. That is, if a significant fraction of the SFR is younger than $\sim10$~Myr \citep{Linden2010}, it would imply a part of the ULX population is likely yet to be formed. However, this cannot be directly explored through the X-ray data presented in this work. In future work, thanks to the availability of good-quality MUSE data, we will determine the star-formation history in different regions of the galaxy (at $\sim 1-2$'' resolution). This hypothesis is worth exploring in more depth, considering that analysis of Spitzer InfraRed Spectrograph \citep[IRS,][]{Houck2004} data found that the starbursts in local LIRGs have ages of 1$-$4.5 Myr, at least in the nuclear regions \citep{Inami2013}. 

\subsubsection{Metallicity}

The efficiency of ULX production is expected to depend on the metallicity of the stellar population. At low metallicity, line-driven stellar winds are weaker, allowing massive stars to retain a larger fraction of their mass and collapse into more massive black holes \citep{Belczynski2010}. Reduced wind-driven mass and angular momentum losses also favor the formation of tighter binaries that are more likely to undergo stable Roche-lobe overflow, increasing both the number and luminosity of HMXBs that can reach the ULX regime \cite{Linden2010}.

\cite{Luangtip2015} attributed the majority of the ULX deficit in LIRGs to obscuration. However, comparing the luminosity function derived in this work to the more recent results of \cite{Lehmer2021} paints a different picture. Indeed, Fig.~\ref{fig:lum_function2}, shows that the ULX luminosity function of NGC~3256 has a similar normalization to that of  normal galaxies when accounting for high metallicity. Previous measurements of the gas phase metallicity in NGC 3256 report 12 + log(O/H) $\sim$ 9 \citep{Rich2012}, which would correspond to the lowest of the \cite{Lehmer2021} luminosity functions shown in Fig. \ref{fig:lum_function2}. Given how the SFR in the right panel of the figure is likely an overestimation (see Sec.~\ref{sec:SFR}), we can explain the low ULX ratio in NGC~2356 solely with metallicity. We thus conclude we do not need to invoke any obscuration effects to explain the number of ULXs found in NGC~3256. 

Having said that, with such a strong dependence on metallicity, as shown by \cite{Lehmer2021}, one may wonder why the ratio of ULXs per unit SFR in ``normal'' galaxies is so high in the work of \cite{Mapelli2010}. In fact, most of their sample turns out to be of quite low metallicity when compared to NGC~3256 (12 + log(O/H)$\sim$7.5). \cite{Mapelli2010} already suggest a potential metallicity dependence, when separating their sample into two metallicity bins, Z$>$0.2 Z$_\odot$ and $\leq$ 0.2 Z$_\odot$. When fitting their relation between number of ULXs and SFR, they find a higher normalization for the lower metallicity bin. However, the tension between the two fits is only $\sim2\sigma$. They conclude there is only marginal evidence for an anticorrelation between metallicity and the number of ULXs in a galaxy. Therefore, their general value for the ULX to SFR ratio remains high, at 2 ULXs per unit SFR.

We also consider whether the comparison shown in Fig.~\ref{fig:lum_function2} could be biased due to different estimators of metallicity. \cite{Rich2012} measured the gas phase metallicity gradient of NGC~3256, using data from the Wide Field Spectrograph (WiFeS), a dual-beam, image-slicing IFU \citep{Dopita2007,Dopita2010}, on the Australian National University (ANU) 2.3m Telescope. They calculate the oxygen abundance in each spaxel using three different strong-line gas-Phase metallicity calibrations: 1) The empirical calibration of \cite{Pettini2004} (PP04); 2) The theoretical method of \cite{Kewley2002} (KD02); 3) The R$_{23}$ method, which uses the [O III]/[O II] line ratio \citep{McGaugh1991,Kewley2002}. 

The results reported in Fig.~12 of \cite{Rich2012} show metallicity values derived using PP04 and KD02, but rescaled to KD02 in both cases. On the other hand, \cite{Lehmer2021} takes values from a variety of literature sources; but rescales them all to PP04. As such, they are, a priori, not comparable. However, Table 2 of \cite{Rich2012} provides the derived metallicity gradient slopes for all methods used. We use the PP04 equation, which also yields a metallicity range of 8.9$-$9.0, for the whole galaxy\footnote{The difference between the best fits of the two methods is of about $\Delta($12+log(O/H)$\leq) 0.03$.}. Therefore, the metallicity comparison performed here is not significantly biased.

Another potential caveat of this comparison, however, is that NGC~3256 is a late-stage merger with very high SFR, in comparison to the nearby galaxies with moderate SFRs of \cite{Lehmer2021}. The merger can result in gas mixing from different parts of the galaxy, and the high SFR could imply very rapid enrichment of gas-phase metallicity. It is nontrivial to assess how this rapid changes may have impacted the gas-phase metallicity of NGC~3256 from the moment the current starburst started until now. That is, a direct comparison may require stellar metallicity measurements, rather than gas-phase metallicity. In future work, we plan to take advantage of the MUSE data to derive stellar metallicities for the younger stellar populations, and test whether they significantly differ from the gas-phase metallicity.

\subsubsection{Obscuration and source confusion}

In the previous section, we claim that the normalization of the ULX luminosity function of NGC~3256 is comparable to that of ``normal'' galaxies of a similar metallicity. However, there is a strong disagreement between the shapes of the luminosity functions. There is an excess of luminous ULXs, particularly when compared to \cite{Luangtip2015}, as well as a flattening of the luminosity function at low luminosities. This means that the low-luminosity ULX population is likely escaping detection.

There are two possible explanations for this effect. First, we consider a lack of angular resolution (i.e., source confusion). That is, due to the very high SFR of NGC~3256, a number of ULXs could be formed in compact star-forming regions, leading us to misclassify multiple fainter sources as a single, brighter one. This would explain both the higher cutoff of the luminosity function, and the flattening at lower luminosities. This is unlikely to occur in the whole sample, given the number of highly variable ULXs observed. However, sources such as PS12 and PS13, which remain remarkably stable, are bright, and are deeply embedded in a star-forming region toward the central regions of the galaxy, could easily consist of multiple ULXs. 

Another possible explanation is a combination of obscuration (i.e., incompleteness at low luminosities) and intrinsically brighter ULXs. All galaxies but one in \cite{Luangtip2015} have SFR~$<$~20~M$_\odot$~yr$^{-1}$; making it perhaps less likely that the brightest end of the luminosity distribution is reached. The high SFR of NGC~3256, which is not included in the \cite{Luangtip2015} sample, could result in us detecting a higher luminosity population. On the other hand, LIRGs have high obscuration in X-rays \citep{Torres-Alba2018,Ricci2017mnras}, which can make detection of fainter sources particularly complicated. NGC~3256, furthermore, has a high background from the shocked-gas emission, further complicating the task. As mentioned in previous sections, the detection limit of an isolated, unobscured source in the stacked \textit{Chandra} observation is log~L$_X\sim38.1$~erg~s$^{-1}$. However, the majority of ULXs in NGC~3256 are expected to fall along regions where SFR is intense, and therefore obscuration and background effects would play a significant role. This would likely increase the detection limit along the galaxy, in a nontrivial way to account for.

Whether obscuration or source confusion are the dominant effect has different implications for the ULX ratio in NGC~3256. If source confusion were to dominate, then the ``true'' luminosity function would sit higher at low luminosities, but lower at high luminosities. In such a scenario, it could potentially become compatible with the results of \cite{Lehmer2021}. However, if obscuration is the dominant effect, the ``true'' luminosity function would sit higher at low luminosities without decreasing at high luminosities. In such a scenario, the luminosity function would likely sit above the results of \cite{Lehmer2021} for galaxies of similar metallicity. This is more apparent when considering the SFR~=~32~M$_\odot$~yr$^{-1}$, which is reasonable given that we are excluding the nuclei from X-ray analysis. In such a scenario, NGC~3256 could have an excess of ULXs compared to normal galaxies of similar metallicity; rather than a deficit. 

\subsubsection{SFR determinations}\label{sec:SFR_deficit}

The SFR is directly related to the normalization of the curves shown in Fig.~\ref{fig:lum_function1} and \ref{fig:lum_function2}, and significantly impacts the determination of the number of ULXs per unit SFR. The work of \cite{Lehmer2021} uses a combination of Spitzer 24 $\mu$m maps and Galaxy Evolution Explorer \citep[GALEX,][]{Martin2005} ultraviolet flux, applying the \cite{Hao2011} relation, as described in \cite{Lehmer2019}. This approach aims to calculate the intrinsic far-ultraviolet (FUV) luminosity of the galaxy by combining the observed FUV emission with the IR luminosity, which accounts for the fraction of FUV radiation absorbed and re-emitted by dust. In this work, we use an estimate by \cite{Diaz-Santos2017}, which uses the 8$-$1000~$\mu$m flux from the Infrared Astronomical Satellite \citep[IRAS,][]{Neugebauer1984}, and a conversion to SFR using the scaling relation of \cite{Murphy2011}. This results in SFR~=~61.7~$\pm$~1.3 ~M$_\odot$~yr$^{-1}$. 

To ensure an optimal comparison, we estimated the SFR using the formalism of \cite{Hao2011}, with the FUV luminosity derived by \cite{Howell2010}, and a 24 $mu$m flux from \cite{Engelbracht2008}. This yields  SFR~$\sim$~62.6~M$_\odot$~yr$^{-1}$.
The agreement between two different methods based on two different IR bands indicates that we are unlikely to be missing a significant fraction of the total SFR. 

This comparison provides a consistency check on the adopted SFR: although the different calibrations rely on different assumptions, they yield nearly identical values for NGC~3256. Therefore, we expect any systematic effect associated with the choice of SFR estimator to be small when comparing our luminosity function with that of \cite{Lehmer2021}. Both of these IR-based estimates, which include the nuclear starburst (which cannot contribute ULXs to our analysis, due to angular resolution constraints) are an overestimation of the non-nuclear SFR. As such, the luminosity function likely sits above that shown in the right panel of Fig~\ref{fig:lum_function2}.

\subsection{Transient ULXs}\label{sec:transients}

A transient ULX is commonly understood to be a source that is detected as a ULX in at least one epoch, and then significantly drops in brightness, below the ULX threshold, in others. There is no firm luminosity threshold to claim a transient ULX. Some works call for a variability of at least two orders of magnitude \citep{Allak2022}, while others have claimed a transient ULX after a variability of about a factor of $\sim10-30$ \citep{vanHaaften2019,Robba2022}. It is generally understood that the ULX must be `missing' in one epoch, but the non-detection depends on the depth of the observation, making this a flexible definition. As most transient ULXs are discovered serendipitously \citep[e.g.,][]{Strickland2001,Soria2012,Middleton2012,Middleton2013,Pintore2018,Earnshaw2019a,Earnshaw2020,Brightman2020,Walton2021}, a standardized methodology for classification is lacking in the literature. The first (and, as far as we know, only) systematic search for transient ULXs using \textit{Swift}-XRT observations, by \cite{Brightman2023}, found five transient ULXs in a sample of five galaxies, each of which dropped in luminosity by 2$-$3 orders of magnitude. 

In Sec.~\ref{sec:results} we classify PS6, PS11, PS15 and PS16 as transient ULX candidates based on them being compatible with a non-detection and the percent of the posterior above the ULX limit dropping considerably, to between 0$-$10\%. However, it is worth noting that their luminosity upper limits remain high, in all cases above 10$^{38}$~erg~s$^{-1}$, and in some even above the ULX threshold. While this is partly due to a lack of depth, the combination of flexible models and very high background emission (caused by the intense soft emission of the galaxy) is the main cause of the very high upper limits. As such, we cannot firmly claim any of these sources as transient ULXs, rather they are being thought of as transient ULX candidates.

If these four sources are indeed transient ULXs, rather than just variable, the transient ULX fraction observed in NGC~3256 over the period of the observations would be between 0.26$-$0.4 (4/15 to 4/10). The observed transient ULX fraction reported by \cite{Brightman2023} is 0.42 (5/12)\footnote{This is the ratio between the number of transient ULXs detected during the time of the observations, over the total number, which includes persistent ULXs detected over the same period of time. As noted by \cite{Brightman2023}, the rate of transient ULXs to persistent ULXs in galaxies is much higher, since one must account for the total number of transient ULXs that appear and disappear over the lifetime of a persistent ULX.}. As such, the incidence of transient ULXs in LIRGs is potentially similar to that of normal star-forming galaxies. 

A proper comparison, however, requires deeper observations which would ensure the detection threshold lowers enough to discard relatively small variability \citep[below a factor of $\sim$10, often seen also in persistent ULXs;][]{Earnshaw2025} as the cause of non-detections. Despite this, NGC~3256, with four transient ULX candidates over the course of only three observations, shows how LIRGs are ideal laboratories to search for and understand the nature of transient ULXs. Although their rate of transients does not appear to be higher than that of normal galaxies, their high SFRs (and thus ULX incidence) make them ideal targets to efficiently explore this issue. 

\section{Conclusions} \label{sec:conclusions}

In this work, we have analyzed three \chandra observations of the nearby LIRG NGC~3256, a major merger with a SFR$\sim$50-60~M$_\odot$~yr$^{-1}$. We detect a total of 17 X-ray point sources that overlap with the optical extension of the galaxy, two of which we determine to be foreground stars. We also estimate that about two of the remaining ULXs could be background quasars, likely PS1 and PS17, which have the highest distance to the center of the galaxy. We fit the X-ray spectra of each source, for all three epochs, using three different models: a \texttt{powerlaw}, an absorbed \texttt{powerlaw}, and a \texttt{diskpbb} model. We conclude:
\begin{itemize}
    \item At least 10 point sources in the galaxy, and at most 15 (depending on model and criteria used), are classified as ULXs. This makes NGC 3256 one of the most ULX-rich galaxies known.
    \item The ULX-SFR relation for NGC~3256 is in the range of 0.16$-$0.46 per unit SFR. This is in reasonable agreement (factor of ~2) with the results reported for other local LIRGs (0.2 per unit SFR), and far from the 2 per unit SFR reported for normal galaxies. However, we find that when comparing our luminosity function to normal galaxies in the same range of (high) metallicity, our results become consistent with recent literature. As such, we do not need to invoke any obscuration effects to explain the ULX deficit in NGC~3256.
    \item While the rate of ULXs per unit SFR may be fully explained by the effects of metallicity, the shape of the luminosity functions disagrees significantly. The ULX luminosity function of NGC~3256 presents an excess of sources at high luminosities, and a flattening at lower luminosities. This could be an effect of obscuration, where we miss the fainter population because it is easier to suppress. It could also be due to a lack of angular resolution (i.e. source confusion) in our observations. If the former is true, LIRGs could be more efficient at forming ULXs  than normal galaxies (at similar metallicity), the opposite to what had been found in previous studies. 
    \item Four sources are found to be transient ULX candidates, which are undetected in at least one observation. However, the observations are not deep enough to determine firm upper limits at least 2 orders of magnitude below the ULX threshold, making it possible that they are simply variable. If they are transient, the transient-to-persistent ULX ratio found in NGC~3256 is consistent with systematic searches in normal galaxies.
\end{itemize}

Overall, we find evidence that obscuration is not the only relevant factor in determining the low ratio of ULXs per unit SFR in LIRGs, and that the high gas-phase metallicity typical of these galaxies likely plays an even larger role. In future work, we shall use MUSE and HST data to explore potential additional factors, such as the age of the starburst. We shall also determine stellar, rather than gas-phase metallicities, for a more accurate assessment of metallicity at the moment the starburst started. 

\begin{acknowledgements}
{We thank the referee for their helpful comments and suggestions, which helped improve this work. We also} thank B. Lehmer for fruitful discussions on the comparison between SFR determinations. N.T.A. acknowledges funding under NASA contract 80NSSC26K0289. C.M.L. acknowledges the research project was supported by the Hellenic Foundation for Research and Innovation (HFRI) under the "2nd Call for HFRI Research Projects to support Faculty Members $\&$ Researchers" (Project Number: 03382). The work of DS was carried out at the Jet Propulsion Laboratory, California Institute of Technology, under a contract with the National Aeronautics and Space Administration (80NM0018D0004). CR acknowledges support from SNSF Consolidator grant F01-13252 and ANID BASAL project FB210003. DJW acknowledges support from the Science and Technology Facilities Council (STFC; grant code ST/Y001060/1). HPE acknowledges funding under NASA contract NNG08FD60C. ET acknowledges support from the ANID CATA-BASAL program FB210003, and FONDECYT Regular 1241005 and  1250821. DJW acknowledges support from the Science and Technology Facilities Council (STFC; grant code ST/Y001060/1).
This research has made use of data obtained from the \chandra Source Catalog, provided by the \chandra X-ray Center (CXC). This paper employs a list of Chandra datasets, obtained by the Chandra X-ray Observatory, contained in the Chandra Data Collection (CDC) 741~\href{https://doi.org/10.25574/cdc.741} {doi:10.25574/cdc.741}. This research is based on observations made with the NASA/ESA \textit{Hubble Space Telescope} obtained from the Space Telescope Science Institute, which is operated by the Association of Universities for Research in Astronomy, Inc., under NASA contract NAS 5–26555. Based on observations collected at the European Southern Observatory under ESO programme 0106.A-0284 (PI HUSEMANN).

\end{acknowledgements}

\bibliographystyle{aa} % style aa.bst
\bibliography{literature.bib}

@ARTICLE{Pereira-Santaella2015,
       author = {{Pereira-Santaella}, Miguel and {Alonso-Herrero}, Almudena and {Colina}, Luis and {Miralles-Caballero}, Daniel and {P{\'e}rez-Gonz{\'a}lez}, Pablo G. and {Arribas}, Santiago and {Bellocchi}, Enrica and {Cazzoli}, Sara and {D{\'\i}az-Santos}, Tanio and {Piqueras L{\'o}pez}, Javier},
        title = "{Star-formation histories of local luminous infrared galaxies}",
      journal = {\aap},
         year = 2015,
        month = may,
       volume = {577},
          eid = {A78},
        pages = {A78},
          doi = {10.1051/0004-6361/201425359},
archivePrefix = {arXiv},
       eprint = {1502.07965},
 primaryClass = {astro-ph.GA},
       adsurl = {https://ui.adsabs.harvard.edu/abs/2015A&A...577A..78P}
}

@ARTICLE{Cortijo-Ferrero2017,
       author = {{Cortijo-Ferrero}, C. and {Gonz{\'a}lez Delgado}, R.~M. and {P{\'e}rez}, E. and {Cid Fernandes}, R. and {Garc{\'\i}a-Benito}, R. and {Di Matteo}, P. and {S{\'a}nchez}, S.~F. and {de Amorim}, A.~L. and {Lacerda}, E.~A.~D. and {L{\'o}pez Fern{\'a}ndez}, R. and {Tadhunter}, C.},
        title = "{The spatially resolved star formation history of mergers. A comparative study of the LIRGs IC 1623, NGC 6090, NGC 2623, and Mice}",
      journal = {\aap},
         year = 2017,
        month = nov,
       volume = {607},
          eid = {A70},
        pages = {A70},
          doi = {10.1051/0004-6361/201731217},
archivePrefix = {arXiv},
       eprint = {1707.05324},
 primaryClass = {astro-ph.GA},
       adsurl = {https://ui.adsabs.harvard.edu/abs/2017A&A...607A..70C}
}

@ARTICLE{Anastasopoulou2016,
       author = {{Anastasopoulou}, K. and {Zezas}, A. and {Ballo}, L. and {Della Ceca}, R.},
        title = "{A deep Chandra observation of the interacting star-forming galaxy Arp 299}",
      journal = {\mnras},
         year = 2016,
        month = aug,
       volume = {460},
       number = {4},
        pages = {3570-3586},
          doi = {10.1093/mnras/stw1200},
archivePrefix = {arXiv},
       eprint = {1605.07001},
 primaryClass = {astro-ph.GA},
       adsurl = {https://ui.adsabs.harvard.edu/abs/2016MNRAS.460.3570A}
}

@ARTICLE{Clark2005,
       author = {{Clark}, D.~M. and {Christopher}, M.~H. and {Eikenberry}, S.~S. and {Brandl}, B.~R. and {Wilson}, J.~C. and {Carson}, J.~C. and {Henderson}, C.~P. and {Hayward}, T.~L. and {Barry}, D.~J. and {Ptak}, A.~F. and {Colbert}, E.~J.~M.},
        title = "{The Ultraluminous X-Ray Source X-37 Is a Background Quasar in the Antennae Galaxies}",
      journal = {\apjl},
         year = 2005,
        month = oct,
       volume = {631},
       number = {2},
        pages = {L109-L112},
          doi = {10.1086/497333},
archivePrefix = {arXiv},
       eprint = {astro-ph/0509143},
 primaryClass = {astro-ph},
       adsurl = {https://ui.adsabs.harvard.edu/abs/2005ApJ...631L.109C}
}

@ARTICLE{Zezas2006,
       author = {{Zezas}, A. and {Fabbiano}, G. and {Baldi}, A. and {Schweizer}, F. and {King}, A.~R. and {Ponman}, T.~J. and {Rots}, A.~H.},
        title = "{Chandra Monitoring Observations of The Antennae Galaxies. I. Catalog of Source Properties}",
      journal = {\apjs},
         year = 2006,
        month = sep,
       volume = {166},
       number = {1},
        pages = {211-248},
          doi = {10.1086/501526},
       adsurl = {https://ui.adsabs.harvard.edu/abs/2006ApJS..166..211Z}
}

@ARTICLE{Buchner2014,
       author = {{Buchner}, J. and {Georgakakis}, A. and {Nandra}, K. and {Hsu}, L. and {Rangel}, C. and {Brightman}, M. and {Merloni}, A. and {Salvato}, M. and {Donley}, J. and {Kocevski}, D.},
        title = "{X-ray spectral modelling of the AGN obscuring region in the CDFS: Bayesian model selection and catalogue}",
      journal = {\aap},
         year = 2014,
        month = apr,
       volume = {564},
          eid = {A125},
        pages = {A125},
          doi = {10.1051/0004-6361/201322971},
archivePrefix = {arXiv},
       eprint = {1402.0004},
 primaryClass = {astro-ph.HE},
       adsurl = {https://ui.adsabs.harvard.edu/abs/2014A&A...564A.125B}
}

@book{Jeffreys196,
  title     = {Theory of Probability},
  author    = {Jeffreys, Harold},
  edition   = {3rd},
  year      = {1961},
  publisher = {Oxford University Press},
  address   = {Oxford, United Kingdom},
  isbn      = {978-0198531937}
}

@ARTICLE{Engelbracht2008,
       author = {{Engelbracht}, C.~W. and {Rieke}, G.~H. and {Gordon}, K.~D. and {Smith}, J.-D.~T. and {Werner}, M.~W. and {Moustakas}, J. and {Willmer}, C.~N.~A. and {Vanzi}, L.},
        title = "{Metallicity Effects on Dust Properties in Starbursting Galaxies}",
      journal = {\apj},
         year = 2008,
        month = may,
       volume = {678},
       number = {2},
        pages = {804-827},
          doi = {10.1086/529513},
archivePrefix = {arXiv},
       eprint = {0801.1700},
 primaryClass = {astro-ph},
       adsurl = {https://ui.adsabs.harvard.edu/abs/2008ApJ...678..804E}
}

@ARTICLE{Howell2010,
       author = {{Howell}, Justin H. and {Armus}, Lee and {Mazzarella}, Joseph M. and {Evans}, Aaron S. and {Surace}, Jason A. and {Sanders}, David B. and {Petric}, Andreea and {Appleton}, Phil and {Bothun}, Greg and {Bridge}, Carrie and {Chan}, Ben H.~P. and {Charmandaris}, Vassilis and {Frayer}, David T. and {Haan}, Sebastian and {Inami}, Hanae and {Kim}, Dong-Chan and {Lord}, Steven and {Madore}, Barry F. and {Melbourne}, Jason and {Schulz}, Bernhard and {U}, Vivian and {Vavilkin}, Tatjana and {Veilleux}, Sylvain and {Xu}, Kevin},
        title = "{The Great Observatories All-sky LIRG Survey: Comparison of Ultraviolet and Far-infrared Properties}",
      journal = {\apj},
         year = 2010,
        month = may,
       volume = {715},
       number = {1},
        pages = {572-588},
          doi = {10.1088/0004-637X/715/1/572},
archivePrefix = {arXiv},
       eprint = {1004.0985},
 primaryClass = {astro-ph.GA},
       adsurl = {https://ui.adsabs.harvard.edu/abs/2010ApJ...715..572H}
}

@ARTICLE{Murphy2011,
       author = {{Murphy}, E.~J. and {Condon}, J.~J. and {Schinnerer}, E. and {Kennicutt}, R.~C. and {Calzetti}, D. and {Armus}, L. and {Helou}, G. and {Turner}, J.~L. and {Aniano}, G. and {Beir{\~a}o}, P. and {Bolatto}, A.~D. and {Brandl}, B.~R. and {Croxall}, K.~V. and {Dale}, D.~A. and {Donovan Meyer}, J.~L. and {Draine}, B.~T. and {Engelbracht}, C. and {Hunt}, L.~K. and {Hao}, C.-N. and {Koda}, J. and {Roussel}, H. and {Skibba}, R. and {Smith}, J.-D.~T.},
        title = "{Calibrating Extinction-free Star Formation Rate Diagnostics with 33 GHz Free-free Emission in NGC 6946}",
      journal = {\apj},
         year = 2011,
        month = aug,
       volume = {737},
       number = {2},
          eid = {67},
        pages = {67},
          doi = {10.1088/0004-637X/737/2/67},
archivePrefix = {arXiv},
       eprint = {1105.4877},
 primaryClass = {astro-ph.CO},
       adsurl = {https://ui.adsabs.harvard.edu/abs/2011ApJ...737...67M}
}

@ARTICLE{Neugebauer1984,
       author = {{Neugebauer}, G. and {Habing}, H.~J. and {van Duinen}, R. and {Aumann}, H.~H. and {Baud}, B. and {Beichman}, C.~A. and {Beintema}, D.~A. and {Boggess}, N. and {Clegg}, P.~E. and {de Jong}, T. and {Emerson}, J.~P. and {Gautier}, T.~N. and {Gillett}, F.~C. and {Harris}, S. and {Hauser}, M.~G. and {Houck}, J.~R. and {Jennings}, R.~E. and {Low}, F.~J. and {Marsden}, P.~L. and {Miley}, G. and {Olnon}, F.~M. and {Pottasch}, S.~R. and {Raimond}, E. and {Rowan-Robinson}, M. and {Soifer}, B.~T. and {Walker}, R.~G. and {Wesselius}, P.~R. and {Young}, E.},
        title = "{The Infrared Astronomical Satellite (IRAS) Mission.}",
      journal = {\apjl},
         year = 1984,
        month = mar,
       volume = {278},
        pages = {L1-L6},
          doi = {10.1086/184209},
       adsurl = {https://ui.adsabs.harvard.edu/abs/1984ApJ...278L...1N}
}

@ARTICLE{Kalberla2005,
       author = {{Kalberla}, P.~M.~W. and {Burton}, W.~B. and {Hartmann}, Dap and
         {Arnal}, E.~M. and {Bajaja}, E. and {Morras}, R. and
         {P{\"o}ppel}, W.~G.~L.},
        title = "{The Leiden/Argentine/Bonn (LAB) Survey of Galactic HI. Final data release of the combined LDS and IAR surveys with improved stray-radiation corrections}",
      journal = {\aap},
         year = 2005,
        month = sep,
       volume = {440},
       number = {2},
        pages = {775-782},
          doi = {10.1051/0004-6361:20041864},
archivePrefix = {arXiv},
       eprint = {astro-ph/0504140},
 primaryClass = {astro-ph},
       adsurl = {https://ui.adsabs.harvard.edu/abs/2005A&A...440..775K}
}

@ARTICLE{Torres-Alba2018,
       author = {{Torres-Alb{\`a}}, N. and {Iwasawa}, K. and {D{\'\i}az-Santos}, T. and
         {Charmandaris}, V. and {Ricci}, C. and {Chu}, J.~K. and {Sand
        ers}, D.~B. and {Armus}, L. and {Barcos-Mu{\~n}oz}, L. and
         {Evans}, A.~S. and {Howell}, J.~H. and {Inami}, H. and {Linden}, S.~T. and
         {Medling}, A.~M. and {Privon}, G.~C. and {U}, V. and {Yoon}, I.},
        title = "{C-GOALS. II. Chandra observations of the lower luminosity sample of nearby luminous infrared galaxies in GOALS}",
      journal = {\aap},
         year = 2018,
        month = dec,
       volume = {620},
          eid = {A140},
        pages = {A140},
          doi = {10.1051/0004-6361/201834105},
archivePrefix = {arXiv},
       eprint = {1810.02371},
 primaryClass = {astro-ph.GA},
       adsurl = {https://ui.adsabs.harvard.edu/abs/2018A&A...620A.140T}
}

@ARTICLE{Ricci2017mnras,
       author = {{Ricci}, C. and {Bauer}, F.~E. and {Treister}, E. and {Schawinski}, K. and {Privon}, G.~C. and {Blecha}, L. and {Arevalo}, P. and {Armus}, L. and {Harrison}, F. and {Ho}, L.~C. and {Iwasawa}, K. and {Sanders}, D.~B. and {Stern}, D.},
        title = "{Growing supermassive black holes in the late stages of galaxy mergers are heavily obscured}",
      journal = {\mnras},
         year = 2017,
        month = jun,
       volume = {468},
       number = {2},
        pages = {1273-1299},
          doi = {10.1093/mnras/stx173},
archivePrefix = {arXiv},
       eprint = {1701.04825},
 primaryClass = {astro-ph.HE},
       adsurl = {https://ui.adsabs.harvard.edu/abs/2017MNRAS.468.1273R}
}

@ARTICLE{Ueda2014,
       author = {{Ueda}, Yoshihiro and {Akiyama}, Masayuki and {Hasinger}, G{\"u}nther and {Miyaji}, Takamitsu and {Watson}, Michael G.},
        title = "{Toward the Standard Population Synthesis Model of the X-Ray Background: Evolution of X-Ray Luminosity and Absorption Functions of Active Galactic Nuclei Including Compton-thick Populations}",
      journal = {\apj},
         year = 2014,
        month = may,
       volume = {786},
       number = {2},
          eid = {104},
        pages = {104},
          doi = {10.1088/0004-637X/786/2/104},
archivePrefix = {arXiv},
       eprint = {1402.1836},
 primaryClass = {astro-ph.CO},
       adsurl = {https://ui.adsabs.harvard.edu/abs/2014ApJ...786..104U}
}

@ARTICLE{Sutton2013,
       author = {{Sutton}, Andrew D. and {Roberts}, Timothy P. and {Middleton}, Matthew J.},
        title = "{The ultraluminous state revisited: fractional variability and spectral shape as diagnostics of super-Eddington accretion}",
      journal = {\mnras},
         year = 2013,
        month = oct,
       volume = {435},
       number = {2},
        pages = {1758-1775},
          doi = {10.1093/mnras/stt1419},
archivePrefix = {arXiv},
       eprint = {1307.8044},
 primaryClass = {astro-ph.HE},
       adsurl = {https://ui.adsabs.harvard.edu/abs/2013MNRAS.435.1758S}
}

@ARTICLE{Gladstone2009,
       author = {{Gladstone}, Jeanette C. and {Roberts}, Timothy P. and {Done}, Chris},
        title = "{The ultraluminous state}",
      journal = {\mnras},
         year = 2009,
        month = aug,
       volume = {397},
       number = {4},
        pages = {1836-1851},
          doi = {10.1111/j.1365-2966.2009.15123.x},
archivePrefix = {arXiv},
       eprint = {0905.4076},
 primaryClass = {astro-ph.CO},
       adsurl = {https://ui.adsabs.harvard.edu/abs/2009MNRAS.397.1836G}
}

@ARTICLE{Baronchelli,
       author = {{Baronchelli}, L. and {Nandra}, K. and {Buchner}, J.},
        title = "{Relativistic accretion disc reflection in AGN X-ray spectra at z = 0.5-4: a study of four Chandra Deep Fields}",
      journal = {\mnras},
         year = 2020,
        month = nov,
       volume = {498},
       number = {4},
        pages = {5284-5298},
          doi = {10.1093/mnras/staa2684},
archivePrefix = {arXiv},
       eprint = {2008.11761},
 primaryClass = {astro-ph.HE},
       adsurl = {https://ui.adsabs.harvard.edu/abs/2020MNRAS.498.5284B}
}

@ARTICLE{Allak2022,
       author = {{Allak}, S. and {Akyuz}, A. and {Akkaya Oralhan}, {\.I}. and {Avdan}, S. and {Aksaker}, N. and {Vinokurov}, A. and {Soydugan}, F. and {Sonbas}, E. and {Dhuga}, K.~S.},
        title = "{The transient ultraluminous X-ray source, ULX-4, in M51}",
      journal = {\mnras},
         year = 2022,
        month = mar,
       volume = {510},
       number = {3},
        pages = {4355-4369},
          doi = {10.1093/mnras/stab3693},
archivePrefix = {arXiv},
       eprint = {2106.14322},
 primaryClass = {astro-ph.GA},
       adsurl = {https://ui.adsabs.harvard.edu/abs/2022MNRAS.510.4355A}
}

@ARTICLE{Earnshaw2019a,
       author = {{Earnshaw}, Hannah P. and {Grefenstette}, Brian W. and {Brightman}, Murray and {Walton}, Dominic J. and {Barret}, Didier and {F{\"u}rst}, Felix and {Harrison}, Fiona A. and {Heida}, Marianne and {Pike}, Sean N. and {Stern}, Daniel and {Webb}, Natalie A.},
        title = "{A Broadband Look at the Old and New ULXs of NGC 6946}",
      journal = {\apj},
         year = 2019,
        month = aug,
       volume = {881},
       number = {1},
          eid = {38},
        pages = {38},
          doi = {10.3847/1538-4357/ab20cd},
archivePrefix = {arXiv},
       eprint = {1905.03383},
 primaryClass = {astro-ph.HE},
       adsurl = {https://ui.adsabs.harvard.edu/abs/2019ApJ...881...38E}
}

@ARTICLE{Pintore2018,
       author = {{Pintore}, F. and {Belfiore}, A. and {Novara}, G. and {Salvaterra}, R. and {Marelli}, M. and {De Luca}, A. and {Rigoselli}, M. and {Israel}, G. and {Rodriguez}, G. and {Mereghetti}, S. and {Wolter}, A. and {Walton}, D.~J. and {Fuerst}, F. and {Ambrosi}, E. and {Zampieri}, L. and {Tiengo}, A. and {Salvaggio}, C.},
        title = "{A new ultraluminous X-ray source in the galaxy NGC 5907}",
      journal = {\mnras},
         year = 2018,
        month = jun,
       volume = {477},
       number = {1},
        pages = {L90-L95},
          doi = {10.1093/mnrasl/sly048},
archivePrefix = {arXiv},
       eprint = {1806.03633},
 primaryClass = {astro-ph.HE},
       adsurl = {https://ui.adsabs.harvard.edu/abs/2018MNRAS.477L..90P}
}

@ARTICLE{Middleton2012,
       author = {{Middleton}, Matthew J. and {Sutton}, Andrew D. and {Roberts}, Timothy P. and {Jackson}, Floyd E. and {Done}, Chris},
        title = "{The missing link: a low-mass X-ray binary in M31 seen as an ultraluminous X-ray source}",
      journal = {\mnras},
         year = 2012,
        month = mar,
       volume = {420},
       number = {4},
        pages = {2969-2977},
          doi = {10.1111/j.1365-2966.2011.20145.x},
archivePrefix = {arXiv},
       eprint = {1111.1188},
 primaryClass = {astro-ph.HE},
       adsurl = {https://ui.adsabs.harvard.edu/abs/2012MNRAS.420.2969M}
}

@ARTICLE{Soria2012,
       author = {{Soria}, Roberto and {Kuntz}, K.~D. and {Winkler}, P. Frank and {Blair}, William P. and {Long}, Knox S. and {Plucinsky}, Paul P. and {Whitmore}, Bradley C.},
        title = "{The Birth of an Ultraluminous X-Ray Source in M83}",
      journal = {\apj},
         year = 2012,
        month = may,
       volume = {750},
       number = {2},
          eid = {152},
        pages = {152},
          doi = {10.1088/0004-637X/750/2/152},
archivePrefix = {arXiv},
       eprint = {1203.2335},
 primaryClass = {astro-ph.HE},
       adsurl = {https://ui.adsabs.harvard.edu/abs/2012ApJ...750..152S}
}

@ARTICLE{Strickland2001,
       author = {{Strickland}, David K. and {Colbert}, Edward J.~M. and {Heckman}, Timothy M. and {Weaver}, Kimberly A. and {Dahlem}, Michael and {Stevens}, Ian R.},
        title = "{Another Intermediate-Mass Black Hole in a Starburst Galaxy? The Luminous X-Ray Source in NGC 3628 Reappears}",
      journal = {\apj},
         year = 2001,
        month = oct,
       volume = {560},
       number = {2},
        pages = {707-714},
          doi = {10.1086/323055},
archivePrefix = {arXiv},
       eprint = {astro-ph/0107115},
 primaryClass = {astro-ph},
       adsurl = {https://ui.adsabs.harvard.edu/abs/2001ApJ...560..707S}
}

@ARTICLE{Bohn2024,
       author = {{Bohn}, Thomas and {Inami}, Hanae and {Togi}, Aditya and {Armus}, Lee and {Lai}, Thomas S.-Y. and {Barcos-Munoz}, Loreto and {Song}, Yiqing and {Linden}, S.~T. and {Surace}, Jason and {Bianchin}, Marina and {U}, Vivian and {Evans}, Aaron S. and {B{\"o}ker}, Torsten and {Malkan}, Matthew A. and {Larson}, Kirsten L. and {Stierwalt}, Sabrina and {Buiten}, Victorine A. and {Charmandaris}, Vassilis and {Diaz-Santos}, Tanio and {Howell}, Justin H. and {Privon}, George C. and {Ricci}, Claudio and {van der Werf}, Paul P. and {Aalto}, Susanne and {Hayward}, Christopher C. and {Kader}, Justin A. and {Mazzarella}, Joseph M. and {Muller-Sanchez}, Francisco and {Sanders}, David B.},
        title = "{GOALS-JWST: The Warm Molecular Outflows of the Merging Starburst Galaxy NGC 3256}",
      journal = {\apj},
         year = 2024,
        month = dec,
       volume = {977},
       number = {1},
          eid = {36},
        pages = {36},
          doi = {10.3847/1538-4357/ad87d3},
archivePrefix = {arXiv},
       eprint = {2403.14751},
 primaryClass = {astro-ph.GA},
       adsurl = {https://ui.adsabs.harvard.edu/abs/2024ApJ...977...36B}
}

@ARTICLE{Dauser2017,
       author = {{Dauser}, T. and {Middleton}, M. and {Wilms}, J.},
        title = "{Modelling the light curves of ultraluminous X-ray sources as precession}",
      journal = {\mnras},
         year = 2017,
        month = apr,
       volume = {466},
       number = {2},
        pages = {2236-2241},
          doi = {10.1093/mnras/stw3304},
archivePrefix = {arXiv},
       eprint = {1612.05035},
 primaryClass = {astro-ph.HE},
       adsurl = {https://ui.adsabs.harvard.edu/abs/2017MNRAS.466.2236D}
}

@ARTICLE{King2009,
       author = {{King}, A.~R.},
        title = "{Masses, beaming and Eddington ratios in ultraluminous X-ray sources}",
      journal = {\mnras},
         year = 2009,
        month = feb,
       volume = {393},
       number = {1},
        pages = {L41-L44},
          doi = {10.1111/j.1745-3933.2008.00594.x},
archivePrefix = {arXiv},
       eprint = {0811.1473},
 primaryClass = {astro-ph},
       adsurl = {https://ui.adsabs.harvard.edu/abs/2009MNRAS.393L..41K}
}

@ARTICLE{Israel2017,
       author = {{Israel}, Gian Luca and {Belfiore}, Andrea and {Stella}, Luigi and {Esposito}, Paolo and {Casella}, Piergiorgio and {De Luca}, Andrea and {Marelli}, Martino and {Papitto}, Alessandro and {Perri}, Matteo and {Puccetti}, Simonetta and {Castillo}, Guillermo A. Rodr{\'\i}guez and {Salvetti}, David and {Tiengo}, Andrea and {Zampieri}, Luca and {D'Agostino}, Daniele and {Greiner}, Jochen and {Haberl}, Frank and {Novara}, Giovanni and {Salvaterra}, Ruben and {Turolla}, Roberto and {Watson}, Mike and {Wilms}, Joern and {Wolter}, Anna},
        title = "{An accreting pulsar with extreme properties drives an ultraluminous x-ray source in NGC 5907}",
      journal = {Science},
         year = 2017,
        month = feb,
       volume = {355},
       number = {6327},
        pages = {817-819},
          doi = {10.1126/science.aai8635},
archivePrefix = {arXiv},
       eprint = {1609.07375},
 primaryClass = {astro-ph.HE},
       adsurl = {https://ui.adsabs.harvard.edu/abs/2017Sci...355..817I}
}

@ARTICLE{Sutton2012,
       author = {{Sutton}, Andrew D. and {Roberts}, Timothy P. and {Walton}, Dominic J. and {Gladstone}, Jeanette C. and {Scott}, Amy E.},
        title = "{The most extreme ultraluminous X-ray sources: evidence for intermediate-mass black holes?}",
      journal = {\mnras},
         year = 2012,
        month = jun,
       volume = {423},
       number = {2},
        pages = {1154-1177},
          doi = {10.1111/j.1365-2966.2012.20944.x},
archivePrefix = {arXiv},
       eprint = {1203.4100},
 primaryClass = {astro-ph.HE},
       adsurl = {https://ui.adsabs.harvard.edu/abs/2012MNRAS.423.1154S}
}

@ARTICLE{Farrell2009,
       author = {{Farrell}, Sean A. and {Webb}, Natalie A. and {Barret}, Didier and {Godet}, Olivier and {Rodrigues}, Joana M.},
        title = "{An intermediate-mass black hole of over 500 solar masses in the galaxy ESO243-49}",
      journal = {\nat},
         year = 2009,
        month = jul,
       volume = {460},
       number = {7251},
        pages = {73-75},
          doi = {10.1038/nature08083},
archivePrefix = {arXiv},
       eprint = {1001.0567},
 primaryClass = {astro-ph.HE},
       adsurl = {https://ui.adsabs.harvard.edu/abs/2009Natur.460...73F}
}

@ARTICLE{West2023,
       author = {{West}, Lacey and {Garofali}, Kristen and {Lehmer}, Bret D. and {Prestwich}, Andrea and {Eufrasio}, Rafael and {Luangtip}, Wasutep and {Roberts}, Timothy P. and {Zezas}, Andreas},
        title = "{The Large Deficit of HMXB Emission from Luminous Infrared Galaxies: The Case of the Circumnuclear Starburst Ring in NGC 7552}",
      journal = {\apj},
         year = 2023,
        month = jul,
       volume = {952},
       number = {1},
          eid = {22},
        pages = {22},
          doi = {10.3847/1538-4357/acd9aa},
archivePrefix = {arXiv},
       eprint = {2305.19491},
 primaryClass = {astro-ph.HE},
       adsurl = {https://ui.adsabs.harvard.edu/abs/2023ApJ...952...22W}
}

@ARTICLE{Sakamoto2014,
       author = {{Sakamoto}, Kazushi and {Aalto}, Susanne and {Combes}, Francoise and {Evans}, Aaron and {Peck}, Alison},
        title = "{An Infrared-luminous Merger with Two Bipolar Molecular Outflows: ALMA and SMA Observations of NGC 3256}",
      journal = {\apj},
         year = 2014,
        month = dec,
       volume = {797},
       number = {2},
          eid = {90},
        pages = {90},
          doi = {10.1088/0004-637X/797/2/90},
archivePrefix = {arXiv},
       eprint = {1403.7117},
 primaryClass = {astro-ph.GA},
       adsurl = {https://ui.adsabs.harvard.edu/abs/2014ApJ...797...90S}
}

@ARTICLE{Gonzalez-Martin2025,
       author = {{Gonz{\'a}lez-Mart{\'\i}n}, Omaira and {D{\'\i}az-Gonz{\'a}lez}, Daniel J. and {Mart{\'\i}nez-Paredes}, Mariela and {Alonso-Herrero}, Almudena and {L{\'o}pez-Rodr{\'\i}guez}, Enrique and {Garc{\'\i}a-Lorenzo}, Bego{\~n}a and {Ramos Almeida}, Cristina and {Garc{\'\i}a-Bernete}, Ismael and {Esparza-Arredondo}, Donaji and {Hoenig}, Sebastian F. and {Garc{\'\i}a-Burillo}, Santiago and {Packham}, Chris and {Levenson}, Nancy A. and {Labiano}, Alvaro and {Pereira-Santaella}, Miguel and {Combes}, Francoise and {Audibert}, Anelise and {Hicks}, Erin K.~S. and {Zhang}, Lulu and {Bellocchi}, Enrica and {Davies}, Richard I. and {Mu{\~n}oz}, Laura Hermosa and {Imanishi}, Masatoshi and {Ricci}, Claudio and {Stalevski}, Marko},
        title = "{JWST reveals the diversity of nuclear obscuring dust in nearby AGN: nuclear isolation of MIRI/MRS data cubes and continuum spectral fitting}",
      journal = {\mnras},
         year = 2025,
        month = may,
       volume = {539},
       number = {3},
        pages = {2158-2184},
          doi = {10.1093/mnras/staf573},
archivePrefix = {arXiv},
       eprint = {2504.01103},
 primaryClass = {astro-ph.GA},
       adsurl = {https://ui.adsabs.harvard.edu/abs/2025MNRAS.539.2158G}
}

@ARTICLE{Brightman2023,
       author = {{Brightman}, Murray and {Hameury}, Jean-Marie and {Lasota}, Jean-Pierre and {Baldi}, Ranieri D. and {Bruni}, Gabriele and {Cann}, Jenna M. and {Earnshaw}, Hannah and {F{\"u}rst}, Felix and {Heida}, Marianne and {Jaodand}, Amruta and {Lazzarini}, Margaret and {Middleton}, Matthew J. and {Walton}, Dominic J. and {Weaver}, Kimberly A.},
        title = "{A New Sample of Transient Ultraluminous X-Ray Sources Serendipitously Discovered by Swift/XRT}",
      journal = {\apj},
         year = 2023,
        month = jul,
       volume = {951},
       number = {1},
          eid = {51},
        pages = {51},
          doi = {10.3847/1538-4357/acd18a},
archivePrefix = {arXiv},
       eprint = {2305.01693},
 primaryClass = {astro-ph.HE},
       adsurl = {https://ui.adsabs.harvard.edu/abs/2023ApJ...951...51B}
}

@ARTICLE{Walton2021,
       author = {{Walton}, D.~J. and {Heida}, M. and {Bachetti}, M. and {F{\"u}rst}, F. and {Brightman}, M. and {Earnshaw}, H. and {Evans}, P.~A. and {Fabian}, A.~C. and {Grefenstette}, B.~W. and {Harrison}, F.~A. and {Israel}, G.~L. and {Lansbury}, G.~B. and {Middleton}, M.~J. and {Pike}, S. and {Rana}, V. and {Roberts}, T.~P. and {Rodriguez Castillo}, G.~A. and {Salvaterra}, R. and {Song}, X. and {Stern}, D.},
        title = "{A new transient ultraluminous X-ray source in NGC 7090}",
      journal = {\mnras},
         year = 2021,
        month = feb,
       volume = {501},
       number = {1},
        pages = {1002-1012},
          doi = {10.1093/mnras/staa3666},
archivePrefix = {arXiv},
       eprint = {2011.08870},
 primaryClass = {astro-ph.HE},
       adsurl = {https://ui.adsabs.harvard.edu/abs/2021MNRAS.501.1002W}
}

@ARTICLE{Brightman2020,
       author = {{Brightman}, Murray and {Earnshaw}, Hannah and {F{\"u}rst}, Felix and {Harrison}, Fiona A. and {Heida}, Marianne and {Israel}, Gianluca and {Pike}, Sean and {Stern}, Daniel and {Walton}, Dominic J.},
        title = "{Swift Monitoring of M51: A 38 day Superorbital Period for the Pulsar ULX7 and a New Transient Ultraluminous X-Ray Source}",
      journal = {\apj},
         year = 2020,
        month = jun,
       volume = {895},
       number = {2},
          eid = {127},
        pages = {127},
          doi = {10.3847/1538-4357/ab7e2a},
archivePrefix = {arXiv},
       eprint = {1912.04431},
 primaryClass = {astro-ph.HE},
       adsurl = {https://ui.adsabs.harvard.edu/abs/2020ApJ...895..127B}
}

@ARTICLE{Earnshaw2020,
       author = {{Earnshaw}, Hannah P. and {Heida}, Marianne and {Brightman}, Murray and {F{\"u}rst}, Felix and {Harrison}, Fiona A. and {Jaodand}, Amruta and {Middleton}, Matthew J. and {Roberts}, Timothy P. and {Sathyaprakash}, Rajath and {Stern}, Daniel and {Walton}, Dominic J.},
        title = "{The (Re)appearance of NGC 925 ULX-3, a New Transient ULX}",
      journal = {\apj},
         year = 2020,
        month = mar,
       volume = {891},
       number = {2},
          eid = {153},
        pages = {153},
          doi = {10.3847/1538-4357/ab77b8},
archivePrefix = {arXiv},
       eprint = {2001.00642},
 primaryClass = {astro-ph.HE},
       adsurl = {https://ui.adsabs.harvard.edu/abs/2020ApJ...891..153E}
}

@ARTICLE{Nandra1994,
       author = {{Nandra}, K. and {Pounds}, K.~A.},
        title = "{GINGA observations of the X-ray spectra of Seyfert galaxies.}",
      journal = {\mnras},
         year = 1994,
        month = may,
       volume = {268},
        pages = {405-429},
          doi = {10.1093/mnras/268.2.405},
       adsurl = {https://ui.adsabs.harvard.edu/abs/1994MNRAS.268..405N}
}

@ARTICLE{Robba2022,
       author = {{Robba}, A. and {Pinto}, C. and {Pintore}, F. and {Rodriguez}, G. and {Ambrosi}, E. and {Barra}, F. and {Cusumano}, G. and {D'A{\`\i}}, A. and {Del Santo}, M. and {Kosec}, P. and {Marino}, A. and {Middleton}, M. and {Roberts}, T. and {Salvaggio}, C. and {Soria}, R. and {Wolter}, A. and {Walton}, D.},
        title = "{A transient ultraluminous X-ray source in NGC 55}",
      journal = {\mnras},
         year = 2022,
        month = oct,
       volume = {515},
       number = {4},
        pages = {4669-4674},
          doi = {10.1093/mnras/stac2018},
archivePrefix = {arXiv},
       eprint = {2207.09447},
 primaryClass = {astro-ph.HE},
       adsurl = {https://ui.adsabs.harvard.edu/abs/2022MNRAS.515.4669R}
}

@ARTICLE{vanHaaften2019,
       author = {{van Haaften}, Lennart M. and {Maccarone}, Thomas J. and {Rhode}, Katherine L. and {Kundu}, Arunav and {Zepf}, Stephen E.},
        title = "{Discovery of a transient ultraluminous X-ray source in the elliptical galaxy M86}",
      journal = {\mnras},
         year = 2019,
        month = mar,
       volume = {483},
       number = {3},
        pages = {3566-3573},
          doi = {10.1093/mnras/sty3221},
archivePrefix = {arXiv},
       eprint = {1811.10595},
 primaryClass = {astro-ph.HE},
       adsurl = {https://ui.adsabs.harvard.edu/abs/2019MNRAS.483.3566V}
}

@ARTICLE{Harrison2013,
       author = {{Harrison}, Fiona A. and {Craig}, William W. and {Christensen}, Finn E. and {Hailey}, Charles J. and {Zhang}, William W. and {Boggs}, Steven E. and {Stern}, Daniel and {Cook}, W. Rick and {Forster}, Karl and {Giommi}, Paolo and {Grefenstette}, Brian W. and {Kim}, Yunjin and {Kitaguchi}, Takao and {Koglin}, Jason E. and {Madsen}, Kristin K. and {Mao}, Peter H. and {Miyasaka}, Hiromasa and {Mori}, Kaya and {Perri}, Matteo and {Pivovaroff}, Michael J. and {Puccetti}, Simonetta and {Rana}, Vikram R. and {Westergaard}, Niels J. and {Willis}, Jason and {Zoglauer}, Andreas and {An}, Hongjun and {Bachetti}, Matteo and {Barri{\`e}re}, Nicolas M. and {Bellm}, Eric C. and {Bhalerao}, Varun and {Brejnholt}, Nicolai F. and {Fuerst}, Felix and {Liebe}, Carl C. and {Markwardt}, Craig B. and {Nynka}, Melania and {Vogel}, Julia K. and {Walton}, Dominic J. and {Wik}, Daniel R. and {Alexander}, David M. and {Cominsky}, Lynn R. and {Hornschemeier}, Ann E. and {Hornstrup}, Allan and {Kaspi}, Victoria M. and {Madejski}, Greg M. and {Matt}, Giorgio and {Molendi}, Silvano and {Smith}, David M. and {Tomsick}, John A. and {Ajello}, Marco and {Ballantyne}, David R. and {Balokovi{\'c}}, Mislav and {Barret}, Didier and {Bauer}, Franz E. and {Blandford}, Roger D. and {Brandt}, W. Niel and {Brenneman}, Laura W. and {Chiang}, James and {Chakrabarty}, Deepto and {Chenevez}, Jerome and {Comastri}, Andrea and {Dufour}, Francois and {Elvis}, Martin and {Fabian}, Andrew C. and {Farrah}, Duncan and {Fryer}, Chris L. and {Gotthelf}, Eric V. and {Grindlay}, Jonathan E. and {Helfand}, David J. and {Krivonos}, Roman and {Meier}, David L. and {Miller}, Jon M. and {Natalucci}, Lorenzo and {Ogle}, Patrick and {Ofek}, Eran O. and {Ptak}, Andrew and {Reynolds}, Stephen P. and {Rigby}, Jane R. and {Tagliaferri}, Gianpiero and {Thorsett}, Stephen E. and {Treister}, Ezequiel and {Urry}, C. Megan},
        title = "{The Nuclear Spectroscopic Telescope Array (NuSTAR) High-energy X-Ray Mission}",
      journal = {\apj},
         year = 2013,
        month = jun,
       volume = {770},
       number = {2},
          eid = {103},
        pages = {103},
          doi = {10.1088/0004-637X/770/2/103},
archivePrefix = {arXiv},
       eprint = {1301.7307},
 primaryClass = {astro-ph.IM},
       adsurl = {https://ui.adsabs.harvard.edu/abs/2013ApJ...770..103H}
}

@ARTICLE{Gardner2006,
       author = {{Gardner}, Jonathan P. and {Mather}, John C. and {Clampin}, Mark and {Doyon}, Rene and {Greenhouse}, Matthew A. and {Hammel}, Heidi B. and {Hutchings}, John B. and {Jakobsen}, Peter and {Lilly}, Simon J. and {Long}, Knox S. and {Lunine}, Jonathan I. and {McCaughrean}, Mark J. and {Mountain}, Matt and {Nella}, John and {Rieke}, George H. and {Rieke}, Marcia J. and {Rix}, Hans-Walter and {Smith}, Eric P. and {Sonneborn}, George and {Stiavelli}, Massimo and {Stockman}, H.~S. and {Windhorst}, Rogier A. and {Wright}, Gillian S.},
        title = "{The James Webb Space Telescope}",
      journal = {\ssr},
         year = 2006,
        month = apr,
       volume = {123},
       number = {4},
        pages = {485-606},
          doi = {10.1007/s11214-006-8315-7},
archivePrefix = {arXiv},
       eprint = {astro-ph/0606175},
 primaryClass = {astro-ph},
       adsurl = {https://ui.adsabs.harvard.edu/abs/2006SSRv..123..485G}
}

@ARTICLE{Kotilainen1996,
       author = {{Kotilainen}, J.~K. and {Moorwood}, A.~F.~M. and {Ward}, M.~J. and {Forbes}, D.~A.},
        title = "{The nature of the two nuclei in the young merger NGC 3256: an obscured AGN?}",
      journal = {\aap},
         year = 1996,
        month = jan,
       volume = {305},
        pages = {107},
       adsurl = {https://ui.adsabs.harvard.edu/abs/1996A&A...305..107K}
}

@ARTICLE{Stierwalt2013,
       author = {{Stierwalt}, S. and {Armus}, L. and {Surace}, J.~A. and {Inami}, H. and {Petric}, A.~O. and {Diaz-Santos}, T. and {Haan}, S. and {Charmandaris}, V. and {Howell}, J. and {Kim}, D.~C. and {Marshall}, J. and {Mazzarella}, J.~M. and {Spoon}, H.~W.~W. and {Veilleux}, S. and {Evans}, A. and {Sanders}, D.~B. and {Appleton}, P. and {Bothun}, G. and {Bridge}, C.~R. and {Chan}, B. and {Frayer}, D. and {Iwasawa}, K. and {Kewley}, L.~J. and {Lord}, S. and {Madore}, B.~F. and {Melbourne}, J.~E. and {Murphy}, E.~J. and {Rich}, J.~A. and {Schulz}, B. and {Sturm}, E. and {Vavilkin}, T. and {Xu}, K.},
        title = "{Mid-infrared Properties of Nearby Luminous Infrared Galaxies. I. Spitzer Infrared Spectrograph Spectra for the GOALS Sample}",
      journal = {\apjs},
         year = 2013,
        month = may,
       volume = {206},
       number = {1},
          eid = {1},
        pages = {1},
          doi = {10.1088/0067-0049/206/1/1},
archivePrefix = {arXiv},
       eprint = {1302.4477},
 primaryClass = {astro-ph.CO},
       adsurl = {https://ui.adsabs.harvard.edu/abs/2013ApJS..206....1S}
}

@ARTICLE{Wootten2009,
       author = {{Wootten}, Alwyn and {Thompson}, A. Richard},
        title = "{The Atacama Large Millimeter/Submillimeter Array}",
      journal = {IEEE Proceedings},
         year = 2009,
        month = aug,
       volume = {97},
       number = {8},
        pages = {1463-1471},
          doi = {10.1109/JPROC.2009.2020572},
archivePrefix = {arXiv},
       eprint = {0904.3739},
 primaryClass = {astro-ph.IM},
       adsurl = {https://ui.adsabs.harvard.edu/abs/2009IEEEP..97.1463W}
}

@ARTICLE{Michiyama2020A,
       author = {{Michiyama}, Tomonari and {Iono}, Daisuke and {Nakanishi}, Kouichiro and {Ueda}, Junko and {Saito}, Toshiki and {Yamashita}, Takuji and {Bolatto}, Alberto and {Yun}, Min},
        title = "{Star Formation Traced by Optical and Millimeter Hydrogen Recombination Lines and Free-Free Emissions in the Dusty Merging Galaxy NGC 3256{\textemdash}MUSE/VLT and ALMA Synergy}",
      journal = {\apj},
         year = 2020,
        month = jun,
       volume = {895},
       number = {2},
          eid = {85},
        pages = {85},
          doi = {10.3847/1538-4357/ab88a5},
archivePrefix = {arXiv},
       eprint = {2004.06123},
 primaryClass = {astro-ph.GA},
       adsurl = {https://ui.adsabs.harvard.edu/abs/2020ApJ...895...85M}
}

@ARTICLE{Jenkins2004,
       author = {{Jenkins}, L.~P. and {Roberts}, T.~P. and {Ward}, M.~J. and {Zezas}, A.},
        title = "{XMM-Newton observations of the starburst merger galaxies NGC 3256 and NGC 3310}",
      journal = {\mnras},
         year = 2004,
        month = aug,
       volume = {352},
       number = {4},
        pages = {1335-1346},
          doi = {10.1111/j.1365-2966.2004.08025.x},
archivePrefix = {arXiv},
       eprint = {astro-ph/0405285},
 primaryClass = {astro-ph},
       adsurl = {https://ui.adsabs.harvard.edu/abs/2004MNRAS.352.1335J}
}

@ARTICLE{Yamada2021,
       author = {{Yamada}, Satoshi and {Ueda}, Yoshihiro and {Tanimoto}, Atsushi and {Imanishi}, Masatoshi and {Toba}, Yoshiki and {Ricci}, Claudio and {Privon}, George C.},
        title = "{Comprehensive Broadband X-Ray and Multiwavelength Study of Active Galactic Nuclei in 57 Local Luminous and Ultraluminous Infrared Galaxies Observed with NuSTAR and/or Swift/BAT}",
      journal = {\apjs},
         year = 2021,
        month = dec,
       volume = {257},
       number = {2},
          eid = {61},
        pages = {61},
          doi = {10.3847/1538-4365/ac17f5},
archivePrefix = {arXiv},
       eprint = {2107.10855},
 primaryClass = {astro-ph.GA},
       adsurl = {https://ui.adsabs.harvard.edu/abs/2021ApJS..257...61Y}
}

@ARTICLE{Neff2003,
       author = {{Neff}, S.~G. and {Ulvestad}, James S. and {Campion}, S.~D.},
        title = "{Radio Emission Associated with Ultraluminous X-Ray Sources in the Galaxy Merger NGC 3256}",
      journal = {\apj},
         year = 2003,
        month = dec,
       volume = {599},
       number = {2},
        pages = {1043-1048},
          doi = {10.1086/379596},
archivePrefix = {arXiv},
       eprint = {astro-ph/0310499},
 primaryClass = {astro-ph},
       adsurl = {https://ui.adsabs.harvard.edu/abs/2003ApJ...599.1043N}
}

@ARTICLE{Diaz-Santos2017,
       author = {{D{\'\i}az-Santos}, T. and {Armus}, L. and {Charmandaris}, V. and {Lu}, N. and {Stierwalt}, S. and {Stacey}, G. and {Malhotra}, S. and {van der Werf}, P.~P. and {Howell}, J.~H. and {Privon}, G.~C. and {Mazzarella}, J.~M. and {Goldsmith}, P.~F. and {Murphy}, E.~J. and {Barcos-Mu{\~n}oz}, L. and {Linden}, S.~T. and {Inami}, H. and {Larson}, K.~L. and {Evans}, A.~S. and {Appleton}, P. and {Iwasawa}, K. and {Lord}, S. and {Sanders}, D.~B. and {Surace}, J.~A.},
        title = "{A Herschel/PACS Far-infrared Line Emission Survey of Local Luminous Infrared Galaxies}",
      journal = {\apj},
         year = 2017,
        month = sep,
       volume = {846},
       number = {1},
          eid = {32},
        pages = {32},
          doi = {10.3847/1538-4357/aa81d7},
archivePrefix = {arXiv},
       eprint = {1705.04326},
 primaryClass = {astro-ph.GA},
       adsurl = {https://ui.adsabs.harvard.edu/abs/2017ApJ...846...32D}
}

@ARTICLE{Ducci2025,
       author = {{Ducci}, Lorenzo and {Mereghetti}, Sandro and {Pintore}, Fabio and {Allak}, Sinan and {Santangelo}, Andrea and {Sasaki}, Manami and {Kavanagh}, Patrick},
        title = "{Discovery of a 9.67 s Pulsar in an Ultraluminous X-Ray Source in NGC 4631 with XMM-Newton}",
      journal = {\apjl},
         year = 2025,
        month = dec,
       volume = {994},
       number = {2},
          eid = {L38},
        pages = {L38},
          doi = {10.3847/2041-8213/ae1c42},
archivePrefix = {arXiv},
       eprint = {2511.04282},
 primaryClass = {astro-ph.HE},
       adsurl = {https://ui.adsabs.harvard.edu/abs/2025ApJ...994L..38D}
}

@ARTICLE{Earnshaw2025,
       author = {{Earnshaw}, Hannah P. and {Patti}, Gauri and {Brightman}, Murray and {Sathyaprakash}, Rajath and {Walton}, Dominic J. and {F{\"u}rst}, Felix and {Roberts}, Timothy P. and {Harrison}, Fiona A.},
        title = "{The Long-Term Variability of a Population of ULXs Monitored by Chandra}",
      journal = {Astronomische Nachrichten},
         year = 2025,
        month = jan,
       volume = {346},
       number = {1},
          eid = {e20240103},
        pages = {e20240103},
          doi = {10.1002/asna.20240103},
archivePrefix = {arXiv},
       eprint = {2411.07459},
 primaryClass = {astro-ph.HE},
       adsurl = {https://ui.adsabs.harvard.edu/abs/2025AN....34640103E}
}

@ARTICLE{Rana2015,
       author = {{Rana}, Vikram and {Harrison}, Fiona A. and {Bachetti}, Matteo and {Walton}, Dominic J. and {Furst}, Felix and {Barret}, Didier and {Miller}, Jon M. and {Fabian}, Andrew C. and {Boggs}, Steven E. and {Christensen}, Finn C. and {Craig}, William W. and {Grefenstette}, Brian W. and {Hailey}, Charles J. and {Madsen}, Kristin K. and {Ptak}, Andrew F. and {Stern}, Daniel and {Webb}, Natalie A. and {Zhang}, William W.},
        title = "{The Broadband XMM-Newton and NuSTAR X-Ray Spectra of Two Ultraluminous X-Ray Sources in the Galaxy IC 342}",
      journal = {\apj},
         year = 2015,
        month = feb,
       volume = {799},
       number = {2},
          eid = {121},
        pages = {121},
          doi = {10.1088/0004-637X/799/2/121},
archivePrefix = {arXiv},
       eprint = {1401.4637},
 primaryClass = {astro-ph.HE},
       adsurl = {https://ui.adsabs.harvard.edu/abs/2015ApJ...799..121R}
}

@ARTICLE{Pintore2018b,
       author = {{Pintore}, F. and {Zampieri}, L. and {Mereghetti}, S. and {Wolter}, A. and {Rodr{\'\i}guez}, G. and {Israel}, G.~L. and {Esposito}, P. and {Paiano}, S. and {Trinchieri}, G. and {Ochner}, P.},
        title = "{The two ultraluminous X-ray sources in the galaxy NGC 925}",
      journal = {\mnras},
         year = 2018,
        month = sep,
       volume = {479},
       number = {3},
        pages = {4271-4277},
          doi = {10.1093/mnras/sty1766},
archivePrefix = {arXiv},
       eprint = {1806.11444},
 primaryClass = {astro-ph.HE},
       adsurl = {https://ui.adsabs.harvard.edu/abs/2018MNRAS.479.4271P}
}

@ARTICLE{Tranin2024,
       author = {{Tranin}, Hugo and {Webb}, Natalie and {Godet}, Olivier and {Quintin}, Erwan},
        title = "{Statistical study of a large and cleaned sample of ultraluminous and hyperluminous X-ray sources}",
      journal = {\aap},
         year = 2024,
        month = jan,
       volume = {681},
          eid = {A16},
        pages = {A16},
          doi = {10.1051/0004-6361/202244952},
archivePrefix = {arXiv},
       eprint = {2304.11216},
 primaryClass = {astro-ph.HE},
       adsurl = {https://ui.adsabs.harvard.edu/abs/2024A&A...681A..16T}
}

@ARTICLE{Furst2017,
       author = {{F{\"u}rst}, F. and {Walton}, D.~J. and {Stern}, D. and {Bachetti}, M. and {Barret}, D. and {Brightman}, M. and {Harrison}, F.~A. and {Rana}, V.},
        title = "{Spectral Changes in the Hyperluminous Pulsar in NGC 5907 as a Function of Super-orbital Phase}",
      journal = {\apj},
         year = 2017,
        month = jan,
       volume = {834},
       number = {1},
          eid = {77},
        pages = {77},
          doi = {10.3847/1538-4357/834/1/77},
archivePrefix = {arXiv},
       eprint = {1610.00258},
 primaryClass = {astro-ph.HE},
       adsurl = {https://ui.adsabs.harvard.edu/abs/2017ApJ...834...77F}
}

@ARTICLE{Amato2025,
       author = {{Amato}, Roberta and {Quintin}, Erwan and {Tranin}, Hugo and {G{\'u}rpide}, Andr{\'e}s and {Webb}, Natalie and {Godet}, Olivier and {Luca Israel}, Gian and {Imbrogno}, Matteo and {Kammoun}, Elias and {Gupta}, Maitrayee},
        title = "{The population of hyper-luminous X-ray sources as seen by XMM-Newton}",
      journal = {\aap},
         year = 2025,
        month = sep,
       volume = {701},
          eid = {A7},
        pages = {A7},
          doi = {10.1051/0004-6361/202555122},
archivePrefix = {arXiv},
       eprint = {2507.14875},
 primaryClass = {astro-ph.HE},
       adsurl = {https://ui.adsabs.harvard.edu/abs/2025A&A...701A...7A}
}

@ARTICLE{Pintore2025,
       author = {{Pintore}, F. and {Pinto}, C. and {Rodriguez-Castillo}, G. and {Israel}, G.~L. and {Pinciroli Vago}, N.~O. and {Motta}, S. and {Barra}, F. and {Walton}, D.~J. and {Fuerst}, F. and {Kosec}, P. and {Salvaggio}, C. and {Del Santo}, M. and {Wolter}, A. and {Middleton}, M. and {D'A{\`\i}}, A. and {Ambrosi}, E. and {Burderi}, L. and {Imbrogno}, M. and {Salvaterra}, R. and {Robba}, A.},
        title = "{A new pulsating neutron star in the ultraluminous X-ray source NGC 4559 X7?}",
      journal = {\aap},
         year = 2025,
        month = mar,
       volume = {695},
          eid = {A238},
        pages = {A238},
          doi = {10.1051/0004-6361/202453240},
archivePrefix = {arXiv},
       eprint = {2502.10246},
 primaryClass = {astro-ph.HE},
       adsurl = {https://ui.adsabs.harvard.edu/abs/2025A&A...695A.238P}
}

@ARTICLE{Sathyaprakash2019,
       author = {{Sathyaprakash}, R. and {Roberts}, T.~P. and {Walton}, D.~J. and {Fuerst}, F. and {Bachetti}, M. and {Pinto}, C. and {Alston}, W.~N. and {Earnshaw}, H.~P. and {Fabian}, A.~C. and {Middleton}, M.~J. and {Soria}, R.},
        title = "{The discovery of weak coherent pulsations in the ultraluminous X-ray source NGC 1313 X-2}",
      journal = {\mnras},
         year = 2019,
        month = sep,
       volume = {488},
       number = {1},
        pages = {L35-L40},
          doi = {10.1093/mnrasl/slz086},
archivePrefix = {arXiv},
       eprint = {1906.00640},
 primaryClass = {astro-ph.HE},
       adsurl = {https://ui.adsabs.harvard.edu/abs/2019MNRAS.488L..35S}
}

@ARTICLE{Carpano2018,
       author = {{Carpano}, S. and {Haberl}, F. and {Maitra}, C. and {Vasilopoulos}, G.},
        title = "{Discovery of pulsations from NGC 300 ULX1 and its fast period evolution}",
      journal = {\mnras},
         year = 2018,
        month = may,
       volume = {476},
       number = {1},
        pages = {L45-L49},
          doi = {10.1093/mnrasl/sly030},
archivePrefix = {arXiv},
       eprint = {1802.10341},
 primaryClass = {astro-ph.HE},
       adsurl = {https://ui.adsabs.harvard.edu/abs/2018MNRAS.476L..45C}
}

@ARTICLE{Lira2002,
       author = {{Lira}, P. and {Ward}, M. and {Zezas}, A. and {Alonso-Herrero}, A. and {Ueno}, S.},
        title = "{Chandra observations of the luminous infrared galaxy NGC 3256}",
      journal = {\mnras},
         year = 2002,
        month = feb,
       volume = {330},
       number = {2},
        pages = {259-278},
          doi = {10.1046/j.1365-8711.2002.05014.x},
archivePrefix = {arXiv},
       eprint = {astro-ph/0109198},
 primaryClass = {astro-ph},
       adsurl = {https://ui.adsabs.harvard.edu/abs/2002MNRAS.330..259L}
}

@software{Buchner2016bxa,
       author = {{Buchner}, Johannes},
        title = "{BXA: Bayesian X-ray Analysis}",
 howpublished = {Astrophysics Source Code Library, record ascl:1610.011},
         year = 2016,
        month = oct,
          eid = {ascl:1610.011},
archivePrefix = {ascl},
       eprint = {1610.011},
       adsurl = {https://ui.adsabs.harvard.edu/abs/2016ascl.soft10011B}
}

@ARTICLE{Watarai2000,
       author = {{Watarai}, Ken-ya and {Fukue}, Jun and {Takeuchi}, Mitsuru and {Mineshige}, Shin},
        title = "{Galactic Black-Hole Candidates Shining at the Eddington Luminosity}",
      journal = {\pasj},
         year = 2000,
        month = feb,
       volume = {52},
        pages = {133},
          doi = {10.1093/pasj/52.1.133},
       adsurl = {https://ui.adsabs.harvard.edu/abs/2000PASJ...52..133W}
}

@ARTICLE{Kubota2005,
       author = {{Kubota}, Aya and {Ebisawa}, Ken and {Makishima}, Kazuo and {Nakazawa}, Kazuhiro},
        title = "{On the Temperature Profile of Radiatively Efficient Geometrically Thin Disks in Black Hole Binaries with the ASCA GIS}",
      journal = {\apj},
         year = 2005,
        month = oct,
       volume = {631},
       number = {2},
        pages = {1062-1071},
          doi = {10.1086/432900},
archivePrefix = {arXiv},
       eprint = {astro-ph/0505338},
 primaryClass = {astro-ph},
       adsurl = {https://ui.adsabs.harvard.edu/abs/2005ApJ...631.1062K}
}

@ARTICLE{Mineshige1994,
       author = {{Mineshige}, Shin and {Hirano}, Akira and {Kitamoto}, Shunji and {Yamada}, Tatsuya T. and {Fukue}, Jun},
        title = "{Time-dependent Disk Accretion in X-Ray Nova MUSCAE 1991}",
      journal = {\apj},
         year = 1994,
        month = may,
       volume = {426},
        pages = {308},
          doi = {10.1086/174065},
       adsurl = {https://ui.adsabs.harvard.edu/abs/1994ApJ...426..308M}
}

@ARTICLE{Cash1979,
       author = {{Cash}, W.},
        title = "{Parameter estimation in astronomy through application of the likelihood ratio.}",
      journal = {\apj},
         year = 1979,
        month = mar,
       volume = {228},
        pages = {939-947},
          doi = {10.1086/156922},
       adsurl = {https://ui.adsabs.harvard.edu/abs/1979ApJ...228..939C}
}

@INCOLLECTION{Buchner2023,
       author = {{Buchner}, Johannes and {Boorman}, Peter},
        title = "{Statistical Aspects of X-ray Spectral Analysis}",
    booktitle = {Handbook of X-ray and Gamma-ray Astrophysics},
         year = 2023,
          eid = {150},
        pages = {150},
          doi = {10.1007/978-981-16-4544-0_175-1},
       adsurl = {https://ui.adsabs.harvard.edu/abs/2023hxga.book..150B}
}

@ARTICLE{Kaastra2016,
       author = {{Kaastra}, J.~S. and {Bleeker}, J.~A.~M.},
        title = "{Optimal binning of X-ray spectra and response matrix design}",
      journal = {\aap},
         year = 2016,
        month = mar,
       volume = {587},
          eid = {A151},
        pages = {A151},
          doi = {10.1051/0004-6361/201527395},
archivePrefix = {arXiv},
       eprint = {1601.05309},
 primaryClass = {astro-ph.IM},
       adsurl = {https://ui.adsabs.harvard.edu/abs/2016A&A...587A.151K}
}

@ARTICLE{HI4PI2016,
       author = {{HI4PI Collaboration} and {Ben Bekhti}, N. and {Fl{\"o}er}, L. and {Keller}, R. and {Kerp}, J. and {Lenz}, D. and {Winkel}, B. and {Bailin}, J. and {Calabretta}, M.~R. and {Dedes}, L. and {Ford}, H.~A. and {Gibson}, B.~K. and {Haud}, U. and {Janowiecki}, S. and {Kalberla}, P.~M.~W. and {Lockman}, F.~J. and {McClure-Griffiths}, N.~M. and {Murphy}, T. and {Nakanishi}, H. and {Pisano}, D.~J. and {Staveley-Smith}, L.},
        title = "{HI4PI: A full-sky H I survey based on EBHIS and GASS}",
      journal = {\aap},
         year = 2016,
        month = oct,
       volume = {594},
          eid = {A116},
        pages = {A116},
          doi = {10.1051/0004-6361/201629178},
archivePrefix = {arXiv},
       eprint = {1610.06175},
 primaryClass = {astro-ph.GA},
       adsurl = {https://ui.adsabs.harvard.edu/abs/2016A&A...594A.116H}
}

@ARTICLE{Buchner2021UltraNest,
       author = {{Buchner}, Johannes},
        title = "{UltraNest - a robust, general purpose Bayesian inference engine}",
      journal = {The Journal of Open Source Software},
         year = 2021,
        month = apr,
       volume = {6},
       number = {60},
          eid = {3001},
        pages = {3001},
          doi = {10.21105/joss.03001},
archivePrefix = {arXiv},
       eprint = {2101.09604},
 primaryClass = {stat.CO},
       adsurl = {https://ui.adsabs.harvard.edu/abs/2021JOSS....6.3001B}
}

@software{Buchner2016ultranest,
       author = {{Buchner}, Johannes},
        title = "{UltraNest: Pythonic Nested Sampling Development Framework and UltraNest}",
 howpublished = {Astrophysics Source Code Library, record ascl:1611.001},
         year = 2016,
        month = nov,
          eid = {ascl:1611.001},
archivePrefix = {ascl},
       eprint = {1611.001},
       adsurl = {https://ui.adsabs.harvard.edu/abs/2016ascl.soft11001B}
}

@ARTICLE{Luangtip2015,
       author = {{Luangtip}, W. and {Roberts}, T.~P. and {Mineo}, S. and {Lehmer}, B.~D. and {Alexander}, D.~M. and {Jackson}, F.~E. and {Goulding}, A.~D. and {Fischer}, J.~L.},
        title = "{A deficit of ultraluminous X-ray sources in luminous infrared galaxies}",
      journal = {\mnras},
         year = 2015,
        month = jan,
       volume = {446},
       number = {1},
        pages = {470-492},
          doi = {10.1093/mnras/stu2086},
archivePrefix = {arXiv},
       eprint = {1410.1569},
 primaryClass = {astro-ph.HE},
       adsurl = {https://ui.adsabs.harvard.edu/abs/2015MNRAS.446..470L}
}

@ARTICLE{McGaugh1991,
       author = {{McGaugh}, Stacy S.},
        title = "{H II Region Abundances: Model Oxygen Line Ratios}",
      journal = {\apj},
         year = 1991,
        month = oct,
       volume = {380},
        pages = {140},
          doi = {10.1086/170569},
       adsurl = {https://ui.adsabs.harvard.edu/abs/1991ApJ...380..140M}
}

@ARTICLE{Kewley2002,
       author = {{Kewley}, L.~J. and {Dopita}, M.~A.},
        title = "{Using Strong Lines to Estimate Abundances in Extragalactic H II Regions and Starburst Galaxies}",
      journal = {\apjs},
         year = 2002,
        month = sep,
       volume = {142},
       number = {1},
        pages = {35-52},
          doi = {10.1086/341326},
archivePrefix = {arXiv},
       eprint = {astro-ph/0206495},
 primaryClass = {astro-ph},
       adsurl = {https://ui.adsabs.harvard.edu/abs/2002ApJS..142...35K}
}

@ARTICLE{Pettini2004,
       author = {{Pettini}, Max and {Pagel}, Bernard E.~J.},
        title = "{[OIII]/[NII] as an abundance indicator at high redshift}",
      journal = {\mnras},
         year = 2004,
        month = mar,
       volume = {348},
       number = {3},
        pages = {L59-L63},
          doi = {10.1111/j.1365-2966.2004.07591.x},
archivePrefix = {arXiv},
       eprint = {astro-ph/0401128},
 primaryClass = {astro-ph},
       adsurl = {https://ui.adsabs.harvard.edu/abs/2004MNRAS.348L..59P}
}

@ARTICLE{Dopita2010,
       author = {{Dopita}, Michael and {Rhee}, Jonghwan and {Farage}, Catherine and {McGregor}, Peter and {Bloxham}, Gabe and {Green}, Anthony and {Roberts}, Bill and {Neilson}, Jon and {Wilson}, Greg and {Young}, Peter and {Firth}, Peter and {Busarello}, Gianni and {Merluzzi}, Paola},
        title = "{The Wide Field Spectrograph (WiFeS): performance and data reduction}",
      journal = {\apss},
         year = 2010,
        month = jun,
       volume = {327},
       number = {2},
        pages = {245-257},
          doi = {10.1007/s10509-010-0335-9},
archivePrefix = {arXiv},
       eprint = {1002.4472},
 primaryClass = {astro-ph.IM},
       adsurl = {https://ui.adsabs.harvard.edu/abs/2010Ap&SS.327..245D}
}

@ARTICLE{Dopita2007,
       author = {{Dopita}, Michael and {Hart}, John and {McGregor}, Peter and {Oates}, Patrick and {Bloxham}, Gabe and {Jones}, Damien},
        title = "{The Wide Field Spectrograph (WiFeS)}",
      journal = {\apss},
         year = 2007,
        month = aug,
       volume = {310},
       number = {3-4},
        pages = {255-268},
          doi = {10.1007/s10509-007-9510-z},
archivePrefix = {arXiv},
       eprint = {0705.0287},
 primaryClass = {astro-ph},
       adsurl = {https://ui.adsabs.harvard.edu/abs/2007Ap&SS.310..255D}
}

@ARTICLE{Rich2012,
       author = {{Rich}, J.~A. and {Torrey}, P. and {Kewley}, L.~J. and {Dopita}, M.~A. and {Rupke}, D.~S.~N.},
        title = "{An Integral Field Study of Abundance Gradients in nearby Luminous Infrared Galaxies}",
      journal = {\apj},
         year = 2012,
        month = jul,
       volume = {753},
       number = {1},
          eid = {5},
        pages = {5},
          doi = {10.1088/0004-637X/753/1/5},
archivePrefix = {arXiv},
       eprint = {1204.5520},
 primaryClass = {astro-ph.CO},
       adsurl = {https://ui.adsabs.harvard.edu/abs/2012ApJ...753....5R}
}

@ARTICLE{Greene2020,
       author = {{Greene}, Jenny E. and {Strader}, Jay and {Ho}, Luis C.},
        title = "{Intermediate-Mass Black Holes}",
      journal = {\araa},
         year = 2020,
        month = aug,
       volume = {58},
        pages = {257-312},
          doi = {10.1146/annurev-astro-032620-021835},
archivePrefix = {arXiv},
       eprint = {1911.09678},
 primaryClass = {astro-ph.GA},
       adsurl = {https://ui.adsabs.harvard.edu/abs/2020ARA&A..58..257G}
}

@ARTICLE{Pasham2014,
       author = {{Pasham}, Dheeraj R. and {Strohmayer}, Tod E. and {Mushotzky}, Richard F.},
        title = "{A 400-solar-mass black hole in the galaxy M82}",
      journal = {\nat},
         year = 2014,
        month = sep,
       volume = {513},
       number = {7516},
        pages = {74-76},
          doi = {10.1038/nature13710},
archivePrefix = {arXiv},
       eprint = {1501.03180},
 primaryClass = {astro-ph.HE},
       adsurl = {https://ui.adsabs.harvard.edu/abs/2014Natur.513...74P}
}

@ARTICLE{Linden2010,
       author = {{Linden}, T. and {Kalogera}, V. and {Sepinsky}, J.~F. and {Prestwich}, A. and {Zezas}, A. and {Gallagher}, J.~S.},
        title = "{The Effect of Starburst Metallicity on Bright X-ray Binary Formation Pathways}",
      journal = {\apj},
         year = 2010,
        month = dec,
       volume = {725},
       number = {2},
        pages = {1984-1994},
          doi = {10.1088/0004-637X/725/2/1984},
archivePrefix = {arXiv},
       eprint = {1005.1639},
 primaryClass = {astro-ph.CO},
       adsurl = {https://ui.adsabs.harvard.edu/abs/2010ApJ...725.1984L}
}

@ARTICLE{Song2021,
       author = {{Song}, Y. and {Linden}, S.~T. and {Evans}, A.~S. and {Barcos-Mu{\~n}oz}, L. and {Privon}, G.~C. and {Yoon}, I. and {Murphy}, E.~J. and {Larson}, K.~L. and {D{\'\i}az-Santos}, T. and {Armus}, L. and {Mazzarella}, Joseph M. and {Howell}, J. and {Inami}, H. and {Torres-Alb{\`a}}, N. and {U}, V. and {Charmandaris}, V. and {McKinney}, J. and {Kunneriath}, D. and {Momjian}, E.},
        title = "{A Comparison between Nuclear Ring Star Formation in LIRGs and in Normal Galaxies with the Very Large Array}",
      journal = {\apj},
         year = 2021,
        month = aug,
       volume = {916},
       number = {2},
          eid = {73},
        pages = {73},
          doi = {10.3847/1538-4357/ac05c2},
archivePrefix = {arXiv},
       eprint = {2107.00412},
 primaryClass = {astro-ph.GA},
       adsurl = {https://ui.adsabs.harvard.edu/abs/2021ApJ...916...73S}
}

@ARTICLE{Haan2013,
       author = {{Haan}, S. and {Armus}, L. and {Surace}, J.~A. and {Charmandaris}, V. and {Evans}, A.~S. and {Diaz-Santos}, T. and {Melbourne}, J.~L. and {Mazzarella}, J.~M. and {Howell}, J.~H. and {Stierwalt}, S. and {Kim}, D.~C. and {Vavilkin}, T. and {Sanders}, D.~B. and {Petric}, A. and {Murphy}, E.~J. and {Braun}, R. and {Bridge}, C.~R. and {Inami}, H.},
        title = "{The build-up of nuclear stellar cusps in extreme starburst galaxies and major mergers}",
      journal = {\mnras},
         year = 2013,
        month = sep,
       volume = {434},
       number = {2},
        pages = {1264-1286},
          doi = {10.1093/mnras/stt1087},
archivePrefix = {arXiv},
       eprint = {1306.4337},
 primaryClass = {astro-ph.CO},
       adsurl = {https://ui.adsabs.harvard.edu/abs/2013MNRAS.434.1264H}
}

@ARTICLE{Lehmer2021,
       author = {{Lehmer}, Bret D. and {Eufrasio}, Rafael T. and {Basu-Zych}, Antara and {Doore}, Keith and {Fragos}, Tassos and {Garofali}, Kristen and {Kovlakas}, Konstantinos and {Williams}, Benjamin F. and {Zezas}, Andreas and {Santana-Silva}, Luidhy},
        title = "{The Metallicity Dependence of the High-mass X-Ray Binary Luminosity Function}",
      journal = {\apj},
         year = 2021,
        month = jan,
       volume = {907},
       number = {1},
          eid = {17},
        pages = {17},
          doi = {10.3847/1538-4357/abcec1},
archivePrefix = {arXiv},
       eprint = {2011.09476},
 primaryClass = {astro-ph.GA},
       adsurl = {https://ui.adsabs.harvard.edu/abs/2021ApJ...907...17L}
}

@ARTICLE{Belczynski2010,
       author = {{Belczynski}, Krzysztof and {Bulik}, Tomasz and {Fryer}, Chris L. and {Ruiter}, Ashley and {Valsecchi}, Francesca and {Vink}, Jorick S. and {Hurley}, Jarrod R.},
        title = "{On the Maximum Mass of Stellar Black Holes}",
      journal = {\apj},
         year = 2010,
        month = may,
       volume = {714},
       number = {2},
        pages = {1217-1226},
          doi = {10.1088/0004-637X/714/2/1217},
archivePrefix = {arXiv},
       eprint = {0904.2784},
 primaryClass = {astro-ph.SR},
       adsurl = {https://ui.adsabs.harvard.edu/abs/2010ApJ...714.1217B}
}

@ARTICLE{Prestwich2013,
       author = {{Prestwich}, A.~H. and {Tsantaki}, Maria and {Zezas}, A. and {Jackson}, F. and {Roberts}, T.~P. and {Foltz}, R. and {Linden}, T. and {Kalogera}, V.},
        title = "{Ultra-luminous X-Ray Sources in the Most Metal Poor Galaxies}",
      journal = {\apj},
         year = 2013,
        month = jun,
       volume = {769},
       number = {2},
          eid = {92},
        pages = {92},
          doi = {10.1088/0004-637X/769/2/92},
archivePrefix = {arXiv},
       eprint = {1302.6203},
 primaryClass = {astro-ph.HE},
       adsurl = {https://ui.adsabs.harvard.edu/abs/2013ApJ...769...92P}
}

@ARTICLE{Mapelli2011,
       author = {{Mapelli}, M. and {Ripamonti}, E. and {Zampieri}, L. and {Colpi}, M.},
        title = "{Remnants of massive metal-poor stars: Viable engines for ultra-luminous X-ray sources}",
      journal = {Astronomische Nachrichten},
         year = 2011,
        month = may,
       volume = {332},
       number = {4},
        pages = {414},
          doi = {10.1002/asna.201011511},
archivePrefix = {arXiv},
       eprint = {1009.5706},
 primaryClass = {astro-ph.CO},
       adsurl = {https://ui.adsabs.harvard.edu/abs/2011AN....332..414M}
}

@ARTICLE{Ricci2021,
       author = {{Ricci}, C. and {Privon}, G.~C. and {Pfeifle}, R.~W. and {Armus}, L. and {Iwasawa}, K. and {Torres-Alb{\`a}}, N. and {Satyapal}, S. and {Bauer}, F.~E. and {Treister}, E. and {Ho}, L.~C. and {Aalto}, S. and {Ar{\'e}valo}, P. and {Barcos-Mu{\~n}oz}, L. and {Charmandaris}, V. and {Diaz-Santos}, T. and {Evans}, A.~S. and {Gao}, T. and {Inami}, H. and {Koss}, M.~J. and {Lansbury}, G. and {Linden}, S.~T. and {Medling}, A. and {Sanders}, D.~B. and {Song}, Y. and {Stern}, D. and {U}, V. and {Ueda}, Y. and {Yamada}, S.},
        title = "{A hard X-ray view of luminous and ultra-luminous infrared galaxies in GOALS - I. AGN obscuration along the merger sequence}",
      journal = {\mnras},
         year = 2021,
        month = oct,
       volume = {506},
       number = {4},
        pages = {5935-5950},
          doi = {10.1093/mnras/stab2052},
archivePrefix = {arXiv},
       eprint = {2107.10864},
 primaryClass = {astro-ph.GA},
       adsurl = {https://ui.adsabs.harvard.edu/abs/2021MNRAS.506.5935R}
}

@ARTICLE{Chartab2022,
       author = {{Chartab}, Nima and {Cooray}, Asantha and {Ma}, Jingzhe and {Nayyeri}, Hooshang and {Zilliot}, Preston and {Lopez}, Jonathan and {Fadda}, Dario and {Herrera-Camus}, Rodrigo and {Malkan}, Matthew and {Rigopoulou}, Dimitra and {Sheth}, Kartik and {Wardlow}, Julie},
        title = "{Low gas-phase metallicities of ultraluminous infrared galaxies are a result of dust obscuration}",
      journal = {Nature Astronomy},
         year = 2022,
        month = may,
       volume = {6},
        pages = {844-849},
          doi = {10.1038/s41550-022-01679-y},
archivePrefix = {arXiv},
       eprint = {2201.07478},
 primaryClass = {astro-ph.GA},
       adsurl = {https://ui.adsabs.harvard.edu/abs/2022NatAs...6..844C}
}

@ARTICLE{Iwasawa2011,
       author = {{Iwasawa}, K. and {Sanders}, D.~B. and {Teng}, S.~H. and {U}, Vivian and {Armus}, L. and {Evans}, A.~S. and {Howell}, J.~H. and {Komossa}, S. and {Mazzarella}, J.~M. and {Petric}, A.~O. and {Surace}, J.~A. and {Vavilkin}, T. and {Veilleux}, S. and {Trentham}, N.},
        title = "{C-GOALS: Chandra observations of a complete sample of luminous infrared galaxies from the IRAS Revised Bright Galaxy Survey}",
      journal = {\aap},
         year = 2011,
        month = may,
       volume = {529},
          eid = {A106},
        pages = {A106},
          doi = {10.1051/0004-6361/201015264},
archivePrefix = {arXiv},
       eprint = {1103.2755},
 primaryClass = {astro-ph.CO},
       adsurl = {https://ui.adsabs.harvard.edu/abs/2011A&A...529A.106I}
}

@ARTICLE{Piqueras-Lopez2013,
       author = {{Piqueras L{\'o}pez}, J. and {Colina}, L. and {Arribas}, S. and {Alonso-Herrero}, A.},
        title = "{VLT-SINFONI integral field spectroscopy of low-z luminous and ultraluminous infrared galaxies. II. 2D extinction structure and distance effects}",
      journal = {\aap},
         year = 2013,
        month = may,
       volume = {553},
          eid = {A85},
        pages = {A85},
          doi = {10.1051/0004-6361/201220991},
archivePrefix = {arXiv},
       eprint = {1304.0894},
 primaryClass = {astro-ph.CO},
       adsurl = {https://ui.adsabs.harvard.edu/abs/2013A&A...553A..85P}
}

@ARTICLE{Downes1998,
       author = {{Downes}, D. and {Solomon}, P.~M.},
        title = "{Rotating Nuclear Rings and Extreme Starbursts in Ultraluminous Galaxies}",
      journal = {\apj},
         year = 1998,
        month = nov,
       volume = {507},
       number = {2},
        pages = {615-654},
          doi = {10.1086/306339},
archivePrefix = {arXiv},
       eprint = {astro-ph/9806377},
 primaryClass = {astro-ph},
       adsurl = {https://ui.adsabs.harvard.edu/abs/1998ApJ...507..615D}
}

@ARTICLE{Herrero-Illana2019,
       author = {{Herrero-Illana}, R. and {Privon}, G.~C. and {Evans}, A.~S. and {D{\'\i}az-Santos}, T. and {P{\'e}rez-Torres}, M. {\'A}. and {U}, V. and {Alberdi}, A. and {Iwasawa}, K. and {Armus}, L. and {Aalto}, S. and {Mazzarella}, J. and {Chu}, J. and {Sanders}, D.~B. and {Barcos-Mu{\~n}oz}, L. and {Charmandaris}, V. and {Linden}, S.~T. and {Yoon}, I. and {Frayer}, D.~T. and {Inami}, H. and {Kim}, D.-C. and {Borish}, H.~J. and {Conway}, J. and {Murphy}, E.~J. and {Song}, Y. and {Stierwalt}, S. and {Surace}, J.},
        title = "{Molecular gas and dust properties of galaxies from the Great Observatories All-sky LIRG Survey}",
      journal = {\aap},
         year = 2019,
        month = aug,
       volume = {628},
          eid = {A71},
        pages = {A71},
          doi = {10.1051/0004-6361/201834088},
archivePrefix = {arXiv},
       eprint = {1907.03854},
 primaryClass = {astro-ph.GA},
       adsurl = {https://ui.adsabs.harvard.edu/abs/2019A&A...628A..71H}
}

@ARTICLE{Middleton2013,
       author = {{Middleton}, Matthew J. and {Miller-Jones}, James C.~A. and {Markoff}, Sera and {Fender}, Rob and {Henze}, Martin and {Hurley-Walker}, Natasha and {Scaife}, Anna M.~M. and {Roberts}, Timothy P. and {Walton}, Dominic and {Carpenter}, John and {Macquart}, Jean-Pierre and {Bower}, Geoffrey C. and {Gurwell}, Mark and {Pietsch}, Wolfgang and {Haberl}, Frank and {Harris}, Jonathan and {Daniel}, Michael and {Miah}, Junayd and {Done}, Chris and {Morgan}, John S. and {Dickinson}, Hugh and {Charles}, Phil and {Burwitz}, Vadim and {Della Valle}, Massimo and {Freyberg}, Michael and {Greiner}, Jochen and {Hernanz}, Margarita and {Hartmann}, Dieter H. and {Hatzidimitriou}, Despina and {Riffeser}, Arno and {Sala}, Gloria and {Seitz}, Stella and {Reig}, Pablo and {Rau}, Arne and {Orio}, Marina and {Titterington}, David and {Grainge}, Keith},
        title = "{Bright radio emission from an ultraluminous stellar-mass microquasar in M 31}",
      journal = {\nat},
         year = 2013,
        month = jan,
       volume = {493},
       number = {7431},
        pages = {187-190},
          doi = {10.1038/nature11697},
archivePrefix = {arXiv},
       eprint = {1212.4698},
 primaryClass = {astro-ph.HE},
       adsurl = {https://ui.adsabs.harvard.edu/abs/2013Natur.493..187M}
}

@ARTICLE{Furst2016,
       author = {{F{\"u}rst}, F. and {Walton}, D.~J. and {Harrison}, F.~A. and {Stern}, D. and {Barret}, D. and {Brightman}, M. and {Fabian}, A.~C. and {Grefenstette}, B. and {Madsen}, K.~K. and {Middleton}, M.~J. and {Miller}, J.~M. and {Pottschmidt}, K. and {Ptak}, A. and {Rana}, V. and {Webb}, N.},
        title = "{Discovery of Coherent Pulsations from the Ultraluminous X-Ray Source NGC 7793 P13}",
      journal = {\apjl},
         year = 2016,
        month = nov,
       volume = {831},
       number = {2},
          eid = {L14},
        pages = {L14},
          doi = {10.3847/2041-8205/831/2/L14},
archivePrefix = {arXiv},
       eprint = {1609.07129},
 primaryClass = {astro-ph.HE},
       adsurl = {https://ui.adsabs.harvard.edu/abs/2016ApJ...831L..14F}
}

@ARTICLE{Rodriguez-Castillo2020,
       author = {{Rodr{\'\i}guez Castillo}, G.~A. and {Israel}, G.~L. and {Belfiore}, A. and {Bernardini}, F. and {Esposito}, P. and {Pintore}, F. and {De Luca}, A. and {Papitto}, A. and {Stella}, L. and {Tiengo}, A. and {Zampieri}, L. and {Bachetti}, M. and {Brightman}, M. and {Casella}, P. and {D'Agostino}, D. and {Dall'Osso}, S. and {Earnshaw}, H.~P. and {F{\"u}rst}, F. and {Haberl}, F. and {Harrison}, F.~A. and {Mapelli}, M. and {Marelli}, M. and {Middleton}, M. and {Pinto}, C. and {Roberts}, T.~P. and {Salvaterra}, R. and {Turolla}, R. and {Walton}, D.~J. and {Wolter}, A.},
        title = "{Discovery of a 2.8 s Pulsar in a 2 Day Orbit High-mass X-Ray Binary Powering the Ultraluminous X-Ray Source ULX-7 in M51}",
      journal = {\apj},
         year = 2020,
        month = may,
       volume = {895},
       number = {1},
          eid = {60},
        pages = {60},
          doi = {10.3847/1538-4357/ab8a44},
archivePrefix = {arXiv},
       eprint = {1906.04791},
 primaryClass = {astro-ph.HE},
       adsurl = {https://ui.adsabs.harvard.edu/abs/2020ApJ...895...60R}
}

@ARTICLE{Cappellari2017,
       author = {{Cappellari}, Michele},
        title = "{Improving the full spectrum fitting method: accurate convolution with Gauss-Hermite functions}",
      journal = {\mnras},
         year = 2017,
        month = apr,
       volume = {466},
       number = {1},
        pages = {798-811},
          doi = {10.1093/mnras/stw3020},
archivePrefix = {arXiv},
       eprint = {1607.08538},
 primaryClass = {astro-ph.GA},
       adsurl = {https://ui.adsabs.harvard.edu/abs/2017MNRAS.466..798C}
}

@ARTICLE{Martin2005,
       author = {{Martin}, D. Christopher and {Fanson}, James and {Schiminovich}, David and {Morrissey}, Patrick and {Friedman}, Peter G. and {Barlow}, Tom A. and {Conrow}, Tim and {Grange}, Robert and {Jelinsky}, Patrick N. and {Milliard}, Bruno and {Siegmund}, Oswald H.~W. and {Bianchi}, Luciana and {Byun}, Yong-Ik and {Donas}, Jose and {Forster}, Karl and {Heckman}, Timothy M. and {Lee}, Young-Wook and {Madore}, Barry F. and {Malina}, Roger F. and {Neff}, Susan G. and {Rich}, R. Michael and {Small}, Todd and {Surber}, Frank and {Szalay}, Alex S. and {Welsh}, Barry and {Wyder}, Ted K.},
        title = "{The Galaxy Evolution Explorer: A Space Ultraviolet Survey Mission}",
      journal = {\apjl},
         year = 2005,
        month = jan,
       volume = {619},
       number = {1},
        pages = {L1-L6},
          doi = {10.1086/426387},
archivePrefix = {arXiv},
       eprint = {astro-ph/0411302},
 primaryClass = {astro-ph},
       adsurl = {https://ui.adsabs.harvard.edu/abs/2005ApJ...619L...1M}
}

@ARTICLE{Lehmer2019,
       author = {{Lehmer}, Bret D. and {Eufrasio}, Rafael T. and {Tzanavaris}, Panayiotis and {Basu-Zych}, Antara and {Fragos}, Tassos and {Prestwich}, Andrea and {Yukita}, Mihoko and {Zezas}, Andreas and {Hornschemeier}, Ann E. and {Ptak}, Andrew},
        title = "{X-Ray Binary Luminosity Function Scaling Relations for Local Galaxies Based on Subgalactic Modeling}",
      journal = {\apjs},
         year = 2019,
        month = jul,
       volume = {243},
       number = {1},
          eid = {3},
        pages = {3},
          doi = {10.3847/1538-4365/ab22a8},
archivePrefix = {arXiv},
       eprint = {1905.05197},
 primaryClass = {astro-ph.GA},
       adsurl = {https://ui.adsabs.harvard.edu/abs/2019ApJS..243....3L}
}

@ARTICLE{Hao2011,
       author = {{Hao}, Cai-Na and {Kennicutt}, Robert C. and {Johnson}, Benjamin D. and {Calzetti}, Daniela and {Dale}, Daniel A. and {Moustakas}, John},
        title = "{Dust-corrected Star Formation Rates of Galaxies. II. Combinations of Ultraviolet and Infrared Tracers}",
      journal = {\apj},
         year = 2011,
        month = nov,
       volume = {741},
       number = {2},
          eid = {124},
        pages = {124},
          doi = {10.1088/0004-637X/741/2/124},
archivePrefix = {arXiv},
       eprint = {1108.2837},
 primaryClass = {astro-ph.CO},
       adsurl = {https://ui.adsabs.harvard.edu/abs/2011ApJ...741..124H}
}

@ARTICLE{Cenarro2001,
       author = {{Cenarro}, A.~J. and {Cardiel}, N. and {Gorgas}, J. and {Peletier}, R.~F. and {Vazdekis}, A. and {Prada}, F.},
        title = "{Empirical calibration of the near-infrared Ca ii triplet - I. The stellar library and index definition}",
      journal = {\mnras},
         year = 2001,
        month = sep,
       volume = {326},
       number = {3},
        pages = {959-980},
          doi = {10.1046/j.1365-8711.2001.04688.x},
archivePrefix = {arXiv},
       eprint = {astro-ph/0109157},
 primaryClass = {astro-ph},
       adsurl = {https://ui.adsabs.harvard.edu/abs/2001MNRAS.326..959C}
}

@ARTICLE{Cappellari2023,
       author = {{Cappellari}, Michele},
        title = "{Full spectrum fitting with photometry in PPXF: stellar population versus dynamical masses, non-parametric star formation history and metallicity for 3200 LEGA-C galaxies at redshift z {\ensuremath{\approx}} 0.8}",
      journal = {\mnras},
         year = 2023,
        month = dec,
       volume = {526},
       number = {3},
        pages = {3273-3300},
          doi = {10.1093/mnras/stad2597},
archivePrefix = {arXiv},
       eprint = {2208.14974},
 primaryClass = {astro-ph.GA},
       adsurl = {https://ui.adsabs.harvard.edu/abs/2023MNRAS.526.3273C}
}

@ARTICLE{Inami2013,
       author = {{Inami}, H. and {Armus}, L. and {Charmandaris}, V. and {Groves}, B. and {Kewley}, L. and {Petric}, A. and {Stierwalt}, S. and {D{\'\i}az-Santos}, T. and {Surace}, J. and {Rich}, J. and {Haan}, S. and {Howell}, J. and {Evans}, A.~S. and {Mazzarella}, J. and {Marshall}, J. and {Appleton}, P. and {Lord}, S. and {Spoon}, H. and {Frayer}, D. and {Matsuhara}, H. and {Veilleux}, S.},
        title = "{Mid-infrared Atomic Fine-structure Emission-line Spectra of Luminous Infrared Galaxies: Spitzer/IRS Spectra of the GOALS Sample}",
      journal = {\apj},
         year = 2013,
        month = nov,
       volume = {777},
       number = {2},
          eid = {156},
        pages = {156},
          doi = {10.1088/0004-637X/777/2/156},
archivePrefix = {arXiv},
       eprint = {1309.4788},
 primaryClass = {astro-ph.CO},
       adsurl = {https://ui.adsabs.harvard.edu/abs/2013ApJ...777..156I}
}

@ARTICLE{Houck2004,
       author = {{Houck}, J.~R. and {Roellig}, T.~L. and {van Cleve}, J. and {Forrest}, W.~J. and {Herter}, T. and {Lawrence}, C.~R. and {Matthews}, K. and {Reitsema}, H.~J. and {Soifer}, B.~T. and {Watson}, D.~M. and {Weedman}, D. and {Huisjen}, M. and {Troeltzsch}, J. and {Barry}, D.~J. and {Bernard-Salas}, J. and {Blacken}, C.~E. and {Brandl}, B.~R. and {Charmandaris}, V. and {Devost}, D. and {Gull}, G.~E. and {Hall}, P. and {Henderson}, C.~P. and {Higdon}, S.~J.~U. and {Pirger}, B.~E. and {Schoenwald}, J. and {Sloan}, G.~C. and {Uchida}, K.~I. and {Appleton}, P.~N. and {Armus}, L. and {Burgdorf}, M.~J. and {Fajardo-Acosta}, S.~B. and {Grillmair}, C.~J. and {Ingalls}, J.~G. and {Morris}, P.~W. and {Teplitz}, H.~I.},
        title = "{The Infrared Spectrograph (IRS) on the Spitzer Space Telescope}",
      journal = {\apjs},
         year = 2004,
        month = sep,
       volume = {154},
       number = {1},
        pages = {18-24},
          doi = {10.1086/423134},
archivePrefix = {arXiv},
       eprint = {astro-ph/0406167},
 primaryClass = {astro-ph},
       adsurl = {https://ui.adsabs.harvard.edu/abs/2004ApJS..154...18H}
}

@ARTICLE{Armus2009,
       author = {{Armus}, L. and {Mazzarella}, J.~M. and {Evans}, A.~S. and {Surace}, J.~A. and {Sanders}, D.~B. and {Iwasawa}, K. and {Frayer}, D.~T. and {Howell}, J.~H. and {Chan}, B. and {Petric}, A. and {Vavilkin}, T. and {Kim}, D.~C. and {Haan}, S. and {Inami}, H. and {Murphy}, E.~J. and {Appleton}, P.~N. and {Barnes}, J.~E. and {Bothun}, G. and {Bridge}, C.~R. and {Charmandaris}, V. and {Jensen}, J.~B. and {Kewley}, L.~J. and {Lord}, S. and {Madore}, B.~F. and {Marshall}, J.~A. and {Melbourne}, J.~E. and {Rich}, J. and {Satyapal}, S. and {Schulz}, B. and {Spoon}, H.~W.~W. and {Sturm}, E. and {U}, V. and {Veilleux}, S. and {Xu}, K.},
        title = "{GOALS: The Great Observatories All-Sky LIRG Survey}",
      journal = {\pasp},
         year = 2009,
        month = jun,
       volume = {121},
       number = {880},
        pages = {559},
          doi = {10.1086/600092},
archivePrefix = {arXiv},
       eprint = {0904.4498},
 primaryClass = {astro-ph.CO},
       adsurl = {https://ui.adsabs.harvard.edu/abs/2009PASP..121..559A}
}

@ARTICLE{Zaw2020,
       author = {{Zaw}, Ingyin and {Rosenthal}, Michael J. and {Katkov}, Ivan Yu. and {Gelfand}, Joseph D. and {Chen}, Yan-Ping and {Greenhill}, Lincoln J. and {Brisken}, Walter and {Noori}, Hind Al},
        title = "{An Accreting, Anomalously Low-mass Black Hole at the Center of Low-mass Galaxy IC 750}",
      journal = {\apj},
         year = 2020,
        month = jul,
       volume = {897},
       number = {2},
          eid = {111},
        pages = {111},
          doi = {10.3847/1538-4357/ab9944},
archivePrefix = {arXiv},
       eprint = {2006.01114},
 primaryClass = {astro-ph.GA},
       adsurl = {https://ui.adsabs.harvard.edu/abs/2020ApJ...897..111Z}
}

@ARTICLE{Emonts2014,
       author = {{Emonts}, B.~H.~C. and {Piqueras-L{\'o}pez}, J. and {Colina}, L. and {Arribas}, S. and {Villar-Mart{\'\i}n}, M. and {Pereira-Santaella}, M. and {Garcia-Burillo}, S. and {Alonso-Herrero}, A.},
        title = "{Outflow of hot and cold molecular gas from the obscured secondary nucleus of NGC 3256: closing in on feedback physics}",
      journal = {\aap},
         year = 2014,
        month = dec,
       volume = {572},
          eid = {A40},
        pages = {A40},
          doi = {10.1051/0004-6361/201423805},
archivePrefix = {arXiv},
       eprint = {1409.4468},
 primaryClass = {astro-ph.GA},
       adsurl = {https://ui.adsabs.harvard.edu/abs/2014A&A...572A..40E}
}

@ARTICLE{Caputi2007,
       author = {{Caputi}, K.~I. and {Lagache}, G. and {Yan}, Lin and {Dole}, H. and {Bavouzet}, N. and {Le Floc'h}, E. and {Choi}, P.~I. and {Helou}, G. and {Reddy}, N.},
        title = "{The Infrared Luminosity Function of Galaxies at Redshifts z = 1 and z \raisebox{-0.5ex}\textasciitilde 2 in the GOODS Fields}",
      journal = {\apj},
         year = 2007,
        month = may,
       volume = {660},
       number = {1},
        pages = {97-116},
          doi = {10.1086/512667},
archivePrefix = {arXiv},
       eprint = {astro-ph/0701283},
 primaryClass = {astro-ph},
       adsurl = {https://ui.adsabs.harvard.edu/abs/2007ApJ...660...97C}
}

@ARTICLE{Magnelli2011,
       author = {{Magnelli}, B. and {Elbaz}, D. and {Chary}, R.~R. and {Dickinson}, M. and {Le Borgne}, D. and {Frayer}, D.~T. and {Willmer}, C.~N.~A.},
        title = "{Evolution of the dusty infrared luminosity function from z = 0 to z = 2.3 using observations from Spitzer}",
      journal = {\aap},
         year = 2011,
        month = apr,
       volume = {528},
          eid = {A35},
        pages = {A35},
          doi = {10.1051/0004-6361/200913941},
archivePrefix = {arXiv},
       eprint = {1101.2467},
 primaryClass = {astro-ph.CO},
       adsurl = {https://ui.adsabs.harvard.edu/abs/2011A&A...528A..35M}
}

@ARTICLE{Larson2016,
       author = {{Larson}, K.~L. and {Sanders}, D.~B. and {Barnes}, J.~E. and {Ishida}, C.~M. and {Evans}, A.~S. and {U}, V. and {Mazzarella}, J.~M. and {Kim}, D.-C. and {Privon}, G.~C. and {Mirabel}, I.~F. and {Flewelling}, H.~A.},
        title = "{Morphology and Molecular Gas Fractions of Local Luminous Infrared Galaxies as a Function of Infrared Luminosity and Merger Stage}",
      journal = {\apj},
         year = 2016,
        month = jul,
       volume = {825},
       number = {2},
          eid = {128},
        pages = {128},
          doi = {10.3847/0004-637X/825/2/128},
archivePrefix = {arXiv},
       eprint = {1605.05417},
 primaryClass = {astro-ph.GA},
       adsurl = {https://ui.adsabs.harvard.edu/abs/2016ApJ...825..128L}
}

@ARTICLE{Sanders1996,
       author = {{Sanders}, D.~B. and {Mirabel}, I.~F.},
        title = "{Luminous Infrared Galaxies}",
      journal = {\araa},
         year = 1996,
        month = jan,
       volume = {34},
        pages = {749},
          doi = {10.1146/annurev.astro.34.1.749},
       adsurl = {https://ui.adsabs.harvard.edu/abs/1996ARA&A..34..749S}
}

@ARTICLE{Mapelli2010,
       author = {{Mapelli}, M. and {Ripamonti}, E. and {Zampieri}, L. and {Colpi}, M. and {Bressan}, A.},
        title = "{Ultra-luminous X-ray sources and remnants of massive metal-poor stars}",
      journal = {\mnras},
         year = 2010,
        month = oct,
       volume = {408},
       number = {1},
        pages = {234-253},
          doi = {10.1111/j.1365-2966.2010.17048.x},
archivePrefix = {arXiv},
       eprint = {1005.3548},
 primaryClass = {astro-ph.CO},
       adsurl = {https://ui.adsabs.harvard.edu/abs/2010MNRAS.408..234M}
}

@ARTICLE{Walton2022,
       author = {{Walton}, D.~J. and {Mackenzie}, A.~D.~A. and {Gully}, H. and {Patel}, N.~R. and {Roberts}, T.~P. and {Earnshaw}, H.~P. and {Mateos}, S.},
        title = "{A multimission catalogue of ultraluminous X-ray source candidates}",
      journal = {\mnras},
         year = 2022,
        month = jan,
       volume = {509},
       number = {2},
        pages = {1587-1604},
          doi = {10.1093/mnras/stab3001},
archivePrefix = {arXiv},
       eprint = {2110.07625},
 primaryClass = {astro-ph.HE},
       adsurl = {https://ui.adsabs.harvard.edu/abs/2022MNRAS.509.1587W}
}

@ARTICLE{Bachetti2014,
       author = {{Bachetti}, M. and {Harrison}, F.~A. and {Walton}, D.~J. and {Grefenstette}, B.~W. and {Chakrabarty}, D. and {F{\"u}rst}, F. and {Barret}, D. and {Beloborodov}, A. and {Boggs}, S.~E. and {Christensen}, F.~E. and {Craig}, W.~W. and {Fabian}, A.~C. and {Hailey}, C.~J. and {Hornschemeier}, A. and {Kaspi}, V. and {Kulkarni}, S.~R. and {Maccarone}, T. and {Miller}, J.~M. and {Rana}, V. and {Stern}, D. and {Tendulkar}, S.~P. and {Tomsick}, J. and {Webb}, N.~A. and {Zhang}, W.~W.},
        title = "{An ultraluminous X-ray source powered by an accreting neutron star}",
      journal = {\nat},
         year = 2014,
        month = oct,
       volume = {514},
       number = {7521},
        pages = {202-204},
          doi = {10.1038/nature13791},
archivePrefix = {arXiv},
       eprint = {1410.3590},
 primaryClass = {astro-ph.HE},
       adsurl = {https://ui.adsabs.harvard.edu/abs/2014Natur.514..202B}
}

@ARTICLE{Mezcua2015,
       author = {{Mezcua}, M. and {Roberts}, T.~P. and {Lobanov}, A.~P. and {Sutton}, A.~D.},
        title = "{The powerful jet of an off-nuclear intermediate-mass black hole in the spiral galaxy NGC 2276}",
      journal = {\mnras},
         year = 2015,
        month = apr,
       volume = {448},
       number = {2},
        pages = {1893-1899},
          doi = {10.1093/mnras/stv143},
archivePrefix = {arXiv},
       eprint = {1501.04897},
 primaryClass = {astro-ph.GA},
       adsurl = {https://ui.adsabs.harvard.edu/abs/2015MNRAS.448.1893M}
}

@ARTICLE{Mezcua2018,
       author = {{Mezcua}, M. and {Kim}, M. and {Ho}, L.~C. and {Lonsdale}, C.~J.},
        title = "{The extended radio jet of an off-nuclear low-mass AGN in NGC 5252}",
      journal = {\mnras},
         year = 2018,
        month = oct,
       volume = {480},
       number = {1},
        pages = {L74-L78},
          doi = {10.1093/mnrasl/sly130},
archivePrefix = {arXiv},
       eprint = {1807.03792},
 primaryClass = {astro-ph.GA},
       adsurl = {https://ui.adsabs.harvard.edu/abs/2018MNRAS.480L..74M}
}

@ARTICLE{Davis2011,
       author = {{Davis}, Shane W. and {Narayan}, Ramesh and {Zhu}, Yucong and {Barret}, Didier and {Farrell}, Sean A. and {Godet}, Olivier and {Servillat}, Mathieu and {Webb}, Natalie A.},
        title = "{The Cool Accretion Disk in ESO 243-49 HLX-1: Further Evidence of an Intermediate-mass Black Hole}",
      journal = {\apj},
         year = 2011,
        month = jun,
       volume = {734},
       number = {2},
          eid = {111},
        pages = {111},
          doi = {10.1088/0004-637X/734/2/111},
archivePrefix = {arXiv},
       eprint = {1104.2614},
 primaryClass = {astro-ph.HE},
       adsurl = {https://ui.adsabs.harvard.edu/abs/2011ApJ...734..111D}
}

@ARTICLE{Chilingarian2018,
       author = {{Chilingarian}, Igor V. and {Katkov}, Ivan Yu. and {Zolotukhin}, Ivan Yu. and {Grishin}, Kirill A. and {Beletsky}, Yuri and {Boutsia}, Konstantina and {Osip}, David J.},
        title = "{A Population of Bona Fide Intermediate-mass Black Holes Identified as Low-luminosity Active Galactic Nuclei}",
      journal = {\apj},
         year = 2018,
        month = aug,
       volume = {863},
       number = {1},
          eid = {1},
        pages = {1},
          doi = {10.3847/1538-4357/aad184},
archivePrefix = {arXiv},
       eprint = {1805.01467},
 primaryClass = {astro-ph.GA},
       adsurl = {https://ui.adsabs.harvard.edu/abs/2018ApJ...863....1C}
}

@ARTICLE{Kaaret2017,
       author = {{Kaaret}, Philip and {Feng}, Hua and {Roberts}, Timothy P.},
        title = "{Ultraluminous X-Ray Sources}",
      journal = {\araa},
         year = 2017,
        month = aug,
       volume = {55},
       number = {1},
        pages = {303-341},
          doi = {10.1146/annurev-astro-091916-055259},
archivePrefix = {arXiv},
       eprint = {1703.10728},
 primaryClass = {astro-ph.HE},
       adsurl = {https://ui.adsabs.harvard.edu/abs/2017ARA&A..55..303K}
}

@ARTICLE{Fabbiano1987,
       author = {{Fabbiano}, G. and {Trinchieri}, G.},
        title = "{X-Ray Observations of Spiral Galaxies. II. Images and Spectral Parameters of 13 Galaxies}",
      journal = {\apj},
         year = 1987,
        month = apr,
       volume = {315},
        pages = {46},
          doi = {10.1086/165113},
       adsurl = {https://ui.adsabs.harvard.edu/abs/1987ApJ...315...46F}
}

@ARTICLE{Long1981,
       author = {{Long}, K.~S. and {Dodorico}, S. and {Charles}, P.~A. and {Dopita}, M.~A.},
        title = "{Observations of the X-ray sources in the nearby SC galaxy M 33.}",
      journal = {\apjl},
         year = 1981,
        month = jun,
       volume = {246},
        pages = {L61-L64},
          doi = {10.1086/183553},
       adsurl = {https://ui.adsabs.harvard.edu/abs/1981ApJ...246L..61L}
}

@ARTICLE{King2023,
       author = {{King}, Andrew and {Lasota}, Jean-Pierre and {Middleton}, Matthew},
        title = "{Ultraluminous X-ray sources}",
      journal = {\nar},
         year = 2023,
        month = jun,
       volume = {96},
          eid = {101672},
        pages = {101672},
          doi = {10.1016/j.newar.2022.101672},
archivePrefix = {arXiv},
       eprint = {2302.10605},
 primaryClass = {astro-ph.HE},
       adsurl = {https://ui.adsabs.harvard.edu/abs/2023NewAR..9601672K}
}

@ARTICLE{Salvaggio2023,
       author = {{Salvaggio}, Chiara and {Wolter}, A. and {Belfiore}, A. and {Colpi}, M.},
        title = "{The largest bright ULX population in a galaxy: X-ray variability and luminosity function in the Cartwheel ring galaxy}",
      journal = {\mnras},
         year = 2023,
        month = jun,
       volume = {522},
       number = {1},
        pages = {1377-1393},
          doi = {10.1093/mnras/stad943},
archivePrefix = {arXiv},
       eprint = {2303.15502},
 primaryClass = {astro-ph.HE},
       adsurl = {https://ui.adsabs.harvard.edu/abs/2023MNRAS.522.1377S}
}

@ARTICLE{Gao2003,
       author = {{Gao}, Yu and {Wang}, Q. Daniel and {Appleton}, P.~N. and {Lucas}, Ray A.},
        title = "{Nonnuclear Hyper/Ultraluminous X-Ray Sources in the Starbursting Cartwheel Ring Galaxy}",
      journal = {\apjl},
         year = 2003,
        month = oct,
       volume = {596},
       number = {2},
        pages = {L171-L174},
          doi = {10.1086/379598},
archivePrefix = {arXiv},
       eprint = {astro-ph/0309253},
 primaryClass = {astro-ph},
       adsurl = {https://ui.adsabs.harvard.edu/abs/2003ApJ...596L.171G}
}

@ARTICLE{Wolter2004,
       author = {{Wolter}, A. and {Trinchieri}, G.},
        title = "{A thorough study of the intriguing X-ray emission from the Cartwheel ring}",
      journal = {\aap},
         year = 2004,
        month = nov,
       volume = {426},
        pages = {787-796},
          doi = {10.1051/0004-6361:20047110},
archivePrefix = {arXiv},
       eprint = {astro-ph/0407446},
 primaryClass = {astro-ph},
       adsurl = {https://ui.adsabs.harvard.edu/abs/2004A&A...426..787W}
}

@ARTICLE{Gaia2023,
       author = {{Gaia Collaboration} and {Vallenari}, A. and {Brown}, A.~G.~A. and {Prusti}, T. and {de Bruijne}, J.~H.~J. and {Arenou}, F. and {Babusiaux}, C. and {Biermann}, M. and {Creevey}, O.~L. and {Ducourant}, C. and {Evans}, D.~W. and {Eyer}, L. and {Guerra}, R. and {Hutton}, A. and {Jordi}, C. and {Klioner}, S.~A. and {Lammers}, U.~L. and {Lindegren}, L. and {Luri}, X. and {Mignard}, F. and {Panem}, C. and {Pourbaix}, D. and {Randich}, S. and {Sartoretti}, P. and {Soubiran}, C. and {Tanga}, P. and {Walton}, N.~A. and {Bailer-Jones}, C.~A.~L. and {Bastian}, U. and {Drimmel}, R. and {Jansen}, F. and {Katz}, D. and {Lattanzi}, M.~G. and {van Leeuwen}, F. and {Bakker}, J. and {Cacciari}, C. and {Casta{\~n}eda}, J. and {De Angeli}, F. and {Fabricius}, C. and {Fouesneau}, M. and {Fr{\'e}mat}, Y. and {Galluccio}, L. and {Guerrier}, A. and {Heiter}, U. and {Masana}, E. and {Messineo}, R. and {Mowlavi}, N. and {Nicolas}, C. and {Nienartowicz}, K. and {Pailler}, F. and {Panuzzo}, P. and {Riclet}, F. and {Roux}, W. and {Seabroke}, G.~M. and {Sordo}, R. and {Th{\'e}venin}, F. and {Gracia-Abril}, G. and {Portell}, J. and {Teyssier}, D. and {Altmann}, M. and {Andrae}, R. and {Audard}, M. and {Bellas-Velidis}, I. and {Benson}, K. and {Berthier}, J. and {Blomme}, R. and {Burgess}, P.~W. and {Busonero}, D. and {Busso}, G. and {C{\'a}novas}, H. and {Carry}, B. and {Cellino}, A. and {Cheek}, N. and {Clementini}, G. and {Damerdji}, Y. and {Davidson}, M. and {de Teodoro}, P. and {Nu{\~n}ez Campos}, M. and {Delchambre}, L. and {Dell'Oro}, A. and {Esquej}, P. and {Fern{\'a}ndez-Hern{\'a}ndez}, J. and {Fraile}, E. and {Garabato}, D. and {Garc{\'\i}a-Lario}, P. and {Gosset}, E. and {Haigron}, R. and {Halbwachs}, J.-L. and {Hambly}, N.~C. and {Harrison}, D.~L. and {Hern{\'a}ndez}, J. and {Hestroffer}, D. and {Hodgkin}, S.~T. and {Holl}, B. and {Jan{\ss}en}, K. and {Jevardat de Fombelle}, G. and {Jordan}, S. and {Krone-Martins}, A. and {Lanzafame}, A.~C. and {L{\"o}ffler}, W. and {Marchal}, O. and {Marrese}, P.~M. and {Moitinho}, A. and {Muinonen}, K. and {Osborne}, P. and {Pancino}, E. and {Pauwels}, T. and {Recio-Blanco}, A. and {Reyl{\'e}}, C. and {Riello}, M. and {Rimoldini}, L. and {Roegiers}, T. and {Rybizki}, J. and {Sarro}, L.~M. and {Siopis}, C. and {Smith}, M. and {Sozzetti}, A. and {Utrilla}, E. and {van Leeuwen}, M. and {Abbas}, U. and {{\'A}brah{\'a}m}, P. and {Abreu Aramburu}, A. and {Aerts}, C. and {Aguado}, J.~J. and {Ajaj}, M. and {Aldea-Montero}, F. and {Altavilla}, G. and {{\'A}lvarez}, M.~A. and {Alves}, J. and {Anders}, F. and {Anderson}, R.~I. and {Anglada Varela}, E. and {Antoja}, T. and {Baines}, D. and {Baker}, S.~G. and {Balaguer-N{\'u}{\~n}ez}, L. and {Balbinot}, E. and {Balog}, Z. and {Barache}, C. and {Barbato}, D. and {Barros}, M. and {Barstow}, M.~A. and {Bartolom{\'e}}, S. and {Bassilana}, J.-L. and {Bauchet}, N. and {Becciani}, U. and {Bellazzini}, M. and {Berihuete}, A. and {Bernet}, M. and {Bertone}, S. and {Bianchi}, L. and {Binnenfeld}, A. and {Blanco-Cuaresma}, S. and {Blazere}, A. and {Boch}, T. and {Bombrun}, A. and {Bossini}, D. and {Bouquillon}, S. and {Bragaglia}, A. and {Bramante}, L. and {Breedt}, E. and {Bressan}, A. and {Brouillet}, N. and {Brugaletta}, E. and {Bucciarelli}, B. and {Burlacu}, A. and {Butkevich}, A.~G. and {Buzzi}, R. and {Caffau}, E. and {Cancelliere}, R. and {Cantat-Gaudin}, T. and {Carballo}, R. and {Carlucci}, T. and {Carnerero}, M.~I. and {Carrasco}, J.~M. and {Casamiquela}, L. and {Castellani}, M. and {Castro-Ginard}, A. and {Chaoul}, L. and {Charlot}, P. and {Chemin}, L. and {Chiaramida}, V. and {Chiavassa}, A. and {Chornay}, N. and {Comoretto}, G. and {Contursi}, G. and {Cooper}, W.~J. and {Cornez}, T. and {Cowell}, S. and {Crifo}, F. and {Cropper}, M. and {Crosta}, M. and {Crowley}, C. and {Dafonte}, C. and {Dapergolas}, A. and {David}, M. and {David}, P. and {de Laverny}, P. and {De Luise}, F. and {De March}, R.},
        title = "{Gaia Data Release 3. Summary of the content and survey properties}",
      journal = {\aap},
         year = 2023,
        month = jun,
       volume = {674},
          eid = {A1},
        pages = {A1},
          doi = {10.1051/0004-6361/202243940},
archivePrefix = {arXiv},
       eprint = {2208.00211},
 primaryClass = {astro-ph.GA},
       adsurl = {https://ui.adsabs.harvard.edu/abs/2023A&A...674A...1G}
}

@INPROCEEDINGS{Bacon2010,
       author = {{Bacon}, R. and {Accardo}, M. and {Adjali}, L. and {Anwand}, H. and {Bauer}, S. and {Biswas}, I. and {Blaizot}, J. and {Boudon}, D. and {Brau-Nogue}, S. and {Brinchmann}, J. and {Caillier}, P. and {Capoani}, L. and {Carollo}, C.~M. and {Contini}, T. and {Couderc}, P. and {Daguis{\'e}}, E. and {Deiries}, S. and {Delabre}, B. and {Dreizler}, S. and {Dubois}, J. and {Dupieux}, M. and {Dupuy}, C. and {Emsellem}, E. and {Fechner}, T. and {Fleischmann}, A. and {Fran{\c{c}}ois}, M. and {Gallou}, G. and {Gharsa}, T. and {Glindemann}, A. and {Gojak}, D. and {Guiderdoni}, B. and {Hansali}, G. and {Hahn}, T. and {Jarno}, A. and {Kelz}, A. and {Koehler}, C. and {Kosmalski}, J. and {Laurent}, F. and {Le Floch}, M. and {Lilly}, S.~J. and {Lizon}, J.-L. and {Loupias}, M. and {Manescau}, A. and {Monstein}, C. and {Nicklas}, H. and {Olaya}, J.-C. and {Pares}, L. and {Pasquini}, L. and {P{\'e}contal-Rousset}, A. and {Pell{\'o}}, R. and {Petit}, C. and {Popow}, E. and {Reiss}, R. and {Remillieux}, A. and {Renault}, E. and {Roth}, M. and {Rupprecht}, G. and {Serre}, D. and {Schaye}, J. and {Soucail}, G. and {Steinmetz}, M. and {Streicher}, O. and {Stuik}, R. and {Valentin}, H. and {Vernet}, J. and {Weilbacher}, P. and {Wisotzki}, L. and {Yerle}, N.},
        title = "{The MUSE second-generation VLT instrument}",
    booktitle = {Ground-based and Airborne Instrumentation for Astronomy III},
         year = 2010,
       editor = {{McLean}, Ian S. and {Ramsay}, Suzanne K. and {Takami}, Hideki},
       series = {Society of Photo-Optical Instrumentation Engineers (SPIE) Conference Series},
       volume = {7735},
        month = jul,
          eid = {773508},
        pages = {773508},
          doi = {10.1117/12.856027},
archivePrefix = {arXiv},
       eprint = {2211.16795},
 primaryClass = {astro-ph.IM},
       adsurl = {https://ui.adsabs.harvard.edu/abs/2010SPIE.7735E..08B}
}

@ARTICLE{Ohyama2015,
       author = {{Ohyama}, Youichi and {Terashima}, Yuichi and {Sakamoto}, Kazushi},
        title = "{Infrared and X-Ray Evidence of an AGN in the NGC 3256 Southern Nucleus}",
      journal = {\apj},
         year = 2015,
        month = jun,
       volume = {805},
       number = {2},
          eid = {162},
        pages = {162},
          doi = {10.1088/0004-637X/805/2/162},
archivePrefix = {arXiv},
       eprint = {1503.08555},
 primaryClass = {astro-ph.GA},
       adsurl = {https://ui.adsabs.harvard.edu/abs/2015ApJ...805..162O}
}

@ARTICLE{Linden2024,
       author = {{Linden}, Sean T. and {Lai}, Thomas and {Evans}, Aaron S. and {Armus}, Lee and {Larson}, Kirsten L. and {Rich}, Jeffrey A. and {U}, Vivian and {Privon}, George C. and {Inami}, Hanae and {Song}, Yiqing and {Bianchin}, Marina and {Bohn}, Thomas and {Buiten}, Victorine A. and {Sanchez-Garc{\'\i}a}, Maria and {Kader}, Justin and {Lenki{\'c}}, Laura and {Medling}, Anne M. and {B{\"o}ker}, Torsten and {D{\'\i}az-Santos}, Tanio and {Charmandaris}, Vassilis and {Barcos-Mu{\~n}oz}, Loreto and {van der Werf}, Paul and {Stierwalt}, Sabrina and {Aalto}, Susanne and {Appleton}, Philip and {Hayward}, Christopher C. and {Howell}, Justin H. and {Malkan}, Matthew A. and {Mazzarella}, Joseph M. and {Murphy}, Eric J. and {Surace}, Jason},
        title = "{GOALS-JWST: Constraining the Emergence Timescale for Massive Star Clusters in NGC 3256}",
      journal = {\apjl},
         year = 2024,
        month = oct,
       volume = {974},
       number = {2},
          eid = {L27},
        pages = {L27},
          doi = {10.3847/2041-8213/ad7eae},
archivePrefix = {arXiv},
       eprint = {2409.16503},
 primaryClass = {astro-ph.GA},
       adsurl = {https://ui.adsabs.harvard.edu/abs/2024ApJ...974L..27L}
}

@ARTICLE{Kim2013,
       author = {{Kim}, D.-C. and {Evans}, A.~S. and {Vavilkin}, T. and {Armus}, L. and {Mazzarella}, J.~M. and {Sheth}, K. and {Surace}, J.~A. and {Haan}, S. and {Howell}, J.~H. and {D{\'\i}az-Santos}, T. and {Petric}, A. and {Iwasawa}, K. and {Privon}, G.~C. and {Sanders}, D.~B.},
        title = "{Hubble Space Telescope ACS Imaging of the GOALS Sample: Quantitative Structural Properties of Nearby Luminous Infrared Galaxies with L $_{IR}$ > 10$^{11.4}$ L $_{{\ensuremath{\odot}}}$}",
      journal = {\apj},
         year = 2013,
        month = may,
       volume = {768},
       number = {2},
          eid = {102},
        pages = {102},
          doi = {10.1088/0004-637X/768/2/102},
archivePrefix = {arXiv},
       eprint = {1303.3977},
 primaryClass = {astro-ph.CO},
       adsurl = {https://ui.adsabs.harvard.edu/abs/2013ApJ...768..102K}
}

@ARTICLE{Lehmer2015,
       author = {{Lehmer}, B.~D. and {Tyler}, J.~B. and {Hornschemeier}, A.~E. and {Wik}, D.~R. and {Yukita}, M. and {Antoniou}, V. and {Boggs}, S. and {Christensen}, F.~E. and {Craig}, W.~W. and {Hailey}, C.~J. and {Harrison}, F.~A. and {Maccarone}, T.~J. and {Ptak}, A. and {Stern}, D. and {Zezas}, A. and {Zhang}, W.~W.},
        title = "{The 0.3-30 keV Spectra of Powerful Starburst Galaxies: NuSTAR and Chandra Observations of NGC 3256 and NGC 3310}",
      journal = {\apj},
         year = 2015,
        month = jun,
       volume = {806},
       number = {1},
          eid = {126},
        pages = {126},
          doi = {10.1088/0004-637X/806/1/126},
archivePrefix = {arXiv},
       eprint = {1505.00789},
 primaryClass = {astro-ph.GA},
       adsurl = {https://ui.adsabs.harvard.edu/abs/2015ApJ...806..126L}
}

@ARTICLE{Liu2011,
       author = {{Liu}, Jifeng},
        title = "{Chandra ACIS Survey of X-ray Point Sources in 383 Nearby Galaxies. I. The Source Catalog}",
      journal = {\apjs},
         year = 2011,
        month = jan,
       volume = {192},
       number = {1},
          eid = {10},
        pages = {10},
          doi = {10.1088/0067-0049/192/1/10},
archivePrefix = {arXiv},
       eprint = {0811.0804},
 primaryClass = {astro-ph},
       adsurl = {https://ui.adsabs.harvard.edu/abs/2011ApJS..192...10L}
}

@ARTICLE{Evans_csc2024,
       author = {{Evans}, Ian N. and {Evans}, Janet D. and {Mart{\'\i}nez-Galarza}, J. Rafael and {Miller}, Joseph B. and {Primini}, Francis A. and {Azadi}, Mojegan and {Burke}, Douglas J. and {Civano}, Francesca M. and {D'Abrusco}, Raffaele and {Fabbiano}, Giuseppina and {Graessle}, Dale E. and {Grier}, John D. and {Houck}, John C. and {Lauer}, Jennifer and {McCollough}, Michael L. and {Nowak}, Michael A. and {Plummer}, David A. and {Rots}, Arnold H. and {Siemiginowska}, Aneta and {Tibbetts}, Michael S.},
        title = "{The Chandra Source Catalog Release 2 Series}",
      journal = {\apjs},
         year = 2024,
        month = oct,
       volume = {274},
       number = {2},
          eid = {22},
        pages = {22},
          doi = {10.3847/1538-4365/ad6319},
archivePrefix = {arXiv},
       eprint = {2407.10799},
 primaryClass = {astro-ph.HE},
       adsurl = {https://ui.adsabs.harvard.edu/abs/2024ApJS..274...22E}
}

@ARTICLE{Li1983,
  author  = {{Li}, T.-P. and {Ma}, Y.-Q.},
  title   = "{Analysis methods for results in gamma-ray astronomy}",
  journal = {\apj},
  year    = 1983,
  month   = sep,
  volume  = {272},
  pages   = {317-324},
  doi     = {10.1086/161295}
}

\begin{appendix}
\onecolumn
\section{Additional tables}\label{App:Tables}

Table \ref{tab:fit_medians} reports the best fit values of all parameters for each point source, for each observation (as well as the 90\% credible interval derived from the posterior distribution), as described in Sec. \ref{sec:results}.

Table \ref{tab:luminosity} reports, for each point source and each observation, the median value of the luminosity posterior, as well as the percent of the posterior above the ULX threshold, as described in Sec. \ref{sec:results}. Reported luminosities are intrinsic, corrected for obscuration.

\begin{table*}[h!]
\scriptsize
\centering
\caption{Fit medians for all models and observations.}
\label{tab:fit_medians}
\renewcommand{\arraystretch}{1.2}
\begin{tabular}{cccccccccc}
\hline\hline
 &  & \multicolumn{2}{c}{powerlaw} & \multicolumn{3}{c}{Abs. Powerlaw} & \multicolumn{3}{c}{Diskpbb}\\
 & & $\Gamma$ & log(norm) & log(N$\rm_H$) & $\Gamma$ & log(norm) & T$_{\rm in}$ & p & log(norm)  \\
\hline
\multirow{3}{*}{PS1} & 835 & $1.21_{-0.97}^{+1.25}$ & $-6.46_{-0.42}^{\,+0.28}$ & $-1.90_{-u}^{\,+1.34}$ & $1.54_{-u}^{\,+u}$ & $-6.33_{-0.31}^{\,+0.26}$ & $3.74_{-2.86}^{\,+u}$ & $0.71_{-u}^{\,+u}$ & $-5.95_{-1.58}^{\,+2.14}$ \\
 & 3569 & $1.46_{-0.55}^{\,+0.56}$ & $-6.02_{-0.21}^{\,+0.17}$ & $-1.80_{-u}^{\,+1.19}$ & $1.61_{-u}^{\,+0.75}$ & $-5.96_{-0.20}^{\,+0.28}$ & $2.36_{-1.38}^{\,+u}$ & $0.65_{-u}^{\,+0.28}$ & $-5.13_{-2.13}^{\,+1.66}$ \\
 & 16026 & $1.58_{-1.23}^{\,+1.25}$ & $-6.13_{-0.47}^{\,+0.35}$ & $-1.17_{-u}^{\,+1.50}$ & $1.99_{-u}^{\,+0.88}$ & $-5.95_{-0.41}^{\,+0.72}$ & $1.77_{-1.15}^{\,+u}$ & $0.72_{-u}^{\,+u}$ & $-4.66_{-2.45}^{\,+1.85}$ \\
\hline
\noalign{\vspace{2pt}}
\multirow{3}{*}{PS2} & 835 & $2.52_{-1.21}^{\,+1.53}$ & $-6.51_{-0.60}^{\,+0.29}$ & $-1.73_{-u}^{\,+1.14}$ & $2.42_{-1.01}^{\,+u}$ & $-6.41_{-0.44}^{\,+0.33}$ & $0.54_{-0.24}^{\,+3.93}$ & $0.68_{-u}^{\,+u}$ & $-3.22_{-3.86}^{\,+1.10}$ \\
 & 3569 & $2.83_{-0.86}^{\,+0.97}$ & $-6.19_{-0.25}^{\,+0.21}$ & $-1.89_{-u}^{\,+1.16}$ & $2.57_{-0.69}^{\,+u}$ & $-6.12_{-0.25}^{\,+0.25}$ & $0.44_{-0.11}^{\,+0.74}$ & $0.67_{-u}^{\,+u}$ & $-2.55_{-2.10}^{\,+0.51}$ \\
 & 16026 & $2.71_{-1.22}^{\,+1.43}$ & $-5.90_{-0.34}^{\,+0.27}$ & $-1.83_{-u}^{\,+1.20}$ & $2.45_{-0.93}^{\,+u}$ & $-5.83_{-0.35}^{\,+0.29}$ & $0.81_{-0.43}^{\,+5.60}$ & $0.62_{-u}^{\,+0.31}$ & $-3.43_{-3.71}^{\,+1.33}$ \\
\hline
\noalign{\vspace{2pt}}
\multirow{3}{*}{PS3} & 835 & $2.34_{-0.66}^{\,+0.81}$ & $-5.99_{-0.17}^{\,+0.14}$ & $-2.06_{-u}^{\,+1.00}$ & $2.42_{-0.69}^{\,+u}$ & $-5.97_{-0.19}^{\,+0.18}$ & $1.44_{-1.01}^{\,+u}$ & $0.54_{-u}^{\,+0.24}$ & $-4.77_{-2.91}^{\,+2.40}$ \\
 & 3569 & $1.82_{-0.64}^{\,+0.66}$ & $-6.04_{-0.20}^{\,+0.18}$ & $-1.74_{-u}^{\,+1.11}$ & $1.94_{-0.63}^{\,+0.78}$ & $-5.99_{-0.22}^{\,+0.28}$ & $1.22_{-0.58}^{\,+6.24}$ & $0.66_{-u}^{\,+u}$ & $-4.14_{-3.06}^{\,+1.34}$ \\
 & 16026 & $1.95_{-1.04}^{\,+1.13}$ & $-6.06_{-0.39}^{\,+0.30}$ & $-1.79_{-u}^{\,+1.25}$ & $2.06_{-0.88}^{\,+u}$ & $-6.01_{-0.36}^{\,+0.34}$ & $2.18_{-1.44}^{\,+u}$ & $0.64_{-u}^{\,+0.31}$ & $-5.23_{-2.28}^{\,+2.08}$ \\
\hline
\noalign{\vspace{2pt}}
\multirow{3}{*}{PS4} & 835 & $3.46_{-1.04}^{\,+1.35}$ & $-6.49_{-0.42}^{\,+0.27}$ & $-2.10_{-u}^{\,+1.06}$ & $2.69_{-0.67}^{\,+u}$ & $-6.36_{-0.30}^{\,+0.24}$ & $0.36_{-0.09}^{\,+0.15}$ & $0.61_{-u}^{\,+0.31}$ & $-2.45_{-2.17}^{\,+0.41}$ \\
 & 3569 & $1.92_{-0.80}^{\,+0.80}$ & $-6.17_{-0.26}^{\,+0.22}$ & $-1.64_{-u}^{\,+1.10}$ & $2.07_{-0.76}^{\,+u}$ & $-6.10_{-0.27}^{\,+0.31}$ & $0.97_{-0.43}^{\,+4.85}$ & $0.69_{-u}^{\,+u}$ & $-3.85_{-3.31}^{\,+1.14}$ \\
 & 16026 & $2.99_{-2.26}^{\,+2.85}$ & $-6.41_{-3.11}^{\,+0.44}$ & $-1.57_{-u}^{\,+1.69}$ & $2.25_{-1.05}^{\,+u}$ & $-6.32_{-2.61}^{\,+0.49}$ & $0.71_{-0.43}^{\,+u}$ & $0.69_{-u}^{\,+u}$ & $-3.95_{-4.32}^{\,+1.79}$ \\
\hline
\noalign{\vspace{2pt}}
\multirow{3}{*}{PS5} & 835 & $2.36_{-4.68}^{\,+u}$ & $-9.31_{-u}^{\,+2.53}$ & $-0.42_{-u}^{\,+u}$ & $2.08_{-u}^{\,+u}$ & $-7.80_{-u}^{\,+1.80}$ & $0.65_{-u}^{\,+u}$ & $0.74_{-u}^{\,+u}$ & $-6.85_{-u}^{\,+3.95}$ \\
 & 3569 & $1.27_{-0.97}^{\,+1.00}$ & $-6.44_{-0.44}^{\,+0.31}$ & $-0.47_{-u}^{\,+0.68}$ & $1.95_{-u}^{\,+u}$ & $-6.08_{-0.49}^{\,+0.66}$ & $1.85_{-1.14}^{\,+u}$ & $0.77_{-u}^{\,+u}$ & $-4.84_{-2.23}^{\,+1.56}$ \\
 & 16026 & - & - & - & - & - & - & - & - \\
\hline
\noalign{\vspace{2pt}}
\multirow{3}{*}{PS6} & 835 & $3.23_{-3.61}^{\,+4.81}$ & $-7.27_{-u}^{\,+0.71}$ & $-1.62_{-u}^{\,+2.93}$ & $2.31_{-1.12}^{\,+u}$ & $-6.96_{-u}^{\,+0.58}$ & $0.56_{-0.37}^{\,+u}$ & $0.70_{-u}^{\,+u}$ & $-4.57_{-u}^{\,+2.37}$ \\
 & 3569 & $1.52_{-0.32}^{\,+0.32}$ & $-5.49_{-0.11}^{\,+0.10}$ & $-0.67_{-u}^{\,+0.34}$ & $2.10_{-0.69}^{\,+0.71}$ & $-5.22_{-0.29}^{\,+0.30}$ & $1.11_{-0.29}^{\,+2.08}$ & $0.79_{-0.20}^{\,+u}$ & $-3.15_{-2.14}^{\,+0.60}$ \\
 & 16026 & $1.65_{-1.14}^{\,+1.58}$ & $-6.10_{-0.43}^{\,+0.32}$ & $-1.69_{-u}^{\,+1.33}$ & $1.83_{-u}^{\,+0.97}$ & $-6.00_{-0.39}^{\,+0.37}$ & $2.09_{-1.52}^{\,+u}$ & $0.70_{-u}^{\,+u}$ & $-4.87_{-2.38}^{\,+2.09}$ \\
\hline
\noalign{\vspace{2pt}}
\multirow{3}{*}{PS7} & 835 & $1.03_{-0.31}^{\,+0.31}$ & $-5.61_{-0.13}^{\,+0.11}$ & $-0.28_{-0.36}^{\,+0.20}$ & $2.35_{-0.80}^{\,+u}$ & $-5.00_{-0.34}^{\,+0.26}$ & $1.58_{-0.61}^{\,+3.10}$ & $0.85_{-0.19}^{\,+u}$ & $-3.59_{-1.79}^{\,+0.76}$ \\
 & 3569 & $2.32_{-1.63}^{\,+3.56}$ & $-6.04_{-u}^{\,+0.31}$ & $-1.23_{-u}^{\,+0.88}$ & $2.45_{-0.87}^{\,+u}$ & $-5.84_{-1.18}^{\,+0.40}$ & $0.57_{-0.18}^{\,+0.97}$ & $0.73_{-u}^{\,+u}$ & $-2.65_{-2.90}^{\,+0.60}$ \\
 & 16026 & $2.06_{-0.82}^{\,+1.19}$ & $-5.47_{-0.26}^{\,+0.20}$ & $-1.69_{-u}^{\,+1.12}$ & $2.14_{-0.77}^{\,+u}$ & $-5.40_{-0.27}^{\,+0.28}$ & $1.08_{-0.55}^{\,+5.62}$ & $0.64_{-u}^{\,+u}$ & $-3.41_{-3.31}^{\,+1.28}$ \\
\hline
\noalign{\vspace{2pt}}
\multirow{3}{*}{PS8} & 835 & $0.74_{-0.65}^{\,+0.55}$ & $-6.02_{-0.41}^{\,+0.25}$ & $-0.14_{-u}^{\,+0.52}$ & $1.87_{-u}^{\,+u}$ & $-5.36_{-0.49}^{\,+0.59}$ & $2.78_{-1.45}^{\,+u}$ & $0.86_{-0.18}^{\,+u}$ & $-4.65_{-1.55}^{\,+1.11}$ \\
 & 3569 & $1.32_{-0.33}^{\,+0.33}$ & $-5.47_{-0.14}^{\,+0.12}$ & $-0.80_{-u}^{\,+0.46}$ & $1.77_{-u}^{\,+0.85}$ & $-5.25_{-0.26}^{\,+0.35}$ & $1.38_{-0.47}^{\,+3.61}$ & $0.79_{-0.19}^{\,+u}$ & $-3.41_{-2.24}^{\,+0.75}$ \\
 & 16026 & $0.69_{-0.60}^{\,+0.49}$ & $-5.62_{-0.35}^{\,+0.23}$ & $0.34_{-u}^{\,+0.26}$ & $2.29_{-u}^{\,+u}$ & $-4.60_{-0.76}^{\,+0.47}$ & $3.19_{-1.63}^{\,+u}$ & $0.86_{-0.17}^{\,+u}$ & $-4.39_{-1.34}^{\,+1.05}$ \\
\hline
\noalign{\vspace{2pt}}
\multirow{3}{*}{PS9} & 835 & $1.04_{-0.33}^{\,+0.32}$ & $-5.64_{-0.19}^{\,+0.14}$ & $-0.13_{-0.30}^{\,+0.20}$ & $2.60_{-0.69}^{\,+u}$ & $-4.82_{-0.35}^{\,+0.22}$ & $1.32_{-0.35}^{\,+0.86}$ & $0.91_{-0.17}^{\,+u}$ & $-3.27_{-0.86}^{\,+0.49}$ \\
 & 3569 & $1.36_{-0.35}^{\,+0.31}$ & $-5.50_{-0.16}^{\,+0.13}$ & $-0.20_{-0.23}^{\,+0.17}$ & $2.72_{-0.59}^{\,+u}$ & $-4.80_{-0.28}^{\,+0.19}$ & $1.05_{-0.24}^{\,+0.51}$ & $0.90_{-0.19}^{\,+u}$ & $-2.91_{-0.76}^{\,+0.45}$ \\
 & 16026 & $1.43_{-0.49}^{\,+0.53}$ & $-5.34_{-0.23}^{\,+0.16}$ & $-1.00_{-u}^{\,+0.83}$ & $1.74_{-u}^{\,+u}$ & $-5.20_{-0.25}^{\,+0.46}$ & $1.62_{-0.73}^{\,+u}$ & $0.73_{-u}^{\,+u}$ & $-3.69_{-2.46}^{\,+1.13}$ \\
\hline
\noalign{\vspace{2pt}}
\multirow{3}{*}{PS10} & 835 & $1.10_{-0.50}^{\,+0.47}$ & $-6.00_{-0.21}^{\,+0.17}$ & $-1.47_{-u}^{\,+1.03}$ & $1.34_{-u}^{\,+0.70}$ & $-5.90_{-0.19}^{\,+0.33}$ & $2.84_{-1.58}^{\,+u}$ & $0.75_{-0.15}^{\,+u}$ & $-4.99_{-1.74}^{\,+1.30}$ \\
 & 3569 & $0.76_{-1.03}^{\,+0.95}$ & $-6.48_{-0.56}^{\,+0.33}$ & $-1.05_{-u}^{\,+1.37}$ & $1.44_{-u}^{\,+1.08}$ & $-6.18_{-0.31}^{\,+0.67}$ & $3.47_{-2.26}^{\,+u}$ & $0.80_{-0.22}^{\,+u}$ & $-5.53_{-1.46}^{\,+1.46}$ \\
 & 16026 & $2.13_{-0.94}^{\,+1.08}$ & $-5.85_{-0.35}^{\,+0.25}$ & $-1.43_{-u}^{\,+1.17}$ & $2.25_{-0.89}^{\,+u}$ & $-5.77_{-0.35}^{\,+0.40}$ & $0.92_{-0.46}^{\,+5.48}$ & $0.69_{-u}^{\,+u}$ & $-3.50_{-3.31}^{\,+1.25}$ \\
\hline
\noalign{\vspace{2pt}}
\multirow{3}{*}{PS11} & 835 &$1.63_{-0.32}^{\,+0.34}$ & $-5.53_{-0.11}^{\,+0.10}$ & $-1.87_{-u}^{\,+0.97}$ & $1.74_{-0.37}^{\,+0.50}$ & $-5.49_{-0.13}^{\,+0.17}$ & $3.80_{-2.50}^{\,+u}$ & $0.57_{-u}^{\,+0.10}$ & $-5.75_{-1.50}^{\,+1.95}$ \\
 & 3569 & $1.87_{-0.35}^{\,+0.38}$ & $-5.44_{-0.12}^{\,+0.10}$ & $-1.33_{-u}^{\,+0.78}$ & $2.10_{-0.50}^{\,+0.67}$ & $-5.35_{-0.15}^{\,+0.28}$ & $0.86_{-0.24}^{\,+2.61}$ & $0.70_{-u}^{\,+u}$ & $-2.91_{-2.83}^{\,+0.74}$ \\
 & 16026 & $4.21_{-6.44}^{\,+u}$ & $-9.18_{-u}^{\,+2.17}$ & $-0.30_{-u}^{\,+u}$ & $2.04_{-u}^{\,+u}$ & $-8.82_{-u}^{\,+2.59}$ & $0.48_{-u}^{\,+u}$ & $0.74_{-u}^{\,+u}$ & $-7.75_{-u}^{\,+4.52}$ \\
\hline
\noalign{\vspace{2pt}}
\multirow{3}{*}{PS12} & 835 &$2.11_{-0.32}^{\,+0.35}$ & $-5.23_{-0.10}^{\,+0.09}$ & $-1.86_{-u}^{\,+0.91}$ & $2.23_{-0.37}^{\,+0.54}$ & $-5.19_{-0.12}^{\,+0.16}$ & $1.71_{-1.08}^{\,+u}$ & $0.52_{-0.u}^{\,+0.10}$ & $-4.44_{-2.71}^{\,+1.99}$ \\
 & 3569 & $1.88_{-0.27}^{\,+0.25}$ & $-5.04_{-0.09}^{\,+0.08}$ & $-1.45_{-u}^{\,+0.73}$ & $2.04_{-0.33}^{\,+0.58}$ & $-4.98_{-0.12}^{\,+0.21}$ & $1.72_{-0.86}^{\,+u}$ & $0.55_{-0.u}^{\,+0.15}$ & $-4.11_{-2.62}^{\,+1.52}$ \\
 & 16026 & $2.05_{-0.47}^{\,+0.49}$ & $-5.06_{-0.20}^{\,+0.14}$ & $-0.66_{-u}^{\,+0.43}$ & $2.54_{-0.74}^{\,+u}$ & $-4.83_{-0.33}^{\,+0.26}$ & $0.83_{-0.18}^{\,+0.64}$ & $0.73_{-0.u}^{\,+0.u}$ & $-2.44_{-1.33}^{\,+0.40}$ \\
\hline
\noalign{\vspace{2pt}}
\multirow{3}{*}{PS13} & 835 &$1.65_{-0.27}^{\,+0.25}$ & $-5.23_{-0.12}^{\,+0.10}$ & $-0.64_{-0.69}^{\,+0.24}$ & $2.46_{-0.70}^{\,+u}$ & $-4.87_{-0.29}^{\,+0.19}$ & $0.82_{-0.14}^{\,+0.48}$ & $0.80_{-0.19}^{\,+u}$ & $-2.42_{-1.03}^{\,+0.37}$ \\
 & 3569 & $1.65_{-0.28}^{\,+0.29}$ & $-5.14_{-0.12}^{\,+0.10}$ & $-0.46_{-0.35}^{\,+0.19}$ & $2.67_{-0.62}^{\,+u}$ & $-4.68_{-0.26}^{\,+0.17}$ & $0.87_{-0.15}^{\,+0.40}$ & $0.82_{-0.19}^{\,+u}$ & $-2.43_{-0.88}^{\,+0.38}$ \\
 & 16026 & $1.94_{-3.05}^{\,+6.29}$ & $-5.83_{-u}^{\,+0.59}$ & $-0.47_{-u}^{\,+0.63}$ & $2.37_{-1.02}^{\,+u}$ & $-5.17_{-1.65}^{\,+0.50}$ & $0.96_{-0.37}^{\,+2.96}$ & $0.78_{-u}^{\,+u}$ & $-3.05_{-3.81}^{\,+0.90}$ \\
\hline
\noalign{\vspace{2pt}}
\multirow{3}{*}{PS14} & 835 & $0.96_{-3.12}^{\,+7.37}$ & $-8.30_{-u}^{\,+2.33}$ & $0.24_{-u}^{\,+0.37}$ & $2.38_{-u}^{\,+u}$ & $-5.02_{-1.04}^{\,+0.50}$ & $2.10_{-1.91}^{\,+u}$ & $0.84_{-u}^{\,+u}$ & $-5.01_{-u}^{\,+1.60}$ \\
 & 3569 & $1.57_{-4.12}^{\,+u}$ & $-8.87_{-u}^{\,+2.85}$ & $0.15_{-u}^{\,+0.63}$ & $2.29_{-u}^{\,+u}$ & $-5.14_{-3.98}^{\,+0.58}$ & $1.86_{-u}^{\,+u}$ & $0.81_{-u}^{\,+u}$ & $-5.45_{-u}^{\,+2.12}$ \\
 & 16026 & $3.59_{-6.26}^{\,+u}$ & $-9.29_{-u}^{\,+2.21}$ & $-0.32_{-u}^{\,+u}$ & $2.03_{-u}^{\,+u}$ & $-8.96_{-u}^{\,+2.61}$ & $0.58_{-u}^{\,+u}$ & $0.76_{-u}^{\,+u}$ & $-7.62_{-u}^{\,+4.38}$ \\
\hline
\noalign{\vspace{2pt}}
\multirow{3}{*}{PS15} & 835 &$1.20_{-0.34}^{\,+0.36}$ & $-5.75_{-0.14}^{\,+0.12}$ & $-0.71_{-u}^{\,+0.42}$ & $1.71_{-u}^{\,+0.90}$ & $-5.50_{-0.28}^{\,+0.39}$ & $1.55_{-0.53}^{\,+3.74}$ & $0.82_{-0.20}^{\,+u}$ & $-3.77_{-2.14}^{\,+0.72}$ \\
 & 3569 & $1.11_{-0.41}^{\,+0.40}$ & $-5.79_{-0.16}^{\,+0.15}$ & $-0.53_{-u}^{\,+0.44}$ & $1.74_{-u}^{\,+0.91}$ & $-5.48_{-0.31}^{\,+0.45}$ & $1.86_{-0.73}^{\,+4.67}$ & $0.81_{-0.18}^{\,+u}$ & $-4.08_{-1.97}^{\,+0.80}$ \\
 & 16026 & $3.11_{-5.64}^{\,+u}$ & $-9.20_{-u}^{\,+2.13}$ & $-0.04_{-u}^{\,+u}$ & $2.01_{-u}^{\,+u}$ & $-8.72_{-2.08}^{\,+2.47}$ & $0.55_{-u}^{\,+u}$ & $0.78_{-u}^{\,+u}$ & $-8.07_{-u}^{\,+4.94}$ \\
\hline
\noalign{\vspace{2pt}}
\multirow{3}{*}{PS16} & 835 &$1.85_{-0.48}^{\,+0.48}$ & $-5.82_{-0.16}^{\,+0.13}$ & $-0.85_{-u}^{\,+u}$ & $2.38_{-0.78}^{\,+0.55}$ & $-5.62_{-0.27}^{\,+0.28}$ & $0.74_{-0.24}^{\,+1.60}$ & $0.75_{-u}^{\,+u}$ & $-2.94_{-2.41}^{\,+0.76}$ \\
 & 3569 & $2.45_{-0.55}^{\,+0.61}$ & $-5.81_{-0.17}^{\,+0.15}$ & $-1.33_{-u}^{\,+0.71}$ & $2.62_{-0.59}^{\,+u}$ & $-5.72_{-0.20}^{\,+0.26}$ & $0.51_{-0.10}^{\,+0.31}$ & $0.69_{-u}^{\,+u}$ & $-2.46_{-1.08}^{\,+0.42}$ \\
 & 16026 & $5.13_{-7.10}^{\,+u}$ & $-8.00_{-u}^{\,+1.71}$ & $-0.60_{-u}^{\,+u}$ & $2.05_{-u}^{\,+u}$ & $-8.47_{-u}^{\,+2.47}$ & $0.66_{-u}^{\,+u}$ & $0.74_{-u}^{\,+u}$ & $-7.26_{-u}^{\,+4.48}$ \\
\hline
\noalign{\vspace{2pt}}
\multirow{3}{*}{PS17} & 835 &$1.51_{-0.26}^{\,+0.27}$ & $-5.45_{-0.09}^{\,+0.08}$ & $-1.04_{-u}^{\,+0.51}$ & $1.84_{-0.46}^{\,+0.70}$ & $-5.31_{-0.18}^{\,+0.29}$ & $1.13_{-0.35}^{\,+2.70}$ & $0.73_{-u}^{\,+u}$ & $-3.22_{-2.41}^{\,+0.75}$ \\
 & 3569 & $1.81_{-0.32}^{\,+0.31}$ & $-5.40_{-0.10}^{\,+0.09}$ & $-2.12_{-u}^{\,+1.05}$ & $1.88_{-0.34}^{\,+0.39}$ & $-5.37_{-0.12}^{\,+0.13}$ & $3.77_{-2.39}^{\,+u}$ & $0.54_{-u}^{\,+0.07}$ & $-5.80_{-1.49}^{\,+1.88}$ \\
 & 16026 & $1.44_{-0.47}^{\,+0.52}$ & $-5.47_{-0.19}^{\,+0.17}$ & $-0.48_{-u}^{\,+0.42}$ & $2.06_{-u}^{\,+u}$ & $-5.14_{-0.39}^{\,+0.41}$ & $1.16_{-0.35}^{\,+1.86}$ & $0.84_{-0.24}^{\,+u}$ & $-3.14_{-1.84}^{\,+0.63}$ \\
\hline
\noalign{\vspace{2pt}}
\hline
\end{tabular}
\footnotesize{\newline \\ 
{\it Notes:} Models and parameters are as defined in Sec.~\ref{sec:analysis}. Errors reported correspond to the 90\% credible intervals of the posterior. A `u' corresponds to an unconstrained parameter. N$\rm_H$ is in units of $10^{22}$~cm$^{-1}$, and T$\rm_{in}$ is in units of keV.}
\end{table*}

\begin{table*}[h!]
\scriptsize
\centering
\caption{Median luminosity and percent of the posterior above the ULX threshold, for all observations, all sources, and all models.}
\label{tab:luminosity}
\renewcommand{\arraystretch}{1.2}
\begin{tabular}{cc|cc|cc|cc}
\hline\hline
 &  & \multicolumn{2}{c}{Powerlaw} & \multicolumn{2}{c}{Abs. powerlaw} & \multicolumn{2}{c}{Diskpbb}\\
 & & log($L_{\rm 0.3-10 keV}$) & posterior & log($L_{\rm 0.3-10 keV}$) & posterior & log($L_{\rm 0.3-10 keV}$) & posterior  \\
  &  & erg~s$^{-1}$ & \% & erg~s$^{-1}$ & \% & erg~s$^{-1}$ & \% \\
\hline
\multirow{3}{*}{PS1*} & 835 & $38.94_{-0.43}^{+0.39}$ & 40.9 & $38.88_{-0.34}^{+0.31}$ & 25.7 & $38.90_{-0.39}^{+0.35}$ & 30.7 \\
 & 3569 & $39.23_{-0.21}^{+0.20}$ & 97.1 & $39.24_{-0.23}^{+0.17}$ & 97.7 & $39.16_{-0.23}^{+0.21}$ & 87.9  \\
 & 16026 & $39.12_{-0.35}^{+0.35}$ & 73.2 & $39.19_{-0.38}^{+0.52}$ & 79.0 & $39.01_{-0.33}^{+0.38}$ & 51.3 \\
\hline
\noalign{\vspace{2pt}}
\multirow{3}{*}{PS2*} & 835 & $29.63_{-0.38}^{+0.37}$ & 0 & $29.70_{-0.38}^{+0.36}$ & 0 & $29.48_{-0.40}^{+0.39}$ & 0 \\
 & 3569 & $29.92_{-0.27}^{+0.22}$ & 0 & $29.95_{-0.24}^{+0.25}$ & 0 & $29.73_{-0.24}^{+0.23}$ & 0 \\
 & 16026 & $30.25_{-0.31}^{+0.30}$ & 0 & $30.26_{-0.33}^{+0.26}$ & 0 & $30.05_{-0.34}^{+0.32}$ & 0\\
\hline
\noalign{\vspace{2pt}}
\multirow{3}{*}{PS3} & 835 & $39.02_{-0.18}^{+0.19}$ & 58.3 & $39.04_{-0.19}^{+0.21}$ & 64.8 & $38.98_{-0.24}^{+0.26}$ &45.8 \\
 & 3569 & $39.07_{-0.22}^{+0.20}$ & 74.1 & $39.11_{-0.22}^{+0.22}$ & 77.5 & $38.99_{-0.24}^{+0.22}$ & 46.2 \\
 & 16026 & $39.07_{-0.30}^{+0.30}$ & 64.8 & $39.08_{-0.35}^{+0.26}$ & 67.0 & $38.99_{-0.30}^{+0.38}$ & 47.1\\
\hline
\noalign{\vspace{2pt}}
\multirow{3}{*}{PS4*} & 835 & $29.12_{-0.27}^{+0.27}$ & 0 & $29.12_{-0.26}^{+0.27}$ & 0 & $28.93_{-0.31}^{+0.23}$ &0\\
 & 3569 & $29.41_{-0.25}^{+0.24}$ & 0 & $29.45_{-0.29}^{+0.22}$ & 0 & $29.29_{-0.23}^{+0.26}$ & 0 \\
 & 16026 & $29.27_{-0.79}^{+0.90}$ & 0 & $29.23_{-0.94}^{+0.66}$ & 0 & $29.00_{-1.28}^{+0.89}$ & 0 \\
\hline
\noalign{\vspace{2pt}}
\multirow{3}{*}{PS5} & 835 & $37.10_{-u}^{+1.81}$ & 2.9 & $37.34_{-u}^{+1.76}$ & 5.7 & $35.50_{-u}^{+3.29}$& 0.1 \\
 & 3569 & $38.93_{-0.34}^{+0.39}$ & 38.3 & $39.07_{-0.42}^{+0.48}$ & 59.6 & $38.81_{-0.39}^{+0.34}$ & 21.2\\
 & 16026 & - & - & - & - & - & - \\
\hline
\noalign{\vspace{2pt}}
\multirow{3}{*}{PS6} & 835 & $38.08_{-u}^{+0.70}$ & 0.7 & $38.09_{-u}^{+0.77}$ & 1.1 & $37.84_{-u}^{+0.79}$ & 0.1\\
 & 3569 & $39.72_{-0.13}^{+0.12}$ & 100.0 & $39.82_{-0.16}^{+0.23}$ & 100.0 & $39.56_{-0.12}^{+0.15}$ & 100.0 \\
 & 16026 & $39.14_{-0.38}^{+0.34}$ & 75.2 & $39.14_{-0.35}^{+0.32}$ & 76.3 & $39.05_{-0.39}^{+0.38}$ & 57.6 \\
\hline
\noalign{\vspace{2pt}}
\multirow{3}{*}{PS7} & 835 & $39.86_{-0.16}^{+0.13}$ & 100.0 & $40.00_{-0.21}^{+0.22}$ & 100.0 & $39.66_{-0.20}^{+0.21}$ & 100.0 \\
 & 3569 & $39.00_{-u}^{+0.46}$ & 50.2 & $39.18_{-0.48}^{+0.48}$ & 77.6 & $38.96_{-0.32}^{+0.26}$ & 38.4 \\
 & 16026 & $39.62_{-0.25}^{+0.24}$ & 99.7 & $39.67_{-0.27}^{+0.22}$ & 99.9 & $39.52_{-0.33}^{+0.24}$ & 99.1\\
\hline
\noalign{\vspace{2pt}}
\multirow{3}{*}{PS8} & 835 & $39.63_{-0.26}^{+0.23}$ & 99.7 & $39.75_{-0.32}^{+0.41}$ & 99.9 & $39.52_{-0.21}^{+0.24}$ & 99.7\\
 & 3569 & $39.84_{-0.14}^{+0.13}$ & 100.0 & $39.89_{-0.17}^{+0.16}$ & 100.0 & $39.69_{-0.15}^{+0.17}$ & 100.0\\
 & 16026 & $40.08_{-0.20}^{+0.17}$ & 100.0 & $40.41_{-0.45}^{+0.42}$ & 100.0 & $39.96_{-0.20}^{+0.20}$ & 100.0\\
\hline
\noalign{\vspace{2pt}}
\multirow{3}{*}{PS9} & 835 & $39.82_{-0.16}^{+0.15}$ & 100.0 & $40.17_{-0.25}^{+0.25}$ & 100.0 & $39.63_{-0.15}^{+0.14}$ & 100.0\\
 & 3569 & $39.80_{-0.12}^{+0.15}$ & 100.0 & $40.18_{-0.23}^{+0.21}$ & 100.0 & $39.61_{-0.13}^{+0.12}$ & 100.0\\
 & 16026 & $39.91_{-0.17}^{+0.15}$ &100.0 & $39.96_{-0.21}^{+0.24}$ & 100.0 & $39.78_{-0.17}^{+0.23}$ & 100.0\\
\hline
\noalign{\vspace{2pt}}
\multirow{3}{*}{PS10} & 835 & $39.43_{-0.19}^{+0.26}$ & 99.9 & $39.40_{-0.19}^{+0.17}$ & 100.0 & $39.34_{-0.23}^{+0.22}$ & 99.4\\
 & 3569 & $39.17_{-0.37}^{+0.39}$ & 77.0 & $39.09_{-0.40}^{+0.38}$ & 67.3 & $39.04_{-0.32}^{+0.36}$ & 57.6\\
 & 16026 & $39.22_{-0.28}^{+0.24}$ & 90.3 & $39.28_{-0.33}^{+0.30}$ & 93.3 & $39.06_{-0.27}^{+0.29}$ & 64.7\\
\hline
\noalign{\vspace{2pt}}
\multirow{3}{*}{PS11} & 835 & $39.64_{-0.11}^{+0.14}$ & 100.0 & $39.64_{-0.12}^{+0.13}$ & 100.0 & $39.59_{-0.15}^{+0.14}$ & 100.0\\
 & 3569 & $39.65_{-0.11}^{+0.12}$ & 100.0 & $39.70_{-0.16}^{+0.17}$ & 100.0 & $39.49_{-0.14}^{+0.13}$ & 100.0\\
 & 16026 & $37.19_{-u}^{+1.89}$ & 8.5 & $36.28_{-u}^{+2.22}$ & 3.6 & $34.54_{-u}^{+3.89}$ & 0.0\\
\hline
\noalign{\vspace{2pt}}
\multirow{3}{*}{PS12} & 835 & $39.79_{-0.10}^{+0.11}$ & 100.0 & $39.82_{-0.13}^{+0.14}$ & 100.0 & $39.75_{-0.16}^{+0.14}$ & 100.0 \\
 & 3569 & $40.03_{-0.09}^{+0.08}$ & 100.0 & $40.07_{-0.12}^{+0.11}$ & 100.0 & $39.97_{-0.12}^{+0.10}$ & 100.0  \\
 & 16026 & $39.98_{-0.14}^{+0.13}$ & 100.0 & $40.16_{-0.24}^{+0.28}$& 100.0 & $39.82_{-0.15}^{+0.14}$ & 100.0 \\
\hline
\noalign{\vspace{2pt}}
\multirow{3}{*}{PS13} & 835 & $39.92_{-0.11}^{+0.11}$ & 100.0 & $40.11_{-0.20}^{+0.19}$ & 100.0 & $39.75_{-0.10}^{+0.10}$ & 100.0\\
 & 3569 & $40.01_{-0.11}^{+0.10}$ & 100.0 & $40.30_{-0.21}^{+0.18}$  & 100.0 & $39.84_{-0.10}^{+0.09}$ & 100.0\\
 & 16026 &$39.42_{-u}^{+0.54}$ & 60.6 & $39.86_{-0.57}^{+0.59}$ & 93.2 & $39.54_{-0.40}^{+0.30}$ & 92.3\\
\hline
\noalign{\vspace{2pt}}
\multirow{3}{*}{PS14} & 835 & $38.17_{-u}^{+1.77}$ & 38.2 & $39.97_{-0.49}^{+0.59}$ & 95.6 & $39.28_{-u}^{+0.63}$ & 64.2 \\
 & 3569 & $37.59_{-u}^{+2.27}$ & 23.8 & $39.89_{-2.18}^{+0.77}$ & 85.7 & $38.85_{-u}^{+1.01}$ & 46.0\\
 & 16026 & $37.10_{-u}^{+2.45}$ & 8.6 & $36.20_{-u}^{+2.21}$ & 3.4 & $34.80_{-u}^{+3.42}$ & 0.0\\
\hline
\noalign{\vspace{2pt}}
\multirow{3}{*}{PS15} & 835 & $39.63_{-0.17}^{+0.15}$ & 100.0 & $39.67_{-0.19}^{+0.18}$ & 100.0 & $39.48_{-0.17}^{+0.19}$ & 100.0\\
 & 3569 & $39.63_{-0.17}^{+0.17}$ & 100.0 & $39.68_{-0.18}^{+0.24}$ & 100.0 & $39.50_{-0.21}^{+0.18}$ & 100.0\\
 & 16026 & $37.20_{-u}^{+1.73}$ &  6.2 & $36.39_{-u}^{+2.00}$ & 3.9 & $34.48_{-u}^{+3.85}$ & 0.0\\
\hline
\noalign{\vspace{2pt}}
\multirow{3}{*}{PS16} & 835 & $39.27_{-0.17}^{+0.17}$ & 99.6 & $39.40_{-0.20}^{+0.26}$ & 99.9 & $39.10_{-0.15}^{+0.20}$ & 85.0\\
 & 3569 & $39.19_{-0.15}^{+0.15}$ & 97.3 & $39.27_{-0.20}^{+0.25}$ & 98.6 & $39.03_{-0.17}^{+0.15}$ & 61.7\\
 & 16026 & $38.32_{-u}^{+1.70}$ & 29.5 & $36.59_{-u}^{+2.37}$ & 5.8 & $35.35_{-u}^{+3.60}$ & 0.9\\
\hline
\noalign{\vspace{2pt}}
\multirow{3}{*}{PS17*} & 835 & $39.76_{-0.10}^{+0.12}$ & 100.0 & $39.80_{-0.14}^{+0.14}$ & 100.0 & $39.61_{-0.13}^{+0.14}$ & 100.0\\
 & 3569 & $39.70_{-0.10}^{+0.10}$ & 100.0 & $39.71_{-0.10}^{+0.11}$ & 100.0 & $39.68_{-0.10}^{+0.11}$ & 100.0\\
 & 16026 & $39.78_{-0.13}^{+0.19}$ & 100.0 & $39.92_{-0.22}^{+0.32}$ & 100.0 & $39.61_{-0.16}^{+0.17}$ & 100.0\\
\hline
\noalign{\vspace{2pt}}
\hline
\end{tabular}
\footnotesize{\newline \\ 
{\it Columns:} 1) Point Source number. 2) \chandra observation ID. 3,5,7) For the \texttt{powerlaw}, absorbed \texttt{powerlaw} and \texttt{diskpbb} models respectively; median of the luminosity posterior, together with the 90\% credible interval. Sources with a lower bound of the credible interval below $10^{37}$~erg~s$^{-1}$ are compatible with non-detections. 4,6,8) For the \texttt{powerlaw}, absorbed \texttt{powerlaw} and \texttt{diskpbb} models respectively; percent of the luminosity posterior above the ULX threshold (10$^{39}$~erg~s$^{-1}$).\newline
*We note that PS1 and PS17 are potential background quasars, and PS2 and PS4 are confirmed stars within our Galaxy. See Sect. \ref{sec:point_source_comments} for details on how to interpret luminosity values and ULX determination.}
\end{table*}

\FloatBarrier
\section{Additional figures}\label{App:figures}

\begin{figure*}[h!]
        \centering
        \includegraphics[scale=0.47,valign=c]{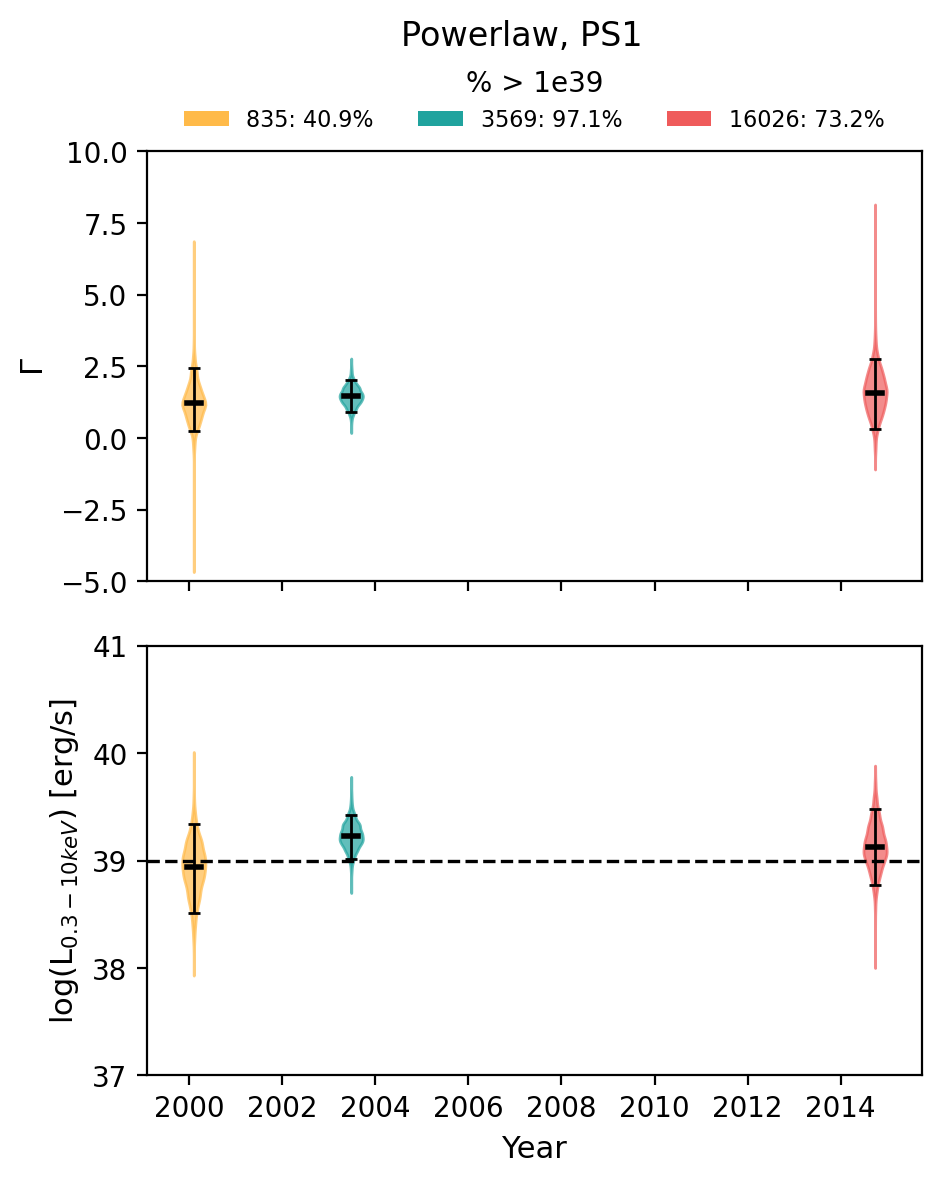}
        \includegraphics[scale=0.47,valign=c]{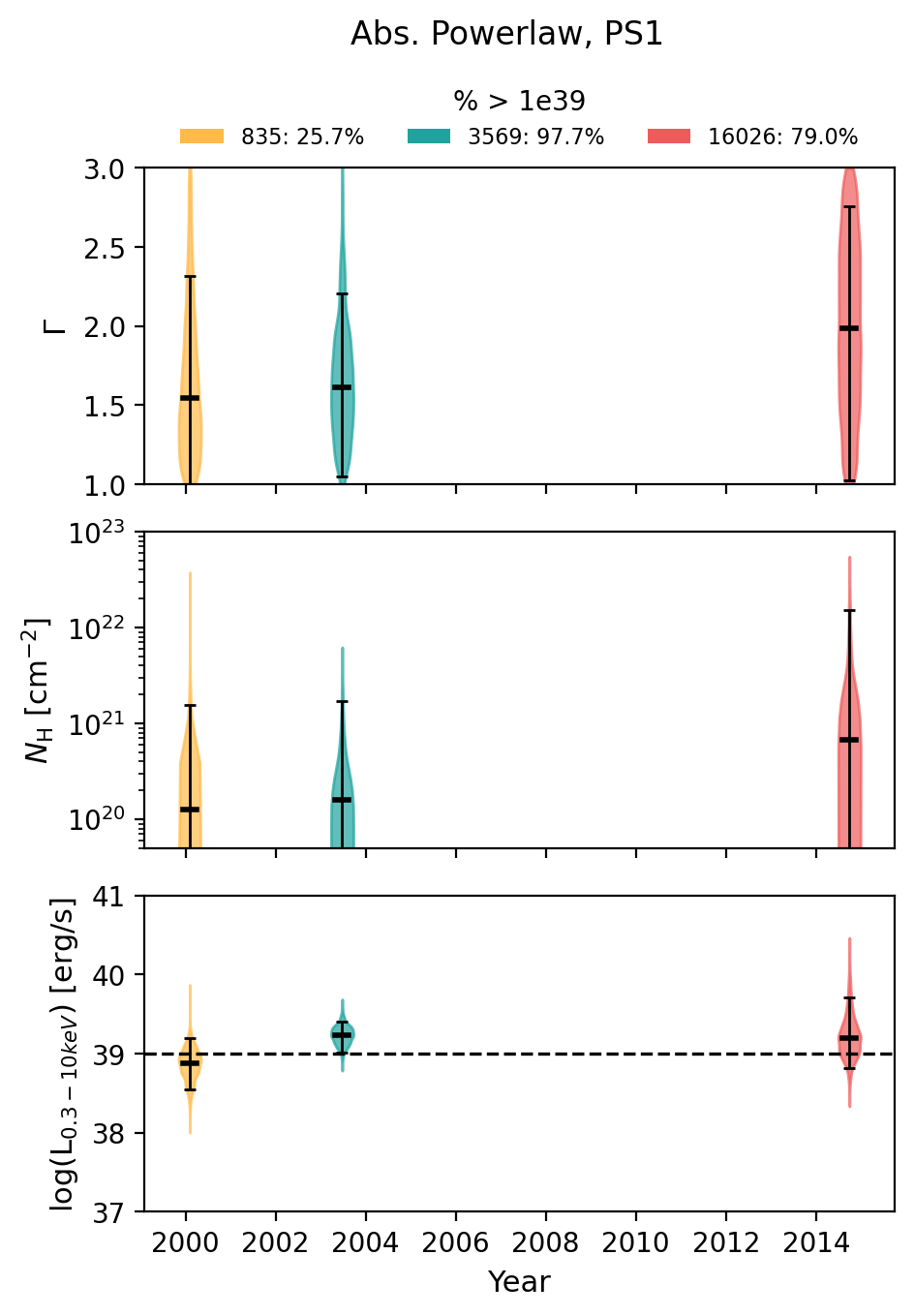  }
        \includegraphics[scale=0.47,valign=c]{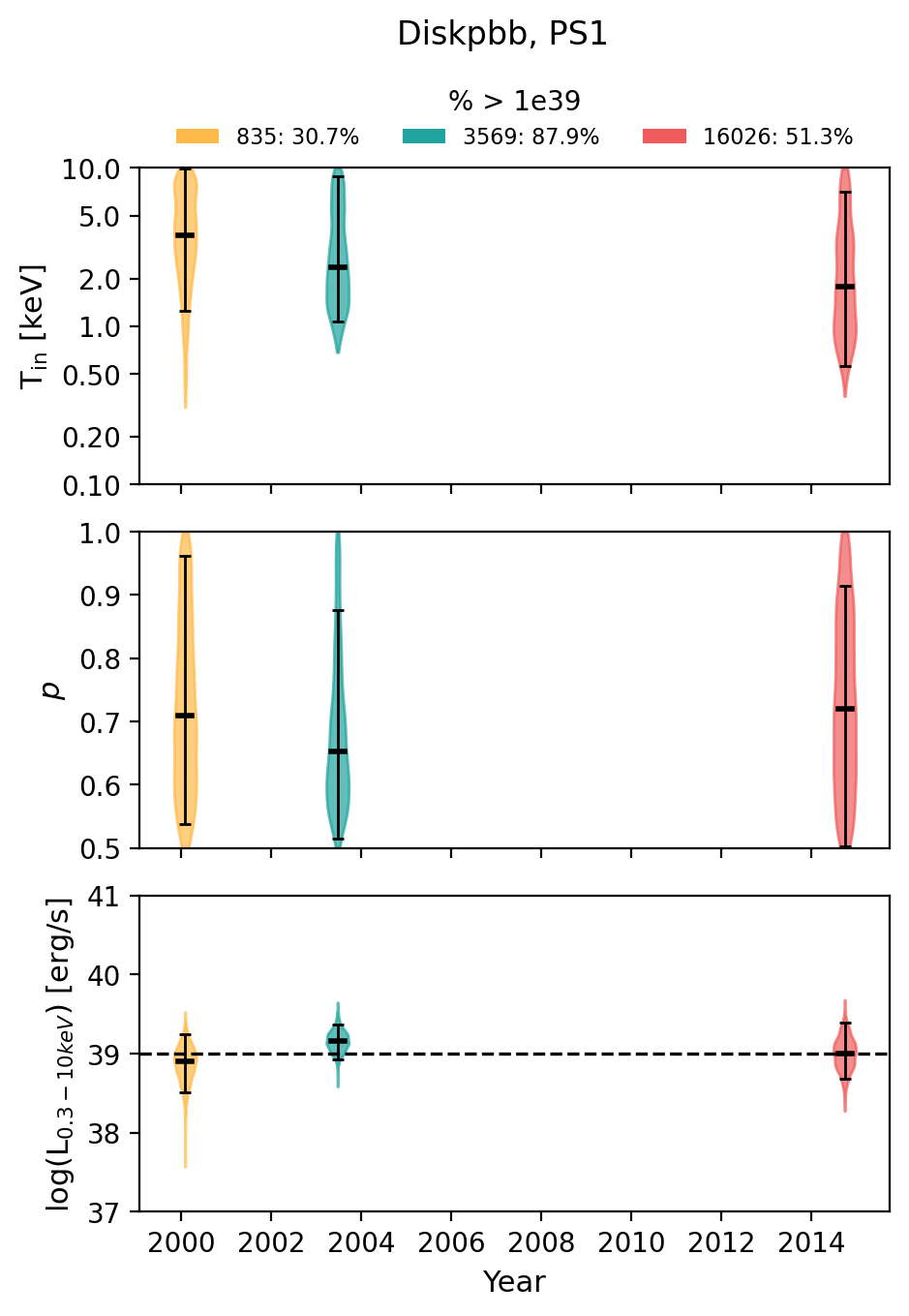}
        \includegraphics[scale=0.47,valign=c]{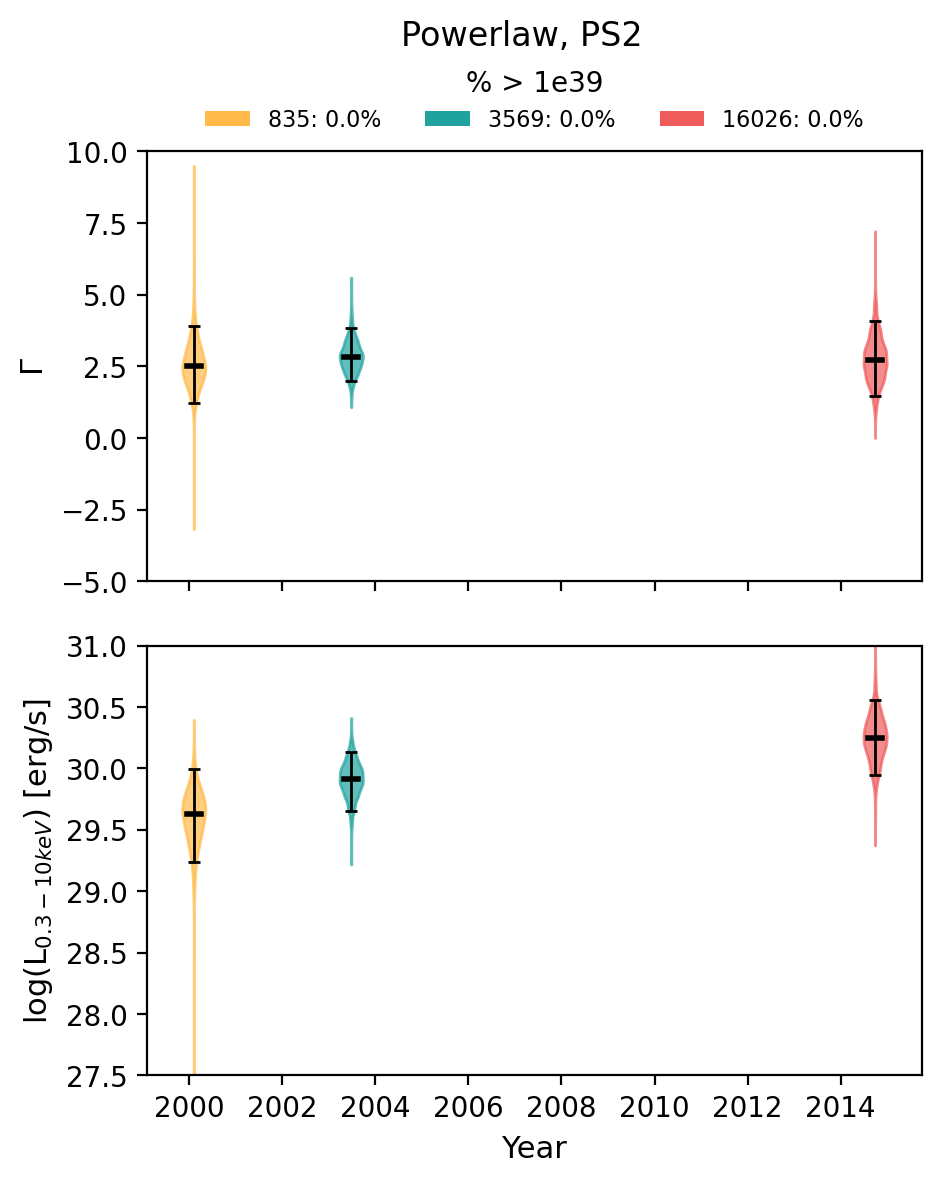}
        \includegraphics[scale=0.47,valign=c]{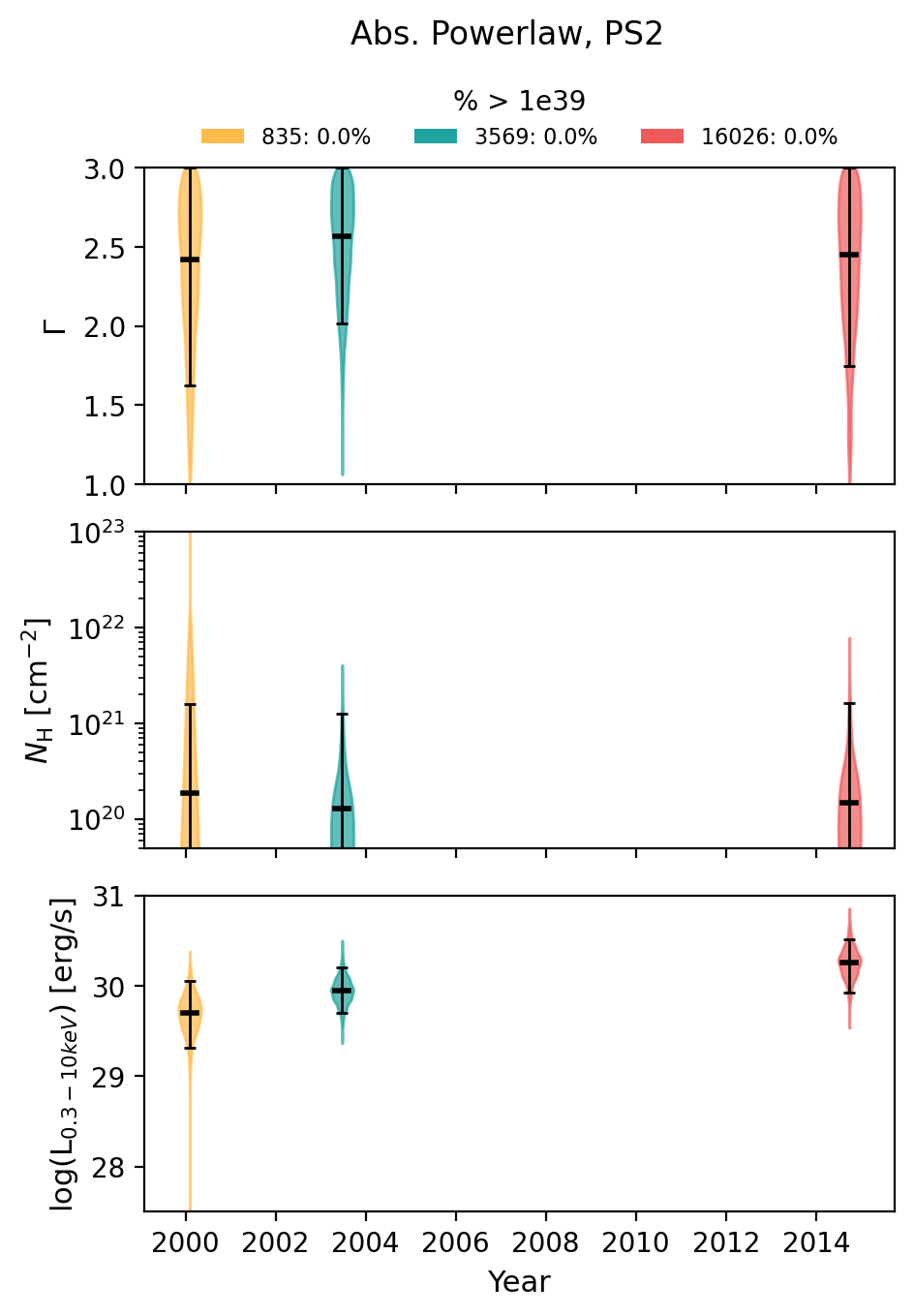  }
        \includegraphics[scale=0.47,valign=c]{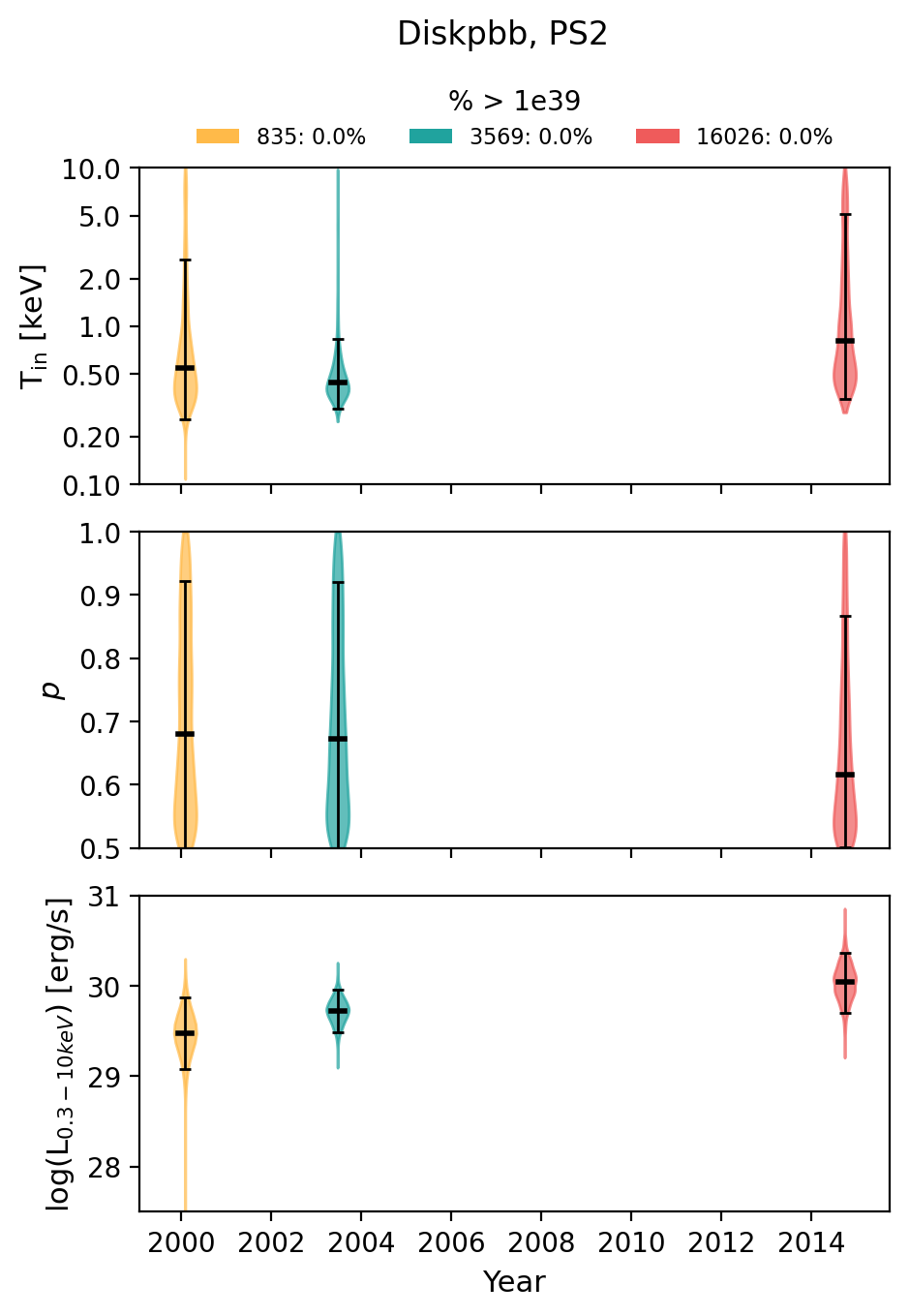}
        \includegraphics[scale=0.47,valign=c]{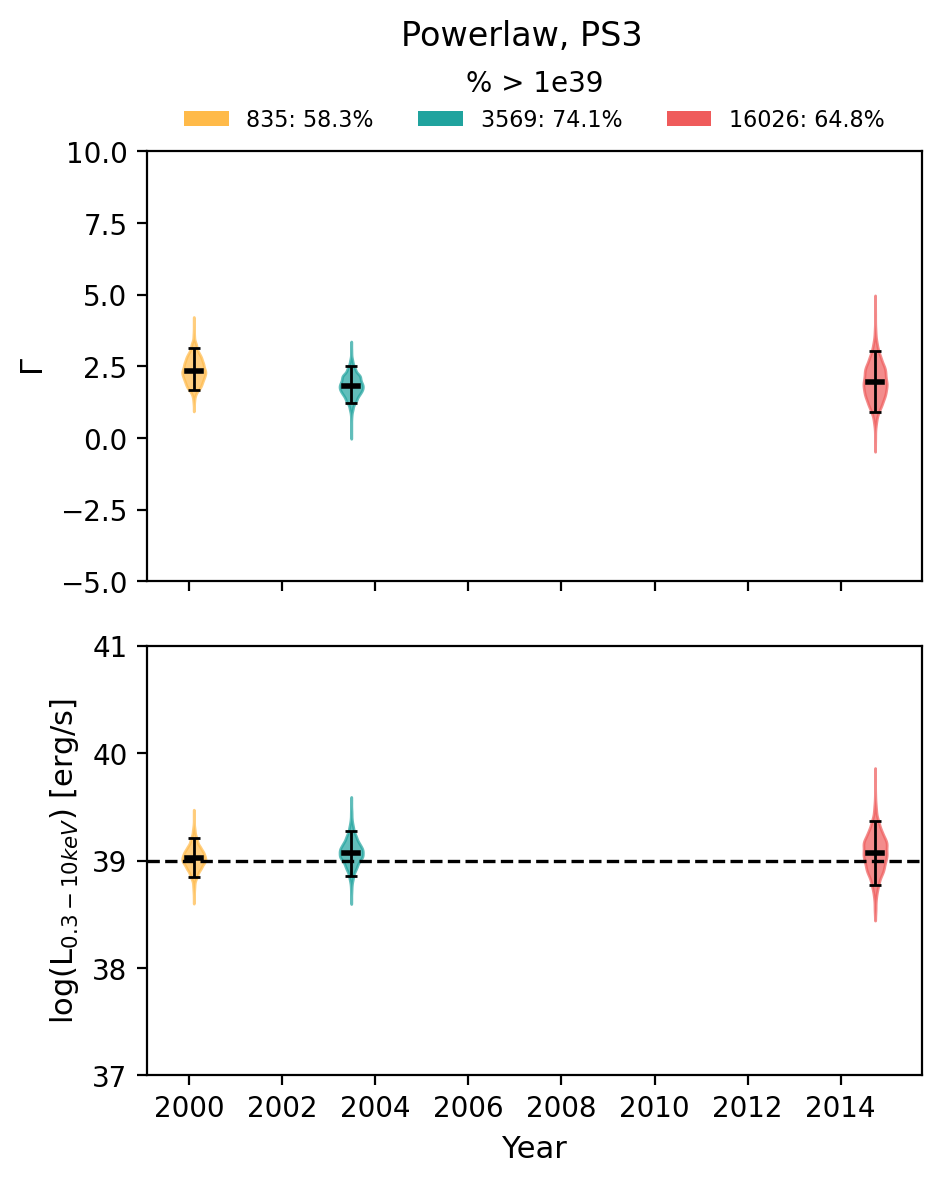}
        \includegraphics[scale=0.47,valign=c]{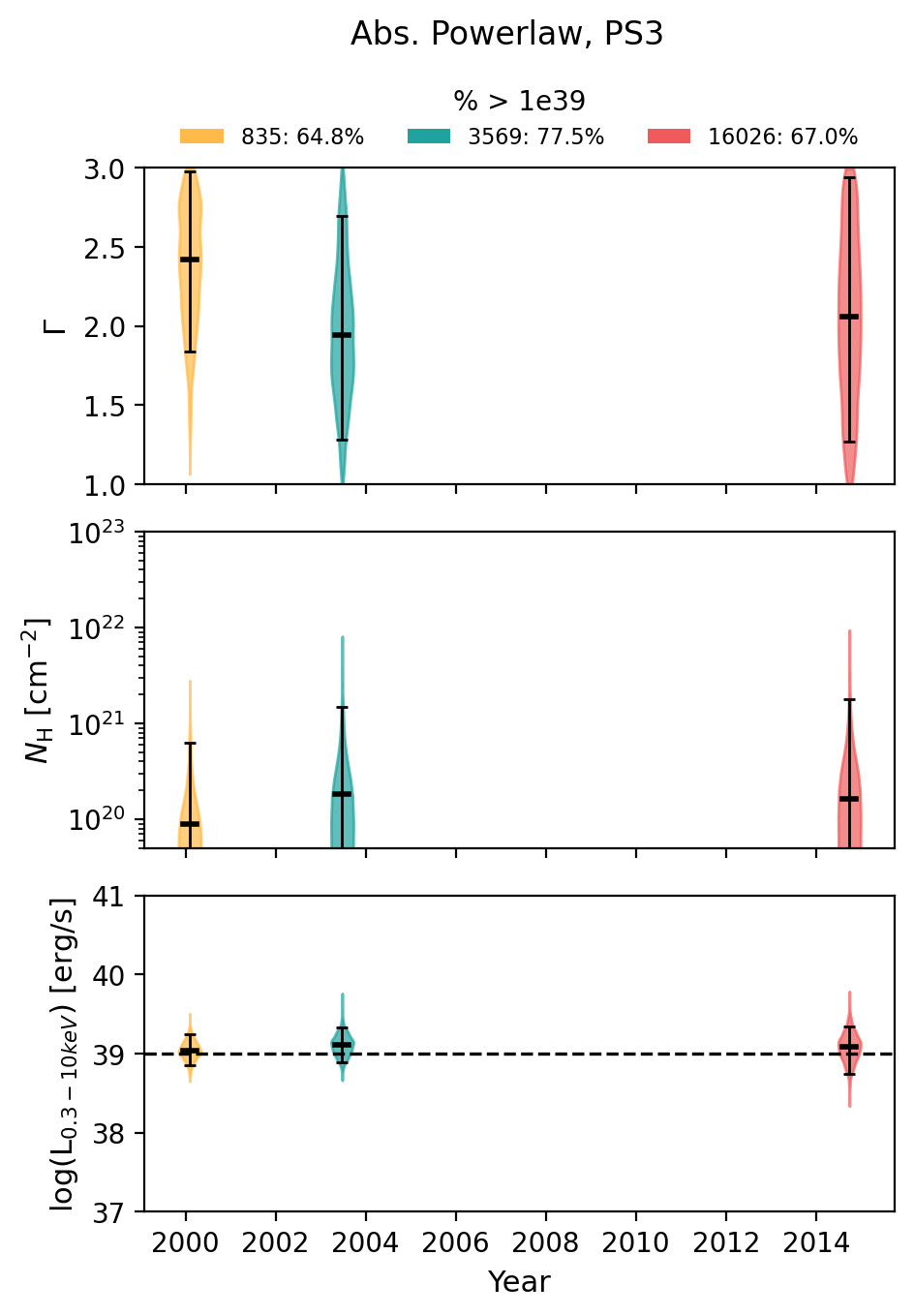  }
        \includegraphics[scale=0.47,valign=c]{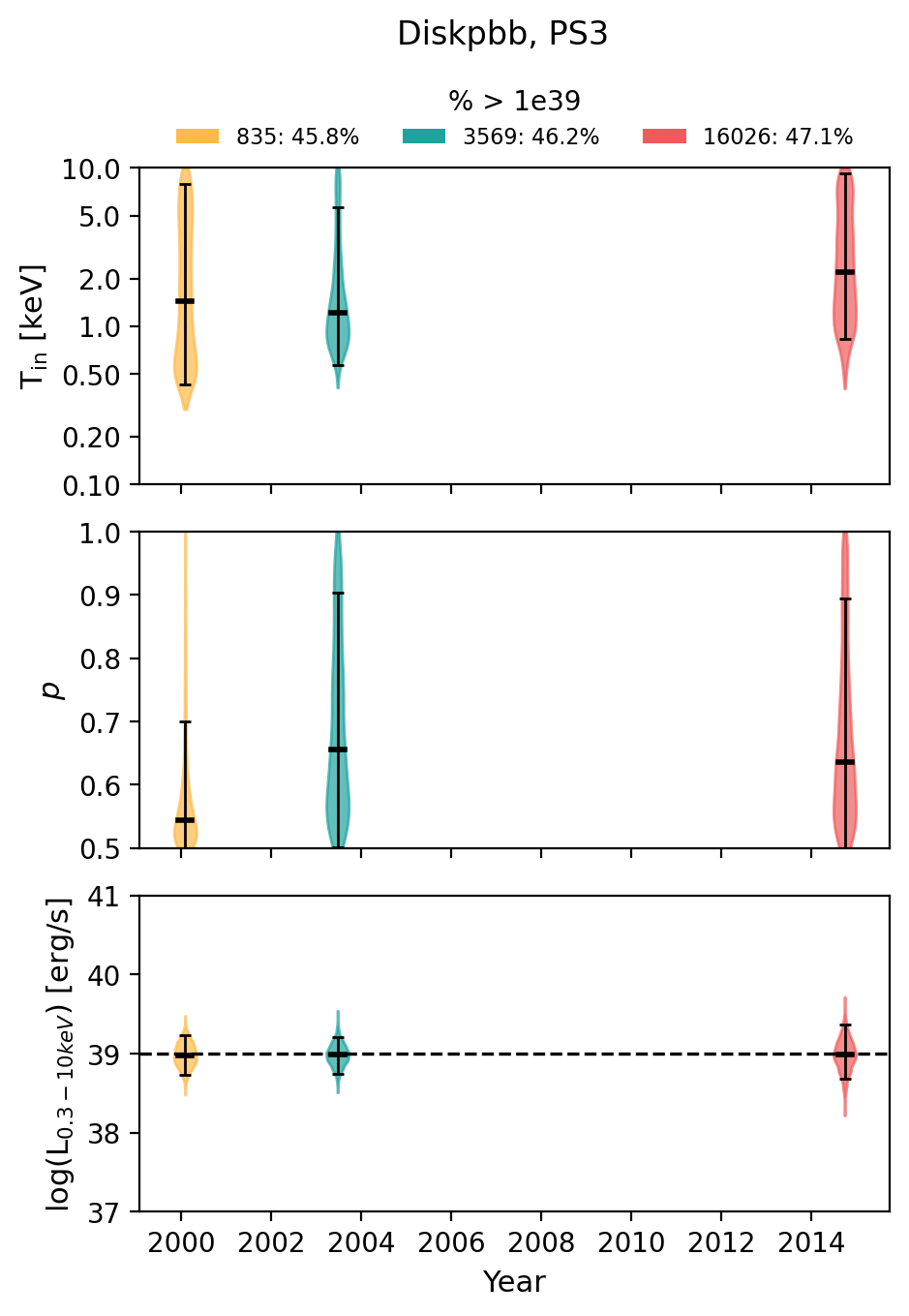}
        \caption{Parameter posterior distributions for the three models used on PS1, PS2 and PS3. See Sect. \ref{App:figures} for details.} \label{fig:app:posteriors_123}
\end{figure*}

\begin{figure*}[h!]
        \centering
        \includegraphics[scale=0.47,valign=c]{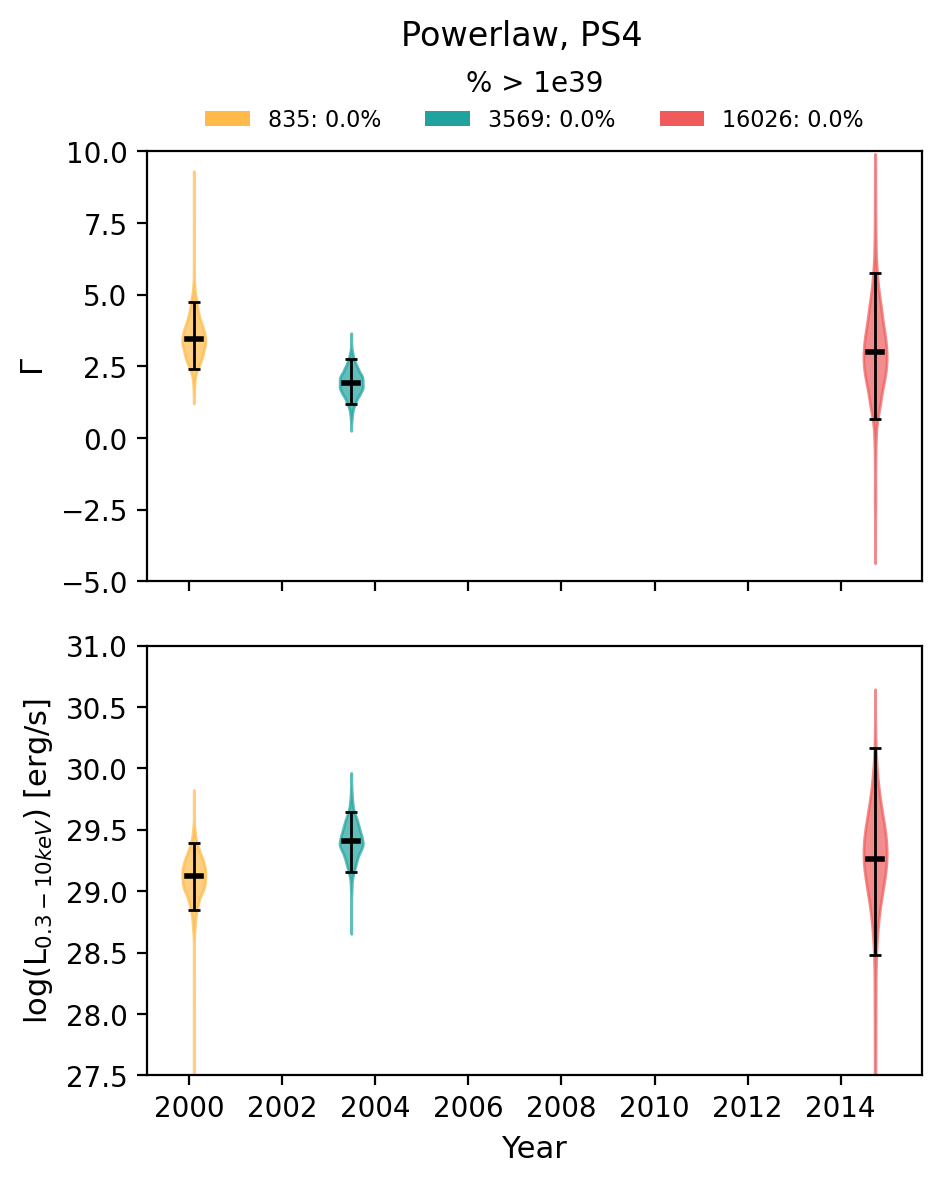}
        \includegraphics[scale=0.47,valign=c]{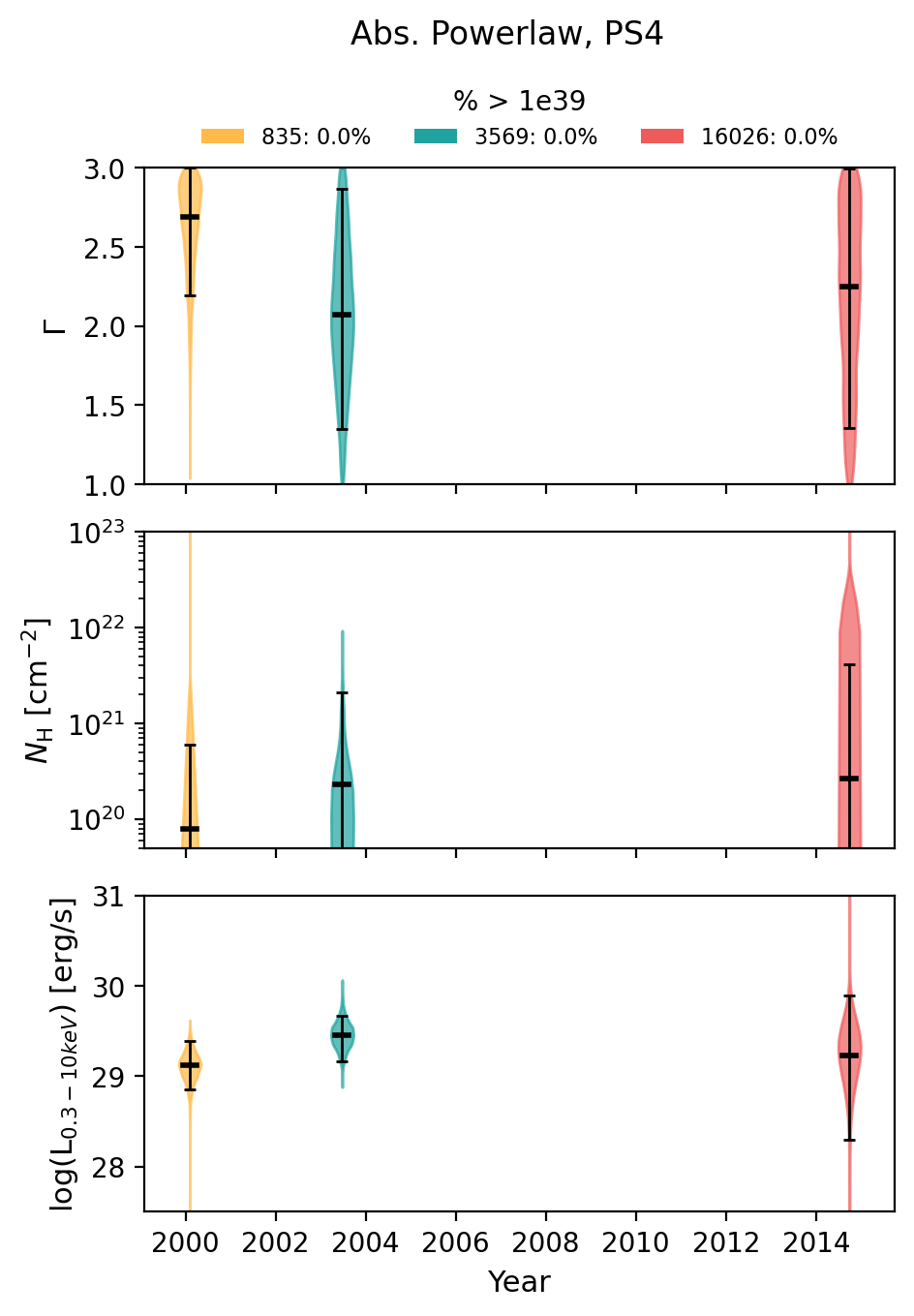  }
        \includegraphics[scale=0.47,valign=c]{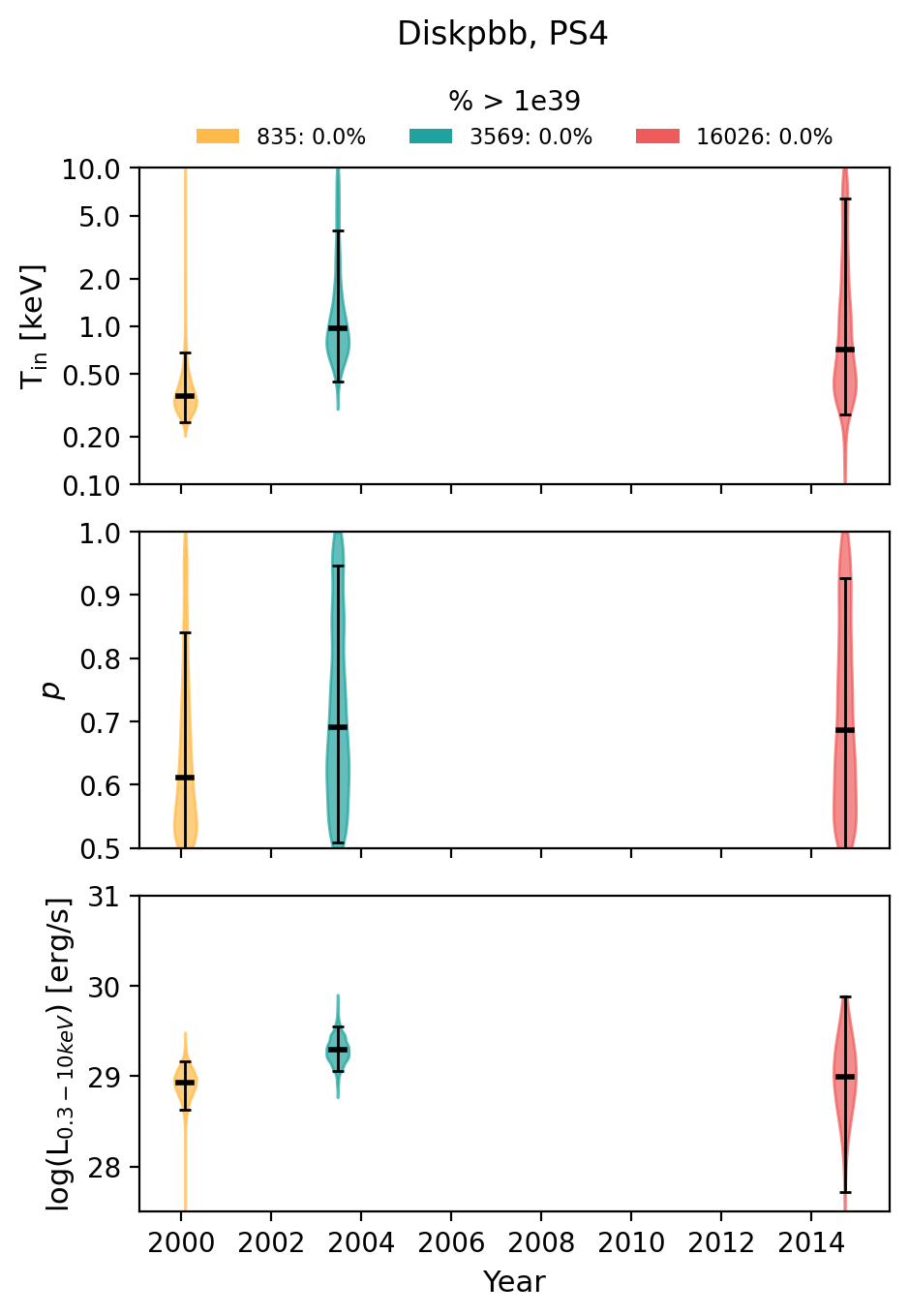}
        \includegraphics[scale=0.47,valign=c]{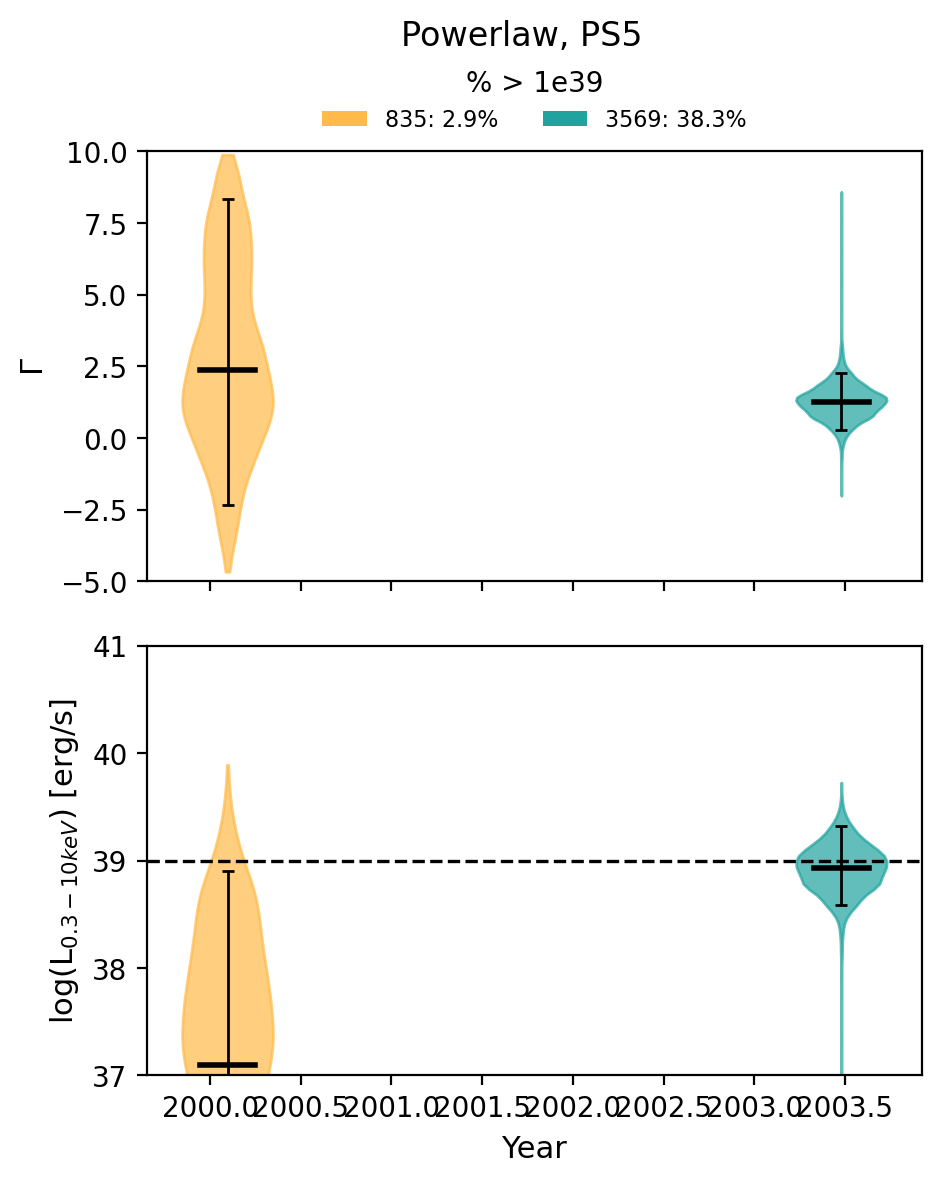}
        \includegraphics[scale=0.47,valign=c]{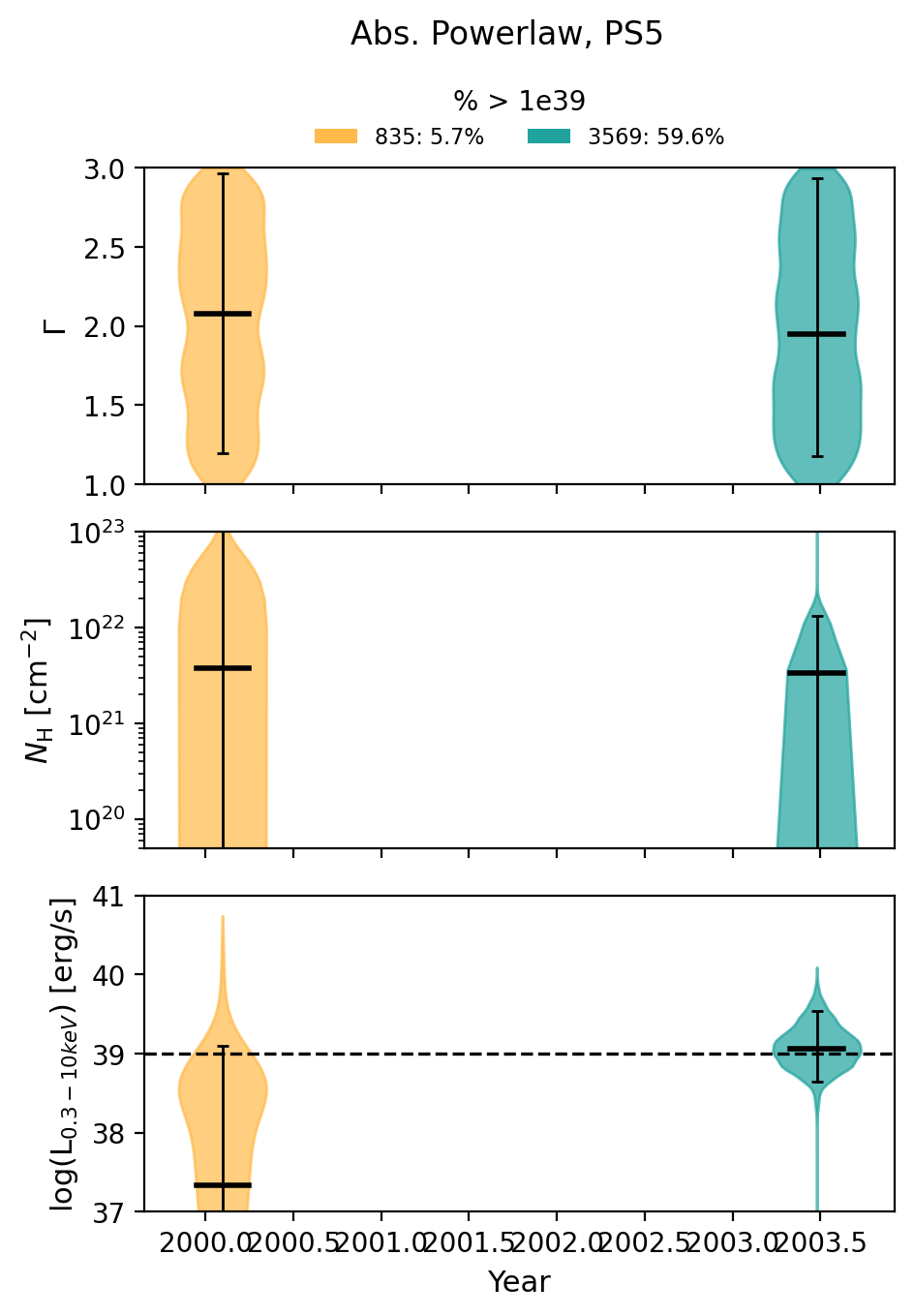  }
        \includegraphics[scale=0.47,valign=c]{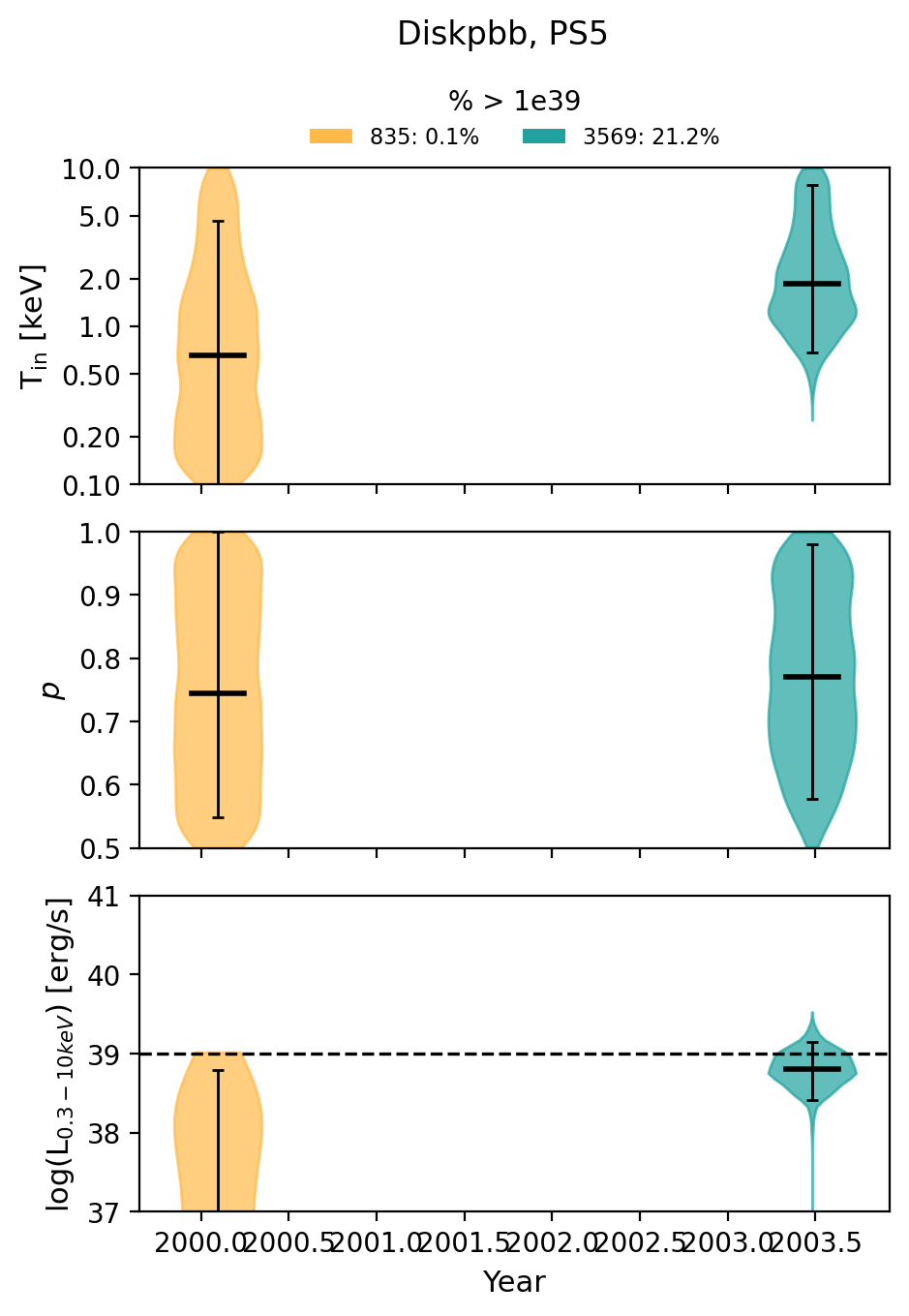}
        \includegraphics[scale=0.47,valign=c]{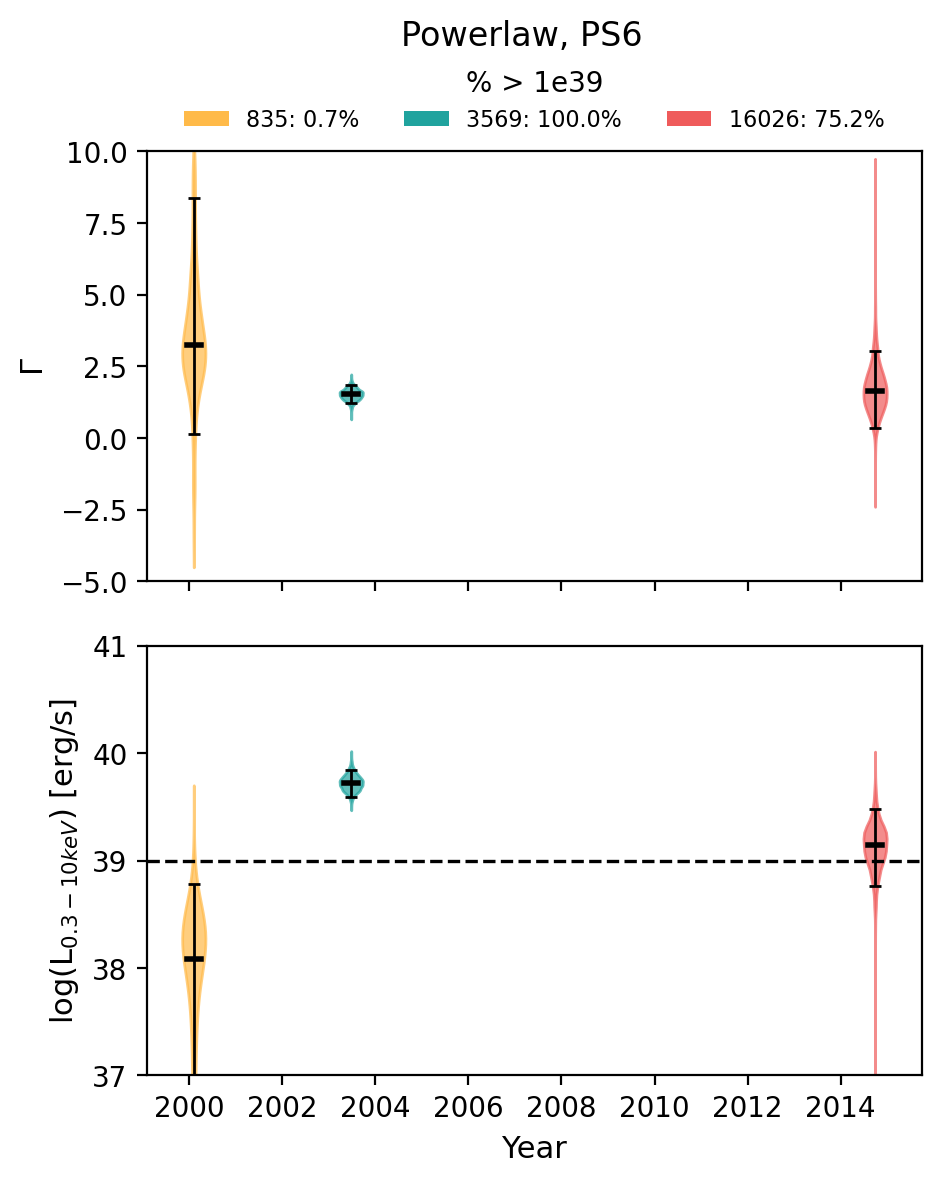}
        \includegraphics[scale=0.47,valign=c]{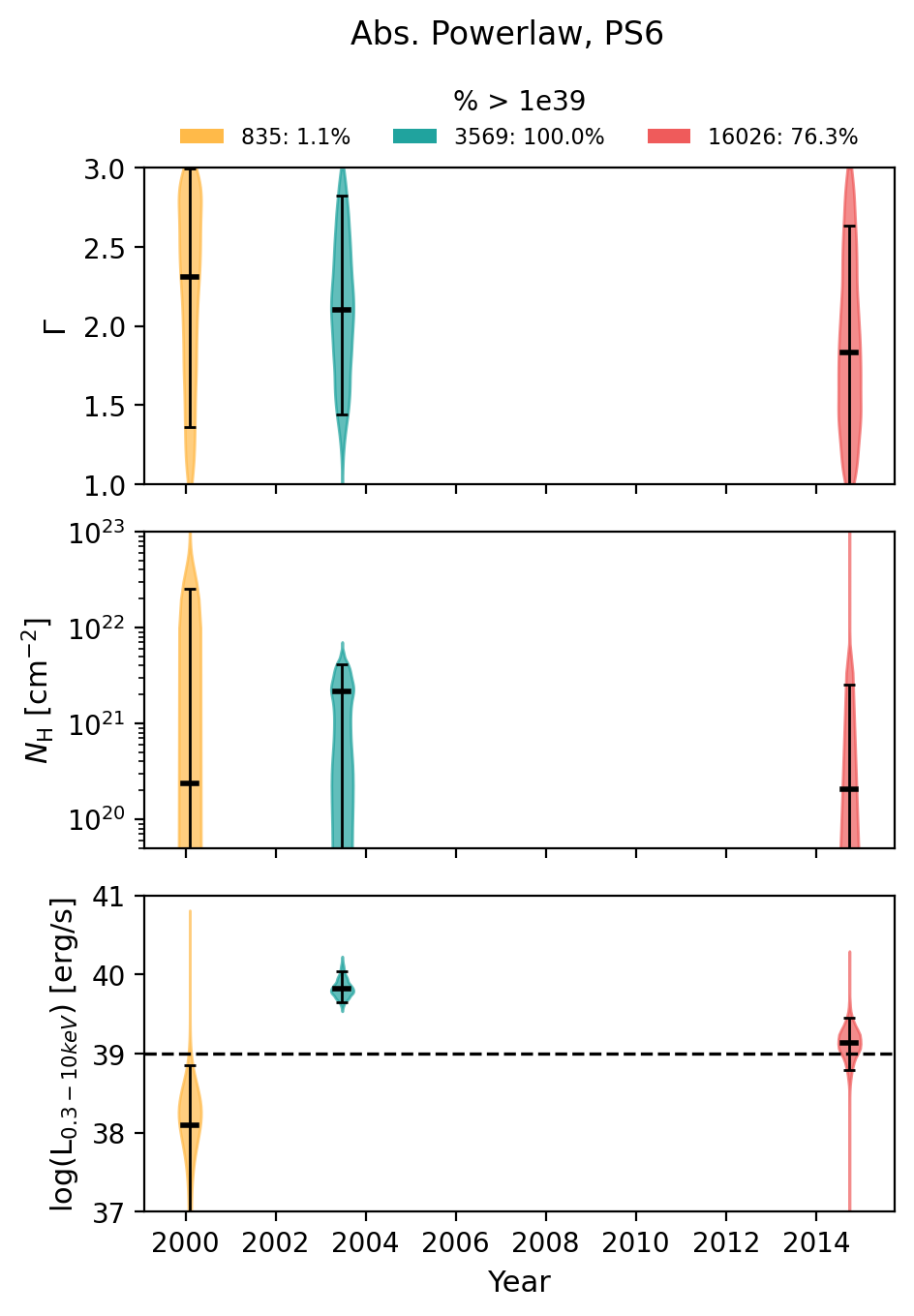  }
        \includegraphics[scale=0.47,valign=c]{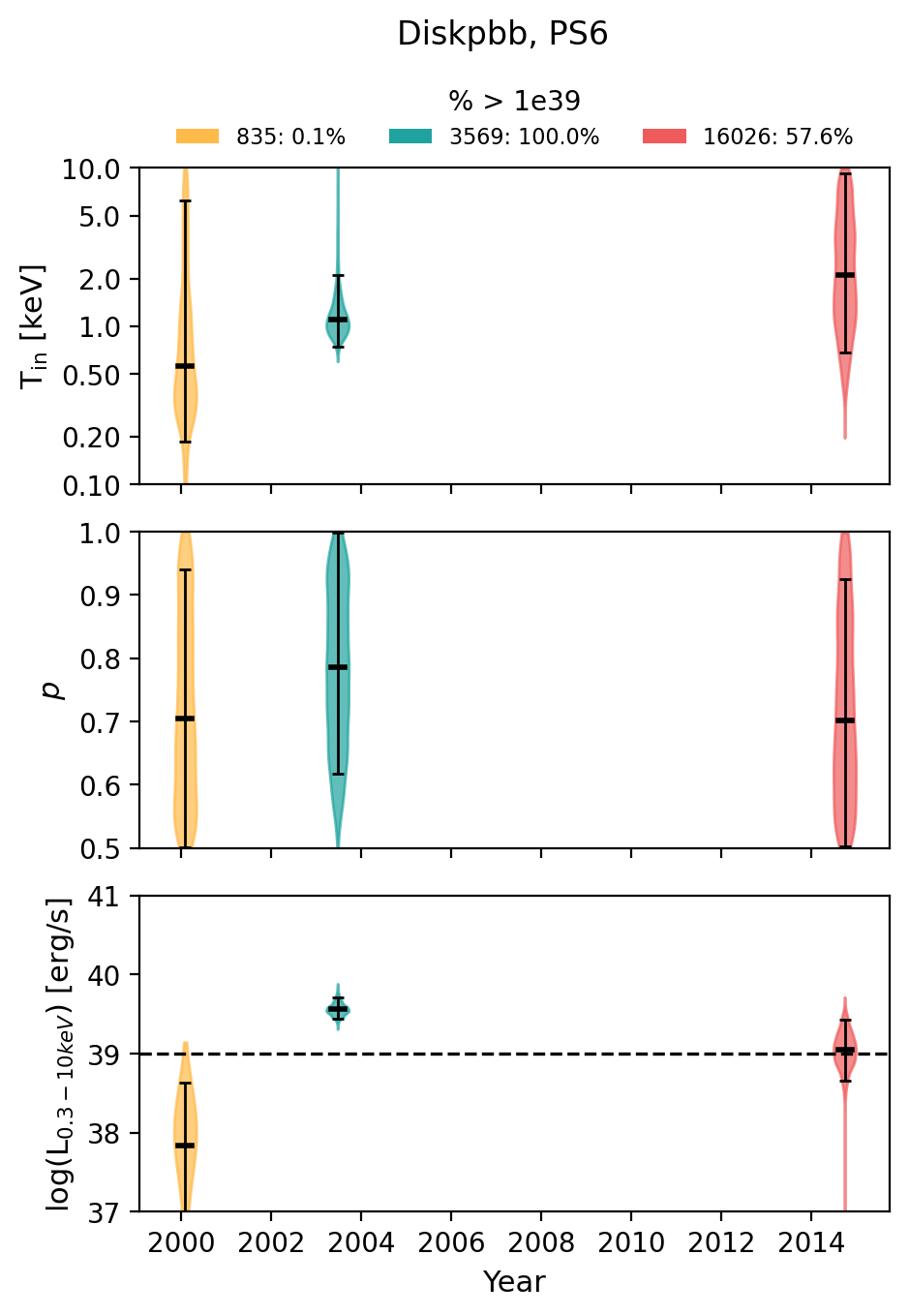}
        \caption{Parameter posterior distributions for the three models used on PS4, PS5 and PS6. See Sect. \ref{App:figures} for details.} \label{fig:app:posteriors_456}
\end{figure*}

\begin{figure*}[h!]
        \centering
        \includegraphics[scale=0.47,valign=c]{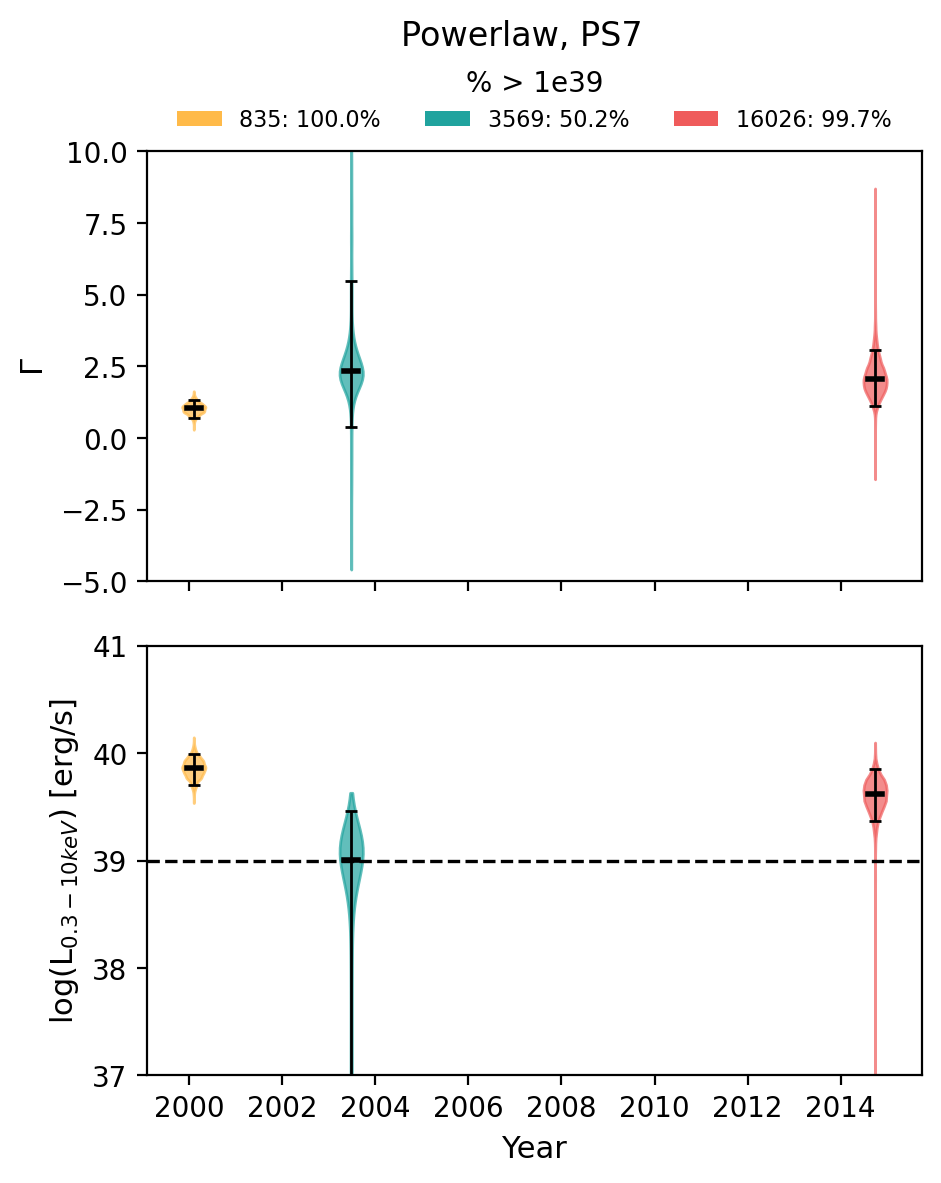}
        \includegraphics[scale=0.47,valign=c]{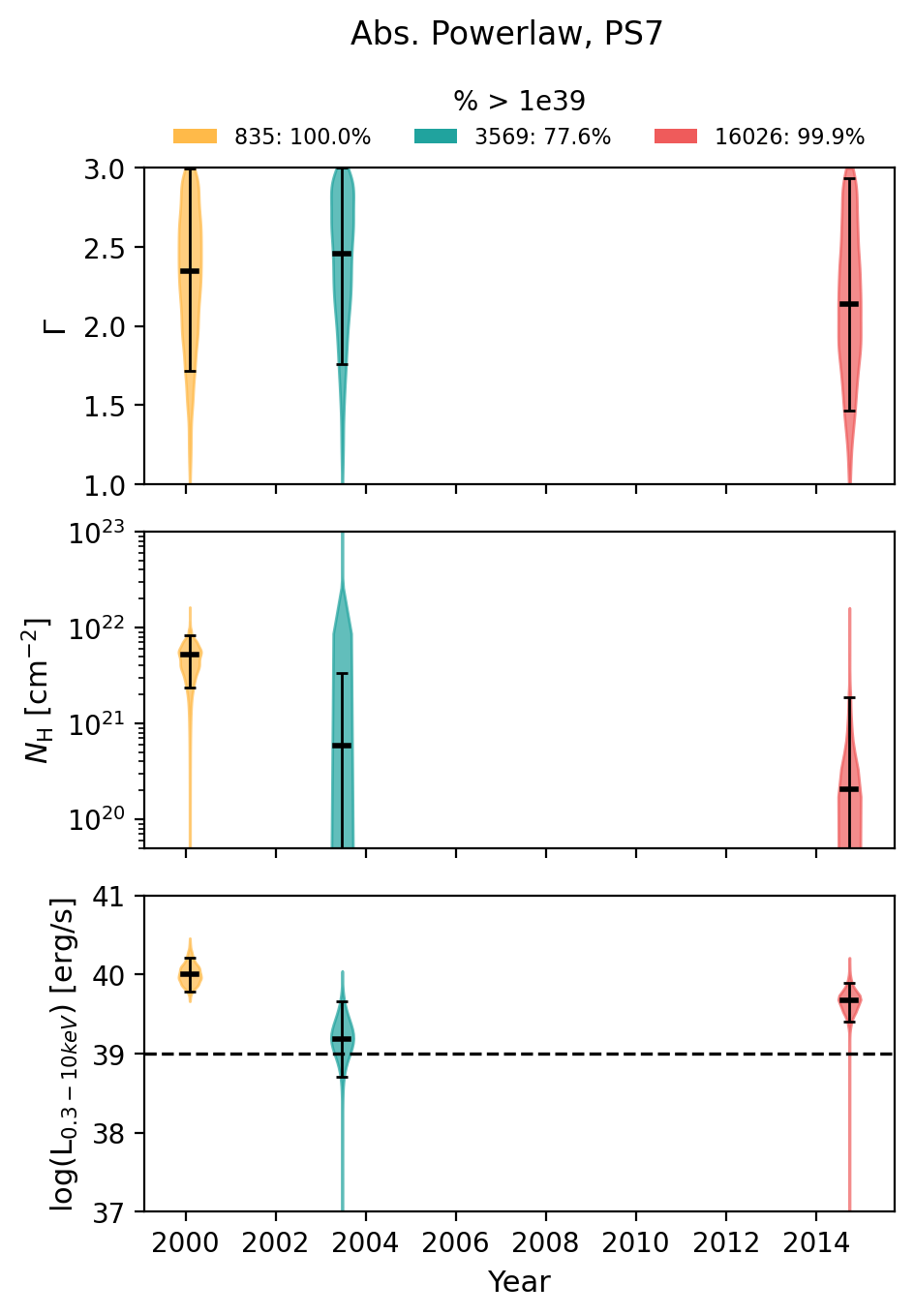  }
        \includegraphics[scale=0.47,valign=c]{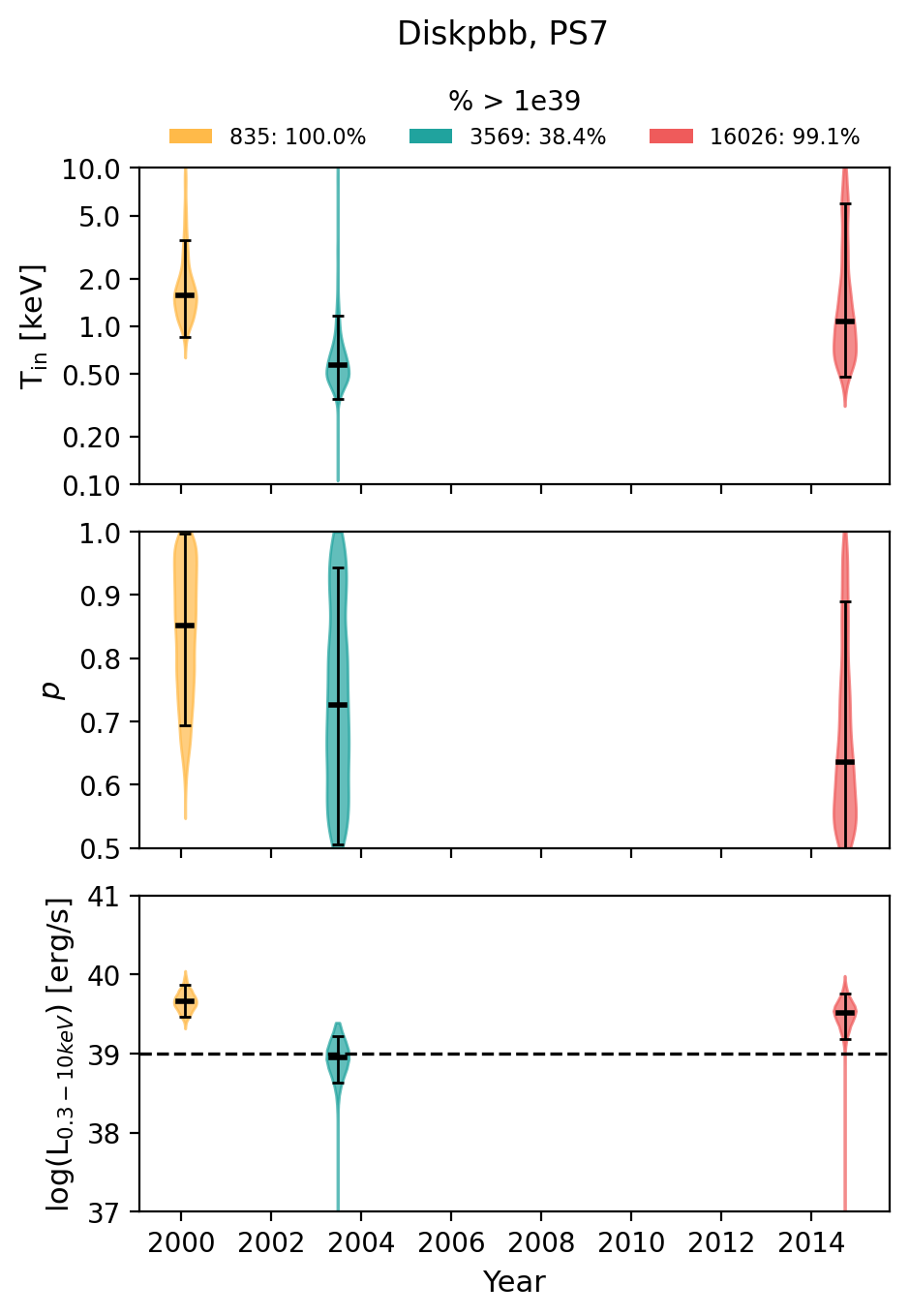}
        \includegraphics[scale=0.47,valign=c]{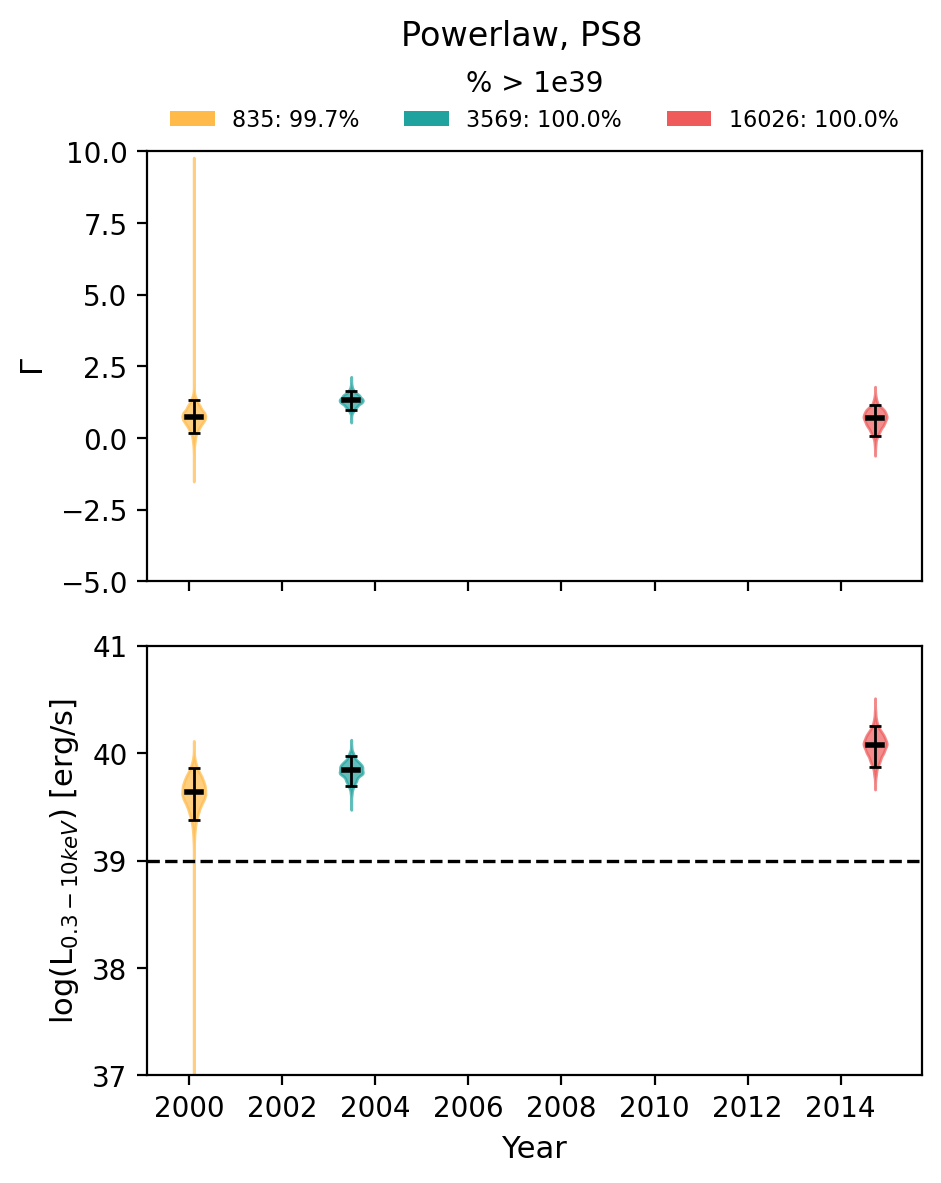}
        \includegraphics[scale=0.47,valign=c]{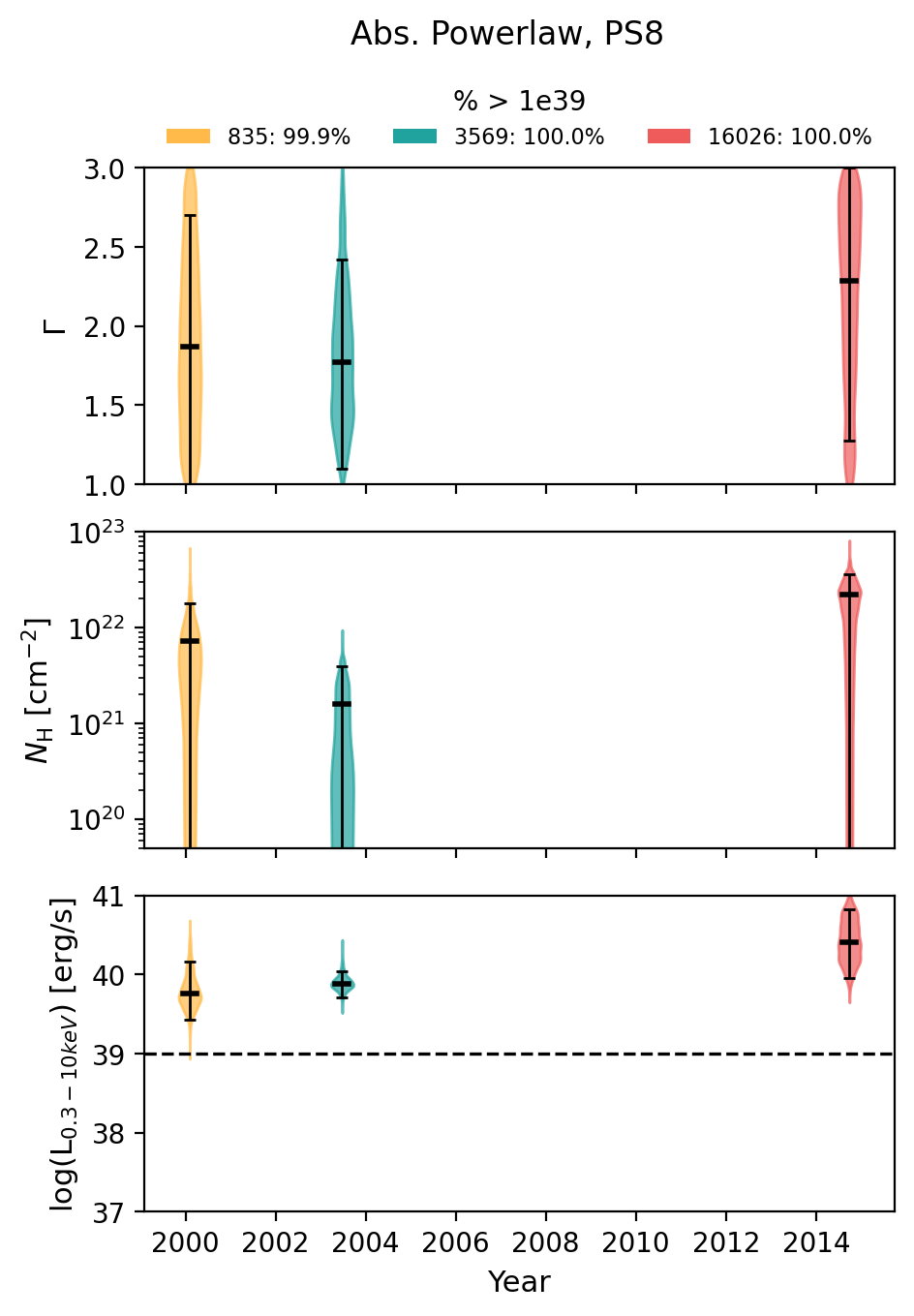  }
        \includegraphics[scale=0.47,valign=c]{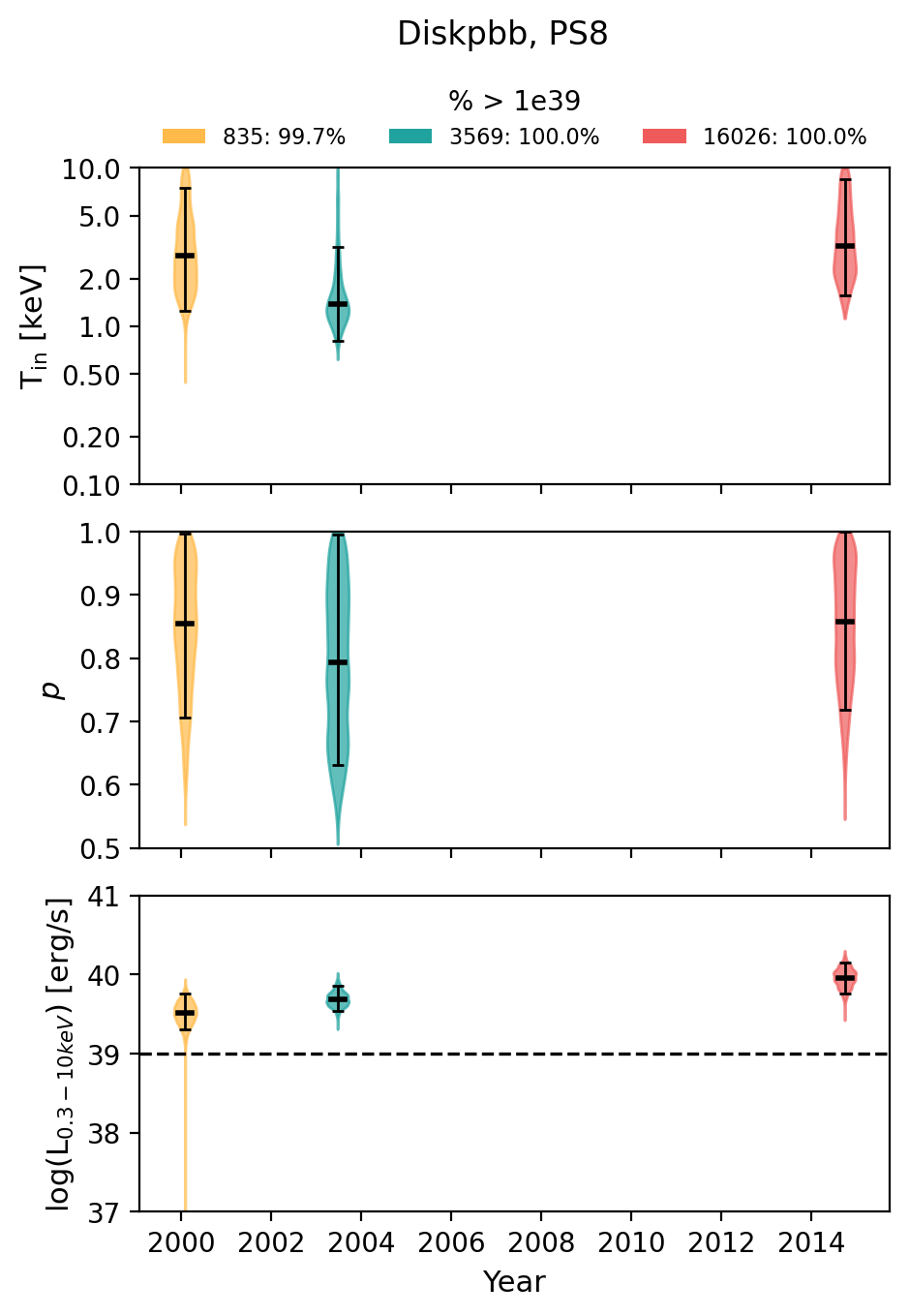}
        \includegraphics[scale=0.47,valign=c]{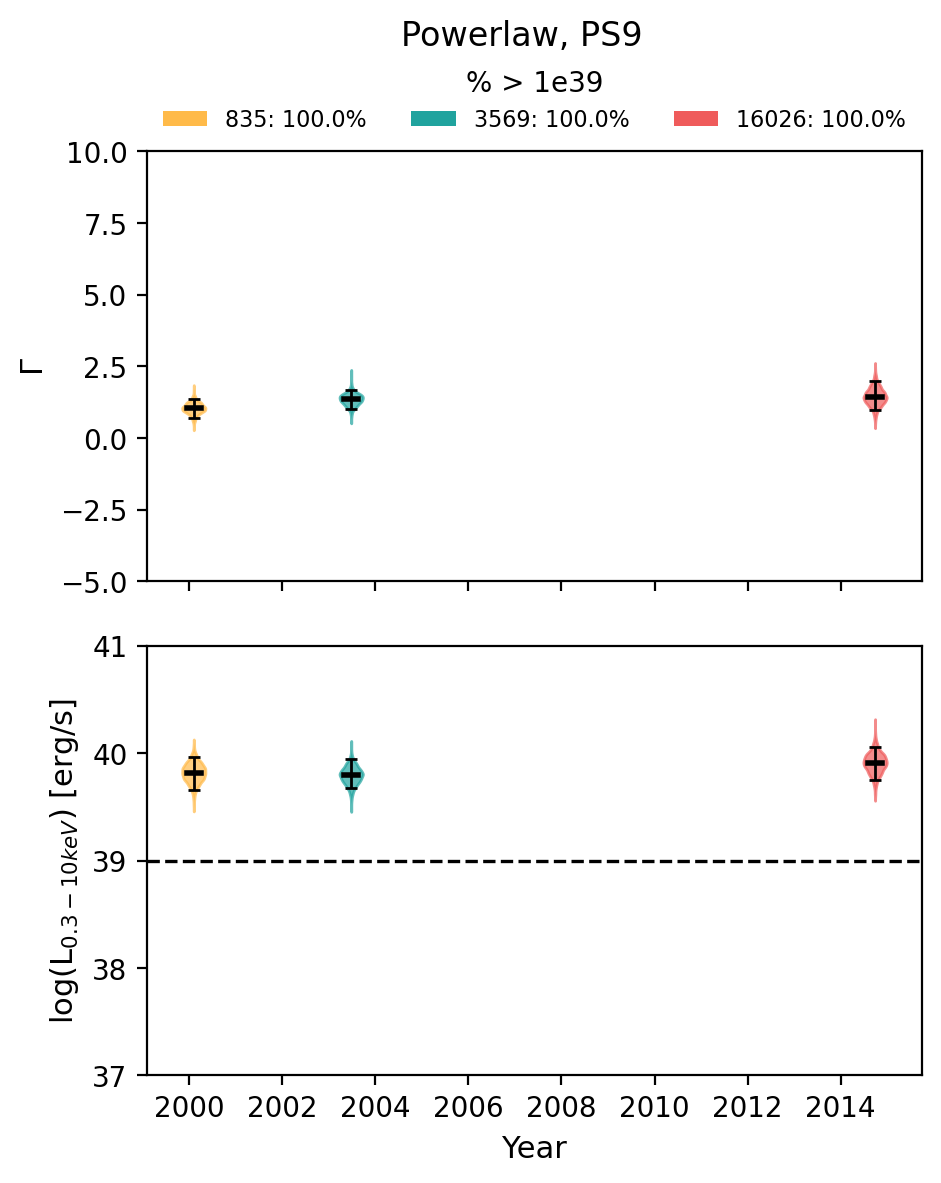}
        \includegraphics[scale=0.47,valign=c]{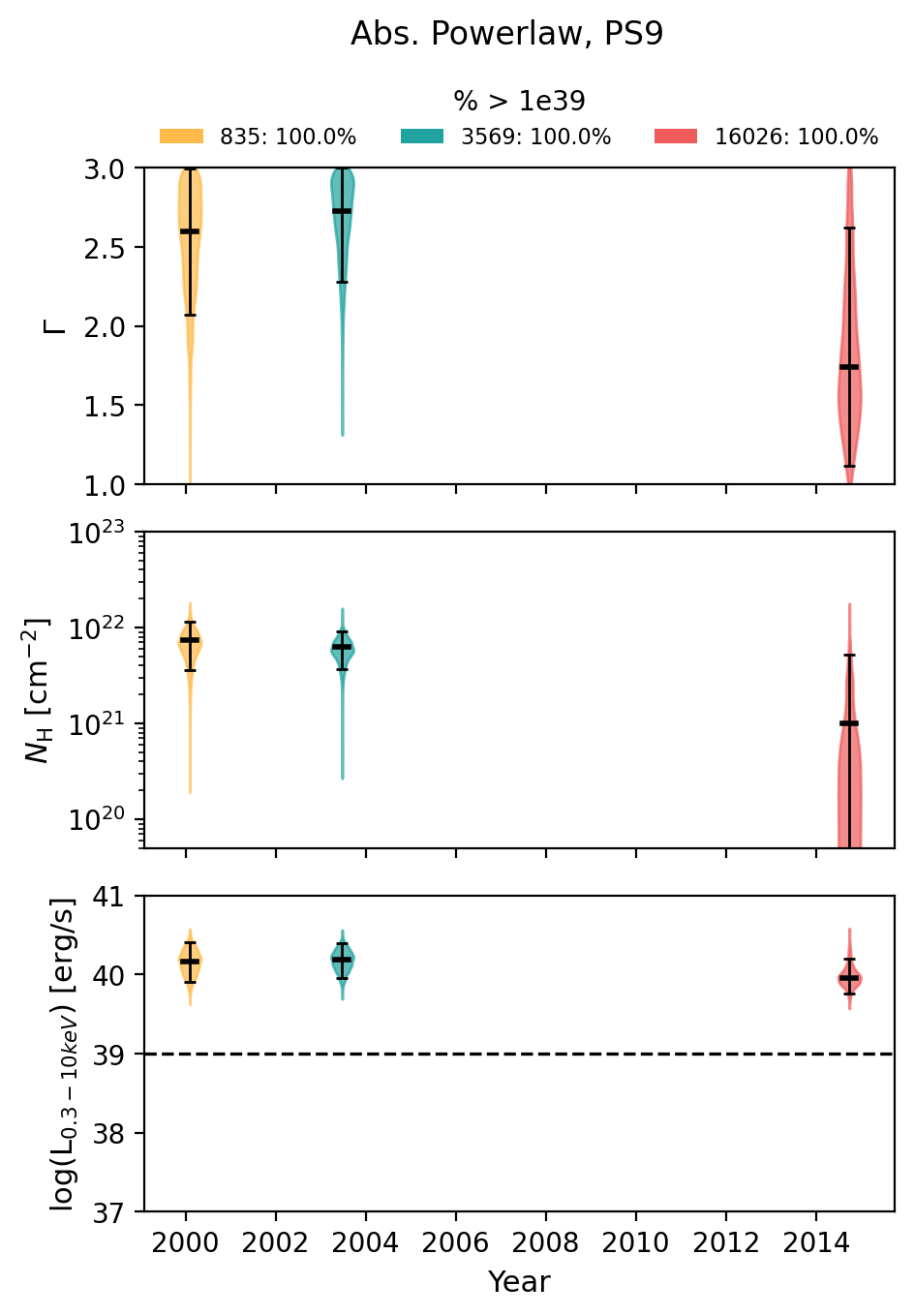  }
        \includegraphics[scale=0.47,valign=c]{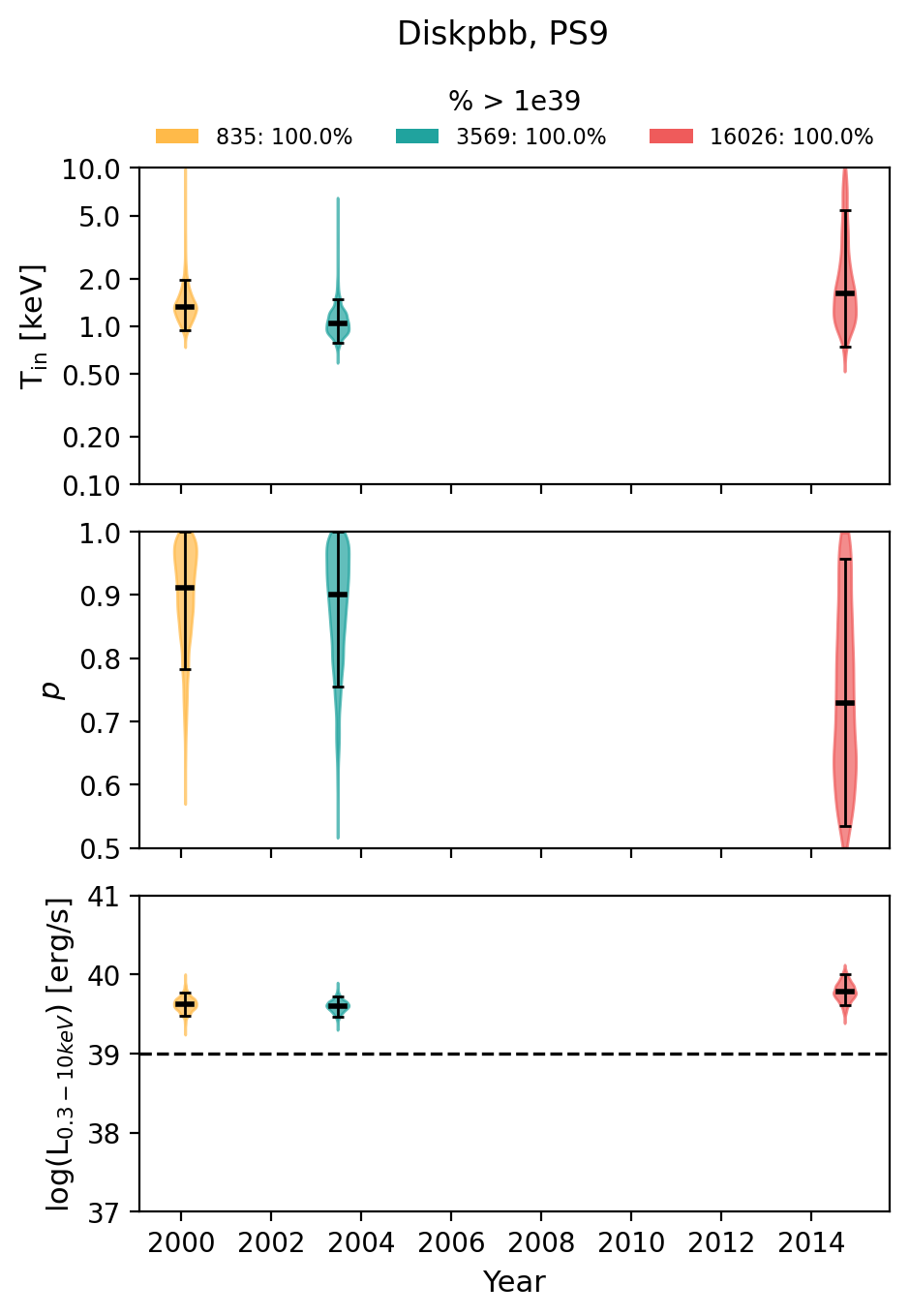}
        \caption{Parameter posterior distributions for the three models used on PS7, PS8 and PS9. See Sect. \ref{App:figures} for details.} \label{fig:app:posteriors_789}
\end{figure*}

\begin{figure*}[h!]
        \centering
        \includegraphics[scale=0.47,valign=c]{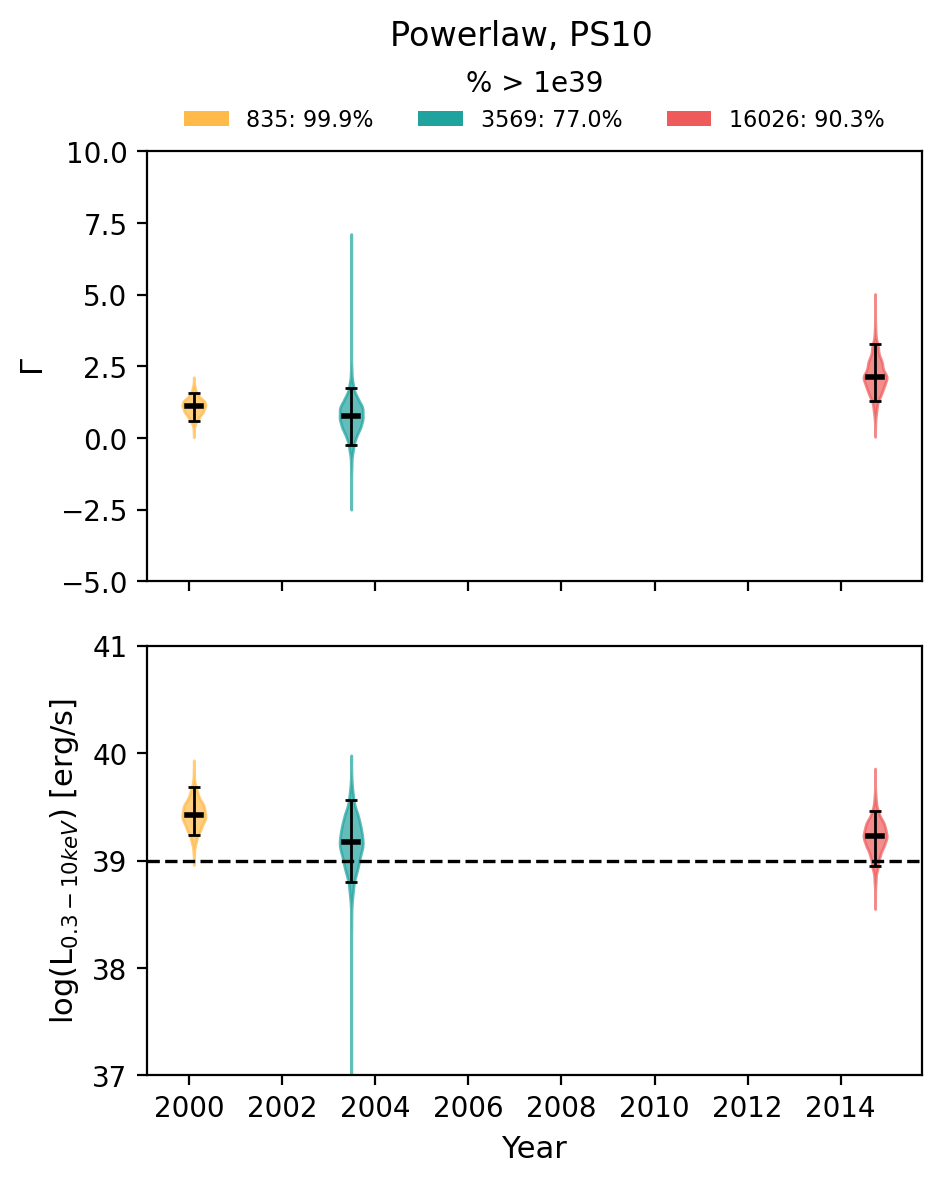}
        \includegraphics[scale=0.47,valign=c]{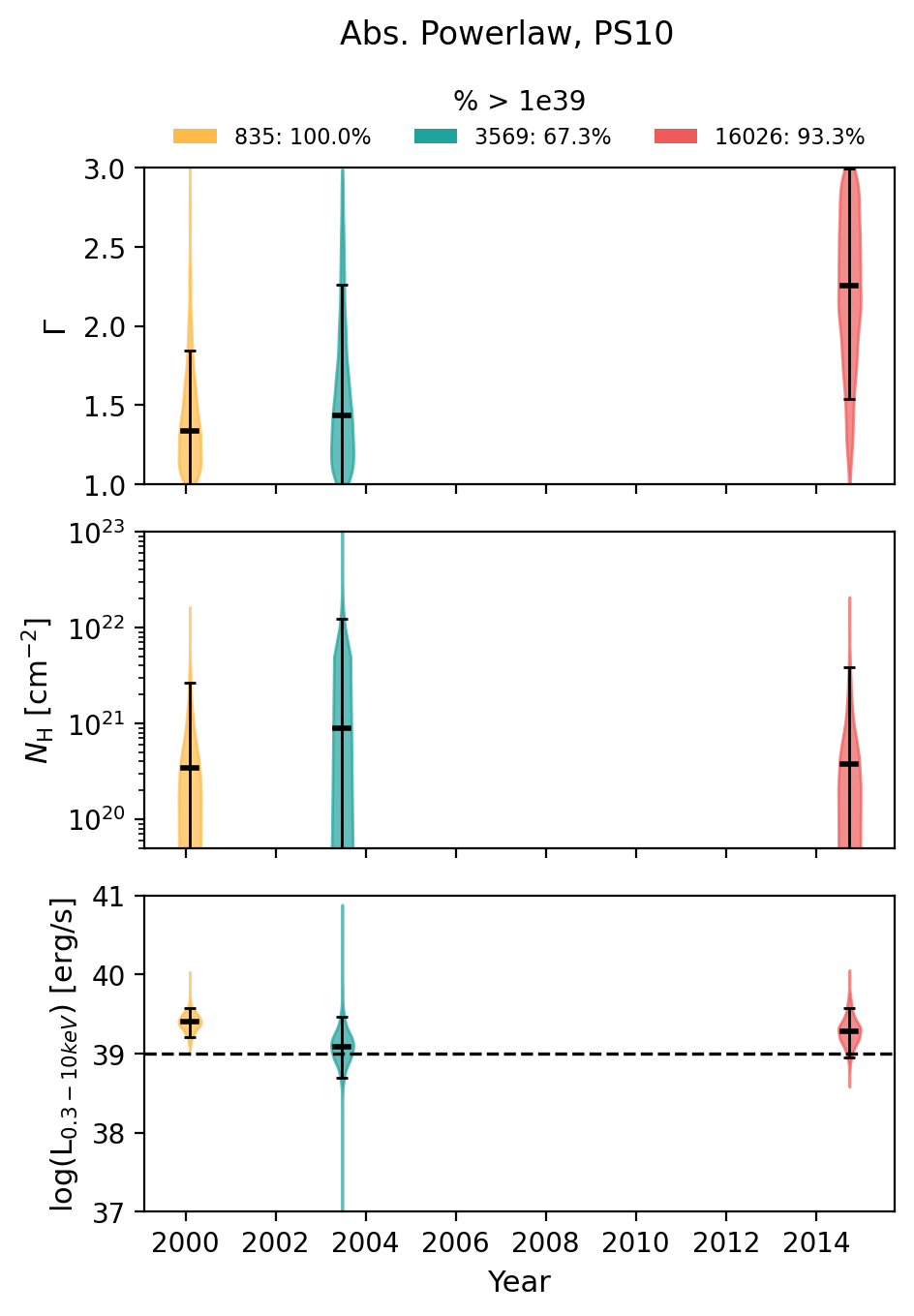  }
        \includegraphics[scale=0.47,valign=c]{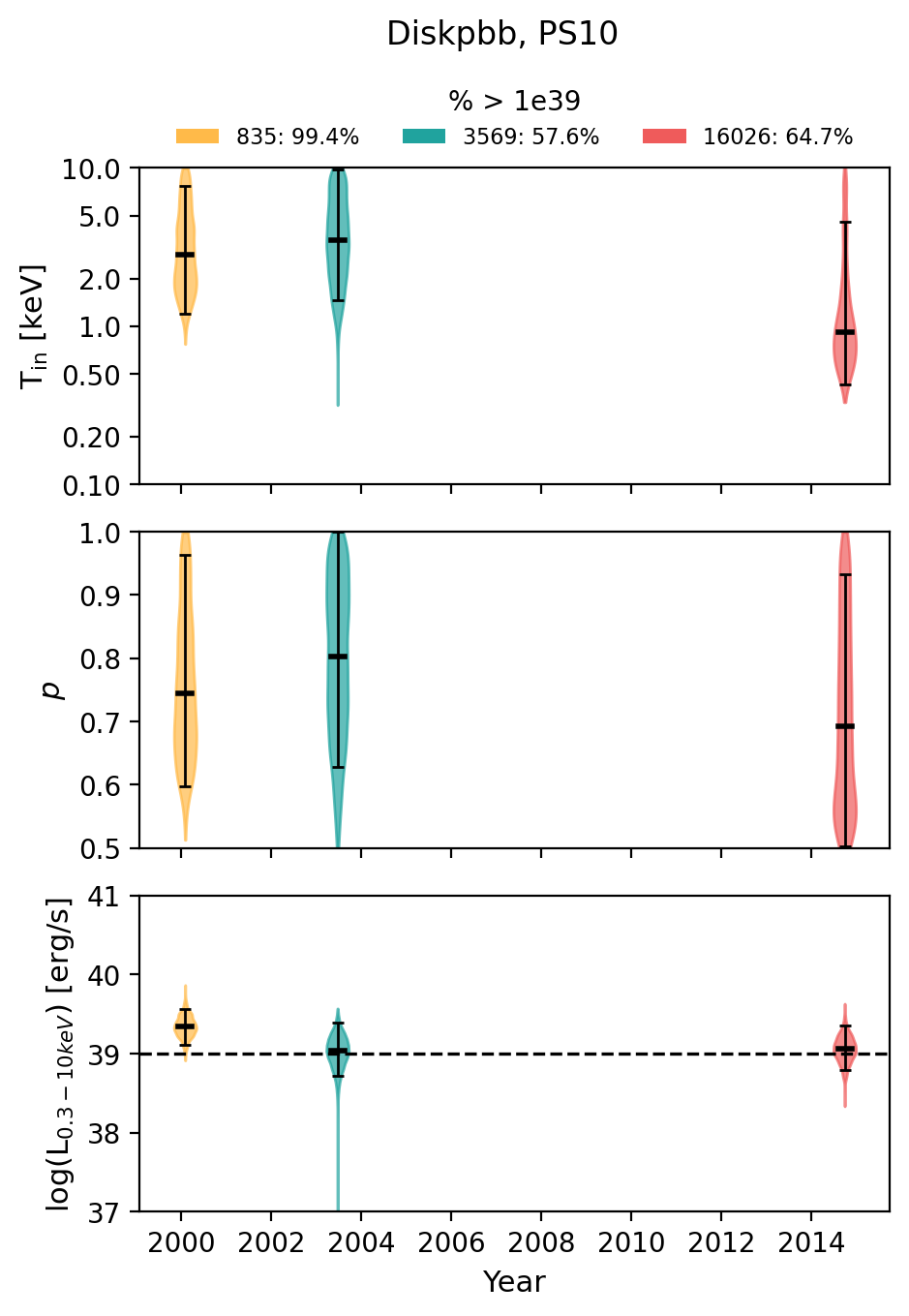}
        \includegraphics[scale=0.47,valign=c]{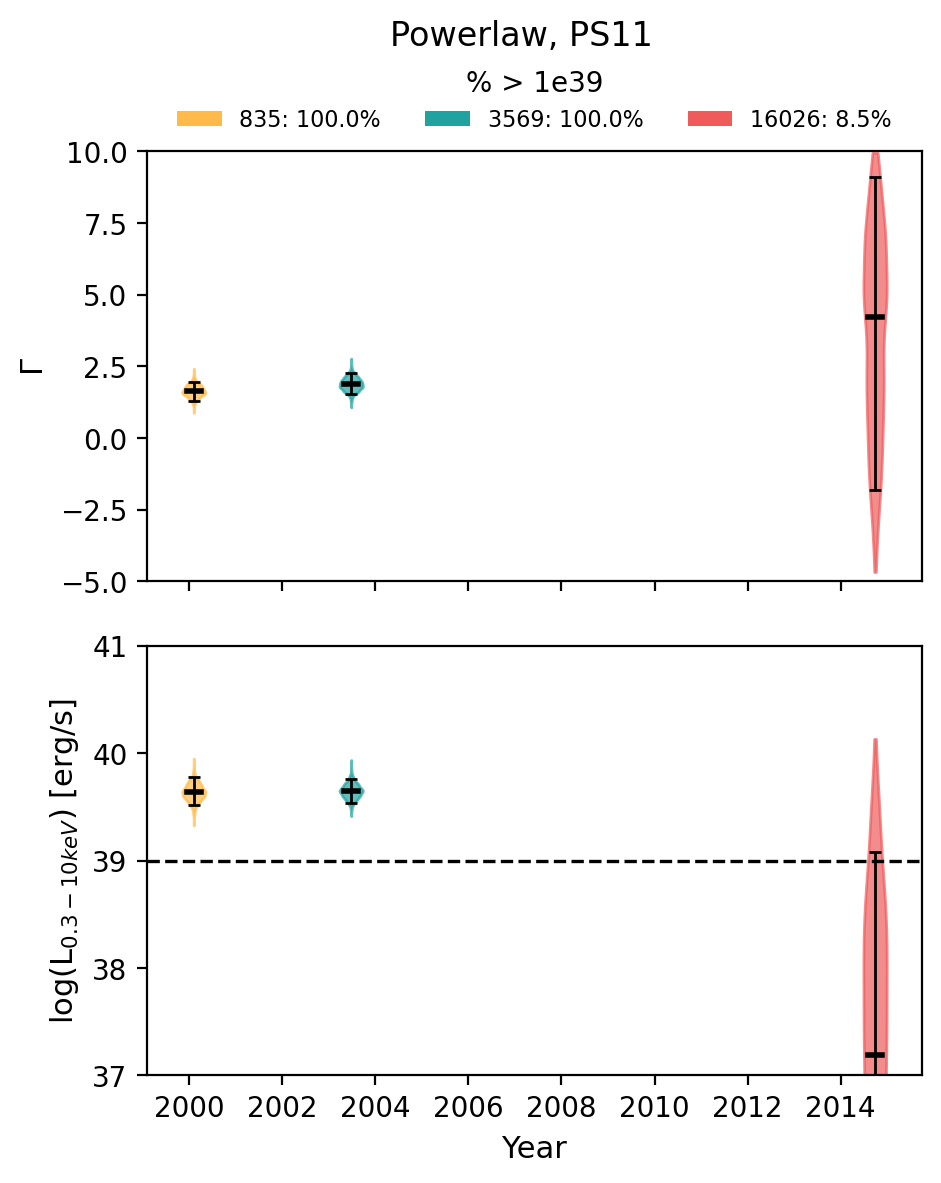}
        \includegraphics[scale=0.47,valign=c]{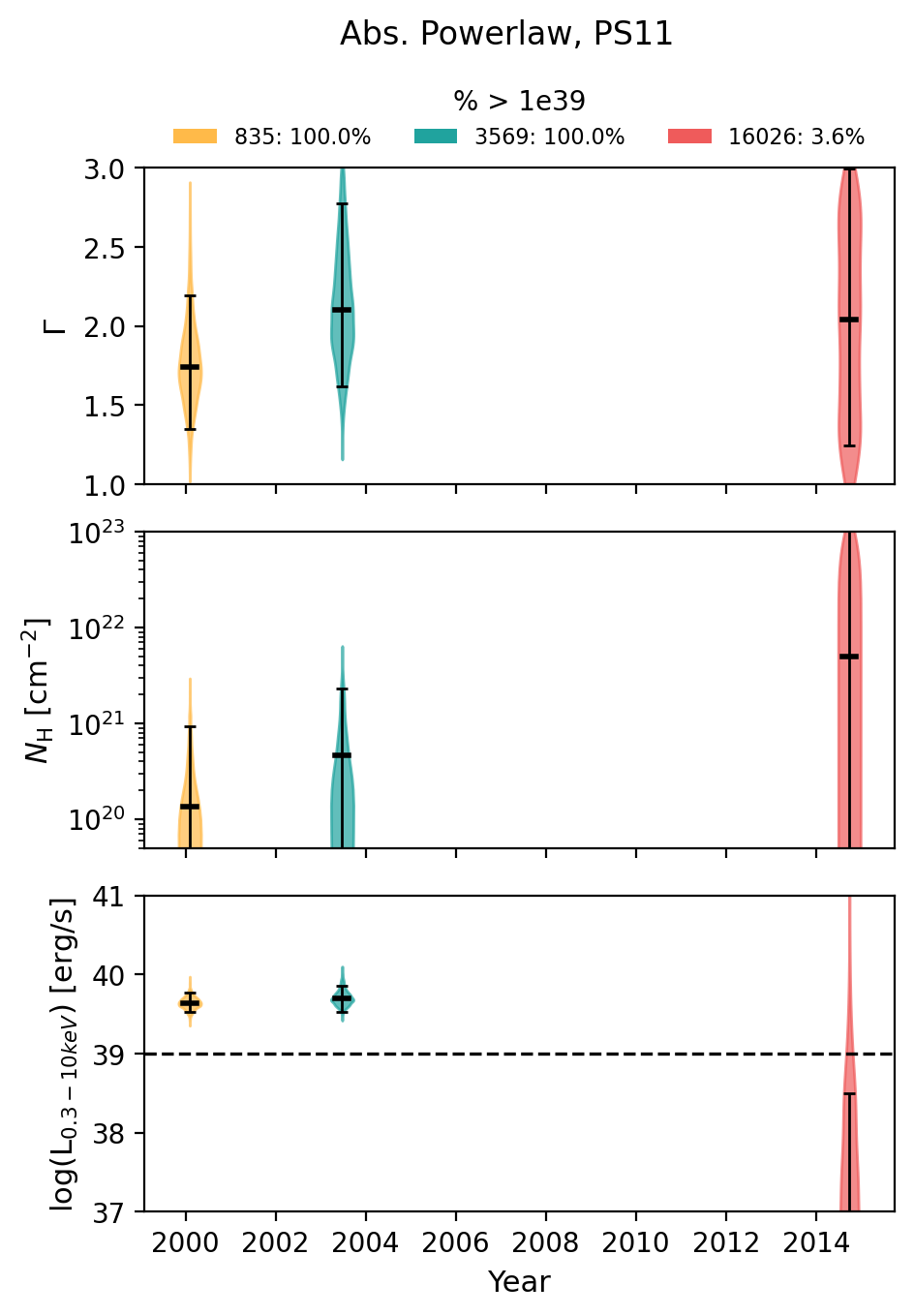  }
        \includegraphics[scale=0.47,valign=c]{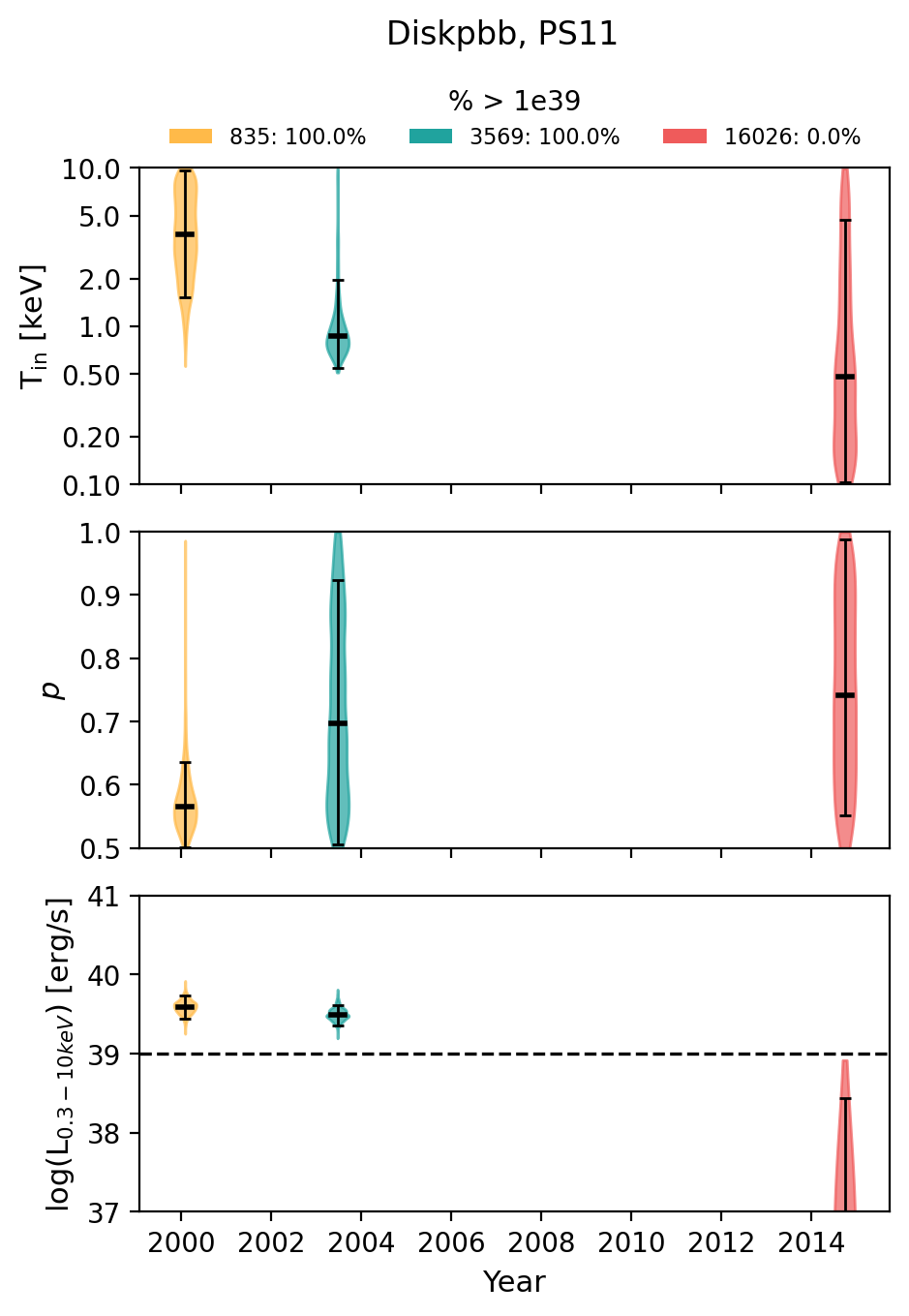}
        \includegraphics[scale=0.47,valign=c]{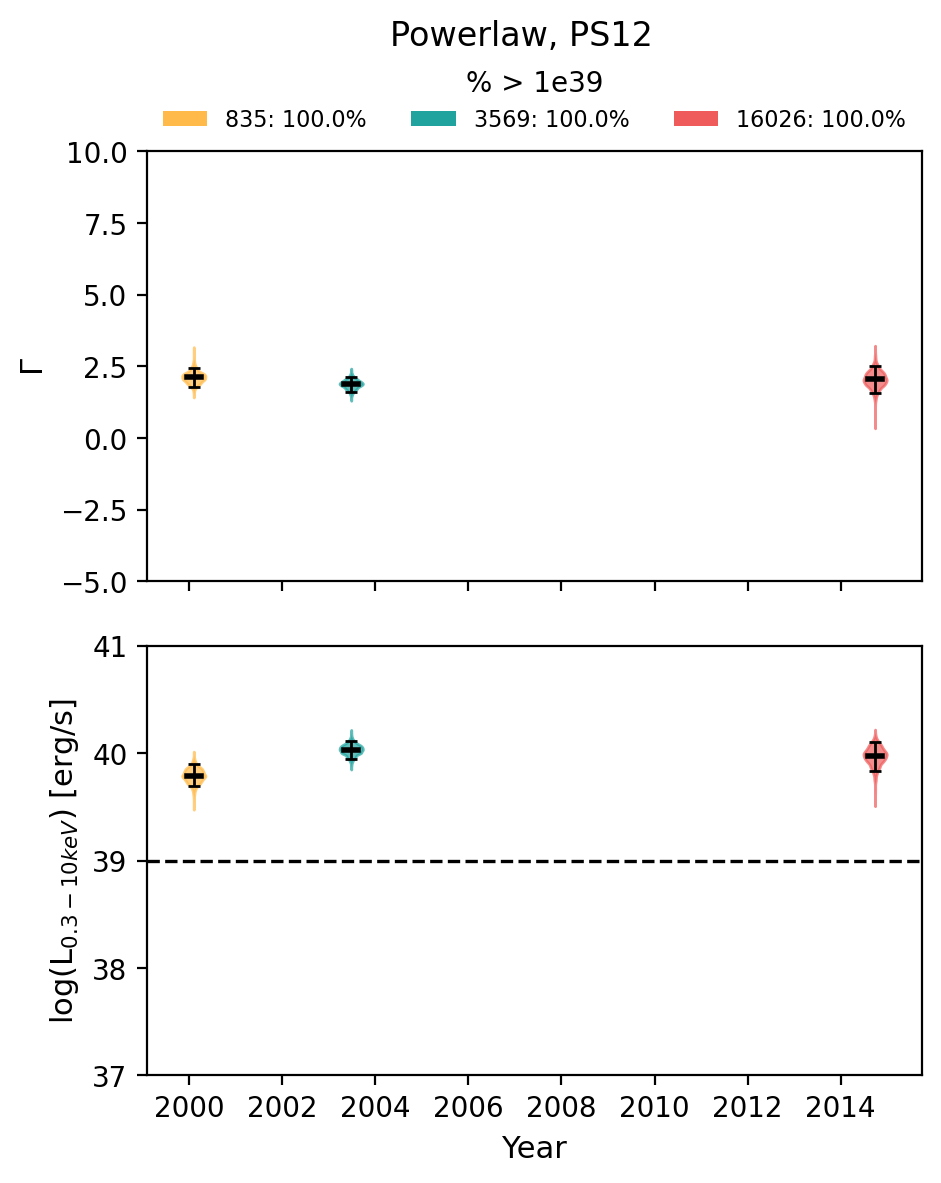}
        \includegraphics[scale=0.47,valign=c]{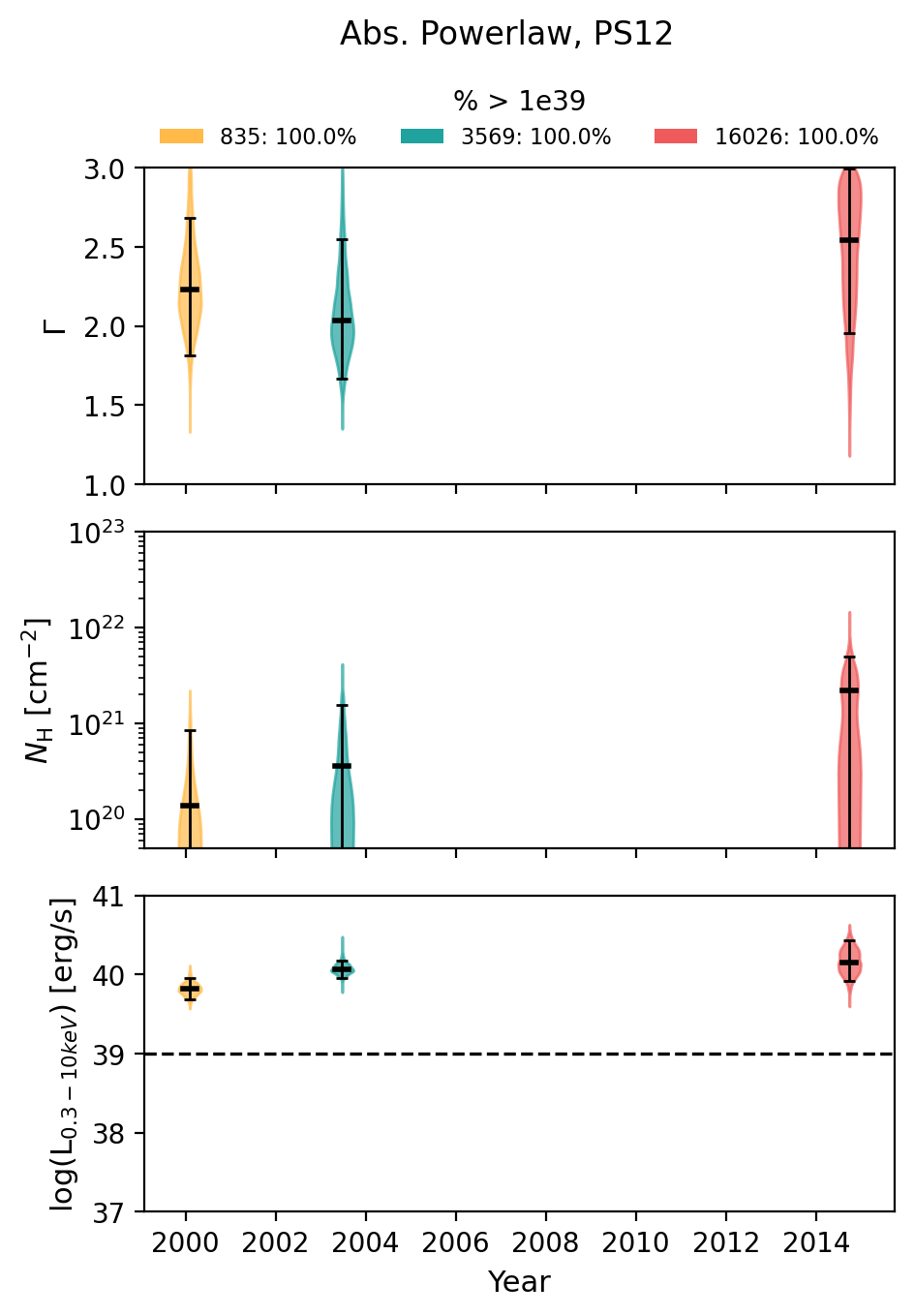  }
        \includegraphics[scale=0.47,valign=c]{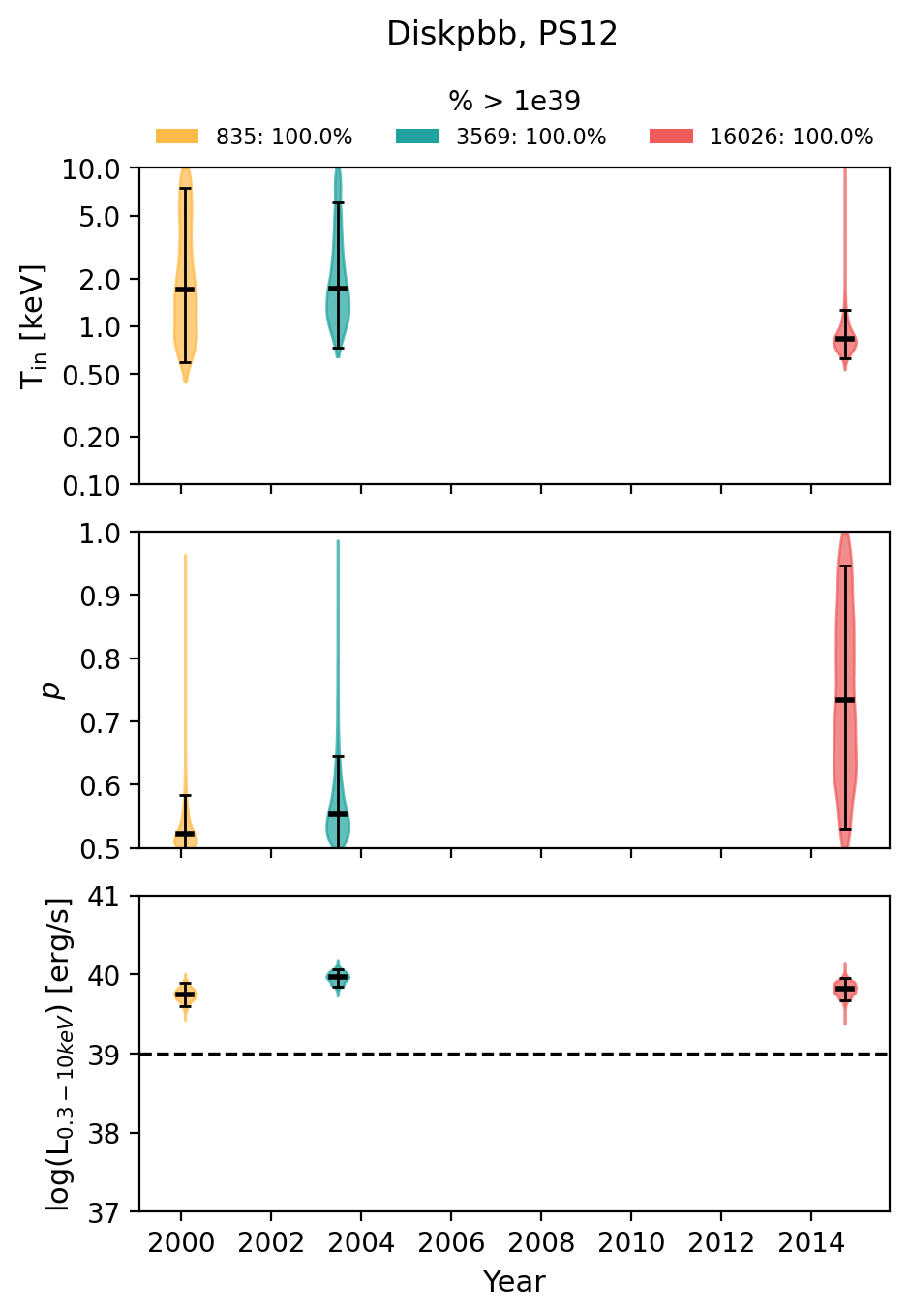}
        \caption{Parameter posterior distributions for the three models used on PS10, PS11 and PS12. See Sect. \ref{App:figures} for details.} \label{fig:app:posteriors_101112}
\end{figure*}

\begin{figure*}[h!]
        \centering
        \includegraphics[scale=0.47,valign=c]{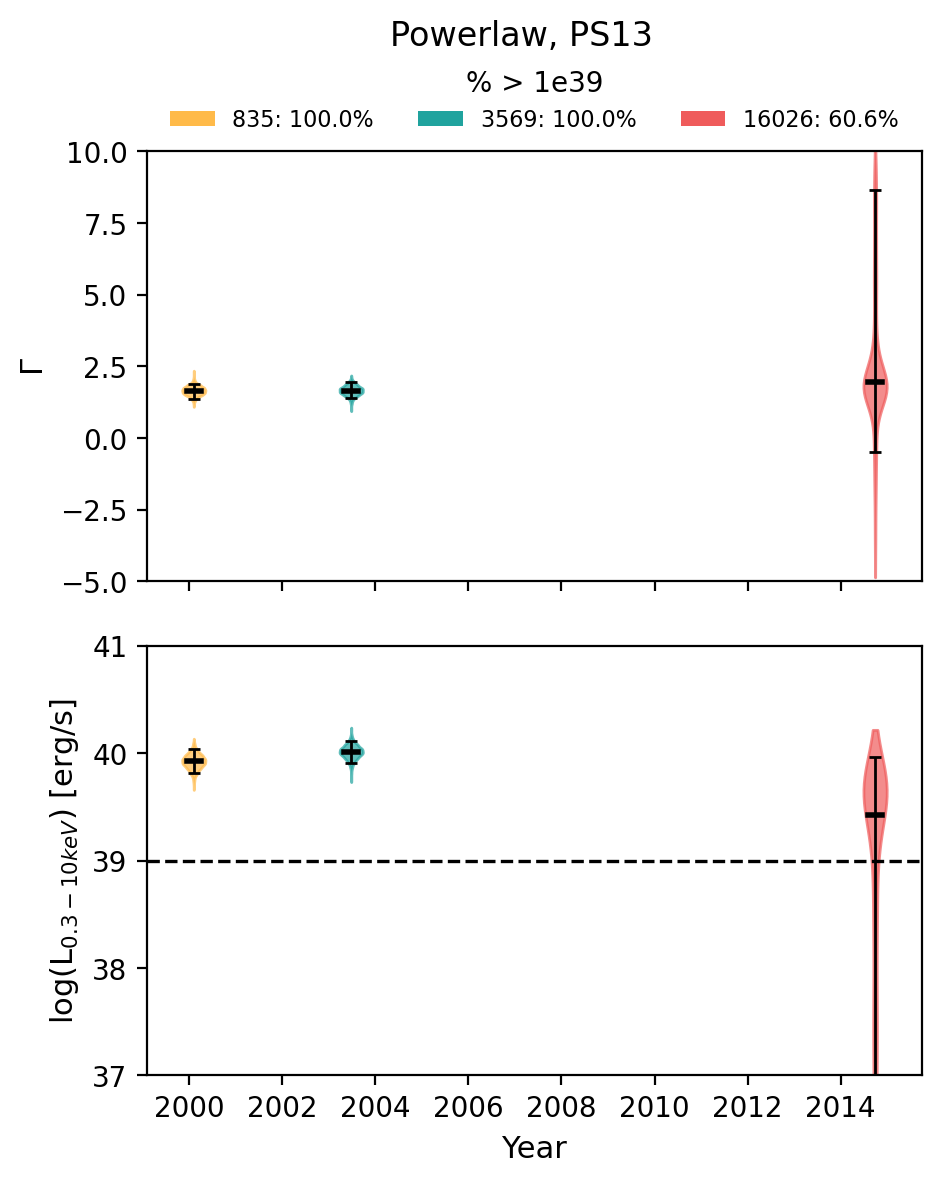}
        \includegraphics[scale=0.47,valign=c]{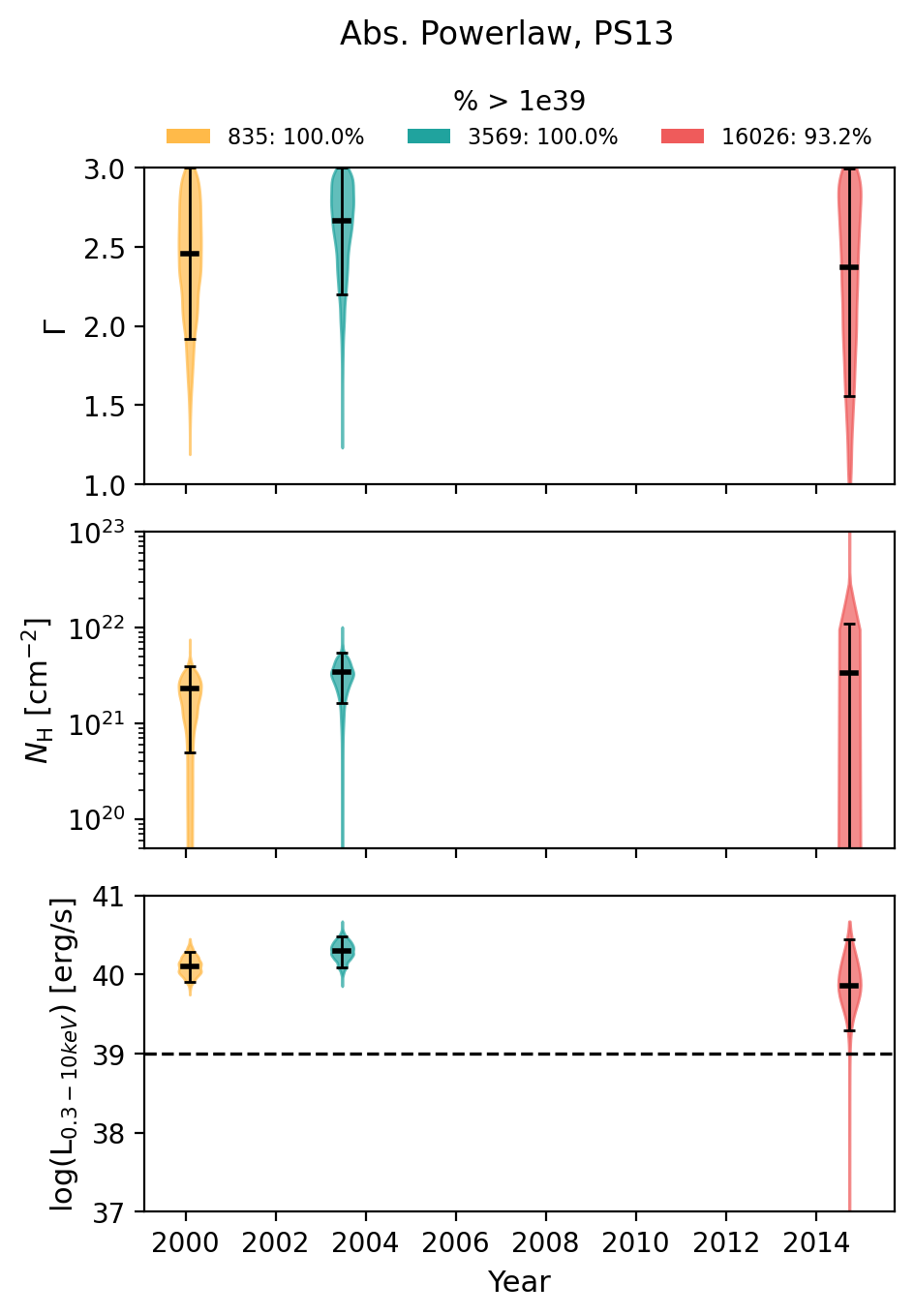  }
        \includegraphics[scale=0.47,valign=c]{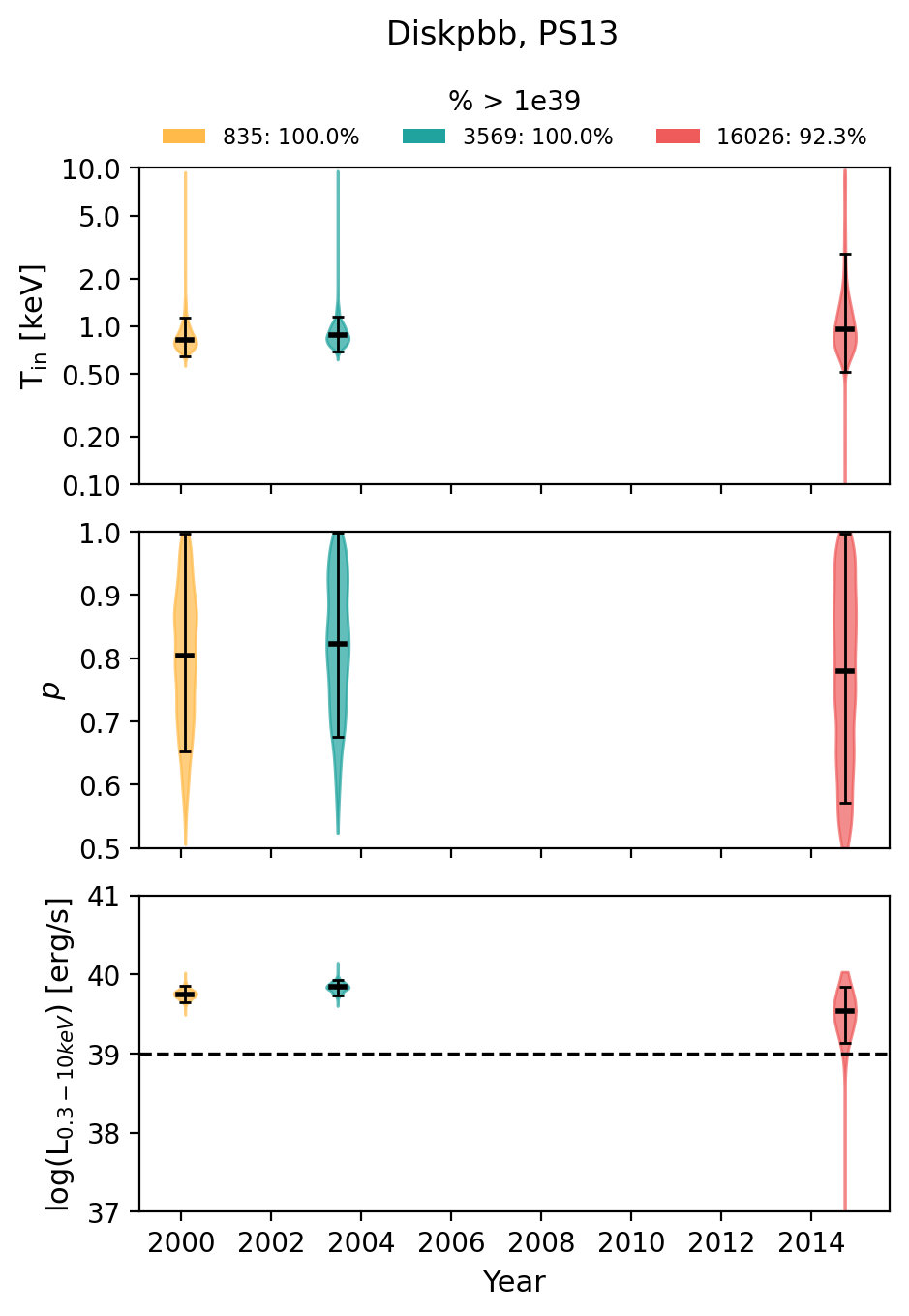}
        \includegraphics[scale=0.47,valign=c]{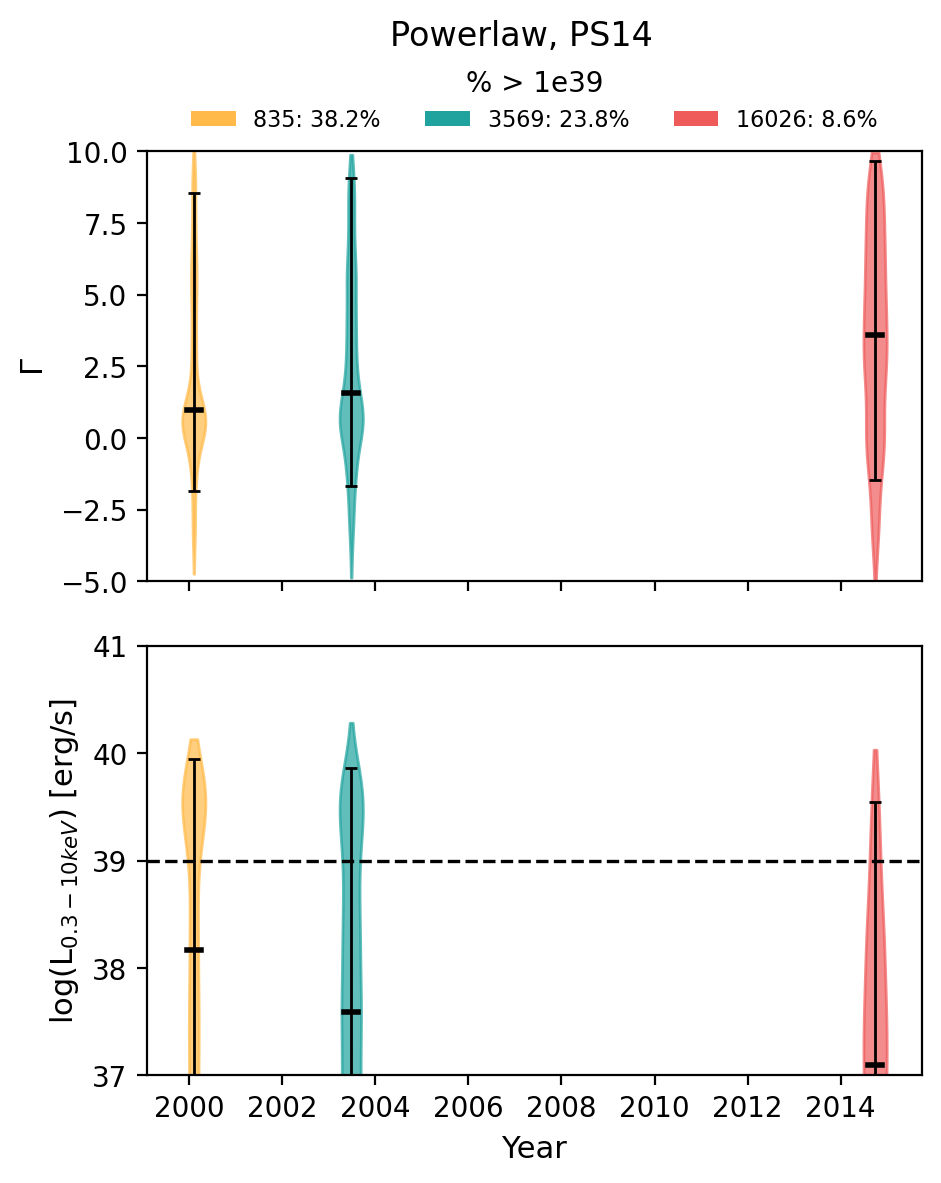}
        \includegraphics[scale=0.47,valign=c]{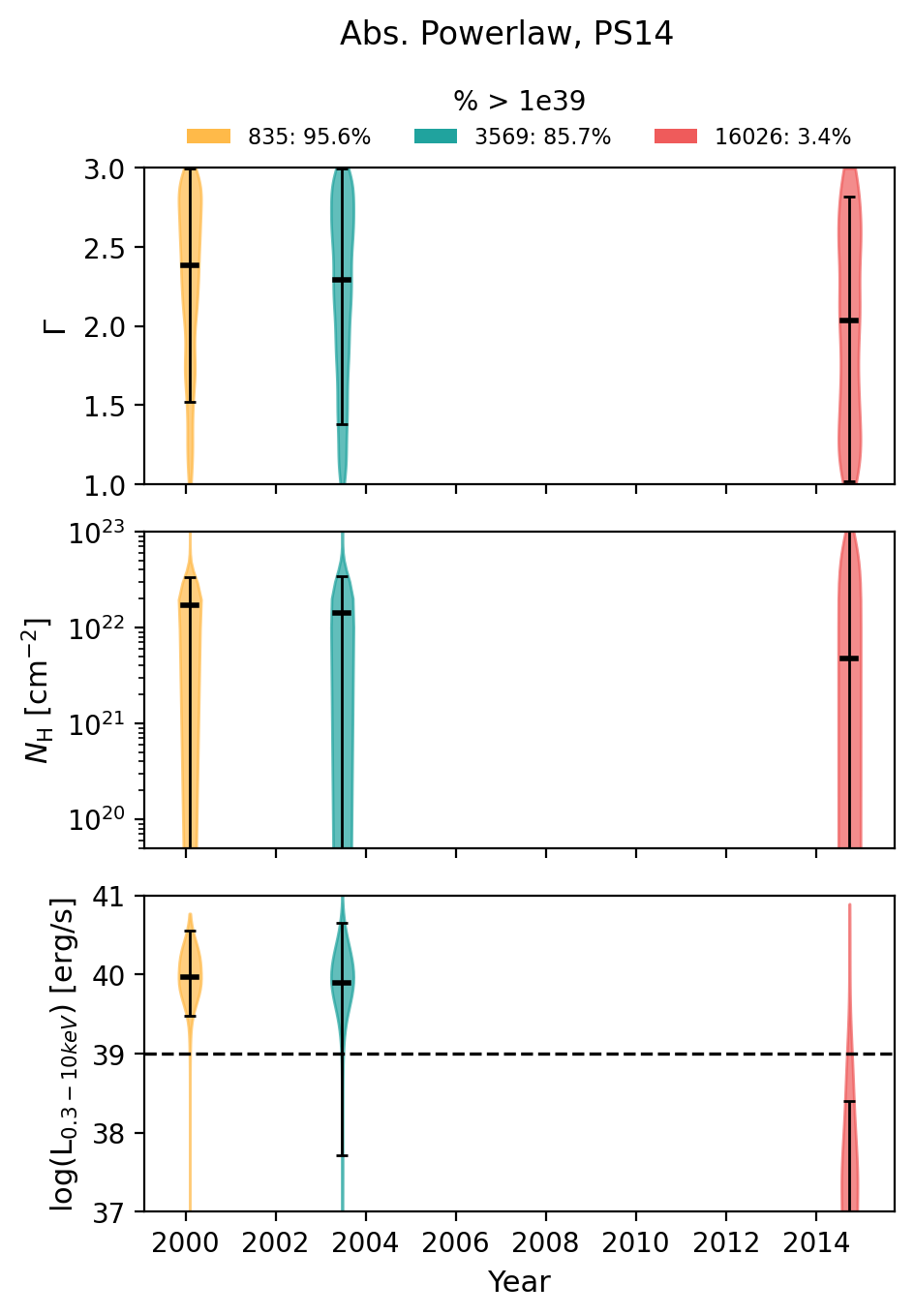  }
        \includegraphics[scale=0.47,valign=c]{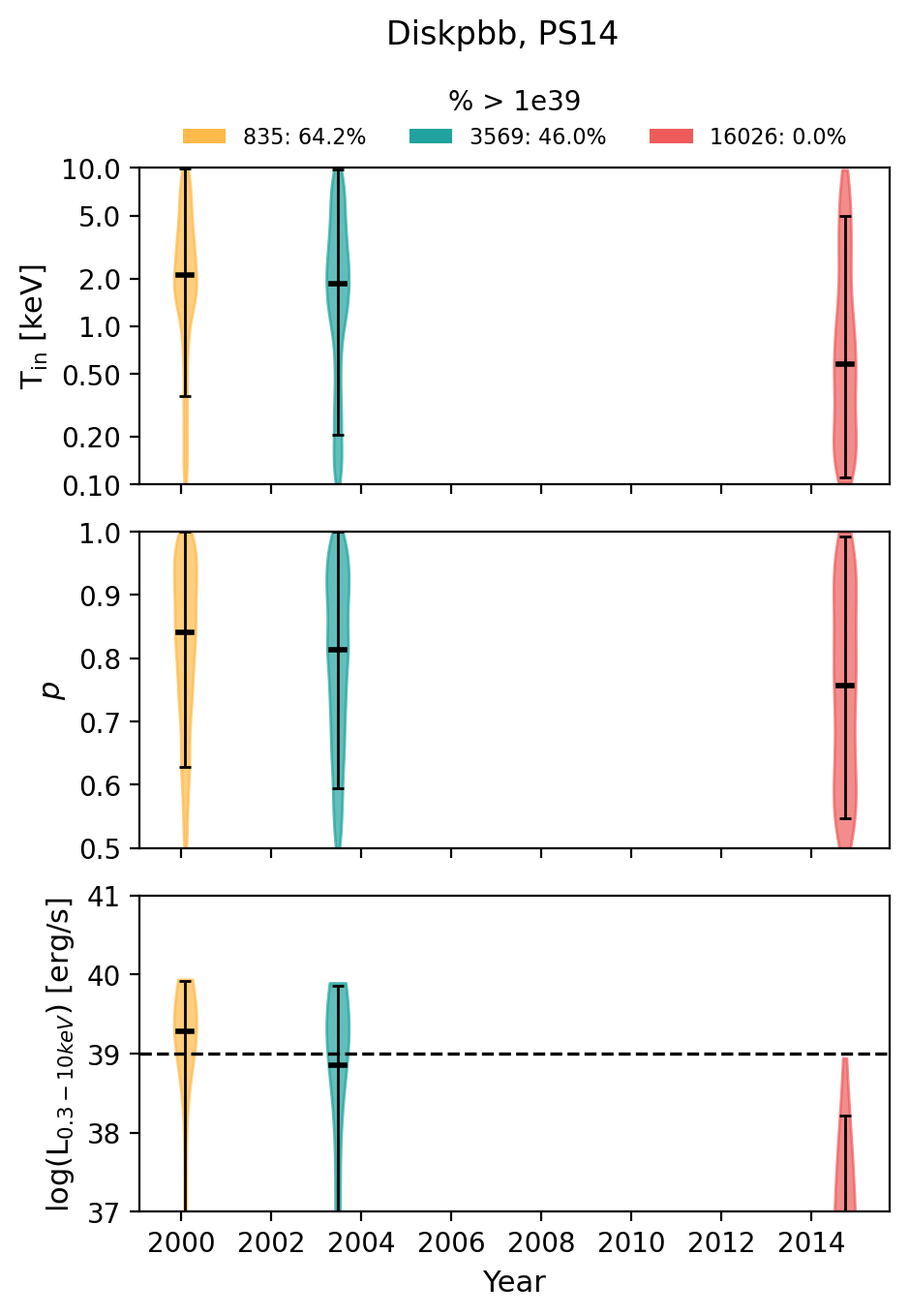}
        \includegraphics[scale=0.47,valign=c]{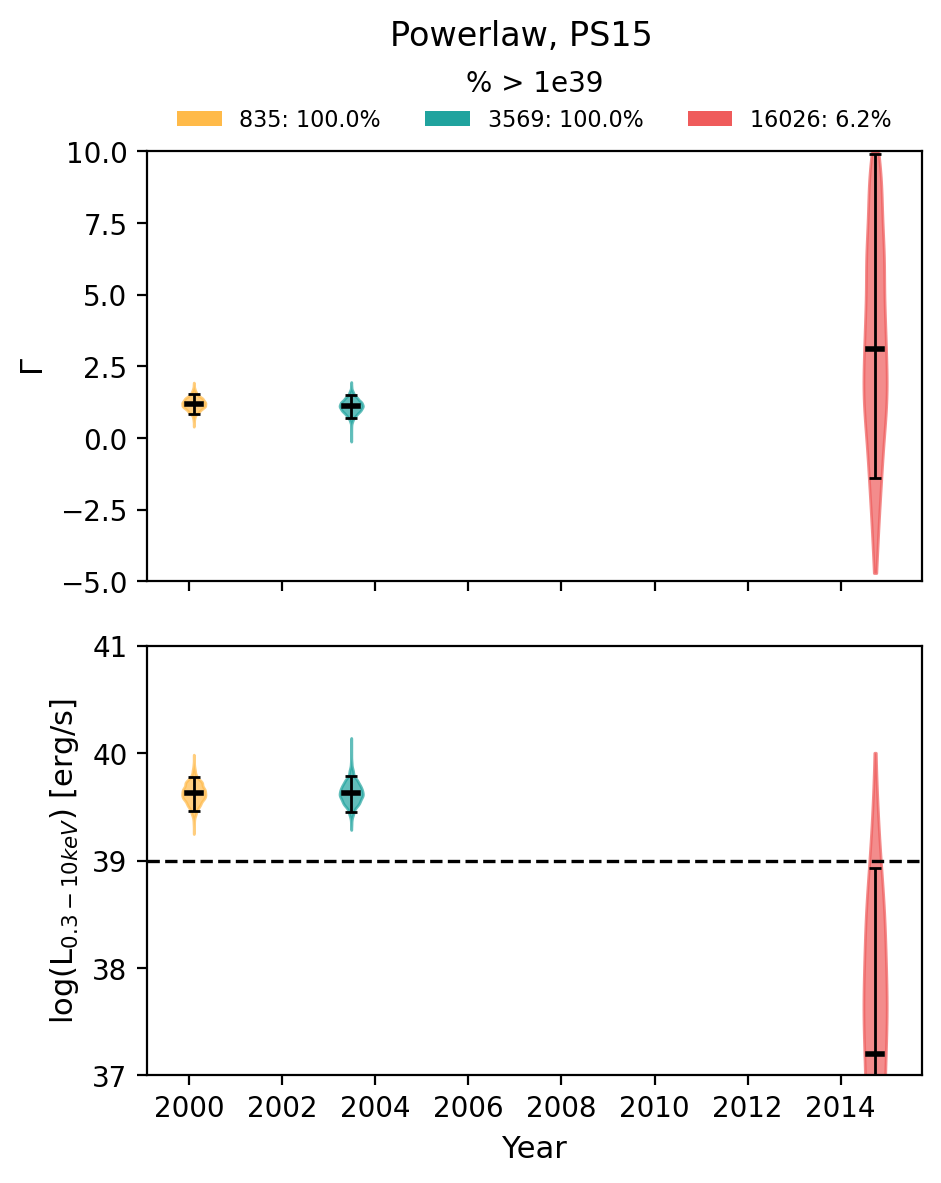}
        \includegraphics[scale=0.47,valign=c]{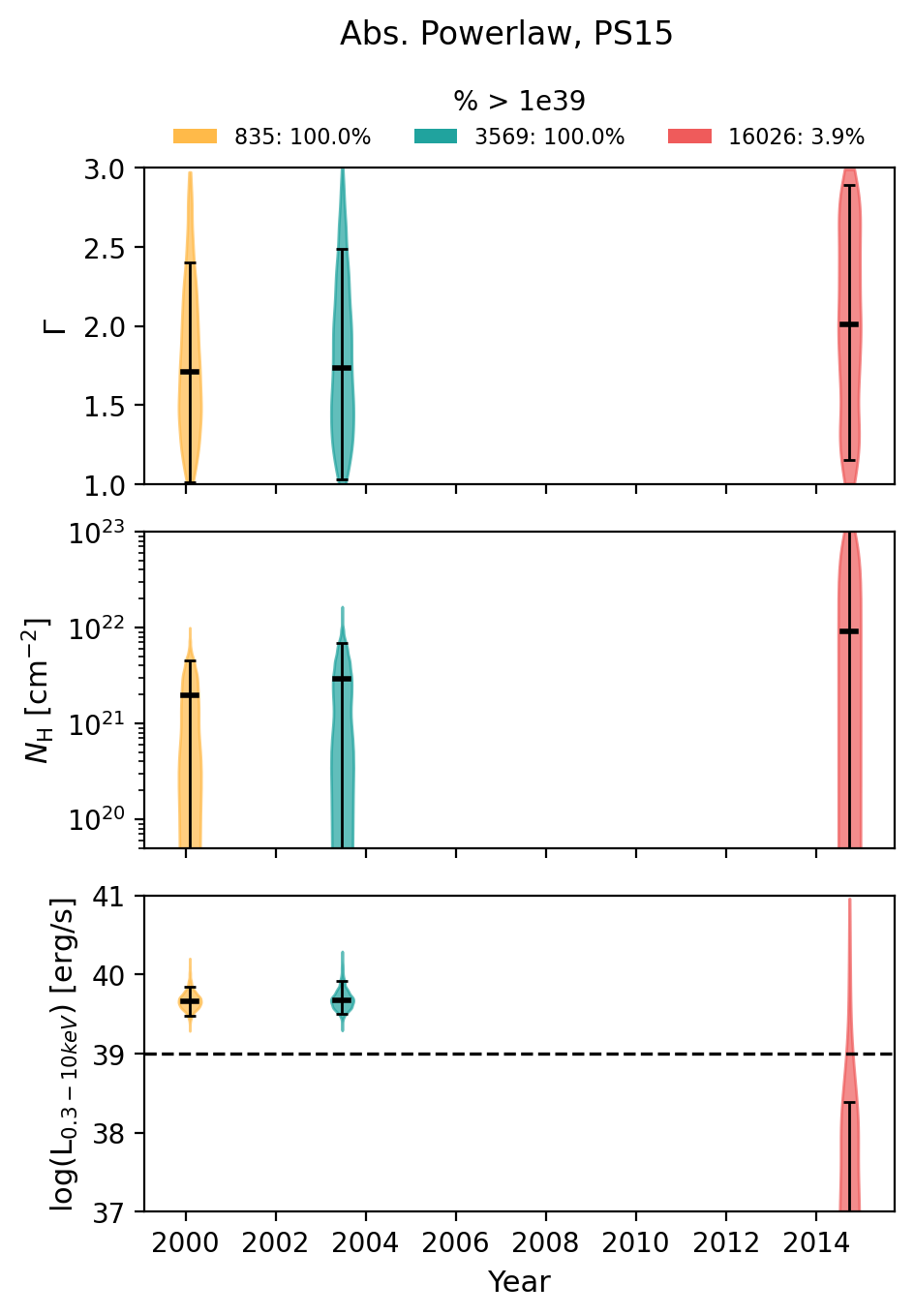  }
        \includegraphics[scale=0.47,valign=c]{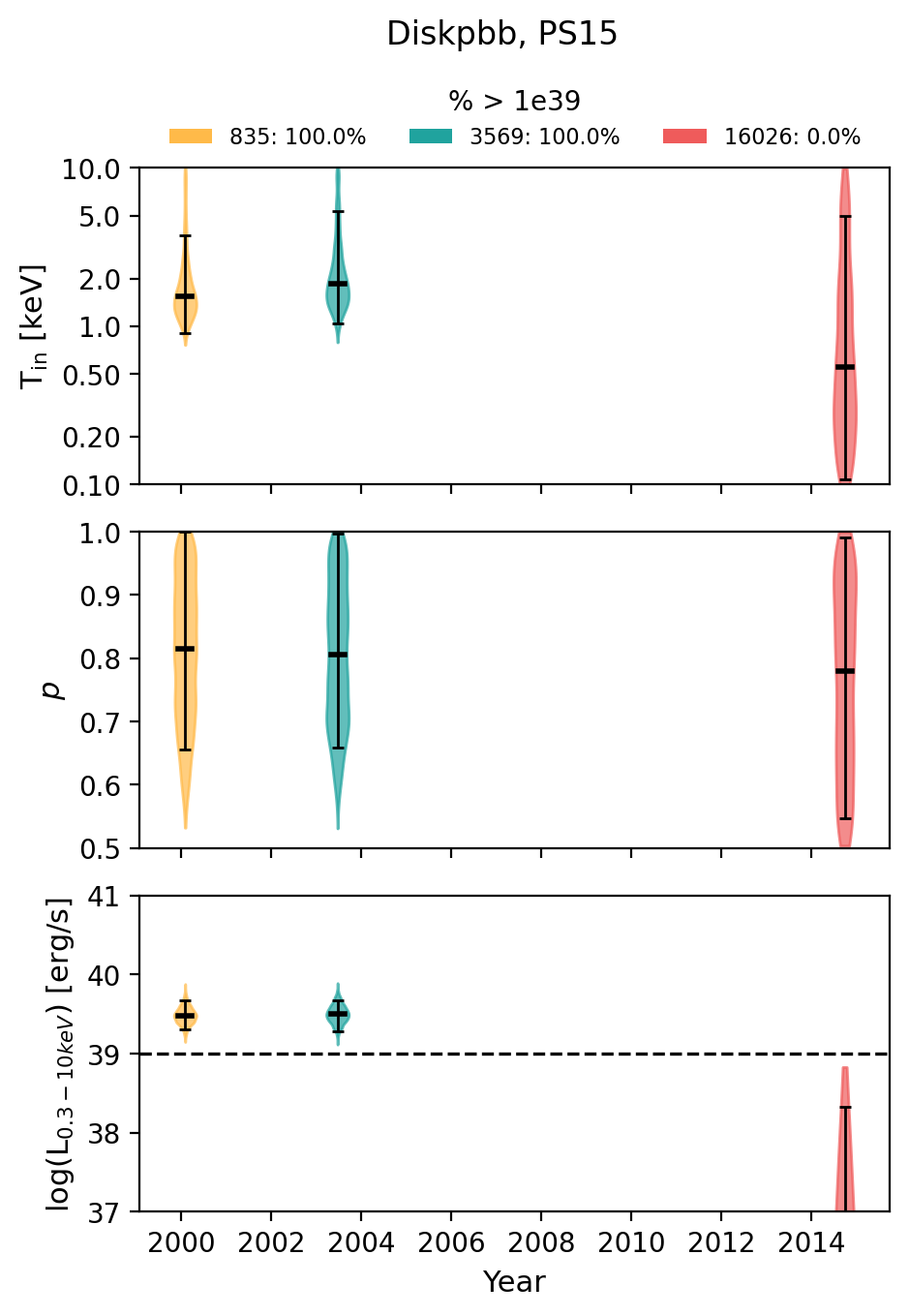}
        \caption{Parameter posterior distributions for the three models used on PS13, PS14 and PS15. See Sect. \ref{App:figures} for details.} \label{fig:app:posteriors_131415}
\end{figure*}

\begin{figure*}[h!]
        \centering
        \includegraphics[scale=0.47,valign=c]{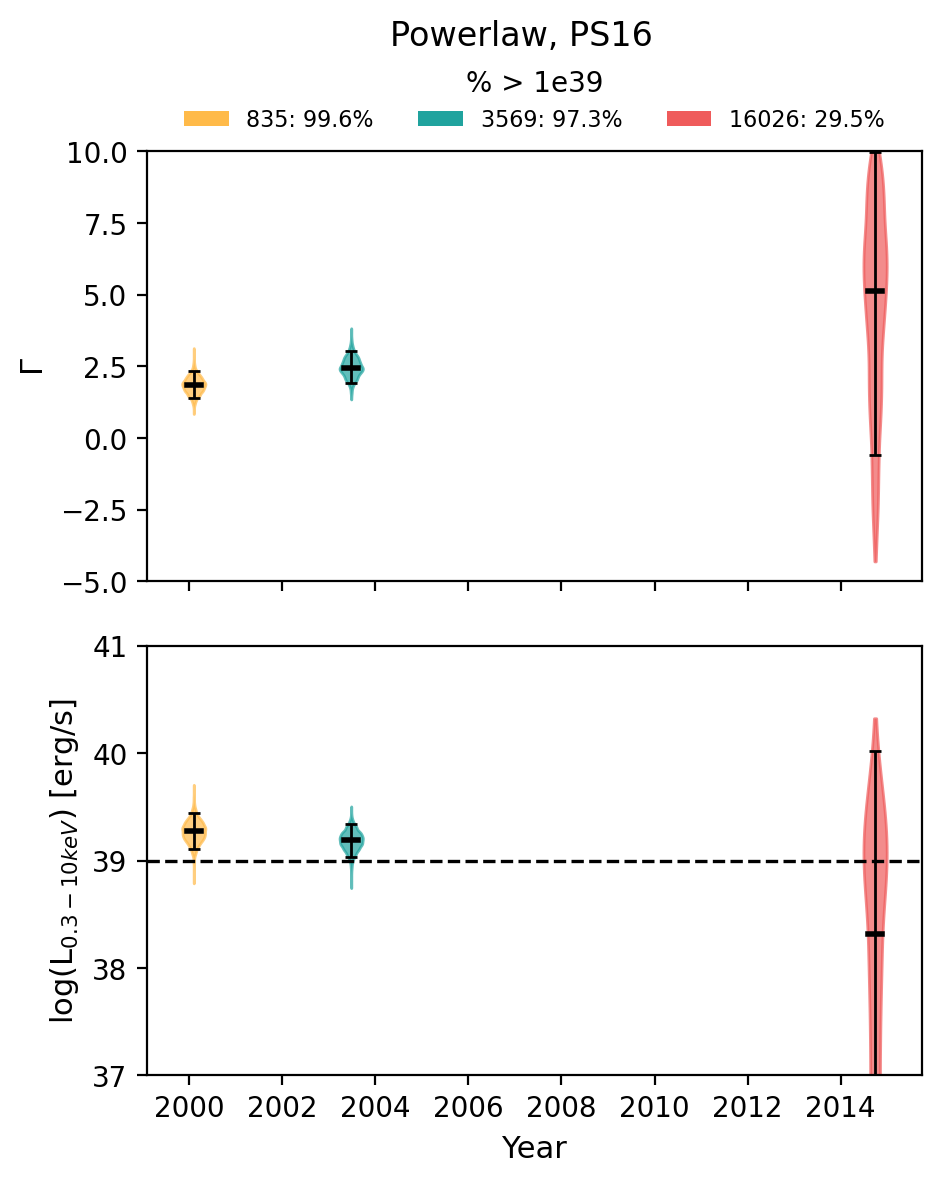}
        \includegraphics[scale=0.47,valign=c]{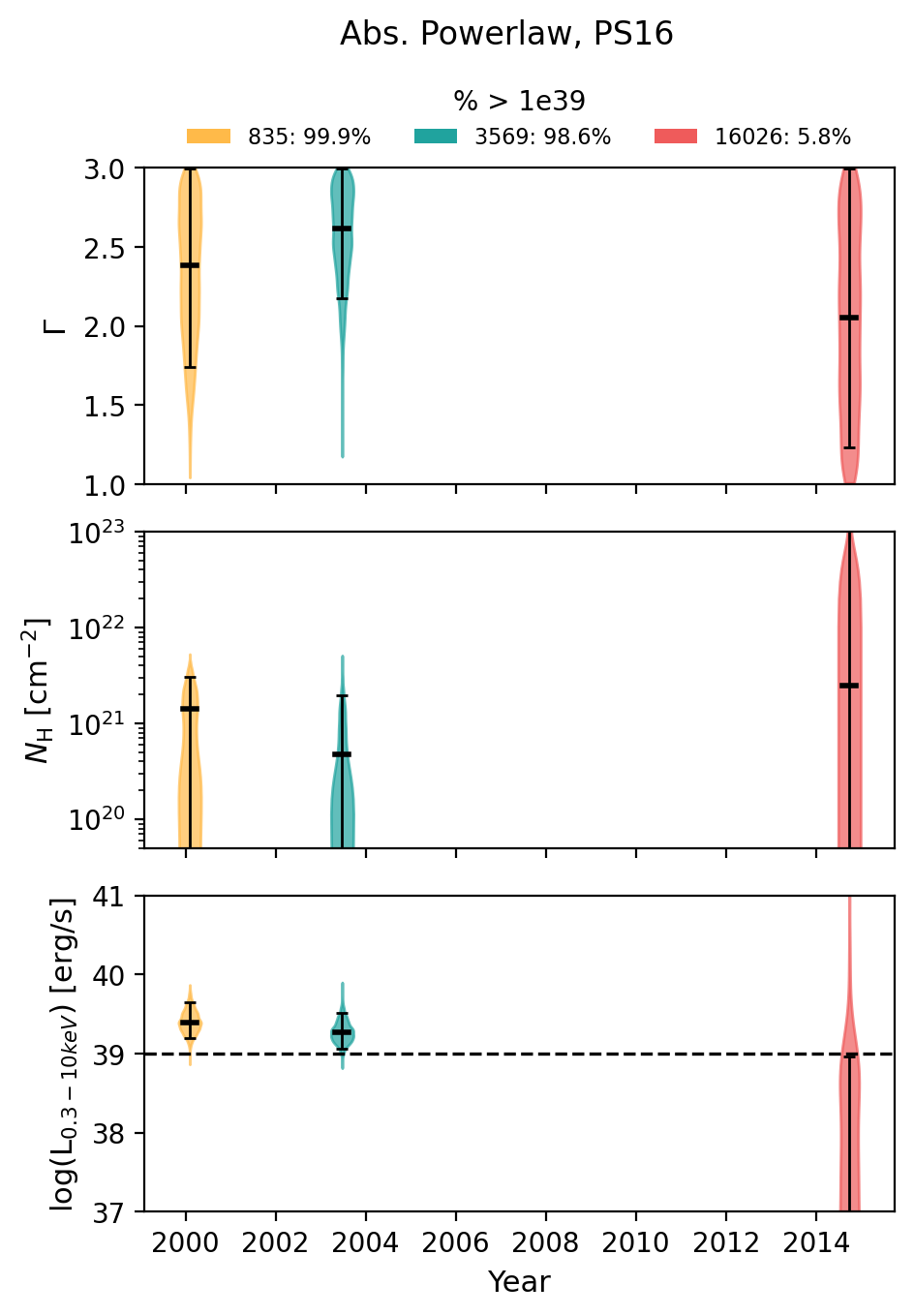  }
        \includegraphics[scale=0.47,valign=c]{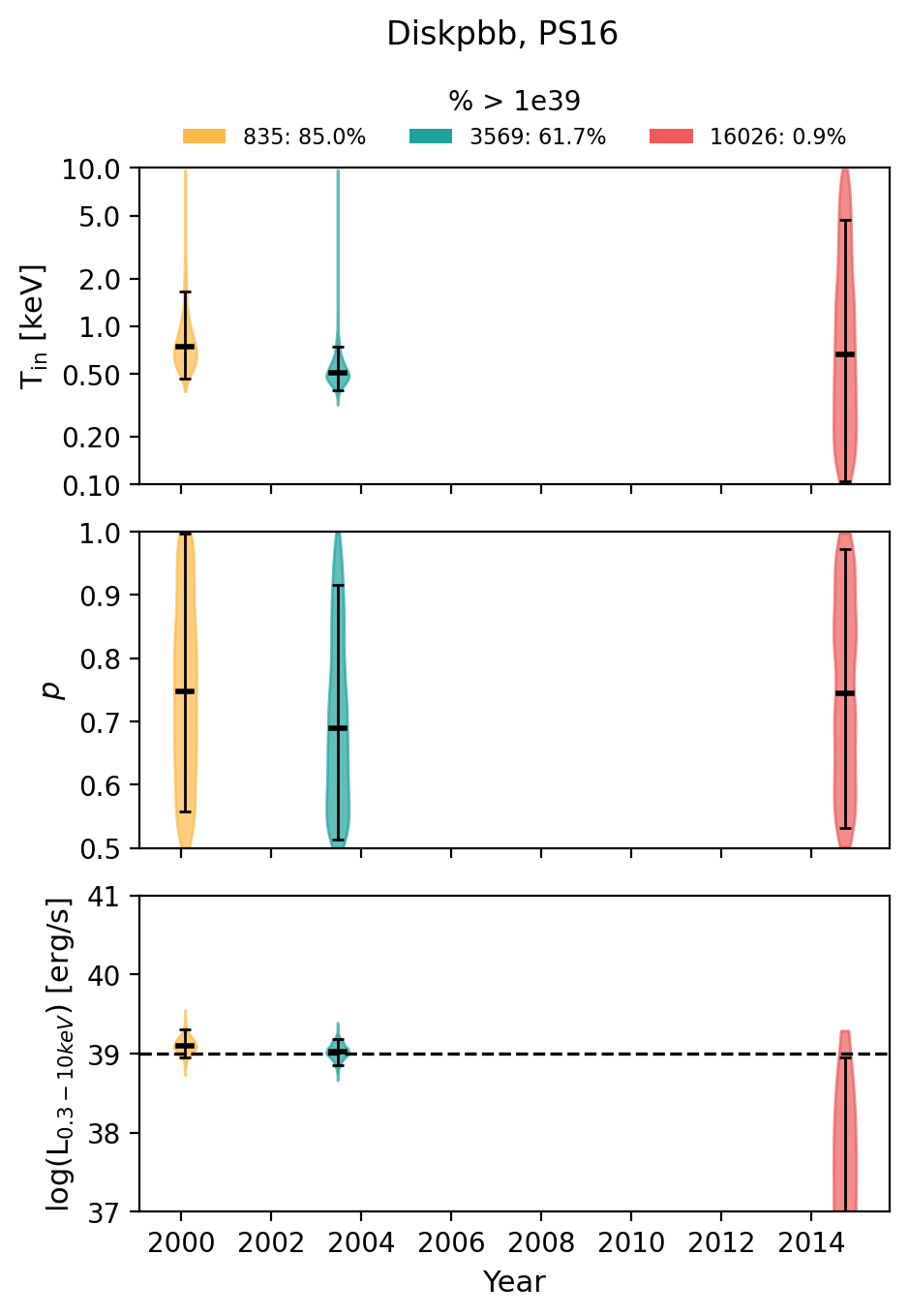}
        \includegraphics[scale=0.47,valign=c]{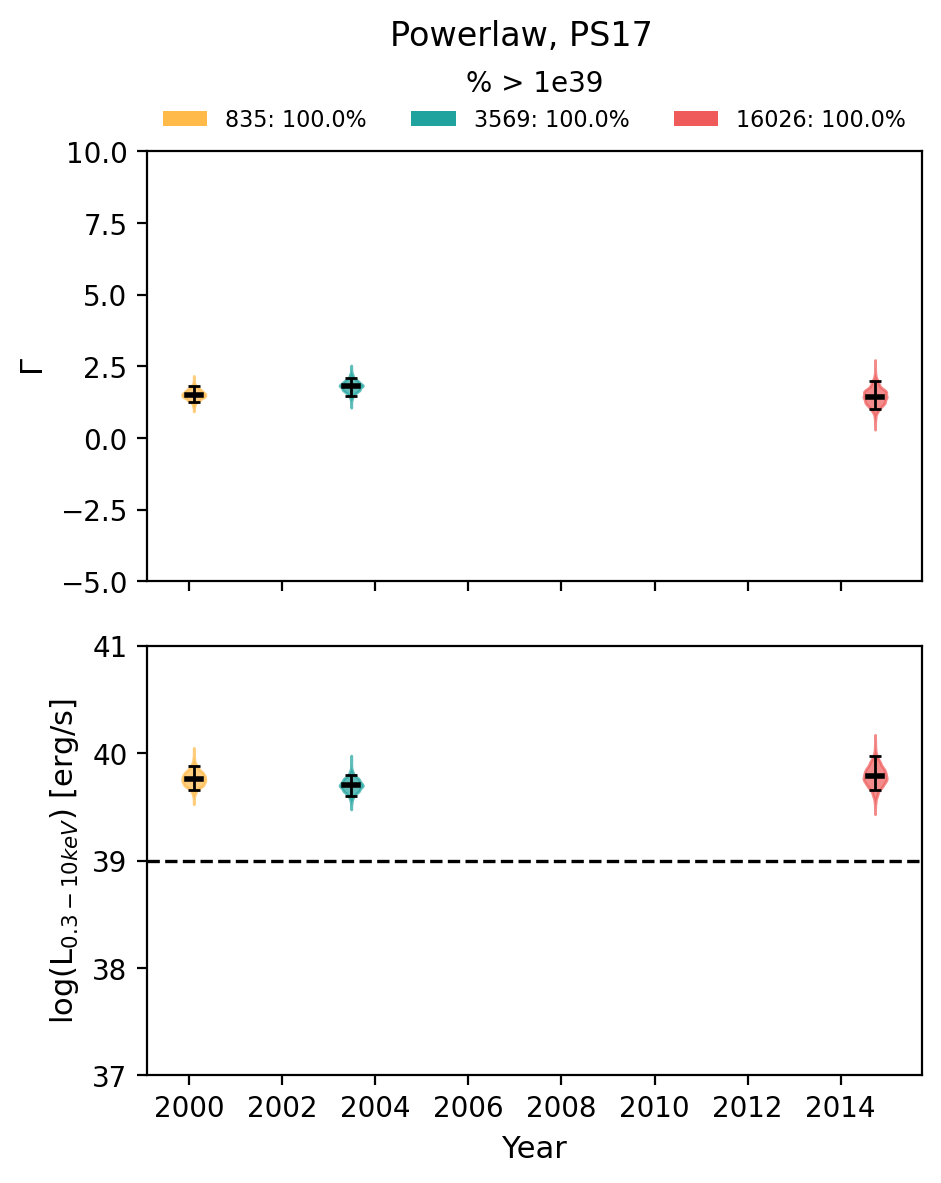}
        \includegraphics[scale=0.47,valign=c]{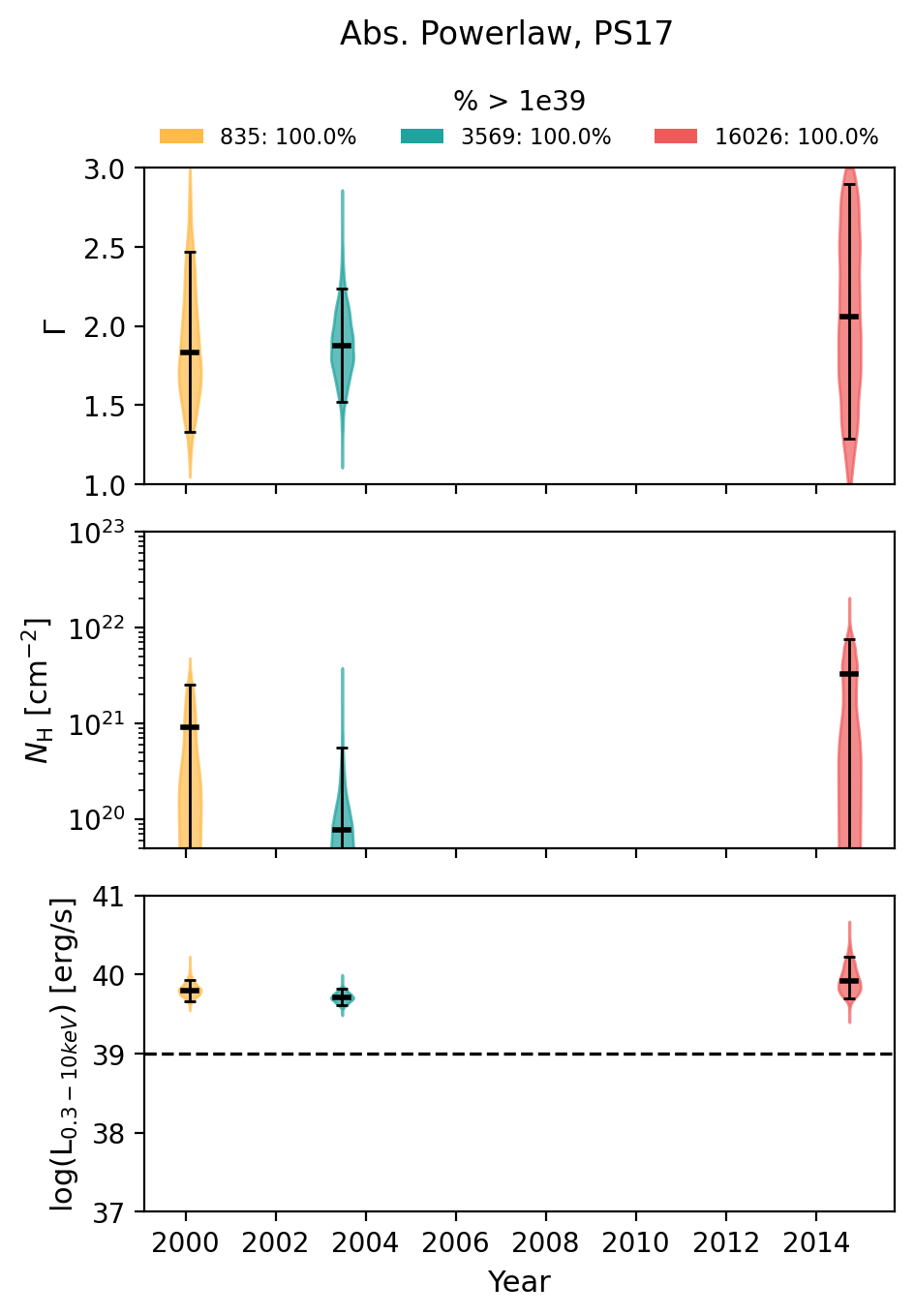  }
        \includegraphics[scale=0.47,valign=c]{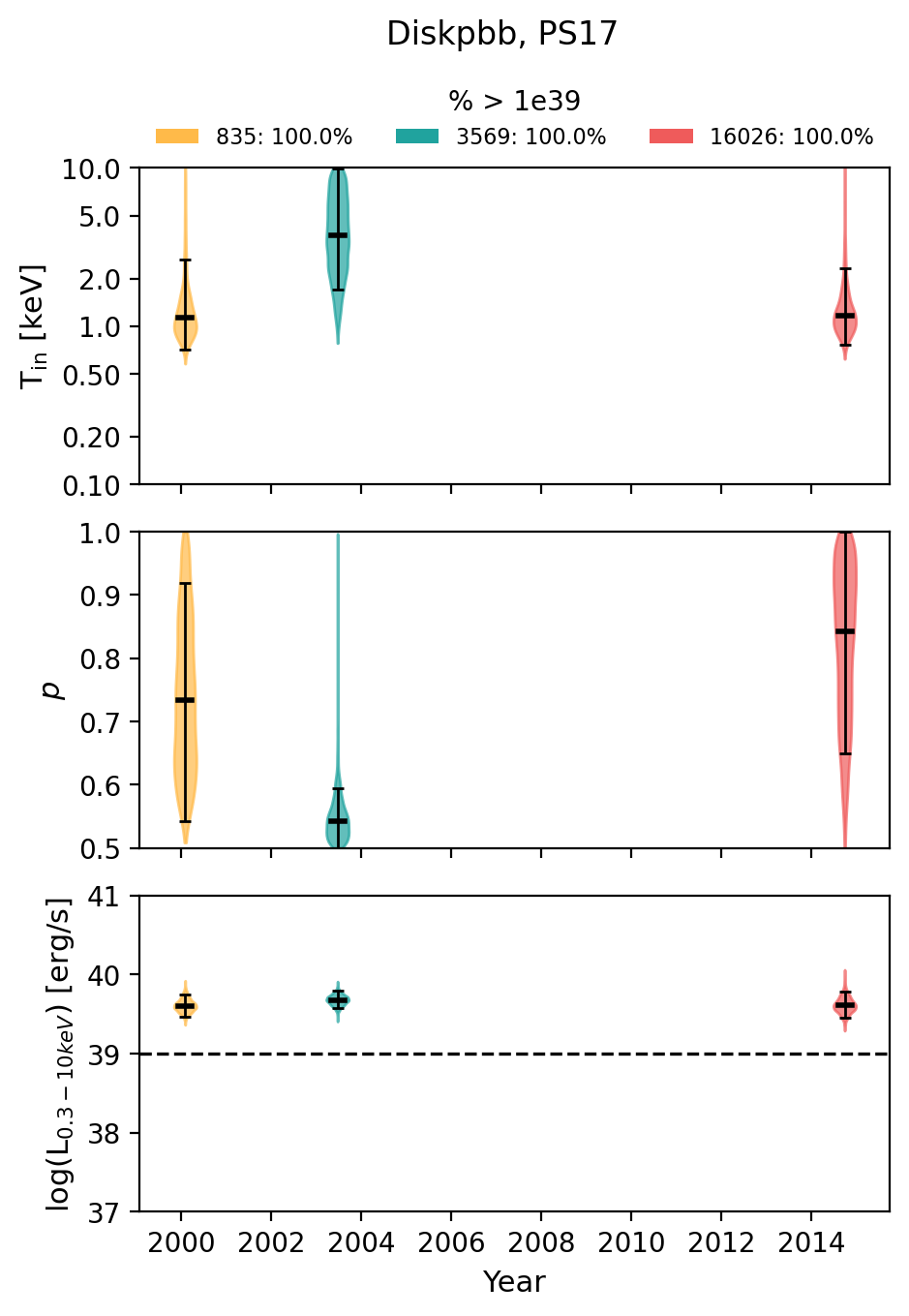}
        \caption{Parameter posterior distributions for the three models used on PS16 and PS17. See Sect. \ref{App:figures} for details.} \label{fig:app:posteriors_1617}
\end{figure*}

This section contains, for all point sources and all three models used, figures depicting the posterior range of all fit parameters except for normalization (Figs \ref{fig:app:posteriors_123}$-$\ref{fig:app:posteriors_1617}). Instead, the luminosity posterior is shown\footnote{Note the violin plots are an approximation of the 1D marginalized posteriors.}. The median value of the posterior is highlighted with a horizontal black line, and the error bars correspond to the 90\% credible interval of the posterior. Error bars reaching below the plot limit should be interpreted as the equivalent of an upper limit. Underneath the title, a legend states the percent of the luminosity posterior above the ULX threshold ($10^{39}$~erg~s$^{-1}$), for each observation.

A ULX with a luminosity compatible with the lower threshold of the plot ($L_{\rm 0.3-10 keV}$=$10^{37}$~erg~s$^{-1}$) is compatible with a non-detection. 

\begin{figure*}[h!]
        \centering
        \includegraphics[scale=0.25,valign=c]{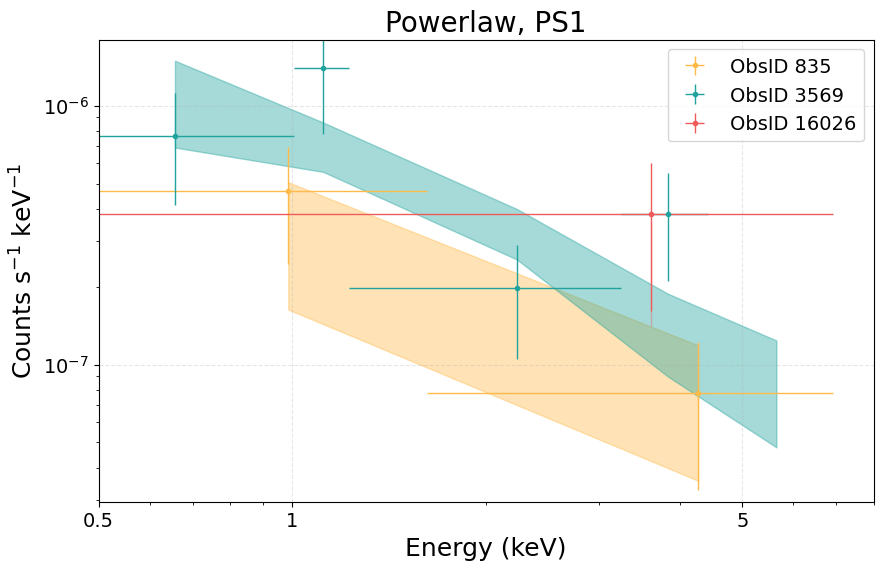}
        \includegraphics[scale=0.25,valign=c]{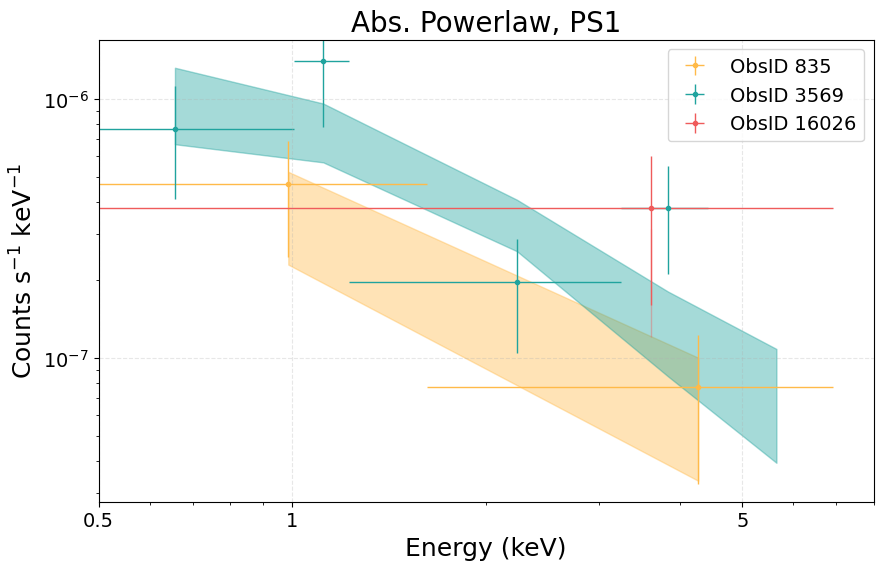}
        \includegraphics[scale=0.25,valign=c]{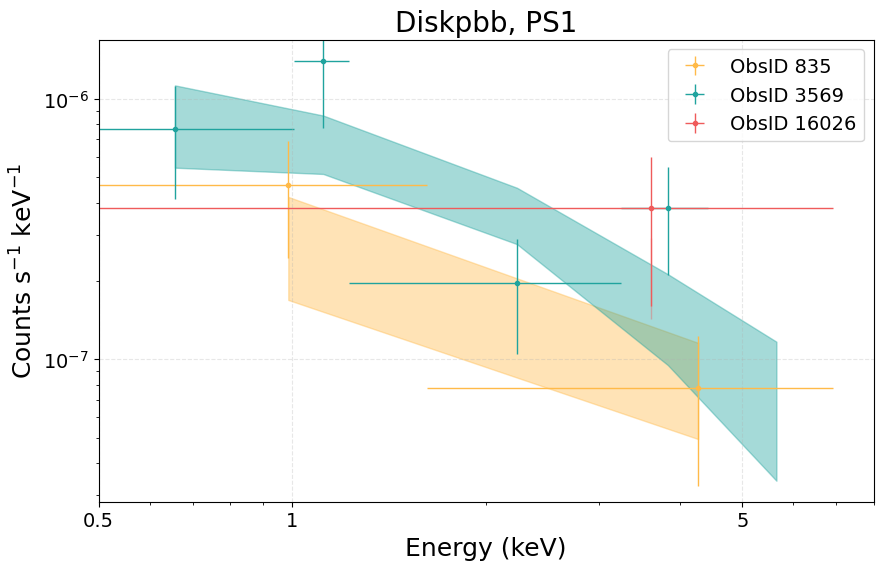}
        \includegraphics[scale=0.25,valign=c]{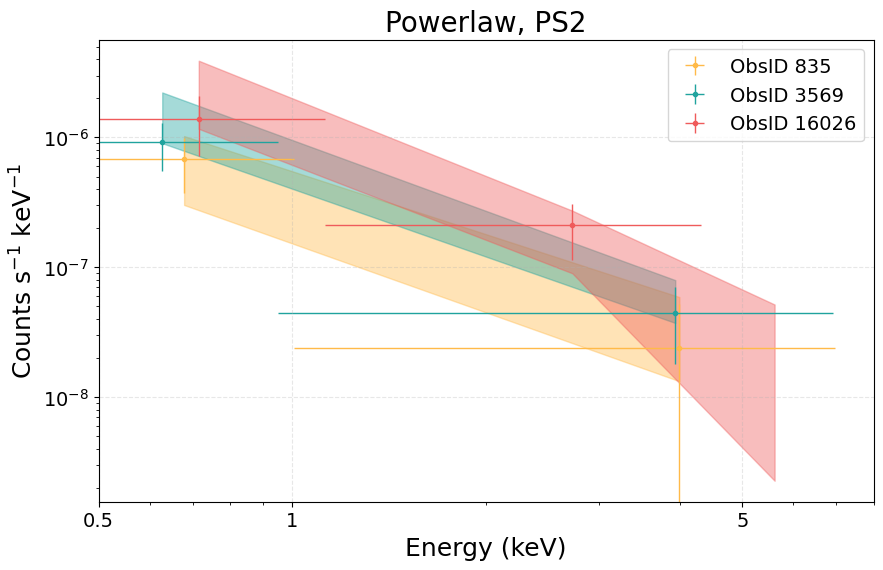}
        \includegraphics[scale=0.25,valign=c]{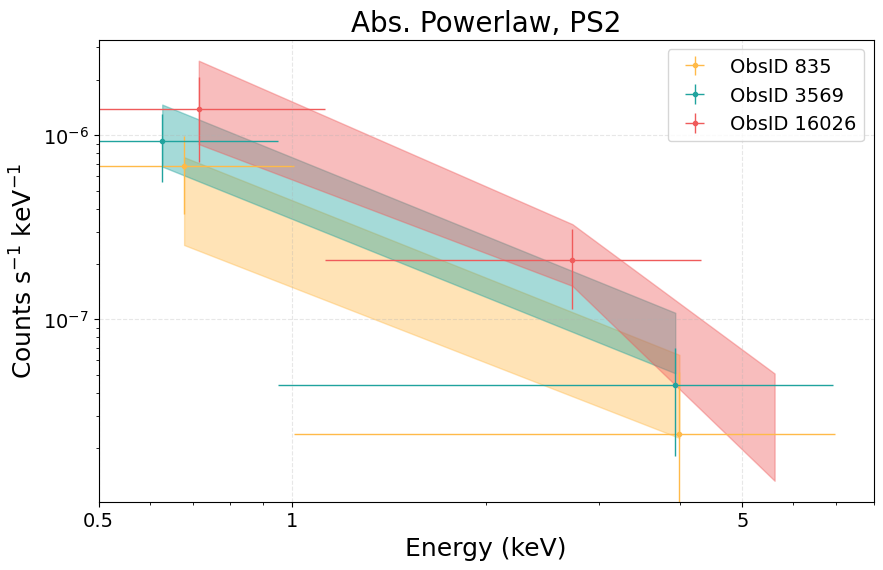}
        \includegraphics[scale=0.25,valign=c]{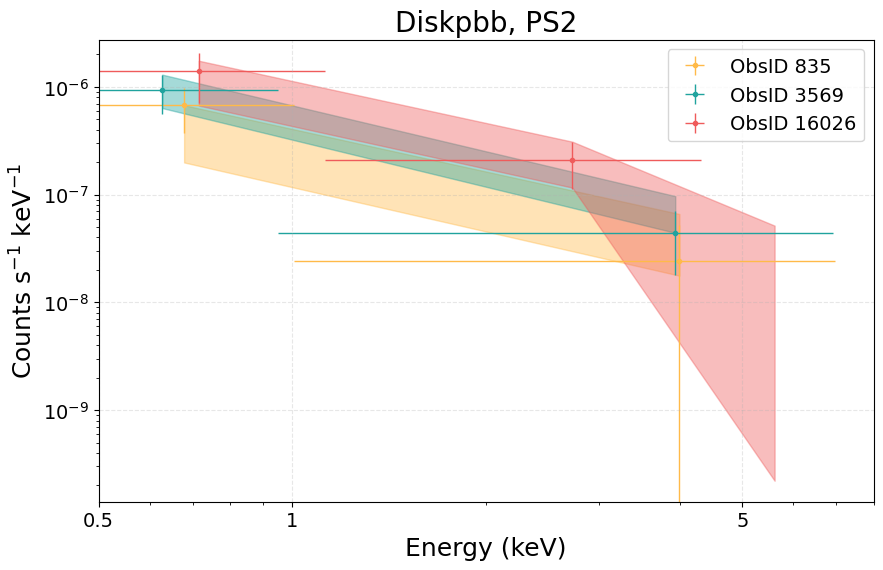}
        \includegraphics[scale=0.25,valign=c]{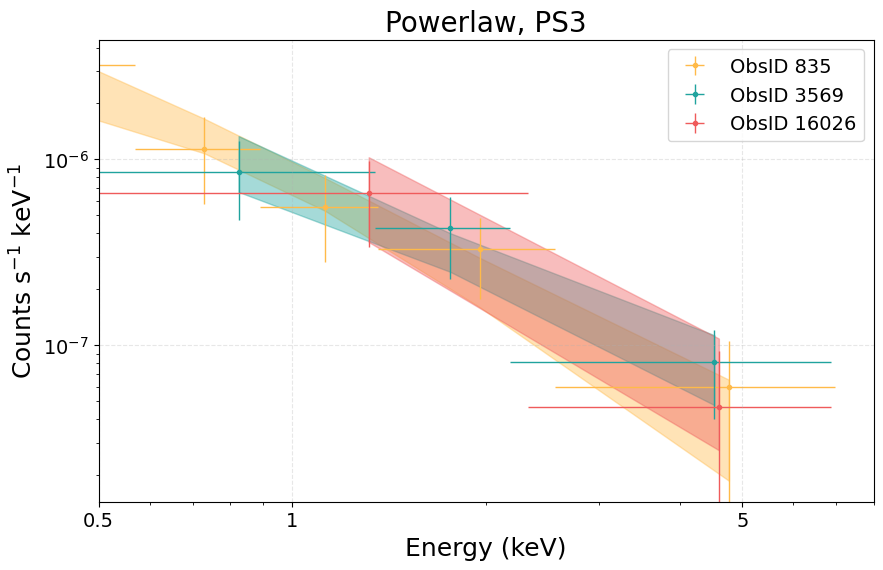}
        \includegraphics[scale=0.25,valign=c]{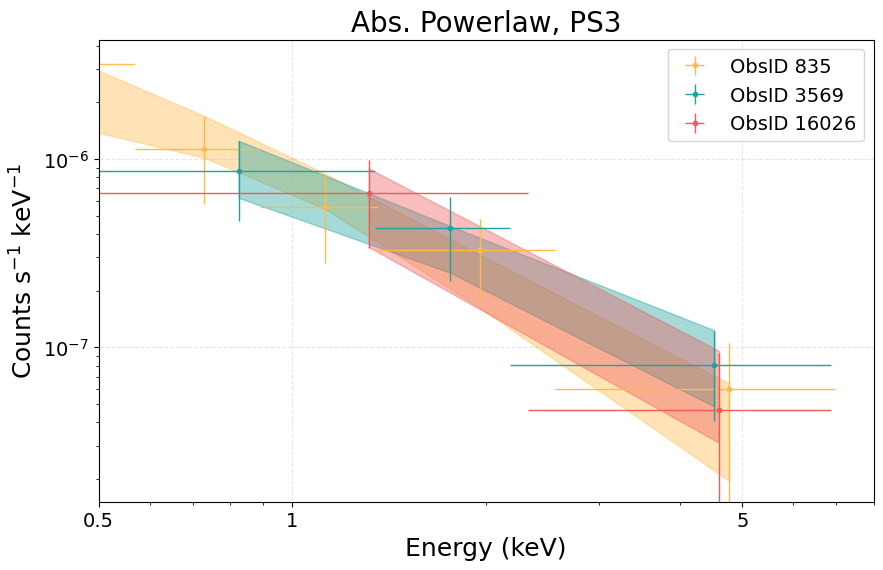}
        \includegraphics[scale=0.25,valign=c]{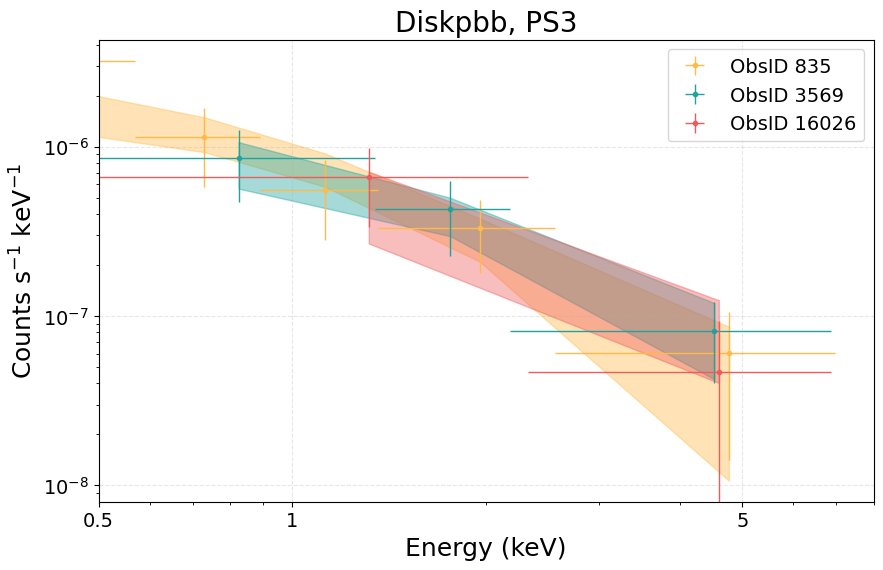}
        \includegraphics[scale=0.25,valign=c]{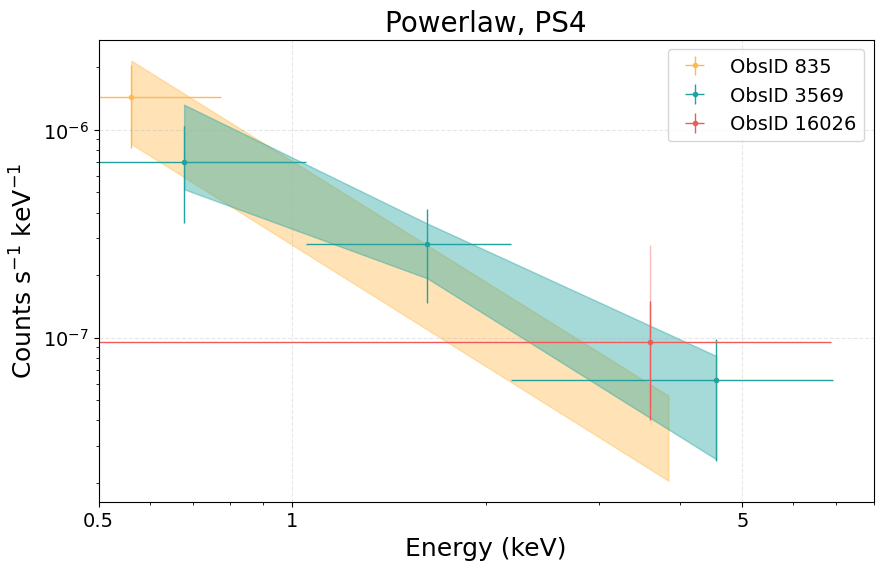}
        \includegraphics[scale=0.25,valign=c]{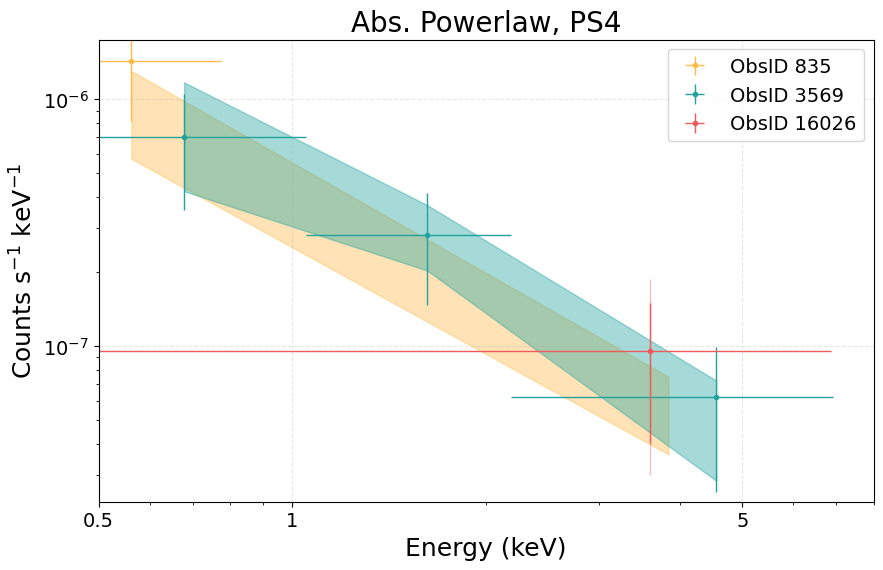}
        \includegraphics[scale=0.25,valign=c]{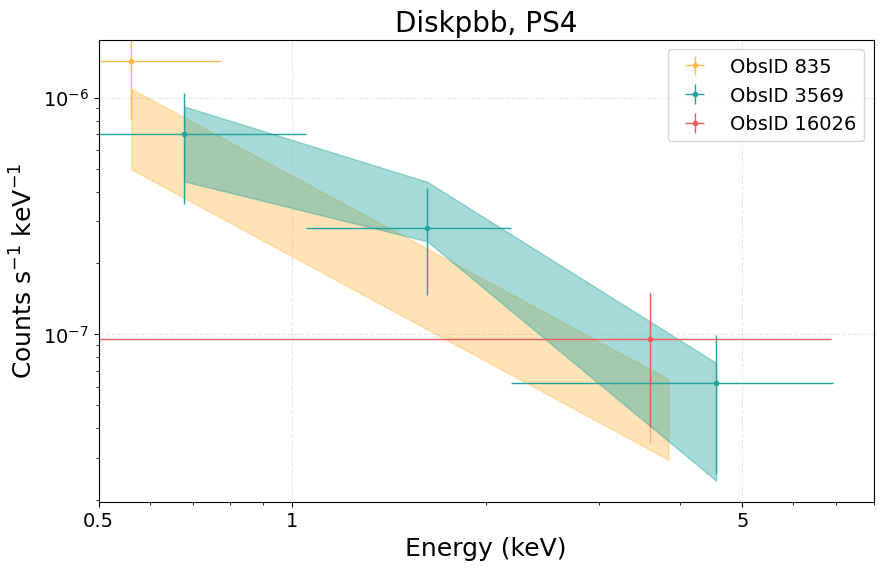}
        \includegraphics[scale=0.25,valign=c]{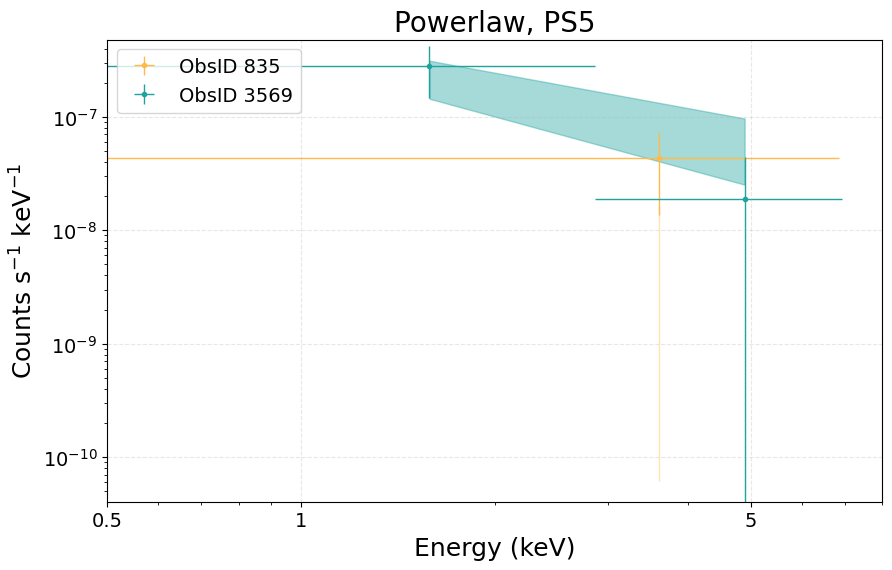}
        \includegraphics[scale=0.25,valign=c]{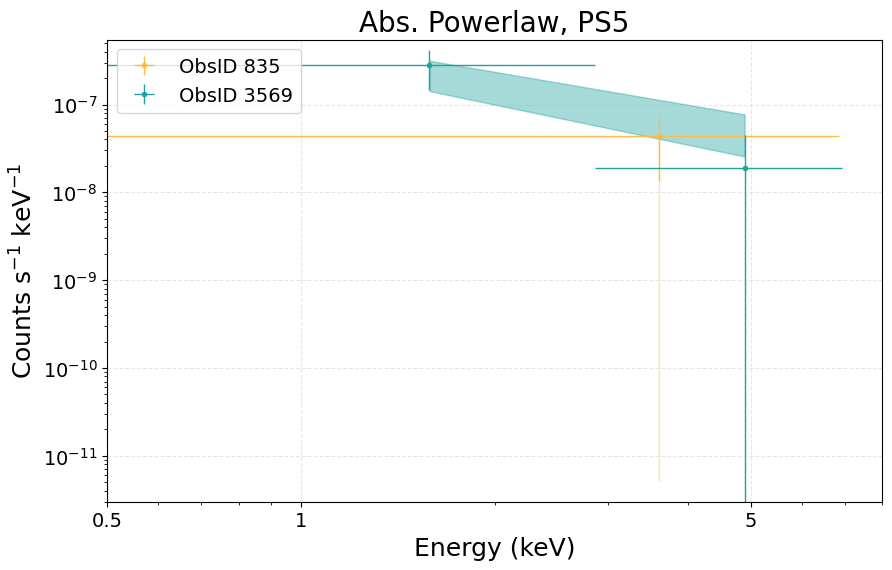}
        \includegraphics[scale=0.25,valign=c]{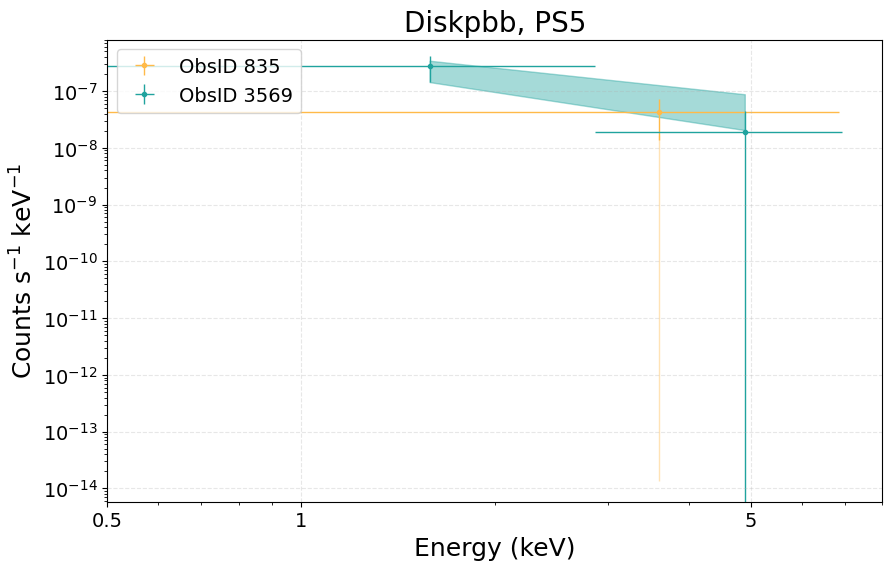}
        \includegraphics[scale=0.25,valign=c]{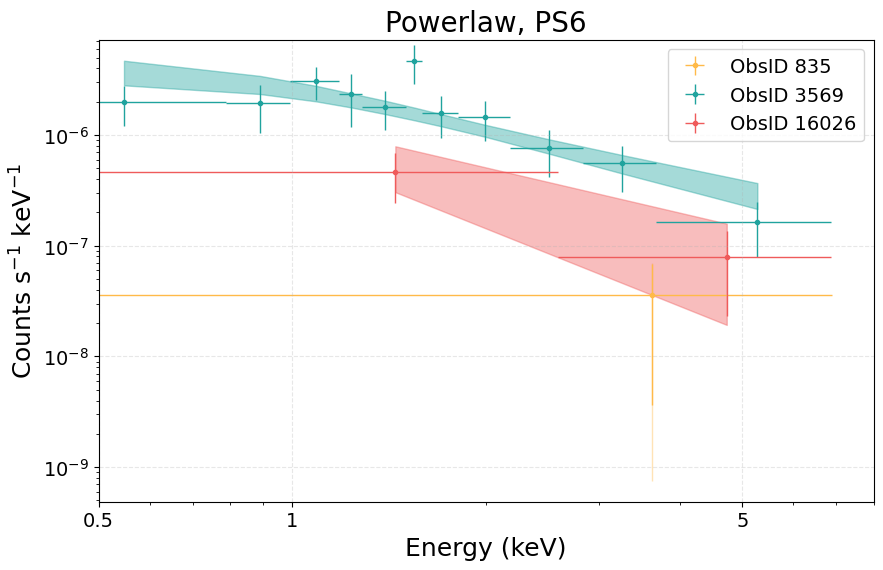}
        \includegraphics[scale=0.25,valign=c]{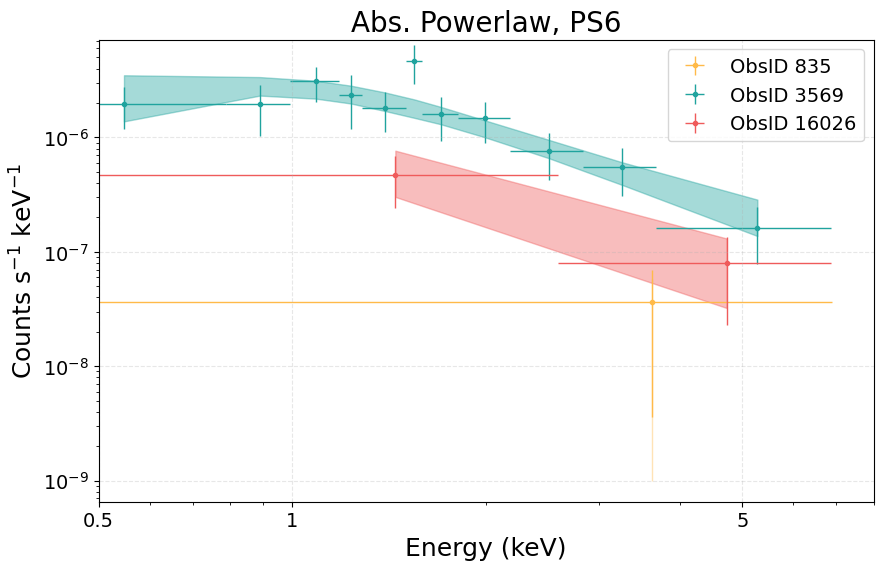}
        \includegraphics[scale=0.25,valign=c]{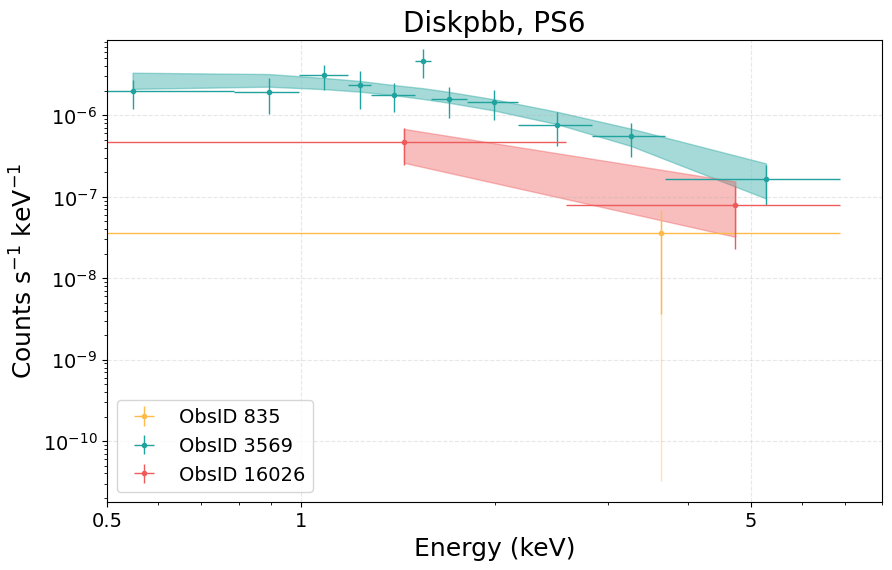}
        \caption{Spectral fits to every observation, for all three models considered in this work, for PS1, PS2, PS3, PS4, PS5 and PS6. See Sect. \ref{App:figures} for details.} \label{fig:app:spectra_panel1}
\end{figure*}

\begin{figure*}[h!]
        \centering
        \includegraphics[scale=0.25,valign=c]{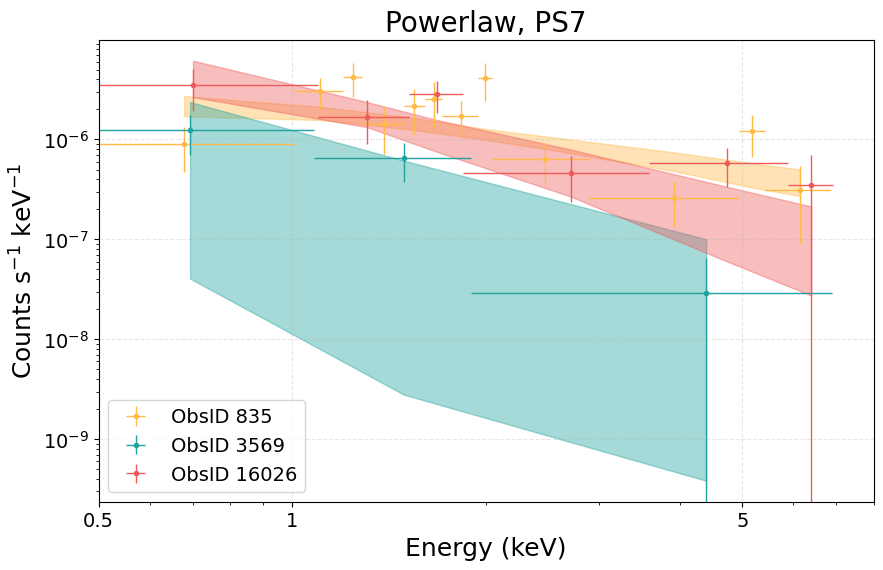}
        \includegraphics[scale=0.25,valign=c]{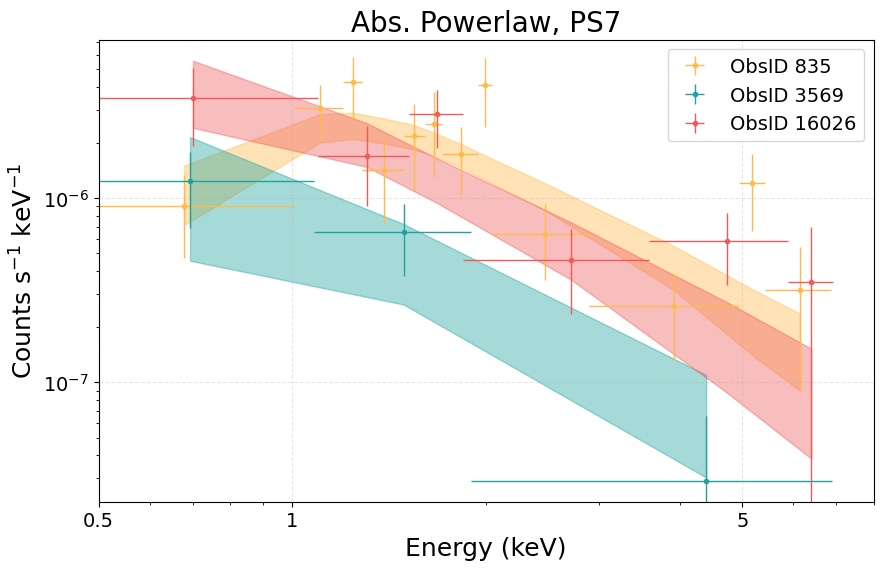}
        \includegraphics[scale=0.25,valign=c]{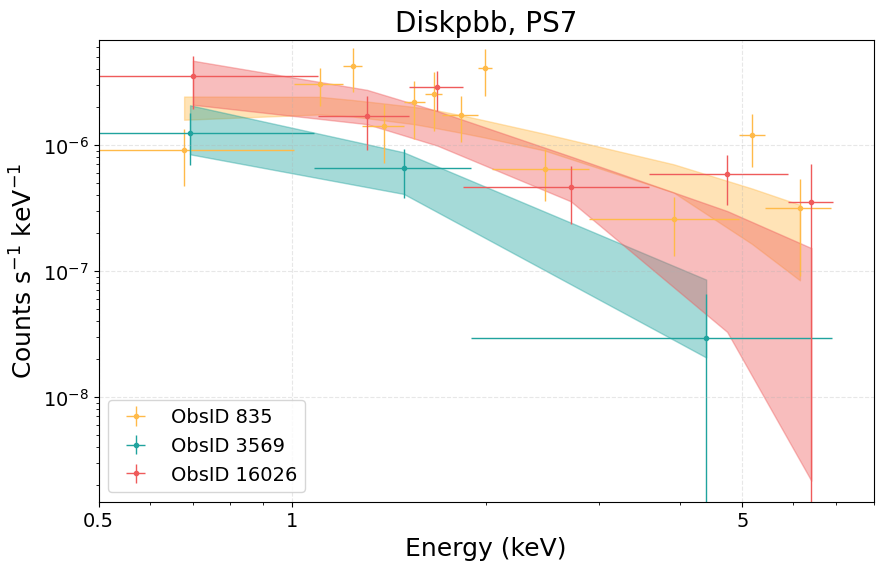}
        \includegraphics[scale=0.25,valign=c]{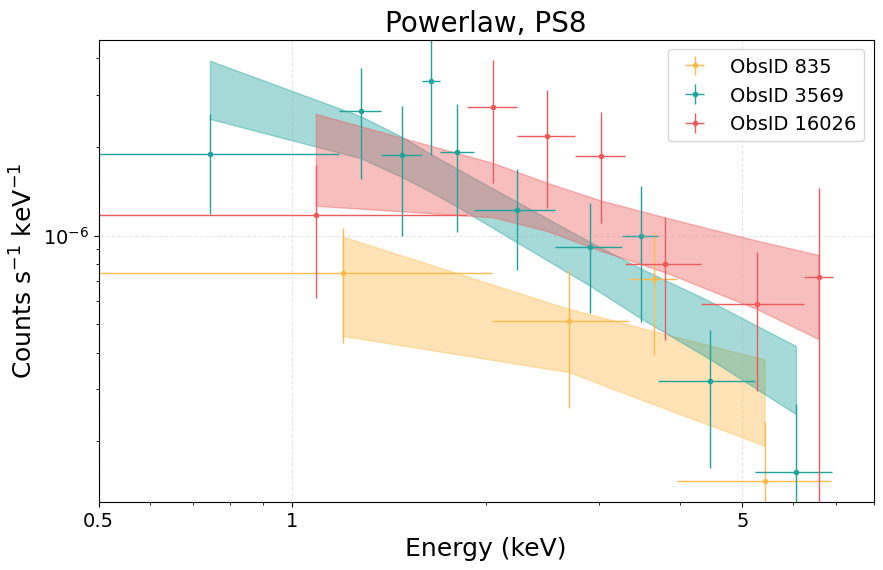}
        \includegraphics[scale=0.25,valign=c]{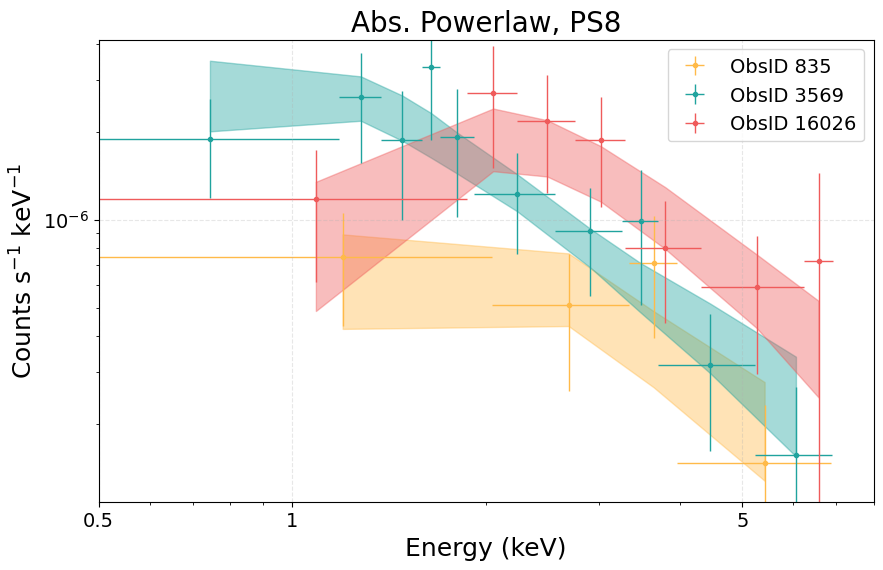}
        \includegraphics[scale=0.25,valign=c]{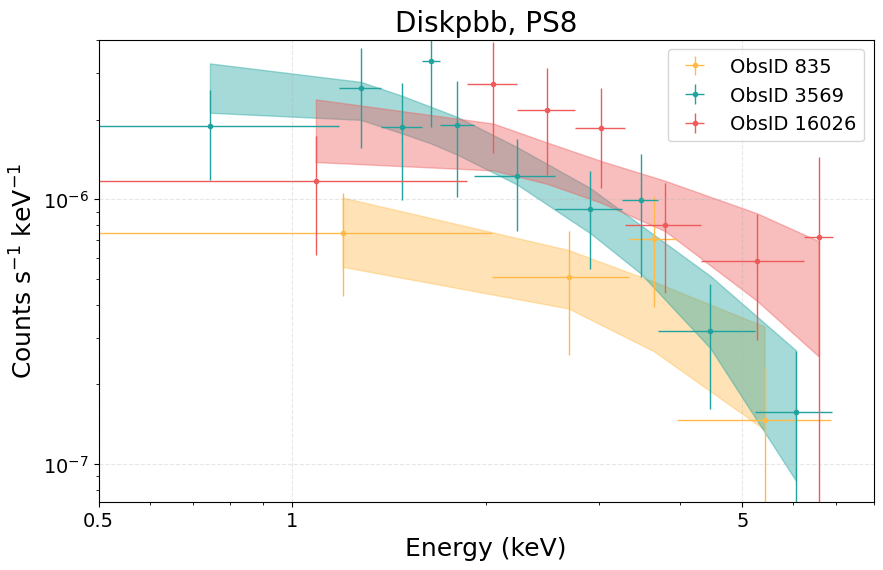}
        \includegraphics[scale=0.25,valign=c]{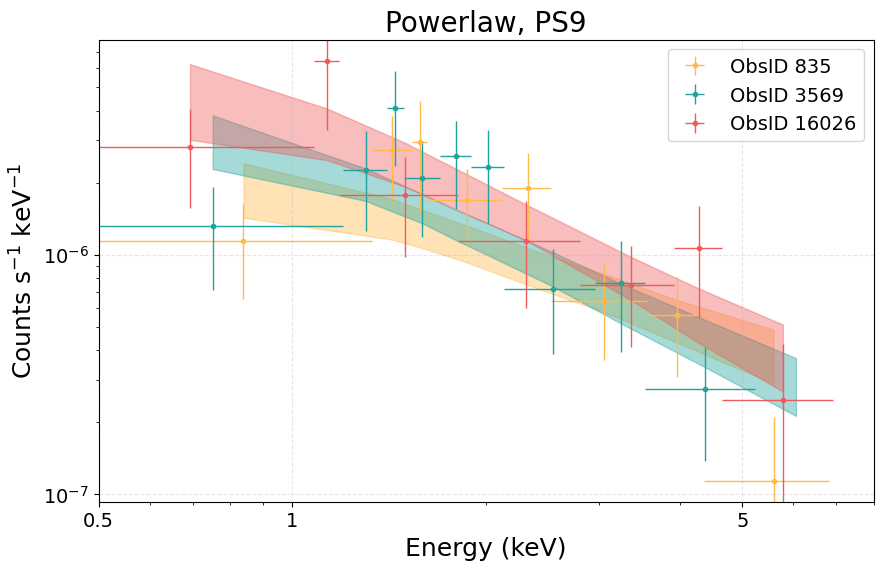}
        \includegraphics[scale=0.25,valign=c]{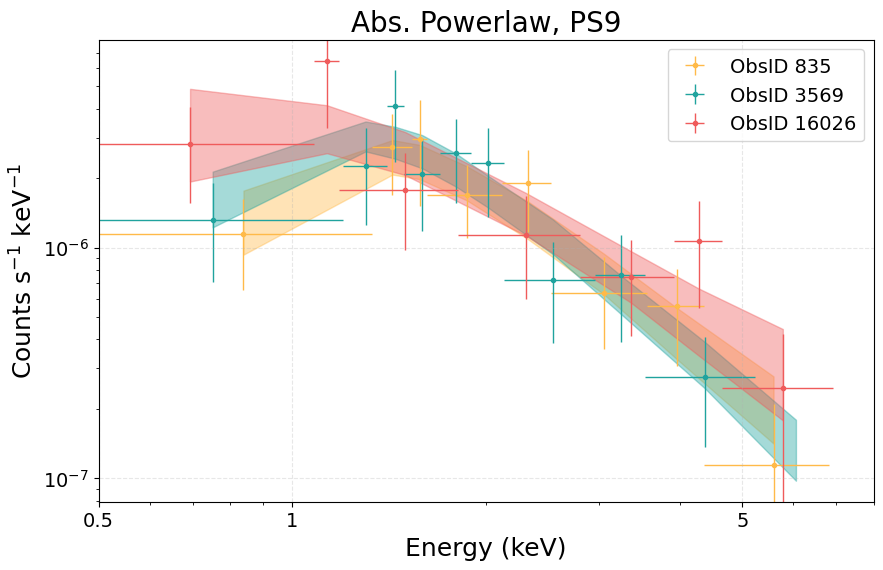}
        \includegraphics[scale=0.25,valign=c]{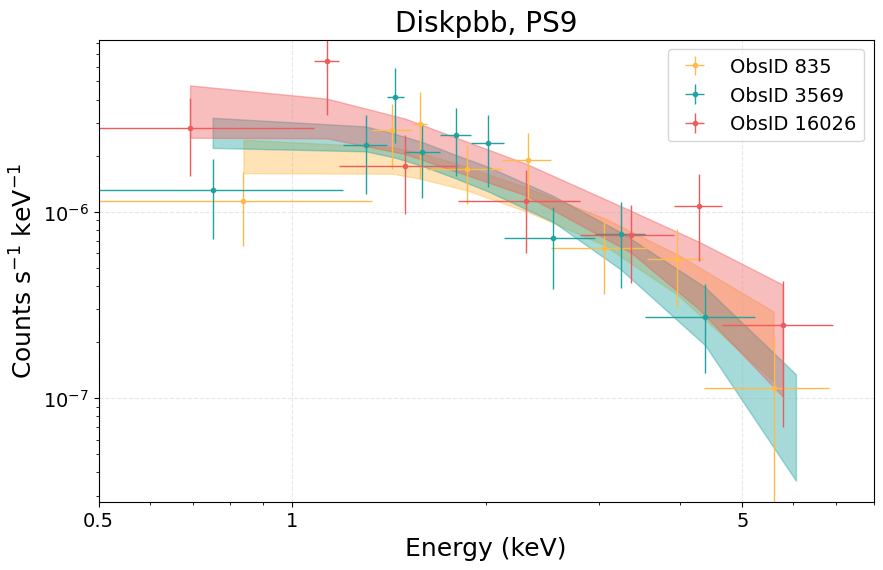}
        \includegraphics[scale=0.25,valign=c]{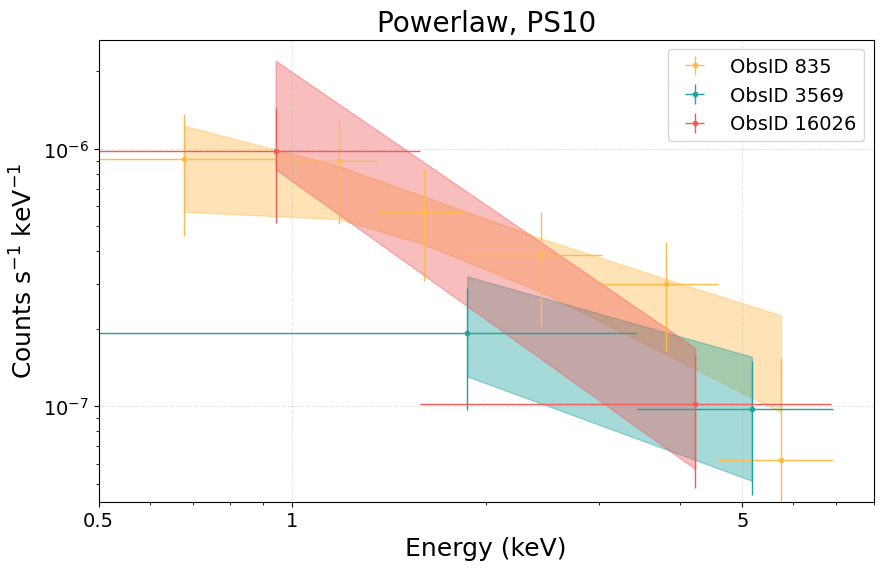}
        \includegraphics[scale=0.25,valign=c]{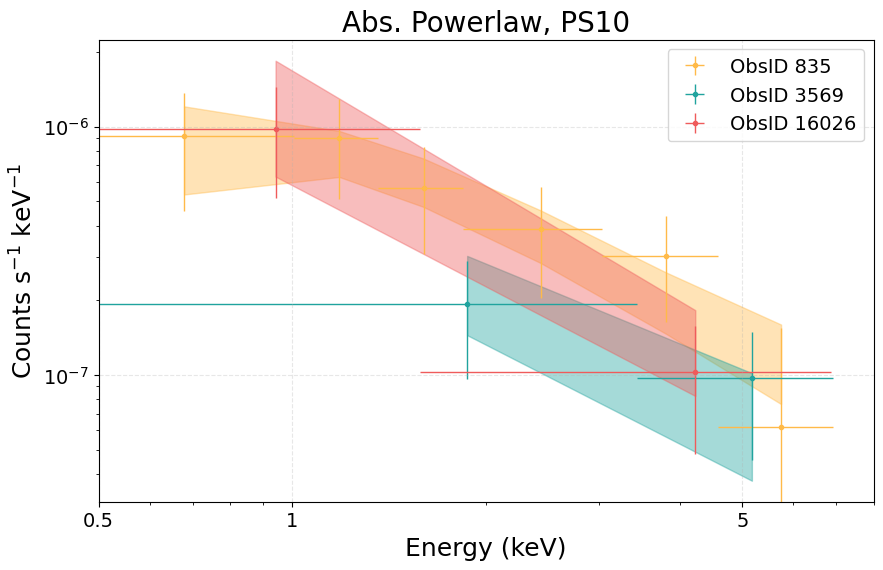}
        \includegraphics[scale=0.25,valign=c]{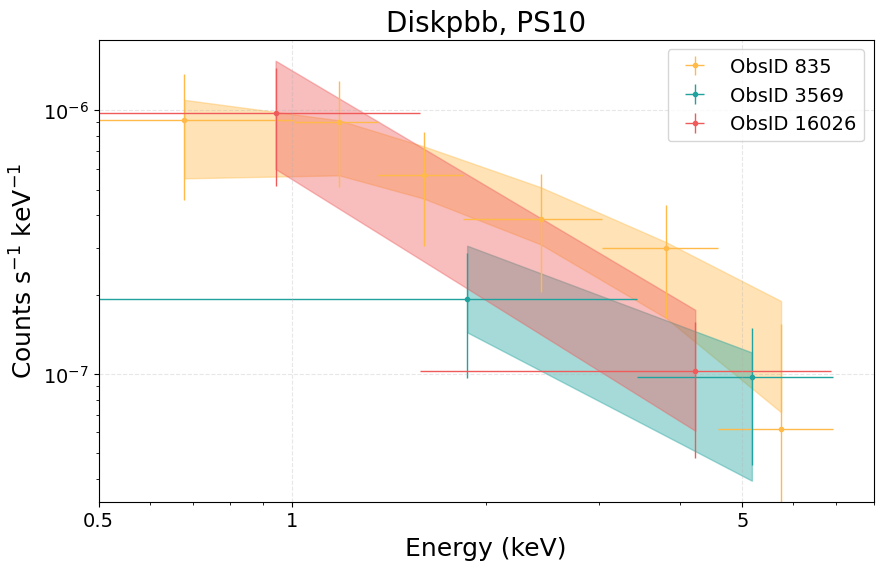}
        \includegraphics[scale=0.25,valign=c]{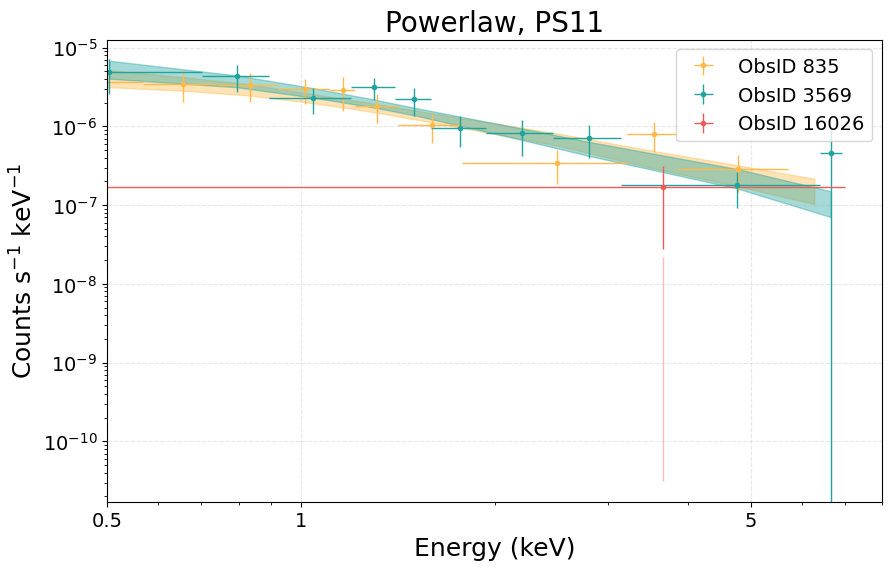}
        \includegraphics[scale=0.25,valign=c]{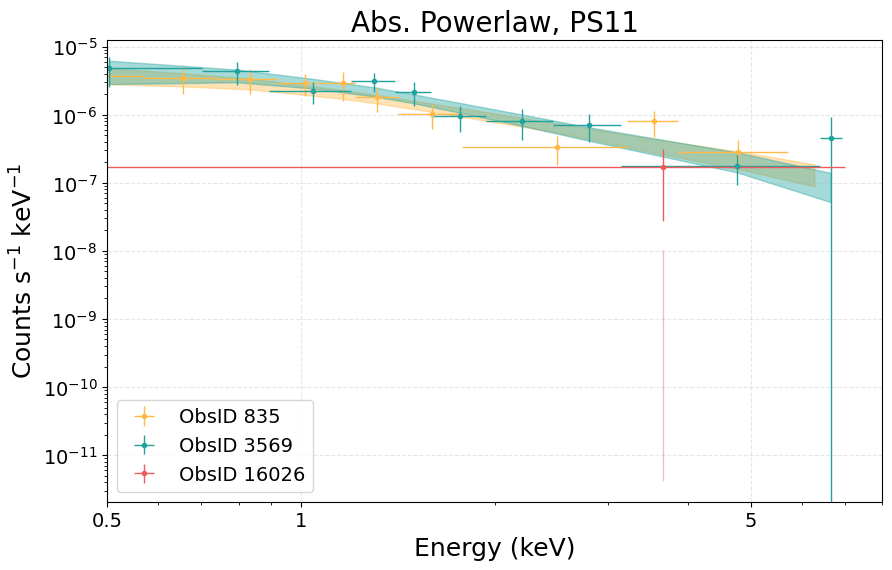}
        \includegraphics[scale=0.25,valign=c]{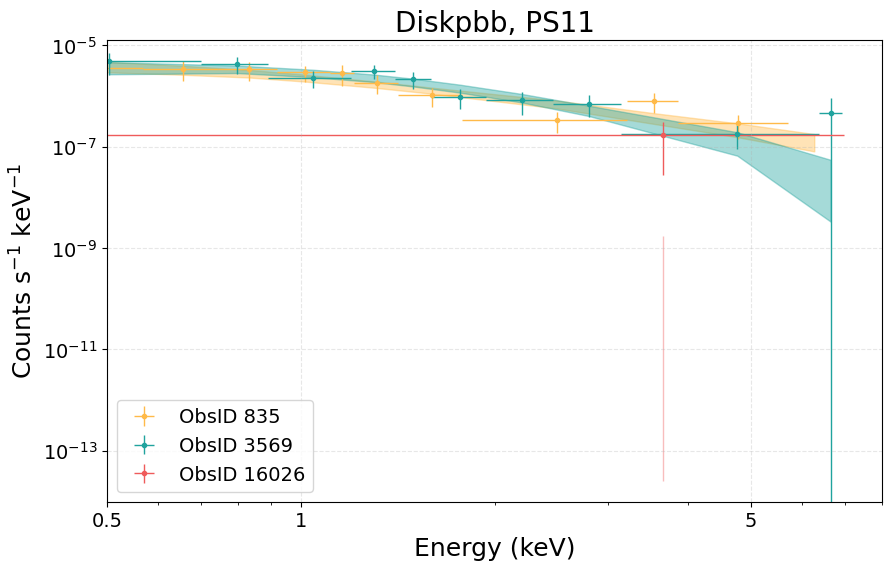}
        \includegraphics[scale=0.25,valign=c]{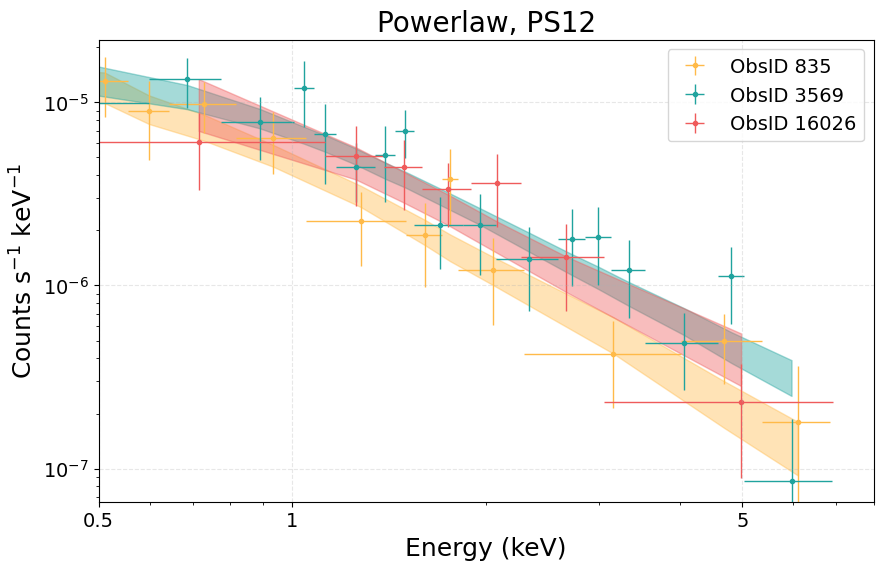}
        \includegraphics[scale=0.25,valign=c]{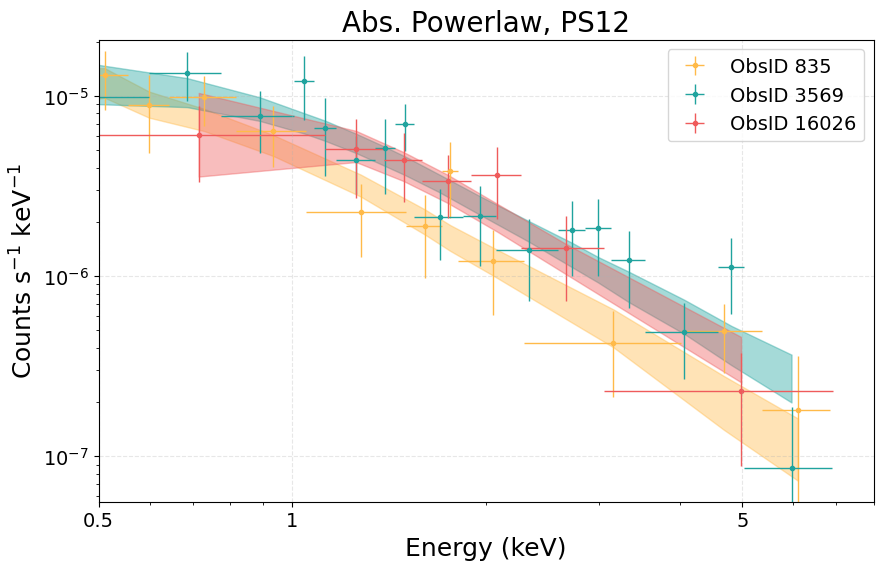}
        \includegraphics[scale=0.25,valign=c]{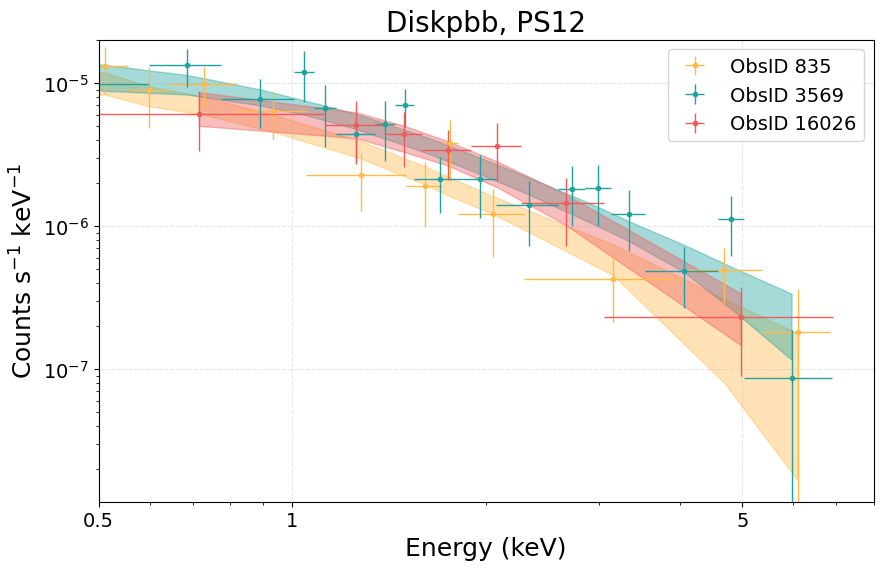}
        \caption{Spectral fits to every observation, for all three models considered in this work, for PS7, PS8, PS9, PS10, PS11 and PS12. See Sect. \ref{App:figures} for details.} \label{fig:app:spectra_panel2}
\end{figure*}

\begin{figure*}[h!]
        \centering
        \includegraphics[scale=0.25,valign=c]{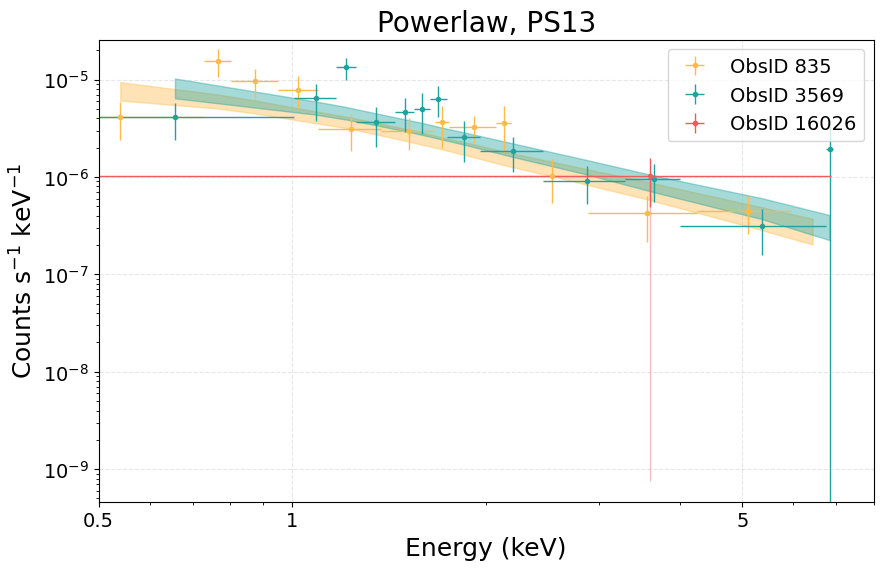}
        \includegraphics[scale=0.25,valign=c]{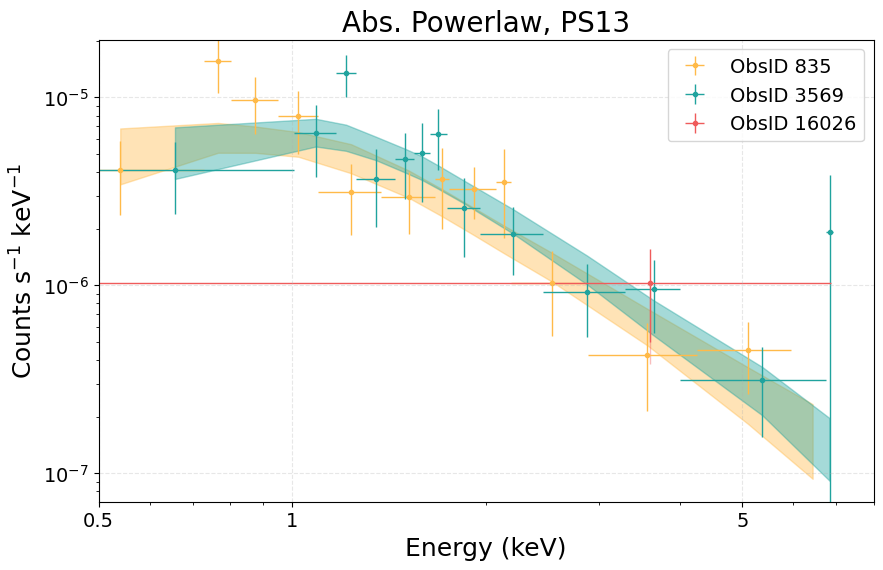}
        \includegraphics[scale=0.25,valign=c]{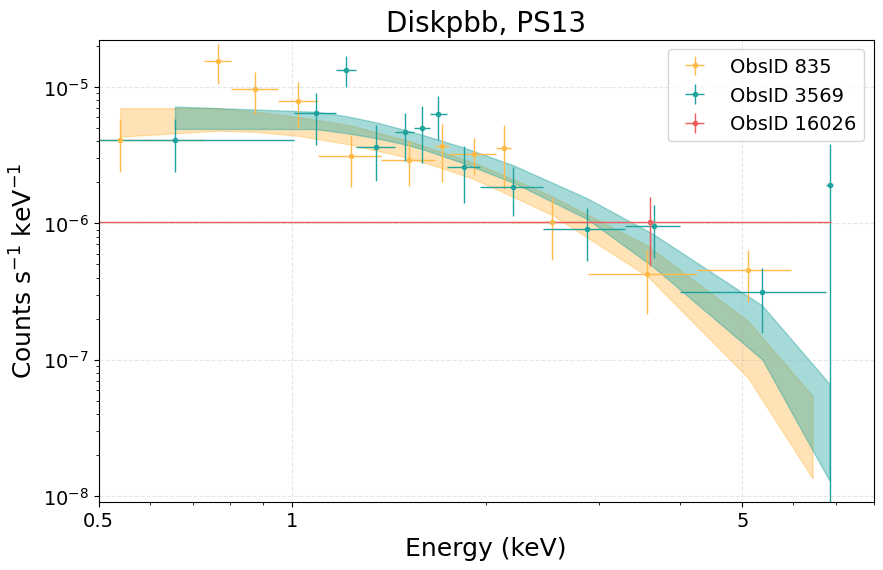}
        \includegraphics[scale=0.25,valign=c]{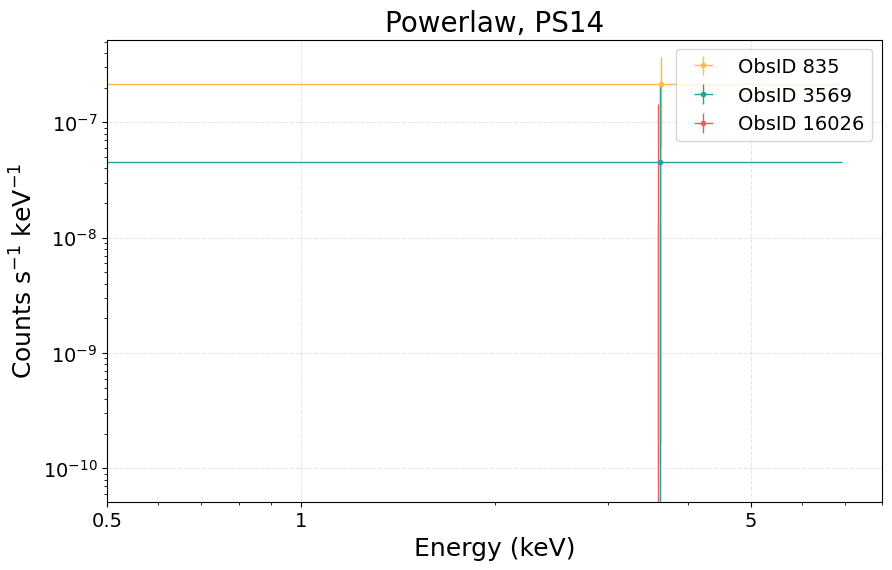}
        \includegraphics[scale=0.25,valign=c]{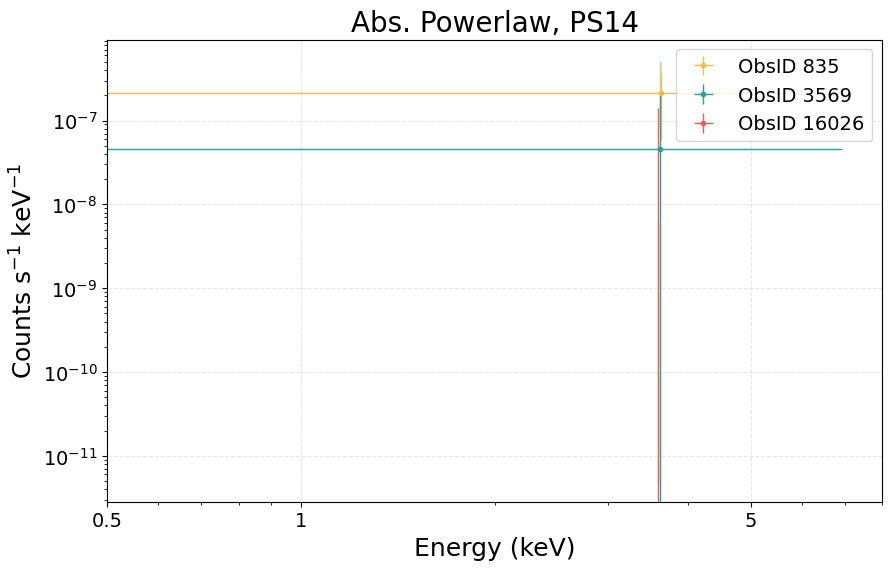}
        \includegraphics[scale=0.25,valign=c]{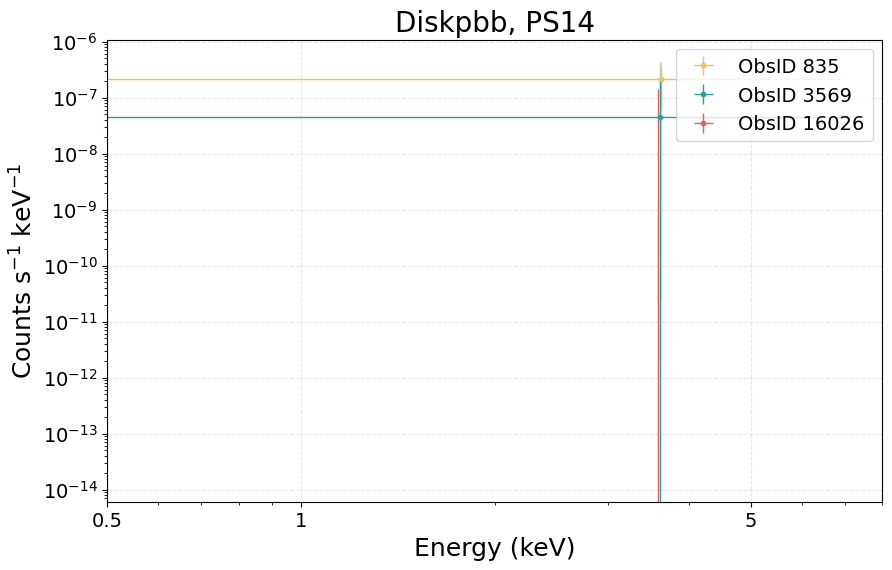}
        \includegraphics[scale=0.25,valign=c]{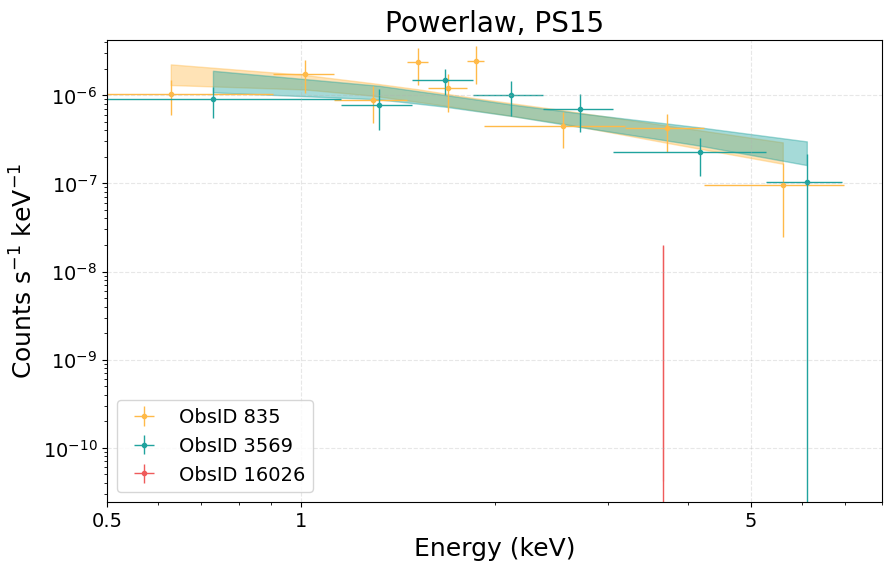}
        \includegraphics[scale=0.25,valign=c]{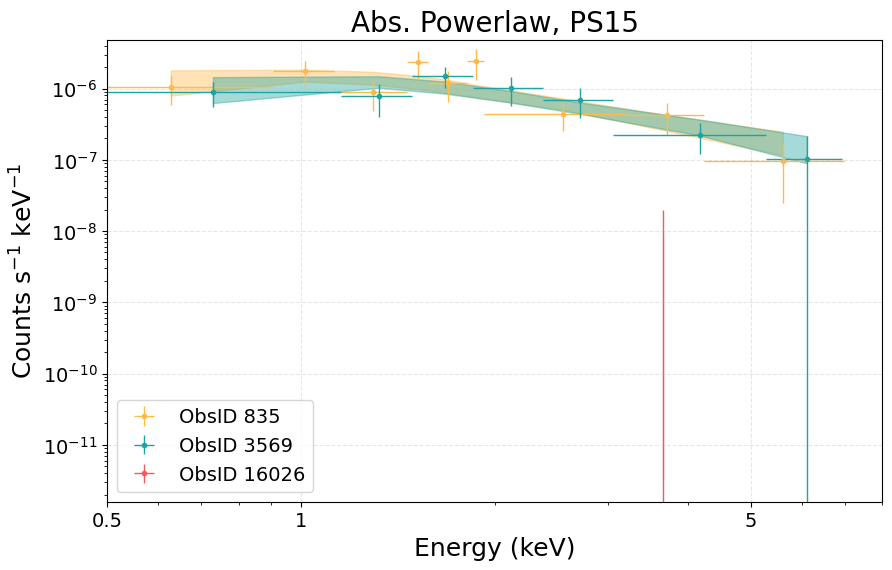}
        \includegraphics[scale=0.25,valign=c]{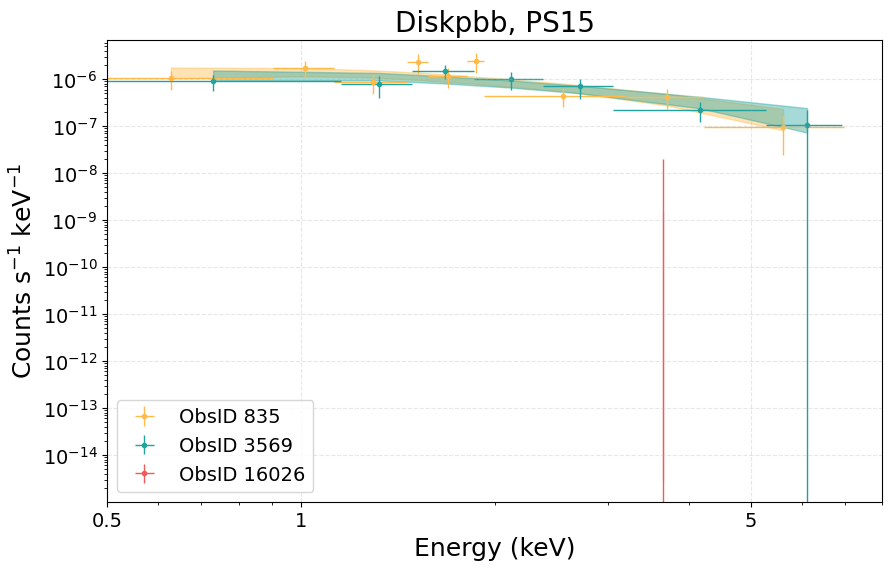}
        \includegraphics[scale=0.25,valign=c]{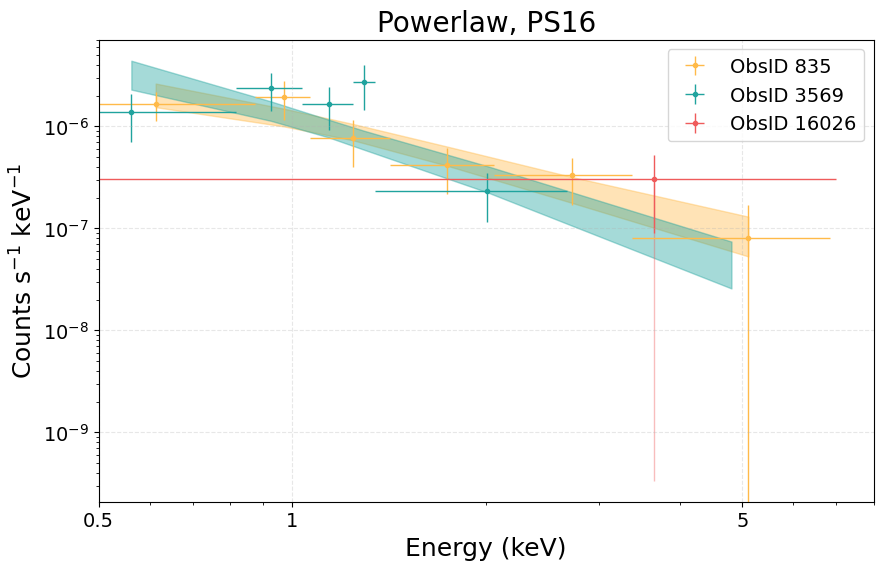}
        \includegraphics[scale=0.25,valign=c]{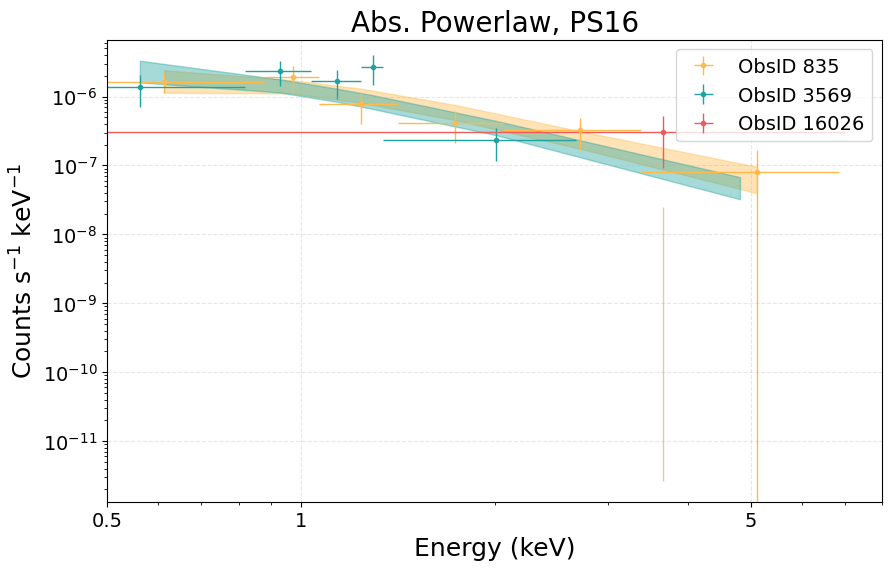}
        \includegraphics[scale=0.25,valign=c]{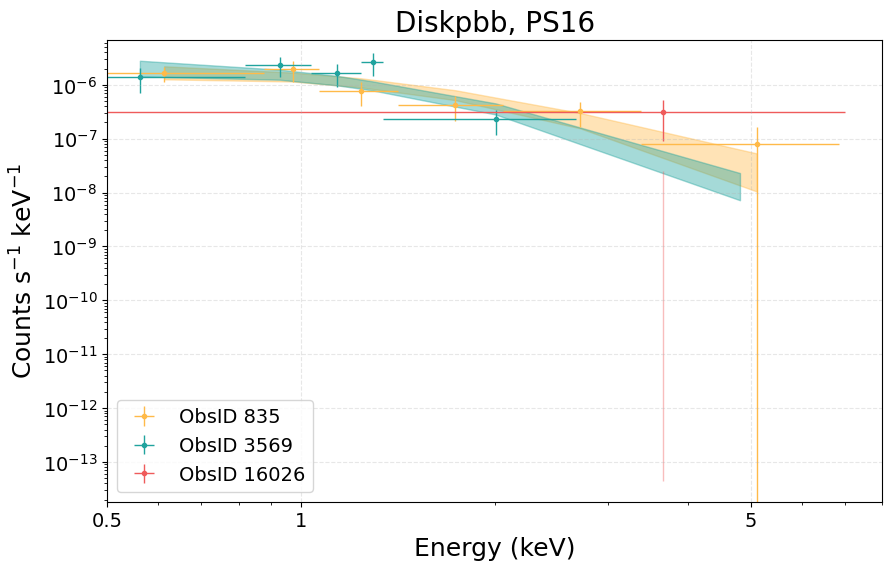}
        \includegraphics[scale=0.25,valign=c]{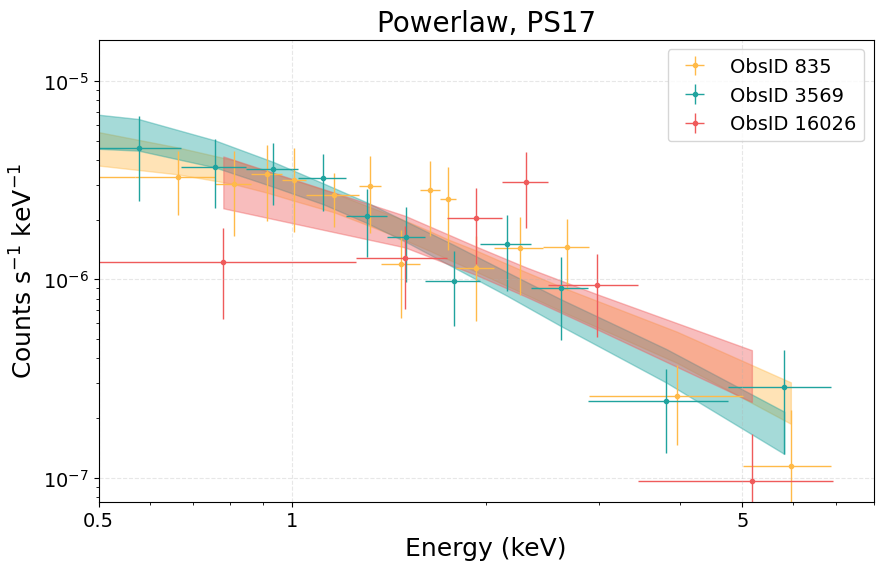}
        \includegraphics[scale=0.25,valign=c]{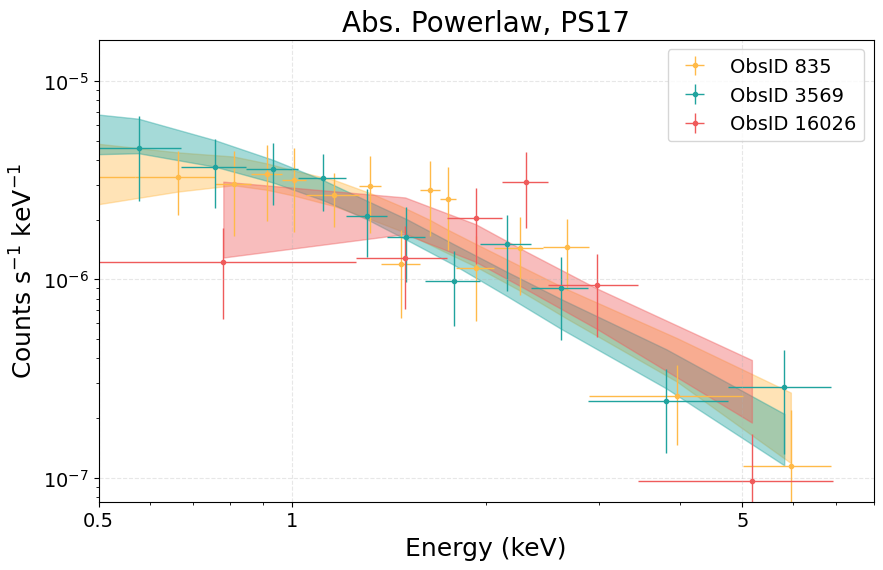}
        \includegraphics[scale=0.25,valign=c]{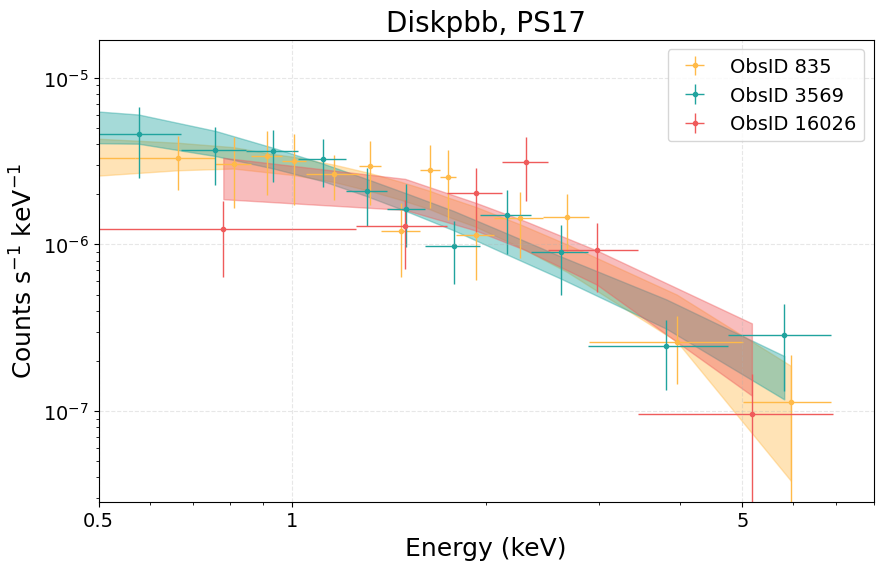}
        \caption{Spectral fits to every observation, for all three models considered in this work, for PS13, PS14, PS15, PS16 and PS17. See Sect. \ref{App:figures} for details.} \label{fig:app:spectra_panel3}
\end{figure*}

This section also contains figures showing the spectral fits to the data of all point sources, for the three models used (Figs \ref{fig:app:spectra_panel1}$-$\ref{fig:app:spectra_panel3}). The spectra are shown in different colors for the three observations available, and the fits are shown as a confidence band, which includes 68\% interquartile range.

\end{appendix}

\end{document}